\documentclass[twocolumn,english,aps,prd,nofootinbib]{revtex4-1}
\usepackage{CJK}
\usepackage{amsfonts}
\usepackage{amsmath}
\usepackage{ dsfont }
\usepackage{hyperref}
\usepackage{amssymb}
\usepackage{xcolor}
\usepackage{xspace}
\usepackage{ragged2e}
\usepackage{relsize}%\mathlarger command 
\usepackage[bottom]{footmisc}

\usepackage{tikz}
\usetikzlibrary{calc} 
\newcommand{\nocontentsline}[3]{}
\newcommand{\tocless}[2]{\bgroup\let\addcontentsline=\nocontentsline#1{#2}\egroup}
\newcommand{\be}{\begin{equation}}
\newcommand{\ee}{\end{equation}}
\newcommand{\bea}{\begin{equation} \begin{aligned}}
\newcommand{\eea}{\end{aligned} \end{equation} }
\newcommand{\bi}{\begin{itemize}}
\newcommand{\ei}{\end{itemize}}

\newcommand{\la}{\lambda}
\renewcommand{\be}{\beta}
\newcommand{\al}{\alpha}
\newcommand{\bpm}{\begin{pmatrix}}
\newcommand{\epm}{\end{pmatrix}}
\newcommand{\eps}{\epsilon}

\newcommand{\lp}{\left(}
\newcommand{\rp}{\right)}

\newcommand{\del}{\partial}

\newcommand{\curl}{\pmb{\nabla} \times }
\newcommand{\Tr}{\text{Tr} \ }

\newcommand{\mbf}[1]{\mathbf{#1}}

\usepackage{color}
\usepackage{graphicx}
\usepackage[space]{grffile}
\usepackage{verbatim}
\usepackage{amsmath}
\usepackage{amssymb}
\usepackage{wasysym}
\usepackage[caption=false]{subfig}
\usepackage{url}
\usepackage{bbold}
\usepackage{slashed}
\usepackage{epstopdf}
\usepackage{braket}
\usepackage{float}
\usepackage[percent]{overpic}

\DeclareRobustCommand{\Sec}[1]{Sec.~\ref{#1}}
\DeclareRobustCommand{\Secs}[2]{Secs.~\ref{#1} and \ref{#2}}
\DeclareRobustCommand{\App}[1]{App.~\ref{#1}}
\DeclareRobustCommand{\Apps}[2]{Apps.~\ref{#1} and \ref{#2}}
\DeclareRobustCommand{\Tab}[1]{Table~\ref{#1}}

\DeclareRobustCommand{\Fig}[1]{Fig.~\ref{#1}}

\DeclareRobustCommand{\Eq}[1]{Eq.~(\ref{#1})}
\DeclareRobustCommand{\Eqs}[2]{Eqs.~(\ref{#1}) and (\ref{#2})}
\DeclareRobustCommand{\Ref}[1]{Ref.~\cite{#1}}
\DeclareRobustCommand{\Refs}[2]{Refs.~\cite{#1, #2}}
\DeclareMathAlphabet\mathbfcal{OMS}{cmsy}{b}{n}

\DeclareRobustCommand{\choose}[2]{#1}

\RequirePackage[normalem]{ulem} %DIF PREAMBLE
\RequirePackage{color}\definecolor{RED}{rgb}{1,0,0}\definecolor{BLUE}{rgb}{0,0,1} %DIF PREAMBLE
\begin{document}

\title{Hofstadter Topology: Non-crystalline Topological Materials at High Flux}

\author{Jonah Herzog-Arbeitman$^{1}$}
\thanks{These authors contributed equally.}
\author{Zhi-Da Song$^{1}$}
\thanks{These authors contributed equally.}
\author{Nicolas Regnault$^{1,2}$}
\author{B. Andrei Bernevig$^{1}$}

\affiliation{$^1$Department of Physics, Princeton University, Princeton, NJ 08544}
\affiliation{$^2$ Laboratoire de Physique de l'Ecole normale sup\'erieure, ENS, Universit\'e PSL, CNRS, Sorbonne Universit\'e, Universit\'e Paris-Diderot, Sorbonne Paris Cit\'e, Paris, France.}

\date{\today}

\begin{abstract}
The Hofstadter problem is the lattice analog of the quantum Hall effect and is the paradigmatic example of topology induced by an applied magnetic field. Conventionally, the Hofstadter problem involves adding $\sim 10^4$ T magnetic fields to a \emph{trivial} band structure. In this work, we show that when a magnetic field is added to an initially \emph{topological} band structure, a wealth of possible phases emerges. Remarkably, we find topological phases which cannot be realized in any crystalline insulators. We prove that threading magnetic flux through a Hamiltonian with nonzero Chern number or Mirror Chern Number enforces a phase transition at fixed filling and that a 2D Hamiltonian with nontrivial Kane-Mele invariant can be classified a 3D TI or 3D weak TI phase in periodic flux. We then study fragile topology protected by the product of two-fold rotation and time-reversal and show that there exists a higher order TI phase where corner modes are pumped by flux. We show that a model of twisted bilayer graphene realizes this phase. Our results rely primarily on the magnetic translation group which exists at rational values of the flux. The advent of Moir\'e lattices renders our work relevant experimentally. Due to the enlarged Moir\'e unit cell, it is possible for laboratory-strength fields to reach one flux per plaquette and allow access to our proposed Hofstadter topological phase. 
\end{abstract}
\maketitle

\tocless\section{Introduction}
\label{sec:hof}

When a two dimensional crystalline lattice in which electrons have a trivial band structure is pierced by a uniform magnetic field, translational symmetry is broken and the energy spectrum develops a complex, fractal structure known as the Hofstadter Butterfly, which hosts a wealth of nontrivial Chern number topology despite the triviality of the original band structure \cite{PhysRevB.14.2239,PhysRevLett.49.405,Harper_1955,2019arXiv190602632N, andreibook,2015NanoL..15.6395W,PhysRevLett.86.147,Dana_1985,PhysRevB.100.245108,2020Sci...367..895D,2013Sci...340..167C, 2010PhRvL.105g6801C}. In this work, we study the Hofstadter problem for an initially \emph{topological} band structure and demonstrate new phases not possible in crystalline insulators. We prove that (1) a nonzero Chern number or mirror Chern number enforces a gapless point in the bulk of the Hofstadter Butterfly and (2) insulators with time-reversal symmetry $\mathcal{T}$ (TRS) and nontrivial $\mathds{Z}_2$ invariant can be considered as either strong or weak 3D topological insulators (TIs) in flux and host gapless edge states. We then study insulators with fragile topology protected by $C_{2z} \mathcal{T}$ symmetry and (3) show that the Hofstadter Hamiltonian can achieve a 3D Higher Order TI (HOTI) phase characterized by corner mode pumping. We then show that a model of twisted bilayer graphene (TBG) realizes the HOTI phase \cite{2018arXiv180710676S}. 

Recently, progress in the manufacture of two dimensional Moir\'e lattices with mesoscale effective unit cells has brought measurements of the Hofstadter Butterfly within reach by enabling access to large fluxes at laboratory-strength magnetic fields \cite{recenthof,moirehofexp,2013Sci...340.1427H,2013Natur.497..594P,2011PNAS..10812233B,2018arXiv180710676S,2010NatNa...5..722D,2018Natur.556...80C,2019Natur.572..101X,Skachkova_2017,AFYoung1,Hunt1427,2011arXiv1101.4712M}. We expect our theoretical predictions to be verifiable in the near future, opening a new field of Hofstadter topology.

First we review the framework for introducing magnetic flux on a lattice using the Peierls substitution \cite{1933ZPhy...80..763P}. We consider a general tight-binding model with unit vectors $\mbf{a}_1, \mbf{a}_2$ whose lattice points we call $\mbf{R}$, with orbitals at $\pmb{\delta}_\al, \al = 1, \dots, N_{orb}$, and hopping elements given by $t_{\al \be}(\mbf{r}-\mbf{r}'), \mbf{r} = \mbf{R} + \pmb{\delta}_{\al},  \mbf{r}' = \mbf{R}' + \pmb{\delta}_{\be}$. The number of occupied bands is $N_{occ}$. We write $c^\dag_{\mbf{R}, \al}$ ($c_{\mbf{R}, \al}$) as the fermion creation (destruction) operator of the $\al$ orbital at position $\mbf{R} + \pmb{\delta}_{\al}$. We find it convenient to work in units where the area of the unit cell, the electron charge $e$, and $\hbar$ are all set to one. By Peierls' substitution, the hoppings acquire a phase $t_{\al \be}(\mbf{r}-\mbf{r}') \to t_{\al \be}(\mbf{r}-\mbf{r}') \exp \left[ i \int^{\mbf{r}}_{\mbf{r}'} \mbf{A} \cdot d\mbf{r} \right]$. The path of integration is a straight line between the orbitals when they are well localized \choose{(see \App{app:Peierlspaths})}{\cite{SM}}. We work in the Landau gauge $\mbf{A}(\mbf{r}) = - \phi \mbf{b}_1 (\mbf{r} \cdot \mbf{b_2})$ where the reciprocal vectors $\mbf{b}_i$ satisfy $\mbf{b}_i \cdot \mbf{a}_j = \delta_{ij}$ and $\phi$ is the flux per unit cell. In this gauge, the hoppings retain translation invariance along $\mbf{a}_1$ but the translation symmetry along $\mbf{a}_2$ is broken. However, at rational values of the flux where $\phi = \frac{2\pi p}{q}$ with $q, p$ coprime, the hoppings recover an extended translational symmetry: $\mbf{r} \to \mbf{r} + q \mbf{a}_2$. In this case, we can diagonalize the Hamiltonian in the $1 \times q$ magnetic unit cell: 
\bea
\label{eq:hofhamdefmain}
H^\phi &= \!\!\! \sum_{k_1, k_2, \al,\be, r_2, r_2'} \!\!\! c^\dag_{k_1,k_2, r_2, \al} [\mathcal{H}^{\phi}(k_1, k_2)]_{r_2,\al, r_2', \be } \ c_{k_1, k_2, r_2', \be} \ .
\eea
Here $r_2', r_2 = 0, \dots, q-1$ are the coordinates of the magnetic unit cell in the $\mbf{a}_2$ direction, $k_1 \in (-\pi,\pi)$ is the momentum along $\mbf{b}_1$, $k_2$ is the momentum along $\mbf{b}_2$ and takes values in $(0, \frac{2\pi}{q})$ due to the enlargement of the magnetic unit cell, and $\mathcal{H}^\phi$ is the single-particle $q N_{orb} \times q N_{orb}$ Hamiltonian which we will refer to as the Hofstadter Hamiltonian. Importantly, the Hofstadter Hamiltonian is periodic in flux up to a unitary transformation: $H^{\phi + \Phi} = U H^\phi U^\dag $, where $\Phi = 2\pi n, n \in \mathbb{N}$ is determined by the condition that \emph{all} closed hopping loops encircle an integer number of flux quanta. In simple models such as nearest-neighbor hopping on the square lattice, $n=1$. We can show \choose{(see \App{app:magper})}{\cite{SM}} that 
\bea
\label{eq:Usinglepart}
U &= \exp \left[ i \sum_{\mbf{R} \al}  \int^{\mbf{R}+ \pmb{\delta}_\al}_{\mbf{r}_0}  \mbf{\tilde{A}}  \cdot d\mbf{r} \,  c^\dag_{\mbf{R},\al} c_{\mbf{R} ,\al} \right], \quad \pmb{\nabla} \times \mbf{\tilde{A}} = \Phi 
\eea
where $\mbf{r}_0$ is the position of a fixed but arbitrary orbital of the Hamiltonian, and the integral may be taken along any sequence of Peierls paths due to the definition of $\Phi$. 

A central feature of the Hofstadter Hamiltonian is the increased periodicity of its Brillouin Zone (BZ) which can be deduced from the magnetic translation group \cite{PhysRev.134.A1602}. As shown in \Eq{eq:hofhamdefmain}, $k_2$ is $2\pi/q$ periodic. Here we show that the energy bands are also $2\pi/q$ periodic along $k_1$. The single-particle magnetic translation operators are
\bea
\label{eqn:T}
T_i(\phi) &= \sum_{\mbf{R} \al} e^{i \int_{\mbf{R} + \pmb{\delta}_\al}^{\mbf{R} + \pmb{\delta}_\al + \mbf{a}_i} \mbf{A} \cdot d\mbf{r} \, + \, i \chi_i(\mbf{R}+\pmb{\delta}_\al )} c^\dag_{\mbf{R}+\mbf{a}_i, \al} c_{\mbf{R},\al}, \\
\eea
where $\chi_i(\mbf{r}) = \phi \, (\mbf{a}_i \times \mbf{r}) \cdot \hat{z} $ has been determined by requiring
$[H^\phi, T_i(\phi)] = 0$ and the integral is taken along a straight-line path \choose{(see \App{app:magtran})}{\cite{SM}}.  While the translation operators commute in the absence of flux, otherwise we find $T_1(\phi) T_2(\phi) = e^{i \phi} T_2(\phi) T_1(\phi) $. However, at rational flux $\phi = \frac{2\pi p}{q}$, we see $[T_1(\phi), T_2^q(\phi)] = 0$. Hence $H^{\phi}, T_1(\phi), $ and $T_2^q(\phi)$ commute and eigenstates may be written as $\ket{m, k_1, k_2}$ with corresponding eigenvalues $\eps_m(\mbf{k}), e^{i k_1}, e^{i q k_2}$, with $m = 1, \dots qN_{orb}$ (see \Eq{eq:hofhamdefmain}). Because $[H^\phi, T_2(\phi)] = 0$, the states $T_2^j(\phi)\ket{m, k_1, k_2}$ also have energy $\eps_m(\mbf{k})$. The $k_1$ momentum of such states is deduced from $T_1(\phi) (T_2^j(\phi) \ket{m, k_1, k_2} ) = e^{i ( k_1+j \phi)} T_2^j(\phi) \ket{m, k_1, k_2}$ and hence they represent the new states at $k_1 + j \phi$ (see \App{app:UTdetails}). Thus we find
\bea
\label{eq:BZper}
T_2^j(\phi) \ket{m, k_1, k_2} \sim \ket{m, k_1 + j \phi, k_2}, \quad j = 0, \dots, q-1 \\
\eea
are all degenerate in energy. Recalling that $k_2 \in (0, \frac{2\pi}{q})$, we conclude that the magnetic BZ has an increased periodicity: $\eps_{n}(\mbf{k}) = \eps_{n}(\mbf{k} + \frac{2\pi}{q} \mbf{b}_i ), \, i = 1,2$. This feature is essential in the following proofs. \\

\tocless\section{Chern Insulators}
\label{sec:cherngap}

\begin{figure*}
 \centering
 \begin{overpic}[height=0.275\textwidth,tics=10,trim =.2cm .4cm 0 1.2cm , clip]{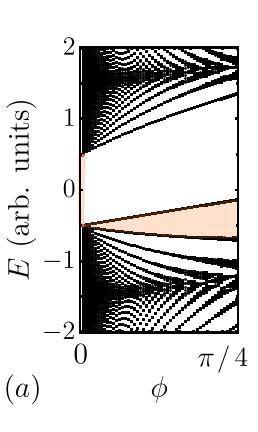}
%\begin{overpic}[width=0.47\textwidth,tics=10]{qsh_mirror_hof.jpg}
\end{overpic} 
\begin{overpic}[height=0.275\textwidth,tics=10,trim =0 0 0 .10cm , clip]{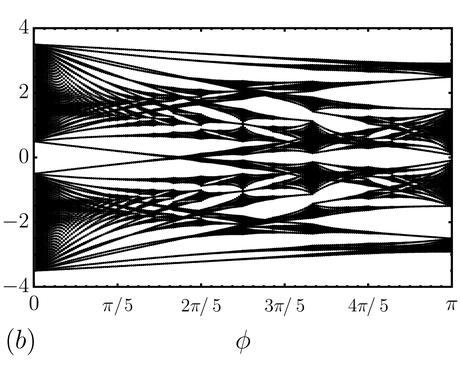}
%\begin{overpic}[width=0.47\textwidth,tics=10]{qsh_mirror_hof.jpg}
\end{overpic} 
\begin{overpic}[height=0.275\textwidth,tics=10,trim =0 0 0 .10cm , clip]{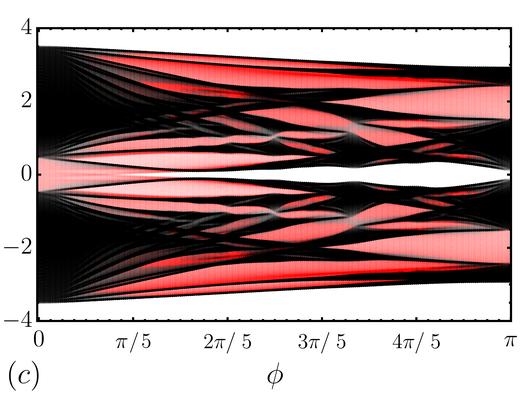}
%\begin{overpic}[width=0.475\textwidth,tics=10]{qsh_edge_hof.jpg}
\end{overpic} 
\caption{(a) We consider a $C=1$ Chern insulator (see \App{app:6}) with the half-filling many-body gap shaded in orange. At zero flux, the gap is $\sim 1$, but when the flux is increased at constant filling, the gap discontinuously closes at infinitesimal $\phi$. Notably, this discontinuity is impossible to realize in \emph{any} crystalline system and relies on the infinitely large magnetic unit cell as $\phi \to 0$. (b) We consider the Hofstadter spectrum of a model $H_{QSH}'$ with nonzero Mirror Chern number and all symmetries broken except for $M_z$ and $\mathcal{T}$ \choose{}{\cite{SM}}. The gap closing at finite $\phi$ is protected by $M_z$; breaking it allows a gap to open in the bulk. (c) We show the Hofstadter Butterfly calculated on a $30\times 30$ lattice for $H_{QSH}''$, a variation where $M_z$ is broken and $\mathcal{T}$ is the only symmetry \choose{}{\cite{SM}}. The bulk spectrum (black) is gapped, but the edge spectrum (red) is gapless at $\phi = 0$. As the flux is increased to $\phi = \pi$, the edge states gap and move into the bulk. We also observe quantum hall states at all fluxes and fillings where $\nu N_{orb} \notin \mathbb{N}$. This is explained by the Diophantine equation, which can be written $\frac{\phi}{2\pi} C = N_{orb} \nu \mod 1$ \cite{Dana_1985}. }
\label{fig:hofs12}
\end{figure*}

As a warmup, we study the Hofstadter Butterfly of a Chern insulator. According to the Streda formula \cite{andreibook}, the filling of a state with fixed nonzero Chern number changes as the flux is increased. In the paradigm of Hofstadter topology, we prove a complementary result: at \emph{fixed} filling, the many-body gap of a Chern insulator has a discontinuity at $\phi =0$ enforced by a mismatch between the Chern number at zero flux and any infinitesimal flux. 

Consider a Hamiltonian $H^{\phi = 0}$ which is gapped with a nonzero Chern number $C^{\phi = 0}$ at filling $\nu = N_{occ} / N_{orb}$. We emphasize that we keep $\nu$ fixed as the flux is increased. Now we choose a flux $\phi = \frac{2\pi p}{q}$ such that $C^{\phi = 0} / q \notin \mathds{Z}$. The magnetic unit cell of $H^\phi$ contains $q N_{orb}$ orbitals and $q N_{occ}$ occupied bands at filling $\nu$. First, we introduce an onsite potential term of overall amplitude $M$ to $H^{\phi= 0}$ that creates an energy splitting between each of the orbitals. For sufficiently large $M$, the model will be split into $N_{orb}$ trivial bands and it will reach a gapped atomic limit for all $\phi$\choose{ (see \App{App:atomiclimit})}{ \cite{SM,PhysRevLett.106.236803,PhysRevB.85.075128}}.

As we tune $M$ from zero to infinity, gap closings occur which eventually cause $H^{\phi = \frac{2\pi p}{q}}$ to undergo a series of phase transitions into a trivial atomic limit \choose{(see \App{App:atomiclimit})}{\cite{SM}} at filling $\nu = q N_{occ} / q N_{orb}$\footnote{Note that the structure of an \emph{onsite} potential is identical for the $q$ unit cells within the magnetic unit cell}, recalling that the change in Chern number is determined locally by the closing bands \cite{andreibook,2013CRPhy..14..816P,Brouder_2007}. If there is gap closing at $\mbf{k}^*$, there must also be an \textit{identical} gap closing at the each of the points $\mbf{k}^* + j \phi \mbf{b}_1 \mod 2\pi, j = 1, \dots q-1$ due to the magnetic BZ periodicity of \Eq{eq:BZper}. Because a multiple of $q$ gap closings separate $H^\phi$ from the trivial atomic limit at large $M$ where the Chern number is zero, it must be that $C^{\phi = \frac{2\pi p}{q} \neq0}  \in q\mathds{Z}$, zero included. Hence we recover the result of \Ref{Dana_1985} using a proof applicable to Sec. III. 

Since we chose $q$ such that $C^{\phi =0}  \notin q \mathds{Z}$, we find by construction that the Chern number has changed during an adiabatic perturbation of $H^\phi$. But this is only possible if the gap closes for $\phi \in [0,\frac{2\pi p}{q}]$. For every $C^{\phi = 0}$, we may choose an arbitrarily large $q$ allowing us to conclude that, at fixed filling, the many-body gap must close immediately when the flux is increased from the fine-tuned point at $\phi =0$ \choose{(see \App{app:chernwilson})}{\cite{SM}}. This enforces a discontinuity in the occupied states as shown for a typical Chern insulator in \Fig{fig:hofs12}a. There is no protected gap closing if $C^{\phi = 0} = 0$ because a vanishing Chern number is possible for all flux at filling $\nu = N_{occ}/N_{orb}$. 

We now extend this result to insulators with a nonzero Mirror Chern number \cite{2006Sci...314.1757B, 2018SciA....4..346S}. Because mirror symmetry $M_z$ is not broken in the presence of flux, $M_z$ remains well-defined at all $\phi$. Then we may block-diagonalize $H^\phi$ at all $\phi$ by its mirror eigenvalues. Each block has a nonzero Chern number at $\phi = 0$, and thus the gap closes immediately at $\phi = 0$ and filling $\nu$ within each individual block. Each block must have a branch of its spectrum connecting its valence and conduction bands. Hence for \emph{any} Fermi energy in the zero-flux gap, there will be a gapless point at \emph{finite} flux in the spectrum of the whole model \choose{ (see \App{app:mirrorchernwilson})}{\cite{SM}}. In \Fig{fig:hofs12}b, we consider the Quantum Spin Hall model $H_{QSH}$ of \Ref{2006Sci...314.1757B} with a nonzero Mirror Chern number.  We show numerical confirmation that although the Chern number is identically zero due to TRS, the gap still closes due to the Mirror Chern number. \\

\tocless\section{Time-Reversal Invariant Insulators}

\label{sec:3DTI}

We show in this section that when a Hofstadter Hamiltonian with spinful TRS $\mathcal{T}$ is topological (in a quantum spin hall state) at $\phi = 0$, it realizes a nontrivial 3D phase where the flux $\phi$ is identified with $k_z$ \cite{2006Sci...314.1757B,2009AIPC.1134...22K,2010NJPh...12f5010R}. Recall that $H^\phi$ is $\Phi = 2 \pi n$ periodic in flux. When $n$ is odd, $H^\phi$ is classified as a 3D TI, and may be a weak TI or 3D TI when $n$ is even. However, it can never be 3D trivial.

The identification of $\phi$ with $k_z$ is deduced from its transformation under $\mathcal{T}$. Because $\mathcal{T}$ is anti-unitary, it flips the sign of $\phi$ in the Peierls substitution and hence
\bea
\mathcal{T}^{-1} \mathcal{H}^{\phi}(\mbf{k})\mathcal{T} &= \mathcal{H}^{-\phi}(-\mbf{k}) \ .
\eea
 Let us first consider the simple case of $\Phi = 2\pi$, i.e. $n = 1$. Then $\phi$ is $2\pi$-periodic and behaves as $k_z$ would in a 3D Hamiltonian. Furthermore, we recall that $H^{\phi + \Phi} = U H^\phi U^\dag$ so from 
 \bea
 \label{eq:mainUT}
 (U \mathcal{T}) H^{\pi} (U \mathcal{T})^\dag = U H^{-\pi} U^\dag = H^{\pi} , 
 \eea
we see that $U \mathcal{T}$ is a symmetry of $H^{\phi= \pi}$. It can be shown that $(U \mathcal{T})^2 = \mathcal{T}^2 = -1$ \choose{(see \App{app:UT}) so $H^{\phi=\pi}$ also has a $\mathds{Z}_2$ topological classification.\footnote{
For pedagogical purposes, we assume that $U$ is diagonal in momentum space. In the generic case, the algebra of $U\mathcal{T}$ and $T_i(\phi)$ acquires a projective phase which leads to an off-diagonal representation of $U\mathcal{T}$ on the magnetic BZ (see \App{app:genz2proof}).
}}{\cite{SM} so $H^{\pi}$ also has a $\mathds{Z}_2$ topological classification. For pedagogical purposes, we assume that $U$ is diagonal in momentum space. In the generic case, the algebra of $U\mathcal{T}$ and $T_i(\phi)$ acquires a projective phase which leads to an off-diagonal representation of $U\mathcal{T}$ on the magnetic BZ \cite{SM}.}

Considering $H^\phi$ as a 3D model with $\mathcal{T}$ symmetry, its topology is characterized by the magnetoelectric polarizability $\theta \in \{0, \pi\}$ where $\pi$ is the nontrivial value of the 3D TI phase. $\theta$ is related to the Pfaffian $\mathds{Z}_2$ invariants by $e^{i \theta} = \delta^{\phi = 0}  \times \delta^{\phi = \pi}$, where $\delta^\phi \in \{-1,1\}$ is the protected by $\mathcal{T}$ ($U \mathcal{T}$) at $\phi = 0$ ($\pi$) \cite{2007PhRvL..98j6803F,2005PhRvL..95n6802K, 2007PhRvL..98j6803F,2011PhRvB..84g5119Y}. Because we assume that $H^{\phi =0}$ is nontrivial, we need only show that $\delta^{\phi=\pi} = +1$ in order to prove $\theta = \pi$. To do so, we introduce the parameter $M$ which tunes the $H^\phi$ to a trivial atomic limit as described in Sec. II. Now consider the magnetic BZ at $\phi=\pi$ with $k_1 \in (-\pi,\pi), k_2 \in (0, \pi)$. \Eq{eq:BZper} requires the BZ to be $\pi$ periodic, $\eps_m(\mbf{k} + \pi \mbf{b}_1) = \eps_m(\mbf{k})$.  As $M \to \infty$, we determine the change in $\delta^{\phi = \pi}$ by counting gap closings in \emph{half} of the magnetic BZ defined by $BZ_{1/2} = k_1 \in(-\pi,\pi), k_2 \in (0, \pi/2)$ \cite{2009PhRvB..79s5322R,2005PhRvL..95n6802K,2006PhRvB..74s5312F}. If there is a gap closing at $\mbf{k}^* \in BZ_{1/2}$, there is a second identical closing at $\mbf{k}^* + \pi \mbf{b}_1 \in BZ_{1/2}$. Each gap closing changes the sign of $\delta^{\phi = \pi}$, so it must be that $\delta^{\pi = \phi} = +1$ because an even number of gap closings occur between $M=0$ and the trivial phase at $M=\infty$. We conclude $\theta = \pi$, proving that the Hofstadter Hamiltonian is a 3D TI. On open boundary conditions (OBC), such a model will pump gapless edge states into the bulk as $\phi$ is increased, as exemplified in \Fig{fig:hofs12}b. There, for a perturbed model with only $\mathcal{T}$ symmertry $H''_{QSH}$  \choose{(see \App{app:QSH})}{\cite{SM}}, we observe gapless edge states for small flux and their disappearance into the bulk. 

The $\pi$ periodicity in the magnetic BZ was crucial to proving that $H^{\phi=\pi}$ is trivial. Generally, if $\Phi = 2\pi n$ then the $U \mathcal{T}$-symmetric point exists at $\Phi/2= n \pi$. When $n$ is odd, the energy spectrum is still $\pi$ periodic along $k_1$, and we conclude that $\delta^{\phi = n\pi} = +1$ \choose{(see \App{app:genz2proof})}{\cite{SM}}. However when $n$ is even, this $\pi$ periodicity is absent so our proof fails, and indeed, $H^\phi$ can be a weak or strong 3D TI (see \App{app:TRissue}). \\

\tocless\section{Fragile Topological Insulators}
\label{sec:fragile}

\begin{figure}
 \centering
\begin{overpic}[width=0.4\textwidth,tics=10,trim =0 0 0 .35cm , clip]{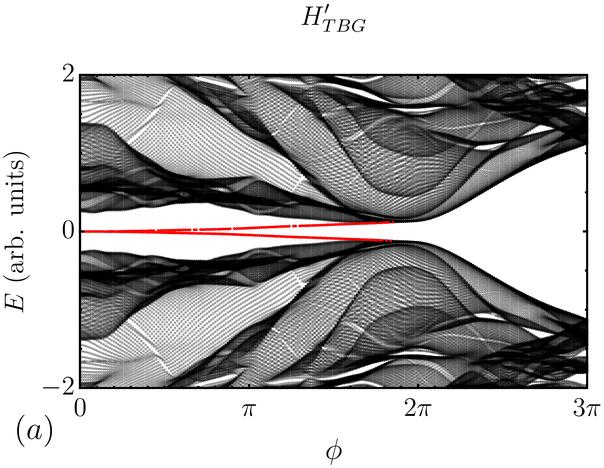}
\end{overpic}  \\
\begin{overpic}[width=0.227\textwidth,tics=10]{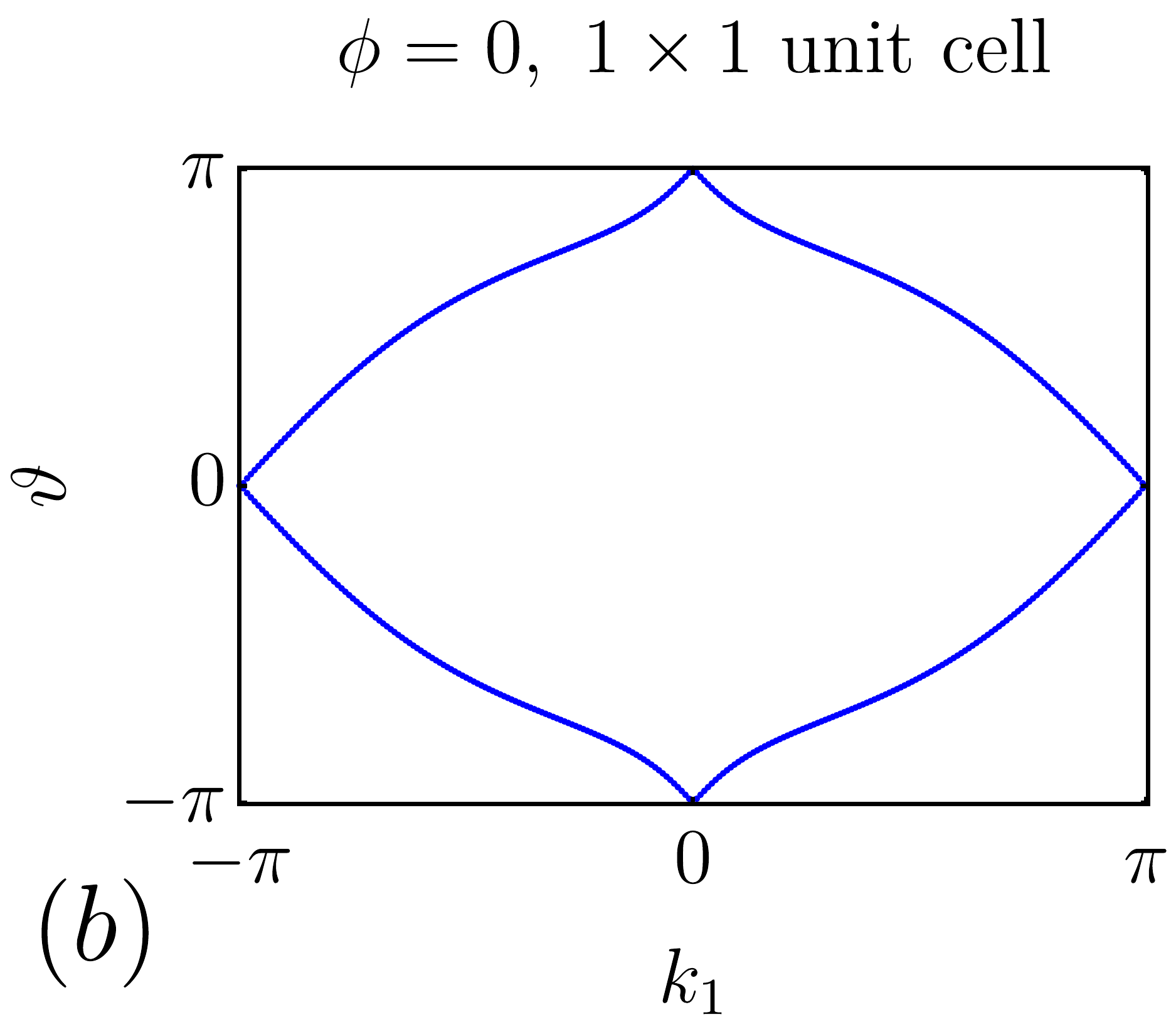}
\end{overpic} 
\begin{overpic}[width=0.245\textwidth,tics=10, trim = 0 .35cm 0 0, clip ]{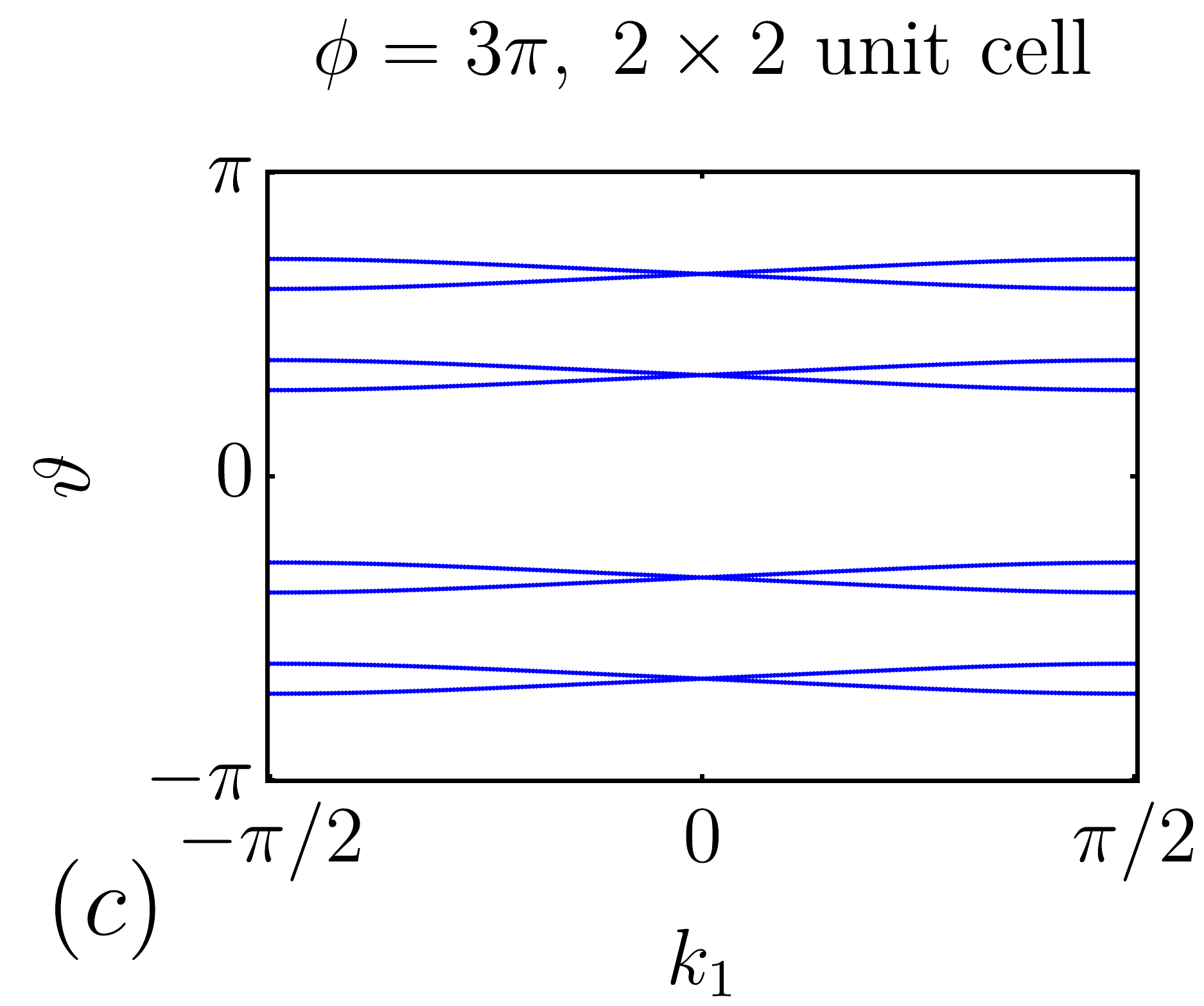}
\end{overpic}
\caption{ (a) The Hofstadter Butterfly is calculated on a $30 \times 30$ lattice for $H_{TBG}'$ which has only $C_{2z} \mathcal{T}$ symmetry \choose{(see \App{app:TBG})}{\cite{SM}}. The corner modes are shown in red over the gapped black bulk and edge spectrum, and pump between the nontrivial $w_2 = 1$ phase at $\phi = 0$ to the trivial phase at $\phi = 3\pi$ where $w_2 = 0$. This model is classified as a HOTI. (b) We observe from the Wilson loop spectrum, with eigenvalues $\exp i \vartheta(k_1) $, that $w^{\phi = 0}_2 = 1$ due to the odd number of crossings at $\vartheta = 0$ and $\vartheta = \pi$ \cite{2019PhRvX...9b1013A}. (c) The Wilson spectrum at $\phi =3\pi$ is calculated in an extended $2\times 2$ unit cell where $U$ is diagonal in momentum space \choose{(see \App{app:C2zTwilson})}{\cite{SM}}. Here, $w_2^{\phi = 3\pi} = 0$ because there are no crossings at $\vartheta = 0$ and $\vartheta = \pi$. 
}
\label{fig:TBGcorner}
\end{figure}

So far we have studied the Hofstadter topology deriving from strong topological 2D phases with a nontrivial  Chern number, the Mirror Chern number, or $\mathds{Z}_2$ index. We now consider a fragile invariant, the second Stiefel-Whitney index $w_2 \in \{0,1\}$ protected by $C_{2z} \mathcal{T}$ (with $(C_{2z} \mathcal{T})^2 = +1$)  \cite{2018arXiv180802482P,2018arXiv180710676S,2019PhRvX...9b1013A,2018arXiv180611116W}. A nontrivial value of $w_2 = 1$ indicates fractional corner states  \cite{2019PhRvX...9b1013A,2019arXiv190503262S,2018arXiv180409719B,PhysRevLett.119.246402,2017Sci...357...61B}, and may be computed in the bulk from the Wilson loop eigenvalues or the nested Wilson loop  \cite{2019PhRvX...9b1013A, 2017PhRvB..96x5115B,Alexandradinata:2012sp}. 

The 3D HOTI phase is characterized by pumping corner states between a $w_2$ nontrivial phase and trivial phase \cite{2018SciA....4..346S,2018arXiv181002373W}. Because $(C_{2z} \mathcal{T})^{-1} \mathcal{H}^{\phi}(\mbf{k}) (C_{2z} \mathcal{T}) = \mathcal{H}^{-\phi}(\mbf{k})$, we can again identify $\phi$ with $k_z$ and use 3D topological invariants to classify $H^\phi$, which we now discuss.

We have assumed that $C_{2z} \mathcal{T}$ is a symmetry of $H^{\phi=0}$ and protects the invariant $w_2^{\phi = 0}$. For a Hofstadter Hamiltonian that has a $\Phi = 2\pi n$ periodicity in flux, the other symmetric point occurs at $\phi = \Phi/2$ where $H^{\Phi/2}$ has the symmetry $U C_{2z} \mathcal{T}$. We can show that $(U C_{2z} \mathcal{T})^2 = \pm 1$ where the sign must be calculated from the Peierls paths \choose{(see \App{app:UCT})}{\cite{SM}}. The Hofstadter topological invariant depends on this sign. If $(U C_{2z} \mathcal{T})^2 = +1$, a nonzero value of $\frac{\theta}{\pi} = w_2^{\phi = 0} - w_2^{\phi = \frac{\Phi}{2}}$ indicates corner state flow. \choose{\footnote{The nontrivial phase with $\theta = \pi$ is a ``strong" symmetry-protected topological phase with corner state pumping. If both $w_2^{\phi = 0} = w_2^{\phi = \Phi/2} = 1$, then $\theta = 0$ but both $H^{\phi = 0}$ and $H^{\phi = \Phi/2}$ have nontrivial corner states, which is a ``weak" 3D fragile state.} \cite{2018arXiv181002373W}}{The nontrivial phase with $\theta = \pi$ is a ``strong" symmetry-protected topological phase with corner state pumping. \cite{endnote2,2018arXiv181002373W,Aroyo:firstpaper,Aroyo:xo5013,hofsymtoappear}}.

If $(U C_{2z} \mathcal{T})^2 = -1$, there is no $w_2$ index at $\phi = \frac{\Phi}{2}$ \cite{2019PhRvX...9b1013A}.  \choose{However, we can diagnose the topology directly with the nested Wilson loop and Kramers' theorem for $(U C_{2z} T)^2 = -1$ (see \App{app:UCTtrivial}), showing there are no protected corner states at $\phi = \Phi/2$ and the Hofstadter HOTI invariant depends only on the zero-field topology, i.e.$ \frac{\theta}{\pi} = w_2^{\phi = 0}$. } {However, we can diagnose the topology directly as a consequence of Kramers' theorem. Because $(U C_{2z} \mathcal{T})^2 = -1$, the states at $C_{2z}$-invariant positions must come in pairs at $\phi = \Phi/2$, and hence can always be removed from these positions without breaking the symmetry. Heuristically, these Kramers' pairs trivialize the bulk, and by the Bulk-Boundary correspondence, the model remains trivial with OBC \cite{2020Sci...367..794S,2018PhRvX...8c1070K,Geier_2018,PhysRevX.9.011012,Khalaf_2018}. This is confirmed explicitly by calculating the nested Wilson loop \cite{SM}. Thus we find that $\frac{\theta}{\pi} = w_2^{\phi = 0}$ only depends on the zero-field topology. }

To exemplify the Hofstadter HOTI phase, we now consider $H_{TBG}$: a 4-band model of TBG with $C_{2x}, C_{3z}, C_{2z}, $ and $\mathcal{T}$ symmetries which possesses fragile Wilson loop winding yielding $w^{\phi = 0}_2 = 1$ \cite{2018arXiv180710676S, 2018arXiv181111786L}. We study a perturbed model $H'_{TBG}$ which has only $C_{2z} \mathcal{T}$ to protect the fragile topology \choose{(see \App{app:TBG})}{\cite{SM,2011PhRvB..83x5132H}}. The Hofstadter Hamiltonian has $\Phi = 6\pi$ and $(U C_{2z} \mathcal{T})^2 = +1$ \choose{(see \App{app:TBG})}{\cite{SM}}. In \Fig{fig:TBGcorner}a, we calculate the Hofstadter Butterfly with OBC and observe the pumping of corner modes (with a gapped bulk and edge) that characterizes a HOTI. We show that $\theta = \pi$ by calculating the $w_2$ indices at $\phi = 0, \Phi/2$ from the Wilson loop spectra shown in \Fig{fig:TBGcorner}b,c\choose{.\footnote{At $\Phi/2$, we have artificially doubled the magnetic unit cell along the $\mbf{a}_1$ direction so that $U C_{2z} \mathcal{T}$ is diagonal in momentum space (see \App{app:C2zTwilson})}}{\cite{endnote3, SM}.} Breaking the $C_{2z}$ and $C_{2x}\mathcal{T}$ symmetries of $H_{TBG}$ (which are \emph{not} true symmetries of TBG) is crucial. Both symmetries are preserved at all $\phi$ and can enforce a bulk gap closing \cite{2017Natur.547..298B,2018PhRvB..97c5139C,PhysRevE.96.023310}, which would disrupt the appearance of the HOTI phase (see \Apps{app:c2xt}{app:symeig}). \\

\tocless\section{Discussion}

The topological phases of the Hofstadter Hamiltonian can be computed in the momentum-flux manifold. We demonstrated that a nonzero Chern number or mirror Chern number enforces a level crossing in the bulk as flux is pumped through the crystal. In analogy to the 3D classifications, we call this a topologically protected Hofstadter semimetal. The Hofstadter topology of a Hamiltonian with a nontrivial $\mathds{Z}_2$ index depended on the flux periodicity $\Phi = 2\pi n$. When $n$ is odd, we proved that the Hofstadter realized a 3D TI phase where flux pumps edge states into the bulk. Finally, we considered fragile topology at zero flux given by nonzero $w_2$ index and found that the topological index of the Hofstadter Hamiltonian depended on the sign of $(UC_{2z} \mathcal{T})^2 = \pm 1$ through the Peierls paths. This is notably different from crystalline systems where $(C_{2z} \mathcal{T})^2 = +1$ with and without SOC. In the strong Hofstadter HOTI phase, realized by a model of TBG, flux pumps corner states into the bulk.

We expect the results of this work to be experimentally verifiable using Moir\'e lattices, which have very large unit cells at small twist angles \cite{2011PNAS..10812233B}. Indeed, after the submission of this work, \Ref{2020arXiv200613963L} observed signatures of fragile Hofstadter topology in a TBG system, and \Ref{2020arXiv200614000B} identified the flux-induced gap closings indicative of a Hofstadter semimetal protected by a valley Chern number in twisted double bilayer graphene. Both experiments show that realistic magnetic fields can probe the Hofstadter phase. \\

\tocless\section{Acknowledgements}
We thank Fang Xie, Biao Lian, Christopher Mora, and Benjamin Wieder for helpful discussions. We also thank one of our referees for pointing out \Ref{Dana_1985}. B. A. B., N.R., and S. Z.-D. were supported by the Department of Energy Grant No. desc0016239, the Schmidt Fund for Innovative Research, Simons Investigator Grant No. 404513, and the Packard Foundation. Further support was provided by the National Science Foundation EAGER Grant No. DMR 1643312, NSF-MRSEC DMR-1420541, BSF Israel US foundation No. 2018226, and ONR No. N00014-20-1-2303.

%%%%%%%%%%%%%%%%%%%%%%%%%%%%%%%%%%
\let\oldaddcontentsline\addcontentsline% Store \addcontentsline
\renewcommand{\addcontentsline}[3]{}% Make \addcontentsline a no-op
%%%
\bibliography{finalbib}
\bibliographystyle{aipnum4-1}
\bibliographystyle{unsrtnat}
%%%
\let\addcontentsline\oldaddcontentsline
%%%%%%%%%%%%%%%%%%%%%%%%%%%%%%%%%%

\onecolumngrid
\clearpage

\appendix

\clearpage
\begin{center}
{\bf Supplementary Appendices for "Hofstadter Topology: Non-crystalline Topological Materials in the Moir\'e Era"}
\end{center}

For readers interested in the substantial additional calculations mentioned in the Main Text, we include a detailed and self-contained Supplementary Materials. For readers interested in specific supporting calculations for the individual sections, we provide a brief outline now. 

Sec. I of the Main Text includes references to \App{app:Peierlspaths} for a discussion of the Peierls paths, \App{app:magper} for a proof of $U$ and the flux periodicity $\Phi = 2\pi n$, \App{app:magtran} for a general definition of the magnetic translation group operators, and \App{app:UTdetails} for a precise discussion of the action of $T_i(\phi)$ on states. 

Sec. II of the Main Text includes references \App{App:gershgorin} to prove the atomic limit of the Hofstadter, \App{app:chernwilson} for detailed examples of the occupied state discontinuity of a Chern insulator, and \App{app:mirrorchernwilson} for comparisons with a Mirror Chern insulator. The QSH model discuss is discussed at length in \App{app:6}. 

Sec III of the Main Text references \App{app:UT} to prove the algebra of $U$ and $\mathcal{T}$, \App{app:Umom} to discuss the off-diagonal representation of $U$ in momentum space, \App{app:wilsonz2} for a Wilson loop proof of the 3D TI phase at $\phi = 2\pi n$ for odd $n$, and exemplifies the weak TI phase for even $n$ in \App{app:TRissue}. 

Sec. IV of the Main Text references \App{app:UCT} to show the Peierls path dependence of the sign of $UC_{2z} \mathcal{T}$, \App{app:w2calc} to discussion the Wilson loop calculation at $\phi = \Phi/2$, \App{app:UCTtrivial} for a proof of the triviality of the nested Wilson loop in the non-crystalline case. Finally, \App{app:7} contains the model of TBG and the calculation of the Wilson loops.

\tableofcontents
\input{Paper_Main3.toc}

\newpage

\section{Features of the Hofstadter Hamiltonian}
\label{app:hof}

In this Appendix, we study Hofstadter Hamiltonian on an arbitrary lattice with arbitrary Peierls paths. We begin by discussing the Peierls substitution (\App{app:Peierlspaths}). Then we prove the periodicity in flux and the gauge invariance of the Hofstadter Hamiltonian (\App{app:magper}). We derive the general form of the magnetic translation operators in \App{app:magtran}. In the remainder of the section, we discuss the momentum space features of the Hofstadter Hamiltonian in a suitable Landau gauge $\mbf{A}(\mbf{r}) = - \phi \mbf{b}_1 (\mbf{r} \cdot \mbf{b}_2)$. This choice of gauge is useful for numerical calculations of the spectrum, but has the disadvantage of generically requiring an \emph{enlarged} magnetic unit cell that arises as an artifact of the gauge choice (\Apps{app:ppA}{app:fluxrat}). We discuss a residual $SL(2, \mathds{Z})$ gauge symmetry associated with the Landau gauge in \App{app:residualgf}. We then construct the Hofstadter Hamiltonian in the Landau gauge (\App{app:hofhamconstruct}) and we give expressions for the embedding matrices which implement the periodicity in flux across the magnetic BZ (\App{sec:embedding}). 

\subsection{Peierls paths}
\label{app:Peierlspaths}

\begin{figure*}
 \centering
{\larger[2] (a)} \includegraphics[width=3cm]{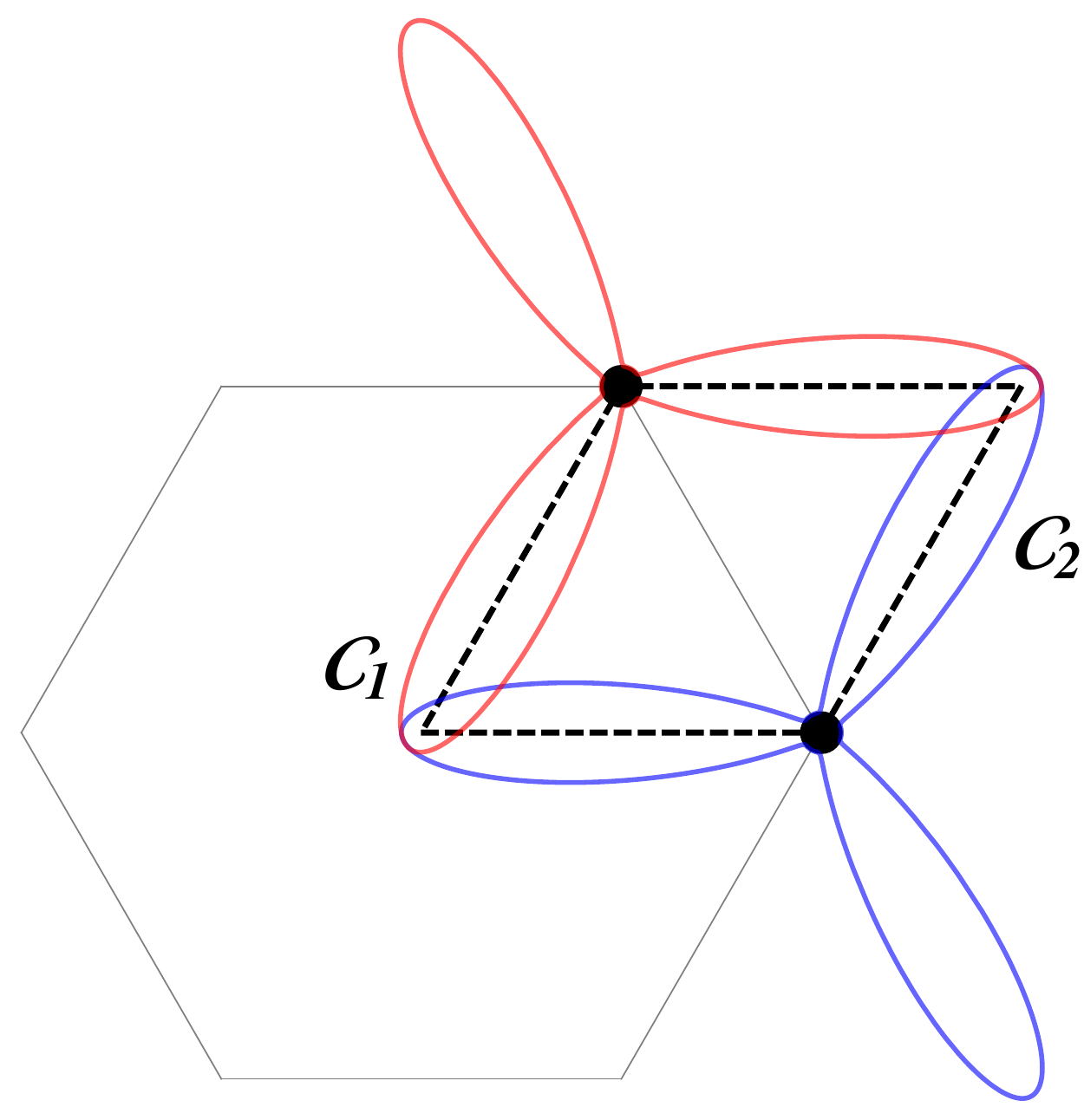} 
{\larger[2] (b)} \includegraphics[width=3.5cm]{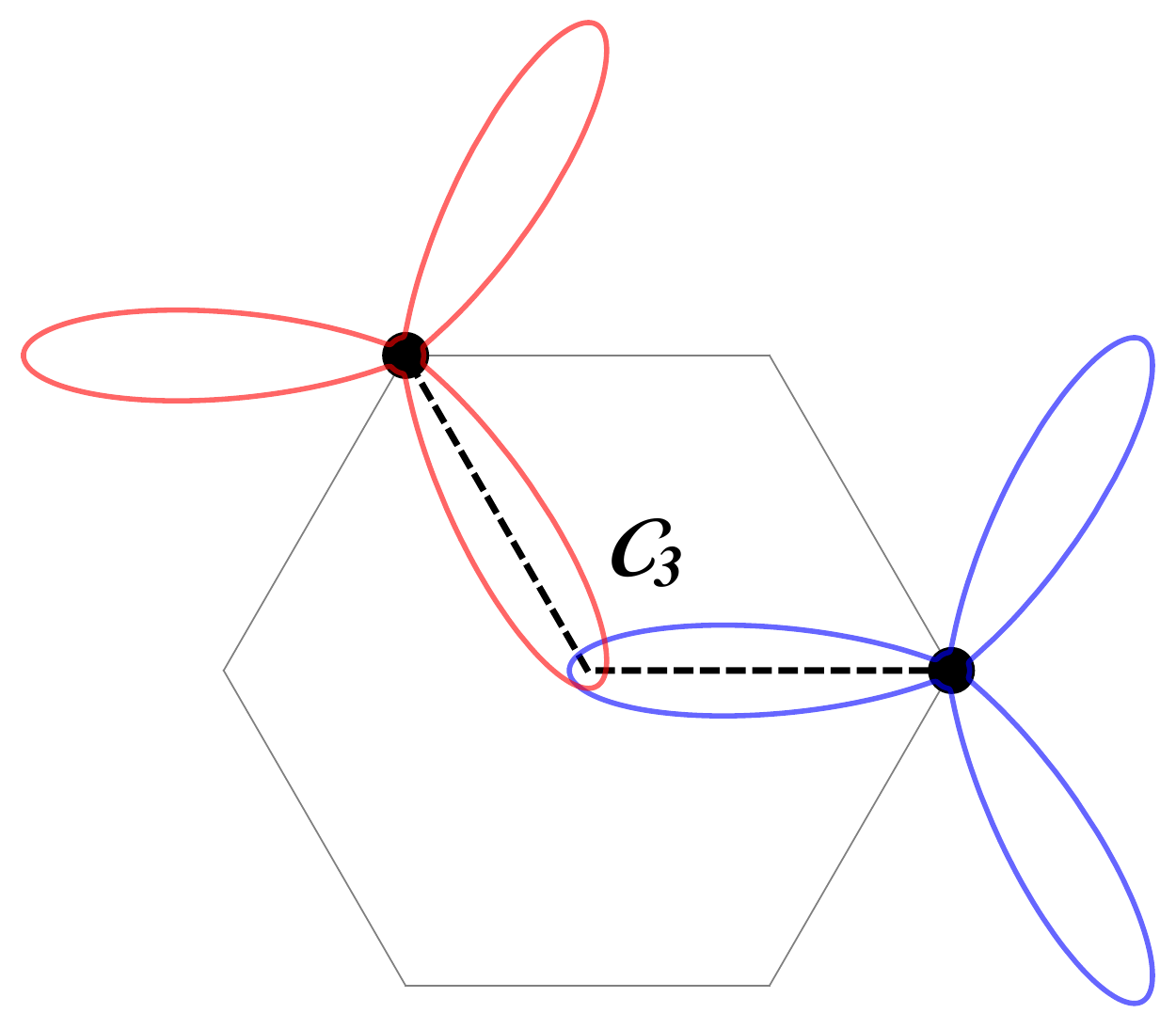} 
{\larger[2] (c)} \includegraphics[width=3.5cm]{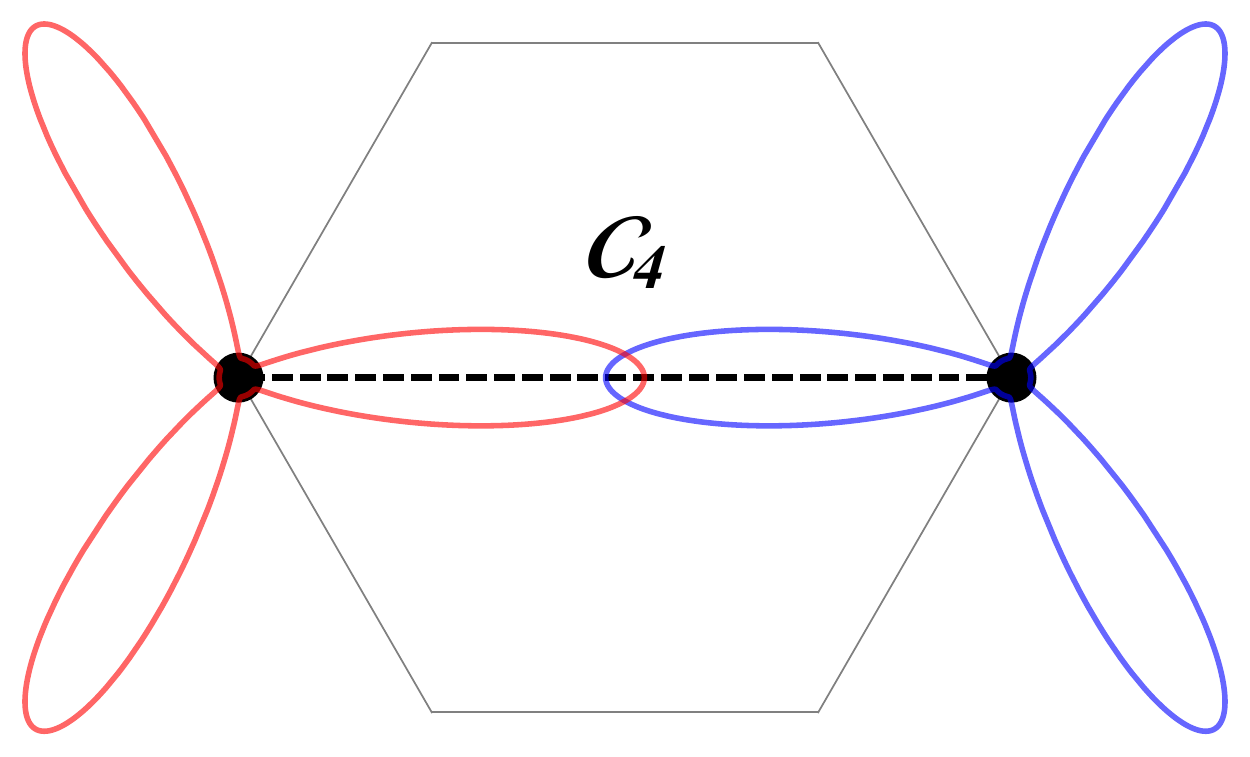} 
{\larger[2] (d)} \includegraphics[width=3.5cm]{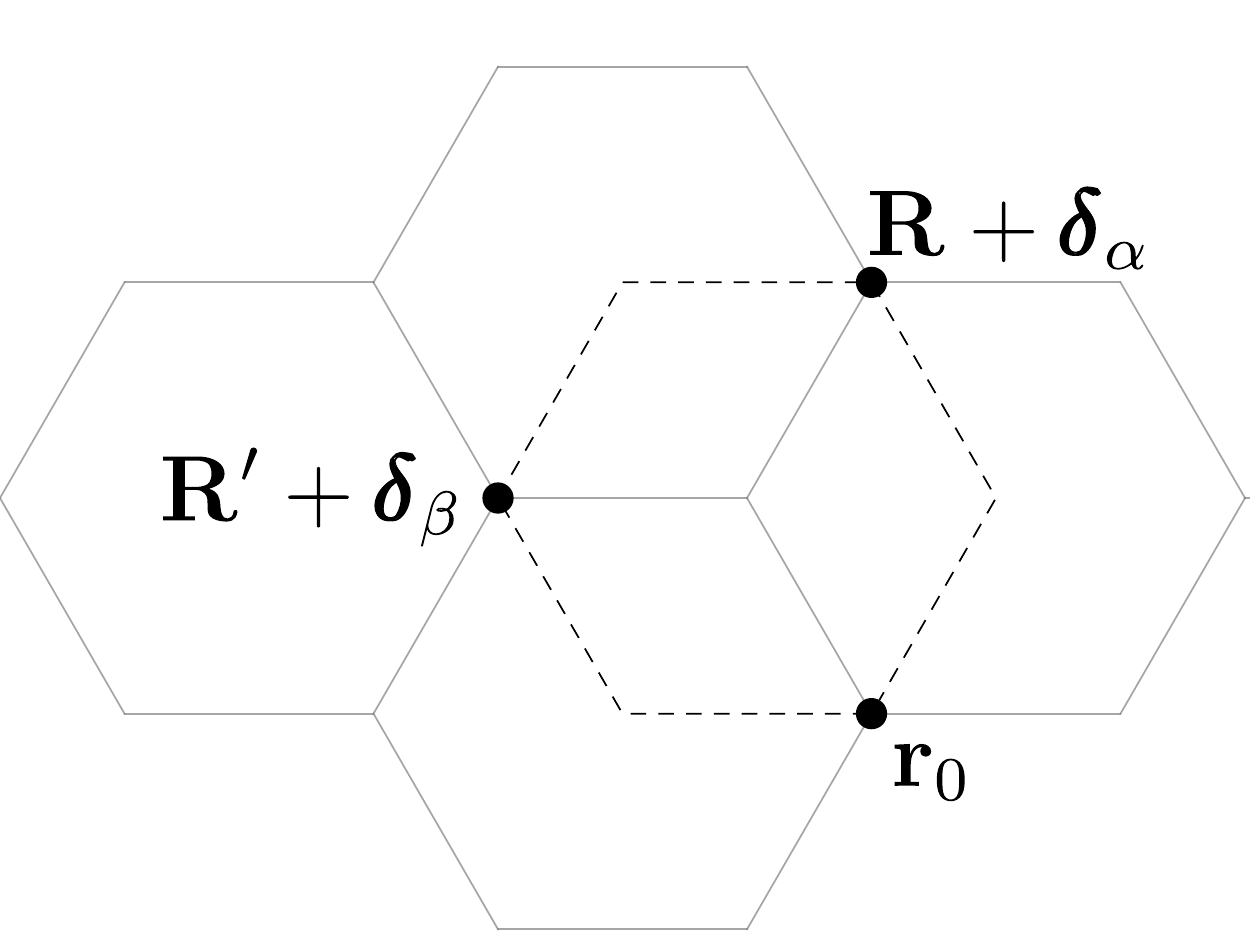} 
\caption{ We sketch the Wannier functions and Peierls paths for three different hoppings in the model of twisted bilayer graphene (introduced fully in \App{app:TBG}). (a) There are two symmetric paths, $\mathcal{C}_1$ and $\mathcal{C}_2$, through the positions of maximum overlap, and thus the Peierls phase is calculated from these two paths \emph{in superposition}. If the Peierls phases along $\mathcal{C}_1, \mathcal{C}_2$ are denoted $\varphi_1,\varphi_2$ respectively, then the hopping $t$ between orbitals becomes $t (e^{i \varphi_1} + e^{i \varphi_2})/2$. (b) There is only one position of maximum overlap at the center of the honeycomb, and the path $\mathcal{C}_3$ (dashed) taken between the orbitals must go through it. (c) There is a straight-line path $\mathcal{C}_4$ (dashed) connecting the orbitals to the center of maximum overlap, matching the conventional Peierls substitution. (d) We show an example of the closed sequence of Peierls paths $\mathcal{C}_{\mbf{r}_0}$ used in \Eq{eq:peierlsdeform0}. When $\phi = \Phi$, any such closed Peierls path encloses a multiple of $2\pi$ flux.  } 
\label{fig:TBGhoppings}
\end{figure*}

To introduce a constant magnetic field to the lattice via the Peierls substitution, we must choose paths $\mathcal{C}_{\mbf{r} \leftarrow \mbf{r}'}$ connecting the orbitals at $\mbf{r} = \mbf{R}+ \pmb{\delta}_\al, \mbf{r}'=\mbf{R}'+ \pmb{\delta}_\be$ where $\mbf{R} =  r_1 \mbf{a}_1 + r_2 \mbf{a}_2, \mbf{R}' = r'_1 \mbf{a}_1 + r'_2 \mbf{a}_2$ with $r_1, r_1', r_2, r_2' \in \mathds{Z} $ and $\pmb{\delta}_\al$ is the position of an orbital $\al$ within the unit cell. Given a path, we calculate the Peierls phases,
\bea
\label{eq:peierlsdefphase}
\varphi_{\mbf{r} \mbf{r}'} &= \int_{\mathcal{C}_{\mbf{r} \leftarrow \mbf{r}'}} \mbf{A} \cdot d\mbf{r},  
\eea
which modify the zero-field hoppings $t_{\al \be}(\mbf{r} - \mbf{r}') \to e^{i \varphi_{\mbf{r} \mbf{r}'} }t_{\al \be}(\mbf{r} - \mbf{r}')  $. Conventionally, the Peierls substitution is for nearest neighbors and consists of straight-line paths between the orbitals.  \Ref{2018arXiv181111786L} discusses Peierls' approximation in more detail and demonstrated that the integral \Eq{eq:peierlsdefphase} should be taken on piecewise straight paths from the orbitals through the points of greatest overlap of the local Wannier functions, possibly in superposition \cite{PhysRev.84.814}. For instance, ``$s$" $\delta$-function-like orbitals on sites  should be connected by a straight-line path since they are centered on the atoms. In a more complicated example, our model of twisted bilayer graphene, the Wannier functions are extended and the paths are not straight but rather are taken through the center of the honeycomb \cite{2018arXiv180710676S,2018arXiv181111786L}. We show this in \Fig{fig:TBGhoppings}. We emphasize that the path $\mathcal{C}_{\mbf{r} \leftarrow \mbf{r}'}$ is physical; different paths lead to a different spectrum, resulting from the orbitals present in the model (see \Ref{2018arXiv181111786L}). Of course, the additional gauge choice made in writing $\mbf{A}$ does affect the individual phases, but does not affect the spectrum.

\subsection{Magnetic Periodicity and Gauge Invariance}
\label{app:magper}
A crucial feature of the Hofstadter Hamiltonian is its periodicity in the flux $\phi$, which we anticipate because the flux dependence enters the Hofstadter Hamiltonian only as a phase. For simplicity, we work in units of length such that the unit cell area $\mbf{a}_1 \times \mbf{a}_2$ is set to $1$. The flux periodicity is given by $\Phi = 2\pi n, n \in \mathbb{N}$ such that taking $\phi \to \phi + \Phi$ leaves the energy spectrum invariant. $n$ is observable (in principle), gauge invariant, and exists so long as the Peierl's paths are commensurate, meaning that all loops along Peierls paths enclose a rational area (given $\mbf{a}_1 \times \mbf{a}_2 = 1$). As we will soon show, $n$ is given by the least common denominator of the fractional area enclosed by all possible loops along the Peierls paths. For example, we consider the model of twisted bilayer graphene. Consulting the hoppings of the twisted bilayer graphene model in \Fig{fig:TBGhoppings}, we see all hoppings must pass through the center of the honeycomb and cannot go along the bonds. Examining the possible closed loops that can be constructed from the Peierls paths in \Fig{fig:TBGhoppings}, we see that all enclose multiples of $1/3$ of a unit cell, so $n=3$. 

We prove the periodicity of $H^\phi$ in flux by constructing the unitary transform $U$ explicitly. Let $\mbf{A} \to \mbf{A} + \mbf{\tilde{A}}$ where the flux of $\mbf{\tilde{A}}$ through the unit cell, denoted $\Omega$, is 
\bea
\pmb{\nabla} \times \mbf{\tilde{A}}  = \Phi 
\eea 
recalling that we take the cross product of 2D vectors to be a scalar, and that the magnetic field is constant. The Peierls phases $\varphi_{\mbf{R} + \pmb{\delta}_\al, \mbf{R}' + \pmb{\delta}_\be}$ acquire the additional contribution $\int^{\mbf{R} + \pmb{\delta}_\al}_{\mbf{R}'+ \pmb{\delta}_\be} \mbf{\tilde{A}} \cdot d\mbf{r}$. By the definition of $n$, all closed line integrals of $\mathbf{\tilde{A}}$ \emph{which are taken along Peierls paths} are pierced by an integer number of flux quanta. In what follows, we assume all integrals are taken along Peierls paths. Let $\mbf{r}_0$ be an arbitrary but fixed orbital of the Hamiltonian that is connected by a sequence of Peierls paths to orbitals at $\mbf{R} + \pmb{\delta}_{\al}$ and $\mbf{R}' + \pmb{\delta}_{\be}$, and let $\mathcal{C}_{\mbf{r}_0}$ be a closed loop along Peierls paths connecting $\mbf{R} + \pmb{\delta}_{\al},\mbf{R}' + \pmb{\delta}_{\be},$ and $\mbf{r}_0$ which we depict an example of in \Fig{fig:TBGhoppings}d. Then at $\phi = \Phi$, we have
\bea
\label{eq:peierlsdeform0}
\oint_{\mathcal{C}_{\mbf{r}_0}} \mbf{\tilde{A}} \cdot d\mbf{r} &= \lp \int_{\mbf{R} + \pmb{\delta}_\al}^{\mbf{R}'+ \pmb{\delta}_\be} +  \int^{\mbf{r}_0}_{\mbf{R}'+ \pmb{\delta}_\be} + \int^{\mbf{R} + \pmb{\delta}_\al}_{\mbf{r}_0} \rp \mbf{\tilde{A}}  \cdot d\mbf{r} = 0 \mod 2\pi 
\eea
from which we conclude
\bea
\label{eq:peierlsdeform}
\int^{\mbf{R} + \pmb{\delta}_\al}_{\mbf{R}'+ \pmb{\delta}_\be}  \mbf{\tilde{A}}  \cdot d\mbf{r} &= \lp \int^{\mbf{r}_0}_{\mbf{R}'+ \pmb{\delta}_\be} + \int^{\mbf{R} + \pmb{\delta}_\al}_{\mbf{r}_0} \rp  \mbf{\tilde{A}}  \cdot d\mbf{r} \mod 2\pi \ .
\eea
This equation shows that the line integral of $\mbf{\tilde{A}}$, generating $\Phi$ flux, taken between two points $\mbf{R} + \pmb{\delta}_\al$ and $\mbf{R}'+ \pmb{\delta}_\be$ along Peierls paths may be deformed to any other point $\mbf{r}_0$ along Peierls paths without changing the value mod $2\pi$. Because of this, whenever integrals in the form \Eq{eq:peierlsdeform} appear in exponentials, the integral is path independent as long as it is taken along Peierls paths. Now we construct the unitary transformation
\bea
\label{eq:hoffluxper}
U^\dag c_{\mbf{R}, \al} U &= e^{i \int^{\mbf{R} + \pmb{\delta}_{\al}}_{\mbf{r}_0}  \mbf{\tilde{A}}  \cdot d\mbf{r}} c_{\mbf{R}, \al}, \qquad U = \exp \ \lp   i \sum_{\mbf{R} \al} c^\dag_{\mbf{R}, \al} c_{\mbf{R}, \al}  \int^{\mbf{R}+ \pmb{\delta}_\al}_{\mbf{r}_0}  \mbf{\tilde{A}}  \cdot d\mbf{r} \rp \ , \\
\eea 
which acts on $c^\dag_{\mbf{R}, \al}$ (resp. $c_{\mbf{R}, \al}$), the fermion creation (resp. destruction) operator of the orbital $\be$ at $\mbf{R} + \pmb{\delta}_\al$.  We note that the path of integration is arbitrary \emph{as long as it is taken along Peierls paths} as per the prior discussion. Using the definition of $U$, we compute
\bea 
U^\dag H^{\phi+\Phi} U &= U^\dag \sum_{\mbf{R} \mbf{R}' \al \be} t_{\al \be}(\mbf{R} + \pmb{\delta}_{\al} - (\mbf{R}' + \pmb{\delta}_{\be})) e^{ i \varphi_{\mbf{R} + \pmb{ \delta}_\al, \mbf{R}' +\pmb{ \delta}_\be}  + i \int^{\mbf{R} + \pmb{\delta}_\al}_{\mbf{R}'+ \pmb{\delta}_\be}  \mbf{\tilde{A}}  \cdot d\mbf{r} } c^\dag_{\mbf{R}, \al} c_{\mbf{R}' ,\be} U \\
&= \sum_{\mbf{R} \mbf{R}' \al \be} t_{\al \be}(\mbf{R} + \pmb{\delta}_{\al} - (\mbf{R}' + \pmb{\delta}_{\be})) e^{ i \varphi_{\mbf{R} +  \pmb{ \delta}_\al, \mbf{R}' +  \pmb{ \delta}_\be} + i\int^{\mbf{R} + \pmb{\delta}_\al}_{\mbf{R}'+ \pmb{\delta}_\be}  \mbf{\tilde{A}}  \cdot d\mbf{r} } U^\dag c^\dag_{\mbf{R}, \al} U U^\dag c_{\mbf{R}', \be} U \\
&= \sum_{\mbf{R} \mbf{R}' \al \be}t_{\al \be}(\mbf{R} + \pmb{\delta}_{\al} - (\mbf{R}' + \pmb{\delta}_{\be})) e^{ i \varphi_{\mbf{R} +  \pmb{ \delta}_\al, \mbf{R}' +  \pmb{ \delta}_\be} + i\int^{\mbf{R} + \pmb{\delta}_\al}_{\mbf{R}'+ \pmb{\delta}_\be}  \mbf{\tilde{A}}  \cdot d\mbf{r} - i\int^{\mbf{r}_0}_{\mbf{R}'+ \pmb{\delta}_\be}  \mbf{\tilde{A}}  \cdot d\mbf{r}  + i \int^{\mbf{R} + \pmb{\delta}_\al}_{\mbf{r}_0}  \mbf{\tilde{A}}  \cdot d\mbf{r} } c^\dag_{\mbf{R}, \al} c_{\mbf{R}', \be} \\
&= \sum_{\mbf{R} \mbf{R}' \al \be} t_{\al \be}(\mbf{R} + \pmb{\delta}_{\al} - (\mbf{R}' + \pmb{\delta}_{\be})) e^{ i \varphi_{\mbf{R} + \pmb{\delta}_\al, \mbf{R}' + \pmb{\delta}_\be} + i\oint_{\mbf{r}_0}  \mbf{\tilde{A}}  \cdot d\mbf{r} } c^\dag_{\mbf{R}, \al} c_{\mbf{R},' \be} \\
&= \sum_{\mbf{R} \mbf{R}' \al \be} t_{\al \be}(\mbf{R} + \pmb{\delta}_{\al} - (\mbf{R}' + \pmb{\delta}_{\be})) e^{ i \varphi_{\mbf{R} + \pmb{\delta}_\al, \mbf{R}' + \pmb{\delta}_\be} } c^\dag_{\mbf{R}, \al} c_{\mbf{R}', \be} \\
&= H^\phi \\
\eea
proving that the Hamiltonian is periodic in $\Phi = 2\pi n$ up to a unitary transformation $U$. We may think of $U$ as a kind of ``embedding matrix" in the flux direction, in analogy to the embedding matrix along $k_z$ of 3D Bloch Hamiltonians \cite{PhysRevLett.113.116403}. 

A very similar proof can be used to show that the Hofstadter Hamiltonian is also gauge-invariant with respect to the electromagnetic field up to a unitary transform. For clarity, we momentarily denote the Hamiltonian's dependence on the gauge field as $H(\mbf{A})$. If we change gauge to $\mbf{A} \to \mbf{A} + \pmb{\nabla} \la$, then we construct the new unitary transformation
\bea
\tilde{U} &= \exp \ \lp   i \sum_{\mbf{R} \al} \lambda(\mbf{R} + \pmb{\delta}_\al) c^\dag_{\mbf{R}, \al} c_{\mbf{R}, \al} \rp 
\eea
and calculate
\bea
\tilde{U}^\dag H(\mbf{A} + \pmb{\nabla} \la) \tilde{U} &= \tilde{U}^\dag \sum_{\mbf{R} \mbf{R}' \al \be} t_{\al \be}(\mbf{r}  - \mbf{r}') e^{ i \varphi_{\mbf{R} + \pmb{\delta}_\al, \mbf{R}' + \pmb{\delta}_\be} + i \int^{\mbf{R} + \pmb{\delta}_\al}_{\mbf{R}'+ \pmb{\delta}_\be} \pmb{\nabla} \la \cdot d\mbf{r} } c^\dag_{\mbf{R}, \al} c_{\mbf{R}', \be} \tilde{U} \\
&= \sum_{\mbf{R} \mbf{R}' \al \be} t_{\al \be}(\mbf{r}  - \mbf{r}') e^{ i \varphi_{\mbf{R} + \pmb{\delta}_\al, \mbf{R}' + \pmb{\delta}_\be} + i\la(\mbf{R}+ \pmb{\delta}_\al) - i \la(\mbf{R}'+ \pmb{\delta}_\be) } \tilde{U}^\dag  c^\dag_{\mbf{R}, \al} \tilde{U} \tilde{U}^\dag c_{\mbf{R},' \be} \tilde{U} \\
&= \sum_{\mbf{R} \mbf{R}' \al \be} t_{\al \be}(\mbf{r}  - \mbf{r}') e^{ i \varphi_{\mbf{R} + \pmb{\delta}_\al, \mbf{R}' + \pmb{\delta}_\be} + i\la(\mbf{R}+ \pmb{\delta}_\al) - i \la(\mbf{R}'+ \pmb{\delta}_\be) - i\la(\mbf{R}+ \pmb{\delta}_\al) + i \la(\mbf{R}' + \pmb{\delta}_\be) } c^\dag_{\mbf{R}, \al} c_{\mbf{R}', \be}  \\
&= H(\mbf{A}) \ . \\
\eea
The transformation $\tilde{U}$ is the straightforward implementation of the $U(1)$ gauge symmetry of continuum electromagnetic to lattice fermions and is an important consistency check on the Peierls substitution. 

\begin{figure*}
 \centering
\includegraphics[width=8cm]{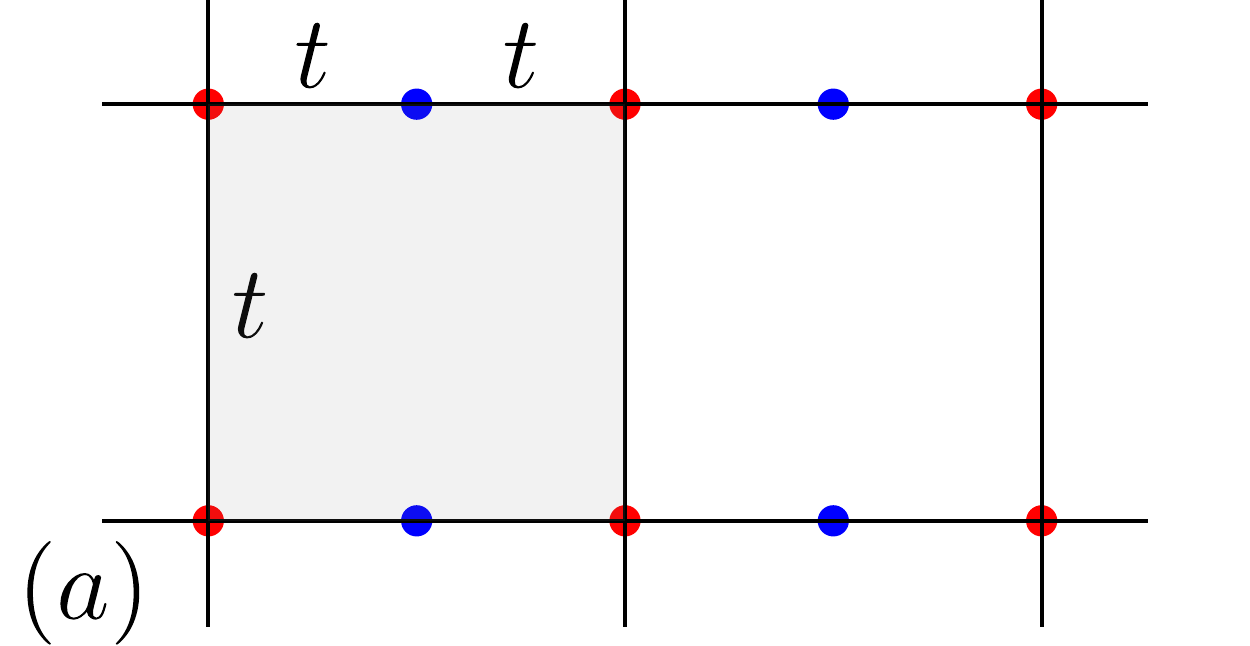} 
\includegraphics[width=8cm]{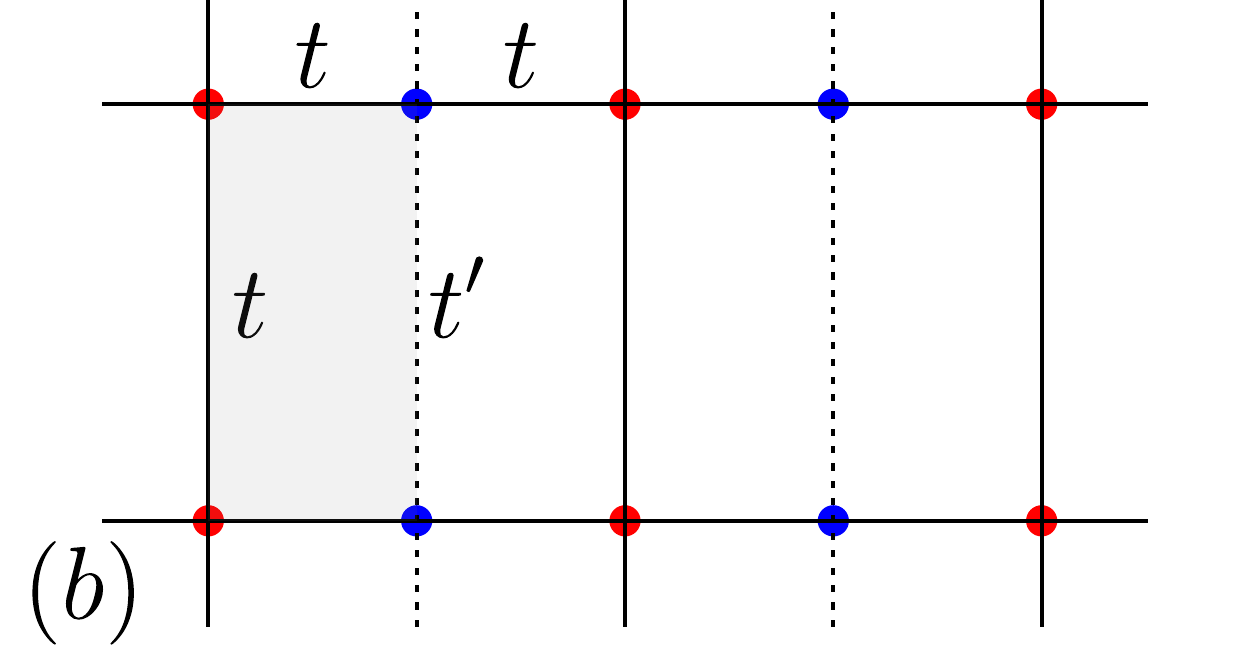} 
\caption{$(a)$ We show a square lattice with atoms at the 1a (red) and 1b (blue) positions. There are hoppings of amplitude $t$ (solid lines) between the 1a atoms separated by $(0,1)$ and also between neighboring 1a and 1b atoms separated by $(1/2,0)$. We take all Peierls paths to be straight lines between the atoms, so the paths are given by the lines shown in the figure. The smallest closed loop along Peierls paths encloses a single unit cell, and hence $n =1, \Phi = 2\pi$.  $(b)$ We now allow the 1b atoms to be connected by a hopping of amplitude $t'$, shown by dotted lines. There is a closed loop that encloses half a unit cell (shaded in gray) by hopping between the 1b positions. In this case $n =2$ and $\Phi = 4\pi$. } 
\label{fig:1blatticeex}
\end{figure*}

We conclude this section with an example of a simple model with $n \neq 1$ (thus $\Phi \neq 2\pi$). Starting from the familiar square lattice Hofstadter model, we add atoms at the 1b position as shown in \Fig{fig:1blatticeex}. If we connect the 1b sites with a hopping $t'$ taken along a straight-line Peierls path (shown with a dotted line in \Fig{fig:1blatticeex}) then there is a closed loop along Peierls paths enclosing half a unit cell, so $n=2$. Accordingly, shifting $\phi \to \phi + 2 \times 2\pi$ leaves the spectrum invariant and we identify $\Phi = 2\pi n = 4\pi$. 

\subsection{The Magnetic Translation Group}
\label{app:magtran}

We now discuss the magnetic translation operators that commute with the Hamiltonian $H^\phi$ in the presence of flux. We will prove that the single-particle operators 
\bea
\label{appeq:T}
T_i(\phi) &= \sum_{\mbf{R} \al} \exp \Big(i \int_{\mbf{R} + \pmb{\delta}_\al}^{\mbf{R} + \pmb{\delta}_\al + \mbf{a}_i} \mbf{A} \cdot d\mbf{r} \, + \, i \chi_i(\mbf{R}+\pmb{\delta}_\al ) \Big) c^\dag_{\mbf{R}+\mbf{a}_i, \al} \ket{0} \bra{0} c_{\mbf{R},\al}, \quad \chi_i(\mbf{r}) = \phi \, \mbf{a}_i \times \mbf{r} \\
\eea
commute with $H^\phi$. We have inserted a projector $\ket{0} \bra{0}$ into \Eq{eqn:T} to arrive at \Eq{appeq:T}, which insures that $T_i^\dag(\phi) c^\dag_{\mbf{R},\pmb{\delta}_{\al}} T_i(\phi)$ is still a single-particle operator. Note that $T_i(\phi)$ is unitary on the single-particle Hilbert space.  The path of integration in \Eq{appeq:T} is taken to be a straight line between $\mbf{R} + \pmb{\delta}_\al$ and $\mbf{R} + \mbf{a}_i + \pmb{\delta}_\al$, although this is \emph{not} necessarily a Peierls path if the local Wannier functions are supported off the orbital sites and the Peierls path is not a straight line (see \App{app:Peierlspaths}). We will prove $[T_i(\phi),H^\phi] = 0$ by showing $T^\dag_i(\phi) H^\phi T_i(\phi) = H^\phi$ in the single-particle Hilbert space. We expand the LHS to find
\bea
\label{eq:THT}
T^\dag_i(\phi) H^\phi T_i(\phi) &=  \sum_{\mbf{R}, \al,\mbf{R}', \be} \exp \lp i \int^{\mbf{R} + \pmb{\delta}_\al}_{\mbf{R}' + \pmb{\delta}_\be} \mbf{A} \cdot d\mbf{r} \rp t_{\al \be}(\mbf{R} + \pmb{\delta}_\al - (\mbf{R}' + \pmb{\delta}_\be) ) T^\dag_i(\phi) c^\dag_{\mbf{R}, \al} T_i(\phi) T_i^\dag(\phi) c_{\mbf{R}', \be}  T_i(\phi) \\
&= \sum_{\mbf{R}, \al,\mbf{R}', \be} \exp \lp {i \int^{\mbf{R} + \mbf{a}_i + \pmb{\delta}_\al}_{\mbf{R}' + \mbf{a}_i + \pmb{\delta}_\be} \mbf{A} \cdot d\mbf{r} - i \int_{\mbf{R} + \pmb{\delta}_\al}^{\mbf{R} + \pmb{\delta}_\al + \mbf{a}_i} \mbf{A} \cdot d\mbf{r} \, - \, i \chi_i(\mbf{R}+\pmb{\delta}_\al )+ i \int_{\mbf{R}' + \pmb{\delta}_\be}^{\mbf{R}' + \pmb{\delta}_\be + \mbf{a}_i} \mbf{A} \cdot d\mbf{r} \, + \, i \chi_i(\mbf{R}'+\pmb{\delta}_\be )} \rp \\
&  \qquad \qquad \times t_{\al \be}(\mbf{R} + \mbf{a}_i + \pmb{\delta}_\al - (\mbf{R}' + \mbf{a}_i + \pmb{\delta}_\be) ) c^\dag_{\mbf{R}, \al} c_{\mbf{R}', \be}  \\
&= \sum_{\mbf{R}, \al,\mbf{R}', \be} \exp \lp {i \lp \int^{\mbf{R} + \mbf{a}_i + \pmb{\delta}_\al}_{\mbf{R}' + \mbf{a}_i + \pmb{\delta}_\be} + \int^{\mbf{R} + \pmb{\delta}_\al}_{\mbf{R} + \pmb{\delta}_\al + \mbf{a}_i} + \int_{\mbf{R}' + \pmb{\delta}_\be}^{\mbf{R}' + \pmb{\delta}_\be + \mbf{a}_i} \rp \mbf{A} \cdot d\mbf{r} \, - \, i \chi_i(\mbf{R}+\pmb{\delta}_\al ) + \, i \chi_i(\mbf{R}'+\pmb{\delta}_\be )} \rp \\&  \qquad \qquad \times t_{\al \be}(\mbf{R}  + \pmb{\delta}_\al - (\mbf{R}' + \pmb{\delta}_\be) ) c^\dag_{\mbf{R}, \al} c_{\mbf{R}', \be} \ . \\
\eea

\begin{figure*}
 \centering
\includegraphics[width=5cm]{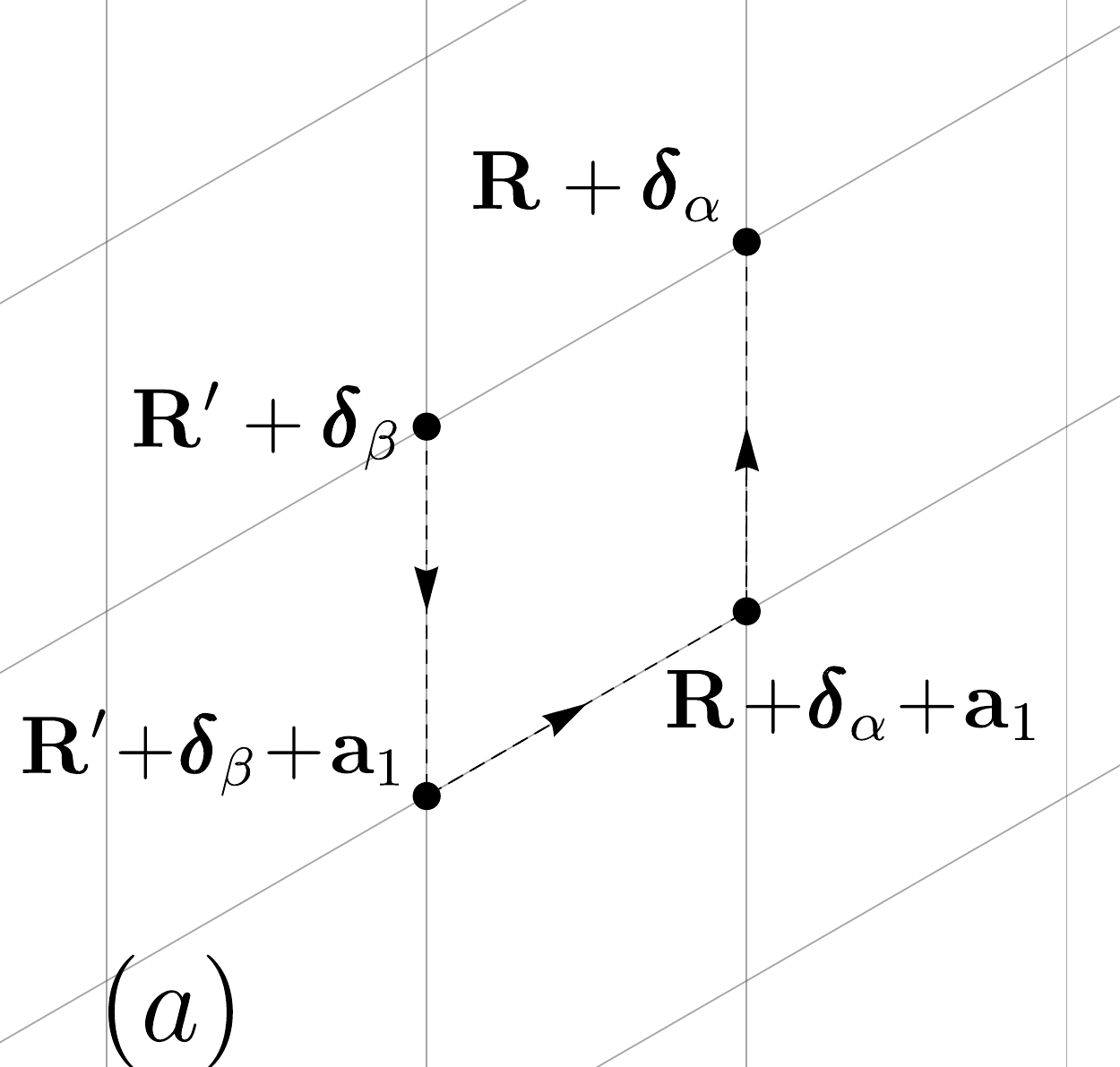}  \qquad 
\includegraphics[width=5cm]{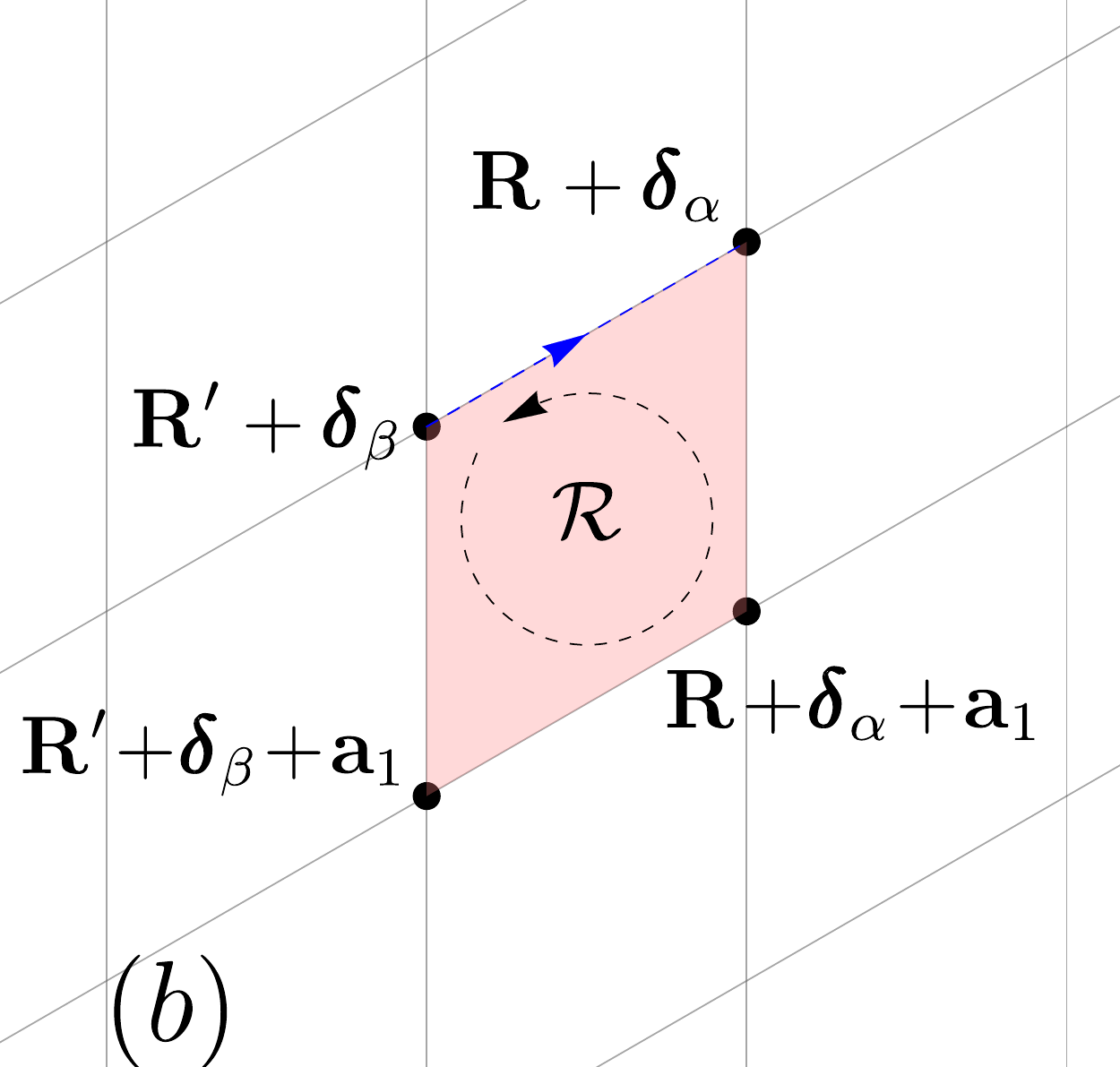} 
\caption{(a) We depict an example of the open path of the line integral on the righthand side of \Eq{eq:intrelation}. (b) Such a path is equivalent to the one on the lefthand side of \Eq{eq:intrelation}, which is a closed path $\mathcal{C}$ encircling $\mathcal{R}$, depicted in pink, with a segment going backwards as shown in blue.} 
\label{fig:Tmagareaenclosed}
\end{figure*}

The path of the open line integral (e.g.  \Fig{fig:Tmagareaenclosed}a) in \Eq{eq:THT} can be rewritten as a closed path that doubles back on itself (e.g.  \Fig{fig:Tmagareaenclosed}b). Written out, we have
\bea
\label{eq:intrelation}
\lp  \int_{\mbf{R}' + \pmb{\delta}_\be}^{\mbf{R}' + \pmb{\delta}_\be + \mbf{a}_i}  + \int^{\mbf{R} + \mbf{a}_i + \pmb{\delta}_\al}_{\mbf{R}' + \mbf{a}_i + \pmb{\delta}_\be} + \int^{\mbf{R} + \pmb{\delta}_\al}_{\mbf{R} + \pmb{\delta}_\al + \mbf{a}_i} \rp \mbf{A} \cdot d\mbf{r} 
&=\lp \int^{\mbf{R} + \mbf{a}_i + \pmb{\delta}_\al}_{\mbf{R}' + \mbf{a}_i + \pmb{\delta}_\be} + \int^{\mbf{R} + \pmb{\delta}_\al}_{\mbf{R} + \pmb{\delta}_\al + \mbf{a}_i} + \int_{\mbf{R}' + \pmb{\delta}_\be}^{\mbf{R}' + \pmb{\delta}_\be + \mbf{a}_i} + \int_{\mbf{R} + \pmb{\delta}_\al}^{\mbf{R}' + \pmb{\delta}_\be} + \int^{\mbf{R} + \pmb{\delta}_\al}_{\mbf{R}' + \pmb{\delta}_\be} \rp \mbf{A} \cdot d\mbf{r} \\
&= \oint_{\mathcal{C}} \mbf{A} \cdot d\mbf{r} + \int^{\mbf{R} + \pmb{\delta}_\al}_{\mbf{R}' + \pmb{\delta}_\be}  \mbf{A} \cdot d\mbf{r} \\
\eea
where the closed loop $\mathcal{C}$ (e.g. the boundary of the pink region in \Fig{fig:Tmagareaenclosed}b) is formed from the straight-line paths $\mbf{R} + \pmb{\delta}_{\al} + \mbf{a}_i \to \mbf{R} + \pmb{\delta}_{\al}$, $\mbf{R}' + \pmb{\delta}_{\be} \to \mbf{R}' + \pmb{\delta}_{\be} + \mbf{a}_i$ of the magnetic translation operator and the Peierls path from $\mbf{R} + \pmb{\delta}_{\al} + \mbf{a}_i \to \mbf{R}' + \pmb{\delta}_{\be} + \mbf{a}_i$. In \Fig{fig:Tmagareaenclosed}, the Peierls path shown in the example happens to be a straight-line path.

\begin{figure}
 \centering
 \includegraphics[width=5.2cm]{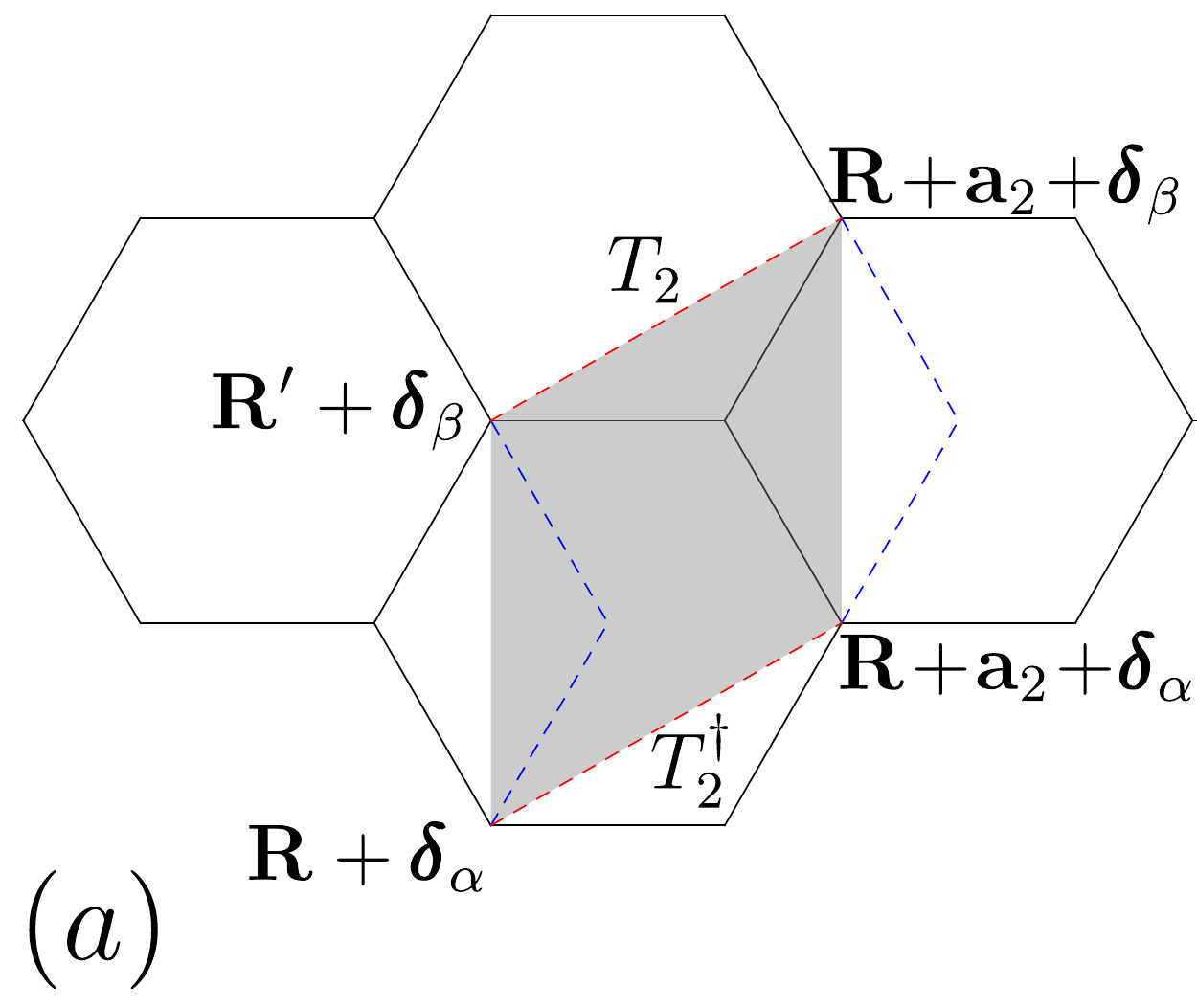} 
\includegraphics[width=5.2cm]{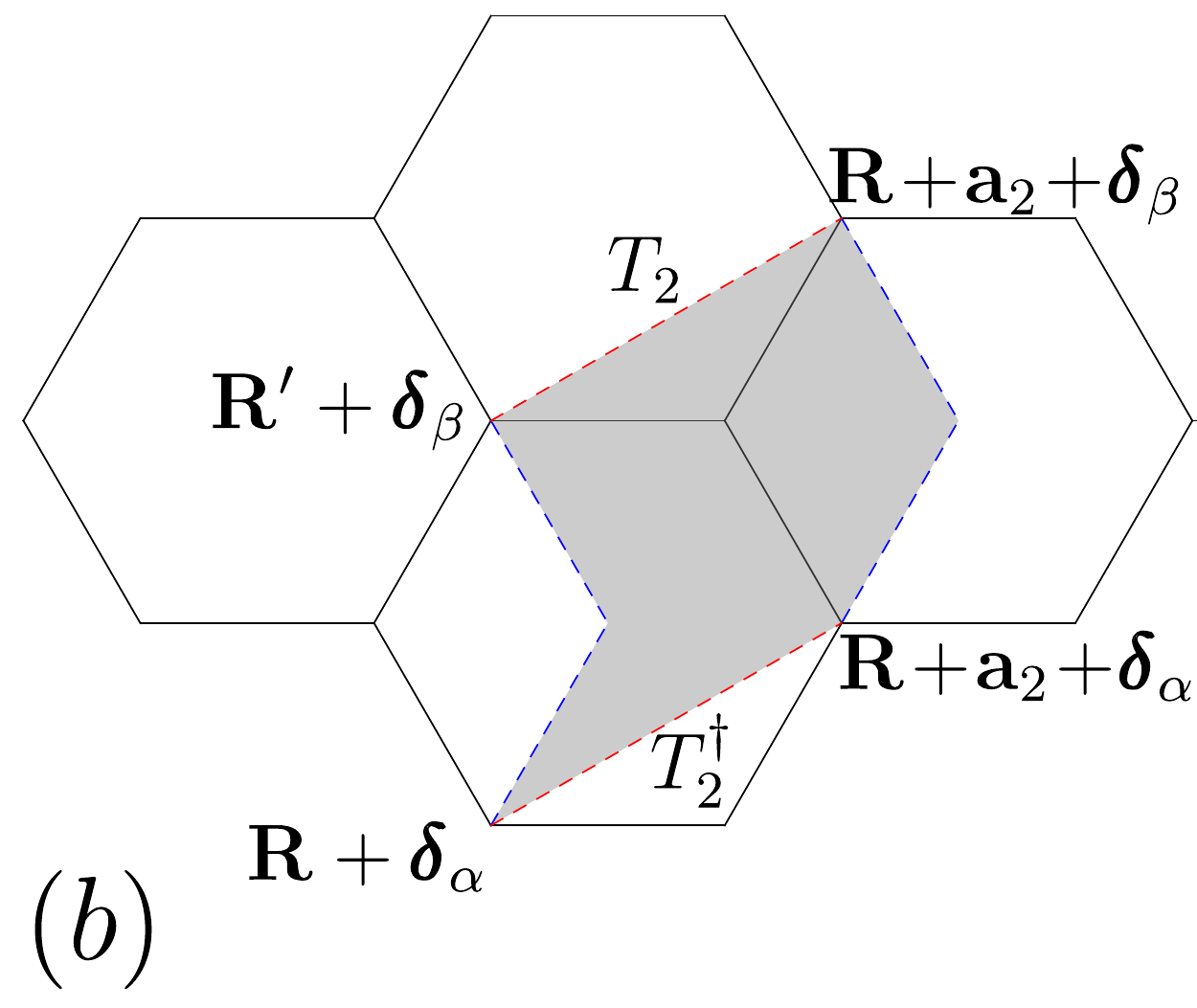}
\includegraphics[width=5.5cm, trim = 0 0cm 0 0, clip]{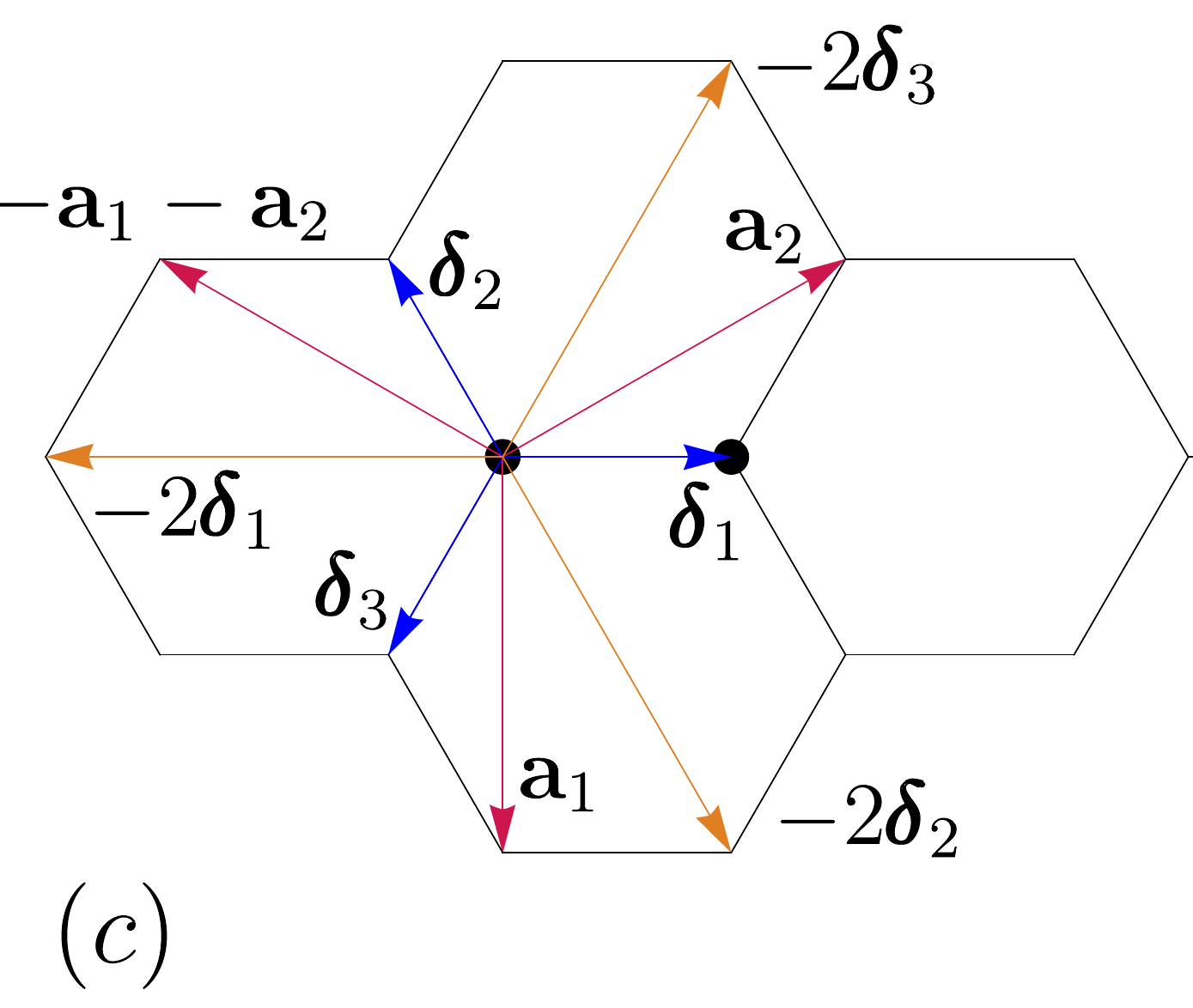}     \\
\caption{We show an example of the area enclosed by the integral in \Eq{eq:stokes} for the Peierls paths of our model of twisted bilayer graphene (as shown in \Fig{fig:TBGhoppings}) and the magnetic translation operator $T_2$. The blue dashed lines are the Peierls' paths $\mathcal{P}_1$ and $|\mathcal{P}_2$ of the hoppings connecting the orbitals, and the red dashed lines show the path of integration in the $T_2$ operator (which is not generically a Peierls path). (a) We depict the area of the parallelogram formed by the vectors $(\mbf{R}' + \pmb{\delta}_\be) - (\mbf{R} + \pmb{\delta}_\al)$ and $\mbf{a}_2$ in grey. (b) We depict the area enclosed by the contiguous integration paths, which has the same value as the area in (a). (c) We depict the lattice vectors $\mbf{a}_1$, the nearest-neighbor vectors $\pmb{\delta}_i$, and the third-nearest neighbor vectors $-2\pmb{\delta}_i$. }
\label{fig:magtranspic}
\end{figure}

We now want to use Stokes' theorem to reduce the closed line integral in \Eq{eq:intrelation} to a surface integral over $\mathcal{R}$, the area enclosed by $\mathcal{C}$. $\mathcal{R}$ is a polygon formed from two parallel sides of length $\mbf{a}_i$ (corresponding to the straight-line paths of integration in the magnetic translation operator $T_i(\phi)$ of \Eq{appeq:T}) and two identical Peierls paths connecting the points $\mbf{R} + \pmb{\delta}_{\al}$ and $\mbf{R}' + \pmb{\delta}_{\be}$ and the points $\mbf{R} + \pmb{\delta}_{\al} +\mbf{a}_i$ and $\mbf{R}' + \pmb{\delta}_{\be} + \mbf{a}_i$, which we call $\mathcal{P}_1$ and $\mathcal{P}_2$ respectively. An example of such a region is shown in \Fig{fig:magtranspic}. If $\mathcal{P}_1$ and $\mathcal{P}_2$ were straight lines, then $\mathcal{R}$ would be a parallelogram with area $(\mbf{R}' + \pmb{\delta}_\be - (\mbf{R} + \pmb{\delta}_\al) ) \times \mbf{a}_i$. In fact, $\mathcal{R}$ still has area $(\mbf{R}' + \pmb{\delta}_\be - (\mbf{R} + \pmb{\delta}_\al) ) \times \mbf{a}_i$ for any (possibly piecewise-straight) Peierls path. This follows because any deviation in area due to $\mathcal{P}_1$ not being straight and is canceled by the same deviation of $\mathcal{P}_2$, as shown for example by the blue dashed lines in \Fig{fig:magtranspic} for the Peierls paths of the twisted bilayer graphene model in \App{app:TBG}. Using this geometrical fact, we find that 
\bea
\label{eq:stokes}
\oint_{\mathcal{C}} \mbf{A} \cdot d\mbf{r} = \phi \int_{\mathcal{R}} dS = \phi \, (\mbf{R}' + \pmb{\delta}_\be - (\mbf{R} + \pmb{\delta}_\al) ) \times \mbf{a}_i \ . \\
\eea

We return to \Eq{eq:THT} and using the results of \Eqs{eq:stokes}{eq:intrelation}, we find
\bea
\label{eq:THTsimp}
T^\dag_i(\phi) H^\phi T_i(\phi) &= \sum_{\mbf{R}, \al,\mbf{R}', \be} \exp \lp {i \int^{\mbf{R} + \pmb{\delta}_\al}_{\mbf{R}' + \pmb{\delta}_\be}  \mbf{A} \cdot d\mbf{r} + i \phi \, (\mbf{R}' + \pmb{\delta}_\be - (\mbf{R} + \pmb{\delta}_\al) ) \times \mbf{a}_i - \, i \chi_i(\mbf{R}+\pmb{\delta}_\al ) + \, i \chi_i(\mbf{R}'+\pmb{\delta}_\be )} \rp \\ &  \qquad \qquad \times t_{\al \be}(\mbf{R}  + \pmb{\delta}_\al - (\mbf{R}' + \pmb{\delta}_\be) ) c^\dag_{\mbf{R}, \al} c_{\mbf{R}', \be} \ . \\
\eea
Now we recall our choice of $\chi_i(\mbf{r}) = \phi \mbf{a}_i \times \mbf{r}$ in \Eq{appeq:T} and notice
\bea
\phi \, (\mbf{R}' + \pmb{\delta}_\be - (\mbf{R} + \pmb{\delta}_\al) ) \times \mbf{a}_i - \,  \chi_i(\mbf{R}+\pmb{\delta}_\al ) + \, \chi_i(\mbf{R}'+\pmb{\delta}_\be ) &=  \phi \, (\mbf{R}' + \pmb{\delta}_\be - (\mbf{R} + \pmb{\delta}_\al) ) \times \mbf{a}_i + \phi  \mbf{a}_i \times (\mbf{R}' + \pmb{\delta}_\be - (\mbf{R} + \pmb{\delta}_\al) )  \\
&= 0 \ . \\
\eea
Hence, we find that \Eq{eq:THTsimp} simplifies to
\bea
T^\dag_i(\phi) H^\phi T_i(\phi) &= \sum_{\mbf{R}, \al,\mbf{R}', \be} \exp \lp i \int^{\mbf{R} + \pmb{\delta}_\al}_{\mbf{R}' + \pmb{\delta}_\be}  \mbf{A} \cdot d\mbf{r} \rp  t_{\al \be}(\mbf{R}  + \pmb{\delta}_\al - (\mbf{R}' + \pmb{\delta}_\be) ) c^\dag_{\mbf{R}, \al} c_{\mbf{R}', \be} \\
&= H^\phi \ . \\
\eea
We have proven that the magnetic translation operators $T_i(\phi)$ commute with $H^\phi$ at all $\phi$. To derive the magnetic translation group algebra, we start from \Eq{appeq:T} and  calculate
\bea
T_1(\phi) T_2(\phi) &=  \sum_{\mbf{R} \al} \exp \lp i \int_{\mbf{R} + \pmb{\delta}_\al + \mbf{a}_2}^{\mbf{R} + \pmb{\delta}_\al + \mbf{a}_1 + \mbf{a}_2} \mbf{A} \cdot d\mbf{r} \, + \, i \chi_1(\mbf{R}+ \mbf{a}_2+\pmb{\delta}_\al ) + i \int_{\mbf{R} + \pmb{\delta}_\al}^{\mbf{R} + \pmb{\delta}_\al + \mbf{a}_2} \mbf{A} \cdot d\mbf{r} \, + \, i \chi_2(\mbf{R}+\pmb{\delta}_\al ) \rp  c^\dag_{\mbf{R}+\mbf{a}_1 + \mbf{a}_2, \al}   c_{\mbf{R},\al}, \\
T_2(\phi) T_1(\phi) &=  \sum_{\mbf{R} \al} \exp \lp i \int_{\mbf{R} + \pmb{\delta}_\al + \mbf{a}_1}^{\mbf{R} + \pmb{\delta}_\al + \mbf{a}_1 + \mbf{a}_2} \mbf{A} \cdot d\mbf{r} \, + \, i \chi_2(\mbf{R}+ \mbf{a}_1+\pmb{\delta}_\al ) + i \int_{\mbf{R} + \pmb{\delta}_\al}^{\mbf{R} + \pmb{\delta}_\al + \mbf{a}_1} \mbf{A} \cdot d\mbf{r} \, + \, i \chi_1(\mbf{R}+\pmb{\delta}_\al ) \rp  c^\dag_{\mbf{R}+\mbf{a}_1 + \mbf{a}_2, \al}   c_{\mbf{R},\al}, \\
\eea
which we use to find that, on the single-particle Hilbert space, 
\bea
\label{eq:t1t2}
(T_2(\phi) T_1\phi))^\dag T_1(\phi) T_2(\phi) &= \sum_{\mbf{R} \al} \exp \Big[ i \int_{\mbf{R} + \pmb{\delta}_\al + \mbf{a}_2}^{\mbf{R} + \pmb{\delta}_\al + \mbf{a}_1 + \mbf{a}_2} \mbf{A} \cdot d\mbf{r} \, + \, i \chi_1(\mbf{R}+ \mbf{a}_2+\pmb{\delta}_\al ) + i \int_{\mbf{R} + \pmb{\delta}_\al}^{\mbf{R} + \pmb{\delta}_\al + \mbf{a}_2} \mbf{A} \cdot d\mbf{r} \, + \, i \chi_2(\mbf{R}+\pmb{\delta}_\al ) \\
& \qquad - i \int_{\mbf{R} + \pmb{\delta}_\al + \mbf{a}_1}^{\mbf{R} + \pmb{\delta}_\al + \mbf{a}_1 + \mbf{a}_2} \mbf{A} \cdot d\mbf{r} \, - \, i \chi_2(\mbf{R}+ \mbf{a}_1+\pmb{\delta}_\al ) - i \int_{\mbf{R} + \pmb{\delta}_\al}^{\mbf{R} + \pmb{\delta}_\al + \mbf{a}_1} \mbf{A} \cdot d\mbf{r} \, - \, i \chi_1(\mbf{R}+\pmb{\delta}_\al )  \Big] c^\dag_{\mbf{R},\al} c_{\mbf{R},\al} \ . \\
\eea
Collecting the integrals, we see
\bea
\label{eq:intunit}
\lp \int_{\mbf{R} + \pmb{\delta}_\al + \mbf{a}_2}^{\mbf{R} + \pmb{\delta}_\al + \mbf{a}_1 + \mbf{a}_2} + \int_{\mbf{R} + \pmb{\delta}_\al}^{\mbf{R} + \pmb{\delta}_\al + \mbf{a}_2} - \int_{\mbf{R} + \pmb{\delta}_\al + \mbf{a}_1}^{\mbf{R} + \pmb{\delta}_\al + \mbf{a}_1 + \mbf{a}_2} - \int_{\mbf{R} + \pmb{\delta}_\al}^{\mbf{R} + \pmb{\delta}_\al + \mbf{a}_1} \rp \mbf{A} \cdot d\mbf{r} &= - \oint_{\text{unit cell}} \mbf{A} \cdot dr \\
&= - \phi \int_{\text{unit cell}} dS \\
&= - \phi \\
\eea
where the minus sign has appeared because the path of integration is clockwise, and we recall that the area of the unit cell is one. Collecting the $\chi_i$ terms from \Eq{eq:t1t2}, we find
\bea
\label{eq:chiunit}
\chi_1(\mbf{R}+ \mbf{a}_2+\pmb{\delta}_\al ) + \chi_2(\mbf{R}+\pmb{\delta}_\al ) - \chi_2(\mbf{R}+ \mbf{a}_1+\pmb{\delta}_\al ) - \chi_1(\mbf{R}+\pmb{\delta}_\al) &= \phi \mbf{a}_1 \times (\mbf{R}+ \mbf{a}_2+\pmb{\delta}_\al - (\mbf{R}+\pmb{\delta}_\al) )  \\
&\qquad + \phi \mbf{a}_2 \times (\mbf{R}+\pmb{\delta}_\al - (\mbf{R}+ \mbf{a}_1+\pmb{\delta}_\al ) ) \\
&= \phi \mbf{a}_1 \times \mbf{a}_2 - \phi \mbf{a}_2 \times \mbf{a}_1 \\
&= 2\phi \mbf{a}_1 \times \mbf{a}_2 \\
&= 2\phi 
\eea
because we have normalized the unit cell to be $\mbf{a}_1 \times \mbf{a}_2 = 1$. Using the results of \Eqs{eq:intunit}{eq:chiunit}, we return to \Eq{eq:t1t2} and find
\bea
\label{eq:t1t2identity}
(T_2(\phi) T_1(\phi))^\dag T_1(\phi) T_2(\phi) &= \sum_{\mbf{R} \al} e^{-i \phi + 2 i \phi} c^\dag_{\mbf{R},\al} c_{\mbf{R},\al}  =  e^{i \phi}
\eea
recalling that $ \sum_{\mbf{R} \al} c^\dag_{\mbf{R},\al} c_{\mbf{R},\al}  = 1$ when acting on single particle states. \Eq{eq:t1t2identity} is the usual single-particle algebra of the magnetic translation operators. 

\subsection{Peierls Phases}
\label{app:ppA}

So far, we have discussed the magnetic translation group in position space which demonstrates the existence of a $1 \times q$ magnetic unit cell at $\phi = \frac{2\pi p}{q}$, $p,q$ coprime. We now discuss a particular gauge choice, a certain Landau gauge, that will makes explicit calculations in momentum space tractable. We work in the Landau gauge $\mbf{A}(\mbf{r}) = - \phi \mbf{b}_1 (\mbf{r} \cdot \mbf{b_2})$ which obeys
\bea
\pmb{\nabla} \times \mbf{A}(\mbf{r}) &= - \phi \mbf{b}_2 \times \mbf{b}_1 = \phi
\eea
where we have used that $\mbf{b}_1 \times \mbf{b}_2 = \mbf{a}_1 \times \mbf{a}_2 = 1$. This choice of gauge is practical. We will see that the Peierls phases do not depend on $r_1$, the coordinate along $\mbf{a}_1$, so the unit cell of the Hofstadter Hamiltonian is only extended along one direction. Thus we may Fourier transform along $\mbf{a}_1$ and obtain an effective 1D chain along $\mbf{a}_2$ which is \emph{not} a priori periodic because of the dependence on $r_2$. However, for a certain choice of \textit{rational} flux (in this specific gauge choice), an extended spatial periodicity reappears and enables the Hamiltonian to be Fourier transformed into momentum space. Importantly, at $\phi = \frac{2\pi p}{q}$, $p,q$ coprime, the Landau gauge does \emph{not} permit a $1 \times q$ unit cell on a general lattice, and may need to be enlarged to a $1 \times q'$ unit cell for $q' > q$. We will discuss this extensively in \App{app:fluxrat}. In \App{app:genz2proof}, we introduce a gauge-invariant formalism where the minimal $1\times q$ magnetic unit cell is manifest. However, the Landau gauge is a much more convenient choice for numerical calculations because we can derive explicit expressions for the Peierls phases. 

First we show that in our Landau gauge, all Peierls phases (\Eq{eq:peierlsdefphase}) must be of the form $\varphi_{\mbf{r} \mbf{r}'} = \phi \, ( \rho_{\mbf{r} \mbf{r}'}  r_2 + \rho'_{\mbf{r} \mbf{r}'} ) $ with $\rho_{\mbf{r} \mbf{r}'} \in \mathbb{Q}$ so long as the orbitals are commensurate, meaning that $\mbf{R}' + \pmb{\delta}_\be - (\mbf{R} + \pmb{\delta}_\al)$ is a rational linear combination of the lattice vectors. As an example of an incommensurate orbitals, consider a lattice with atoms at the positions $r_1 \mbf{a}_1 + r_2 \mbf{a}_2$ and $r_1 \mbf{a}_1 + r_2 \mbf{a}_2 + 1/\sqrt{5} \mbf{a}_1$ which are connected by straight-line Peierls paths. No magnetic unit cell exists on this lattice because there is no periodicity in its Peierls phases. 

Let the Peierls path $\mathcal{C}$ be from $\mbf{R} + \pmb{\delta}_\al$ to $\mbf{R}' + \pmb{\delta}_\be$, taken along the piecewise straight path connecting the points $\mbf{x}_1, \dots, \mbf{x}_N$ (noting that the beginning and ending points are $\mbf{x}_1 = \mbf{R} + \pmb{\delta}_\al$ and $\mbf{x}_N = \mbf{R}' + \pmb{\delta}_\be$ respectively). We now calculate the Peierls phase accumulated along the path $\mathcal{C}$. Breaking up the integral \Eq{eq:peierlsdefphase} along the piecewise-straight path, we find
\bea
\label{eq:peierlssum}
\varphi_{\mbf{x}_1 \mbf{x}_N} &= \int_{\mathcal{C}} \mbf{A} \cdot d \mbf{r} = \sum_{i=1}^{N-1} \int_{\mbf{x}_i}^{\mbf{x}_{i+1}} \mbf{A} \cdot d \mbf{r} \ . \\
\eea
We compute
\bea
\int_{\mbf{x}_i}^{\mbf{x}_{i+1}} \mbf{A} \cdot d \mbf{r} &= -\phi \int_{0}^{1} \mbf{b}_2 \cdot (\mbf{x}_i + t (\mbf{x}_{i+1} - \mbf{x}_i) ) \mbf{b}_1 \cdot (\mbf{x}_{i+1} - \mbf{x}_i) dt \\
&= -\phi \mbf{b}_1 \cdot (\mbf{x}_{i+1} - \mbf{x}_i)  \int_{0}^{1} \lp  \mbf{b}_2 \cdot \mbf{x}_i  + t \mbf{b}_2 \cdot (\mbf{x}_{i+1} - \mbf{x}_i) \rp dt \\
&= -\phi \mbf{b}_1 \cdot (\mbf{x}_{i+1} - \mbf{x}_i) \lp \mbf{b}_2 \cdot \mbf{x}_i  + \frac{1}{2} \mbf{b}_2 \cdot (\mbf{x}_{i+1} - \mbf{x}_i) \rp  \\
&= - \frac{\phi }{2}  \mbf{b}_1 \cdot (\mbf{x}_{i+1} - \mbf{x}_i) \lp \mbf{b}_2 \cdot (\mbf{x}_{i+1} + \mbf{x}_i) \rp \ .  \\
\eea
We work with a fixed reference point (``origin") at $r_1 \mbf{a}_1 + r_2 \mbf{a}_2, \ r_1, r_2 \in \mathds{Z}$ and define $\mbf{t}_i = \mbf{x}_i - ( r_1 \mbf{a}_1 + r_2 \mbf{a}_2) $. Note that because the $\mbf{x}_i$ are assumed commensurate, each $\mbf{t}_i$ is also commensurate. Then we have
\bea
\int_{\mbf{x}_i}^{\mbf{x}_{i+1}} \mbf{A} \cdot d \mbf{r} &= - \phi  \mbf{b}_1 \cdot (\mbf{t}_{i+1} - \mbf{t}_i) \lp \mbf{b}_2 \cdot \big( r_1 \mbf{a}_1 + r_2 \mbf{a}_2 +  \frac{1}{2} (\mbf{t}_{i+1} + \mbf{t}_i)\big)  \rp \\
&= - \phi  \mbf{b}_1 \cdot (\mbf{t}_{i+1} - \mbf{t}_i) \lp r_2  + \frac{\mbf{b}_2}{2} \cdot (\mbf{t}_{i+1} + \mbf{t}_i)  \rp  \\
\eea
using $\mbf{a}_i \cdot \mbf{b}_2 = \delta_{i2}$.  Then by \Eq{eq:peierlssum}, we find
\bea
\label{eq:peierlstel}
\varphi_{\mbf{x}_1 \mbf{x}_N} &= - \phi \sum_{i=1}^{N-1} \mbf{b}_1 \cdot (\mbf{t}_{i+1} - \mbf{t}_i) \lp r_2  + \frac{\mbf{b}_2}{2} \cdot (\mbf{t}_{i+1} + \mbf{t}_i)  \rp  \\
&= - \phi r_2 \left[ \mbf{b}_1 \cdot \sum_{i=1}^{N-1}  (\mbf{t}_{i+1} - \mbf{t}_i)\right] - \phi \left[ \sum_{i=1}^{N-1} \mbf{b}_1 \cdot (\mbf{t}_{i+1} - \mbf{t}_i) \lp \frac{\mbf{b}_2}{2} \cdot (\mbf{t}_{i+1} + \mbf{t}_i)  \rp \right] \\
\eea
We recognize
\bea
\label{eq:peierlsphaseexp}
\rho_{\mbf{r} \mbf{r}'} &= -\sum_{i=1}^{N-1} \mbf{b}_1 \cdot (\mbf{t}_{i+1} - \mbf{t}_i) , \quad \rho'_{\mbf{r} \mbf{r}'} = - \sum_{i=1}^{N-1} \mbf{b}_1 \cdot (\mbf{t}_{i+1} - \mbf{t}_i) \lp \mbf{b}_2 \cdot  \frac{\mbf{t}_{i+1} + \mbf{t}_i}{2} \rp, \\
\eea
Additionally, we observe that our expression for $\rho_{\mbf{r} \mbf{r}'}$ can be simplified because the sum telescopes, canceling term by term:
\bea
\label{eq:peierlsphaseform}
\rho_{\mbf{r} \mbf{r}'} &= -\sum_{i=1}^{N-1} \mbf{b}_1 \cdot (\mbf{t}_{i+1} - \mbf{t}_i) \\
&= - \mbf{b}_1 \cdot (\mbf{t}_{N} - \mbf{t}_1) \\
&= - \mbf{b}_1 \cdot (r_1 \mbf{a}_1 + r_2 \mbf{a}_2 + \mbf{t}_{N} - (r_1 \mbf{a}_1 + r_2 \mbf{a}_2 + \mbf{t}_1) )\\
&= - \mbf{b}_1 \cdot \big( (\mbf{R} + \pmb{\delta}_\al) - (\mbf{R}' + \pmb{\delta}_\be) \big) \ . \\
\eea
and hence $\rho_{\mbf{r} \mbf{r}'}$ is rational because $\mbf{b}_1 \cdot \mbf{a}_i = \delta_{1i}$ and $ (\mbf{R} + \pmb{\delta}_\al) - (\mbf{R}' + \pmb{\delta}_\be) $ is assumed to be a rational linear combination of the lattice vectors because the orbitals are commensurate. \Eq{eq:peierlsphaseform} will be useful as we now determine the magnetic unit cell. We emphasize that the Peierls phases are calculated on open paths and hence are gauge-dependent. 

\subsection{Rationalization of the Flux}
\label{app:fluxrat}

In this section we will demonstrate how to determine the magnetic unit cell in the Landau gauge at rational values of the flux $\phi$. We begin by recalling the well-known procedure on the square lattice. When the Peierls paths are taken along the bonds in the Landau gauge $\mbf{A}(\mbf{r}) = -\phi(y,0)$, only the $x$-directed hoppings acquire nontrivial Peierls phases such that the hopping term becomes $\exp (- \phi y) c^\dag_{x+1,y} c_{x,y}$. For $\phi \neq 0$, the unit cell is broken. However, we can recover a $1 \times q$ unit cell where $y \to y + q$ leaves all the Peierls phases invariant when $\phi$ takes the values $\phi = \frac{2\pi p}{q}$, $p,q$ coprime. We emphasize that this specific magnetic unit cell derives from our Landau gauge choice. Because we can find $p,q$ such that $\phi$ is arbitrarily close to any real number, we are able to form the Hofstadter Hamiltonian in momentum space arbitrarily close to any flux (potentially at the cost of a large magnetic unit cell.) We now consider the magnetic unit cell in a more complicated lattice. It is always possible in principle to choose a $1 \times q$ magnetic unit cell since $[T_1(\phi),T_2^q(\phi)] = [H,T_i(\phi)]=0$. However, since we wish to work \textit{in a specific gauge}, the Landau gauge $\mbf{A}(\mbf{r}) = - \phi \mbf{b}_1 (\mbf{r} \cdot \mbf{b}_2)$, the Peierls phases are not generally periodic in the $1 \times q$ magnetic unit cell. In our Landau gauge, all Peierls phases are in the form $\varphi_{\mbf{r} \mbf{r}'} = \phi \, ( \rho_{\mbf{r} \mbf{r}'}  r_2 + \rho'_{\mbf{r} \mbf{r}'} )$ ( \Eq{eq:peierlsphaseexp}). Under a translation by $T_2^q$ which amounts to taking $r_2 \to r_2 +q$, we have $\varphi_{\mbf{r} \mbf{r}'}  \to \varphi_{\mbf{r} \mbf{r}'}  + 2\pi p \rho_{\mbf{r} \mbf{r}'} $ with $\phi = 2\pi p/q$. Generically, $p \rho_{\mbf{r} \mbf{r}'}$ is not an integer and hence the Peierls phase is \emph{not} periodic in the $1 \times q$ magnetic unit cell. We now show that the Peierls phases \emph{are} periodic for our gauge choice over a $1 \times q'$ magnetic unit cell with $q' \geq q$ if the flux is rationalized to be $\phi = \mu \frac{2\pi p'}{q'}$ where $\mu \in \mathbb{N}$ depends on the orbital positions and the choice of Landau gauge. (Note that $\mu$ depends on the gauge and is unrelated to $n$ and the flux periodicity $\Phi = 2\pi n$.)

Now we give an expression for $\mu$ in the Landau gauge. Let $\mu \in \mathbb{N}$ be the least common denominator (lcd) --- equivalently the greatest common factor of the denominators --- of $\{ \rho_{\mbf{r} \mbf{r'}} \}$ for each hopping of the model. Because $\mbf{R}$ and $\mbf{R}'$ are lattice vectors and $\mbf{b}_1 \cdot \mbf{a}_i = \delta_{i1}$, $\mbf{b}_1 \cdot (\mbf{R}- \mbf{R}')$ is an integer, the rational part of $\rho_{\mbf{r} \mbf{r'}}$ is $ - \mbf{b}_1 \cdot (\pmb{\delta}_\al -  \pmb{\delta}_\be)$ and hence
\bea
\label{eq:mudef}
\mu &=  \text{lcd } \{ \mbf{b}_1 \cdot (\pmb{\delta}_\be - \pmb{\delta}_\al) \} \text{ for all hoppings connecting $\mbf{R} + \pmb{\delta}_\al$ and $\mbf{R}' + \pmb{\delta}_\be$} \ .
\eea
We  seek to determine what (rational) value of the flux $\phi$ permits a spatial periodicity in the Peierls phases \Eq{eq:peierlstel} along $r_2$. For a $1 \times q'$ unit cell, meaning that all Peierls phases at $r_2$ and $r_2 +q'$ are identical modulo $2\pi$, we will prove that we must rationalize the flux to be
\bea
\label{eq:fluxdenommu}
\phi = \mu \frac{2\pi p'}{q'}, \quad p',q' \in \mathds{Z},  \text{ coprime} \ .
\eea
We see that the minimal $1 \times q$ magnetic unit cell at $\phi = \frac{2\pi p}{q}$ given by the magnetic translation group does not generically coincide with the $1 \times q'$ magnetic unit cell which results from choosing the Landau gauge. Throughout this section, we define $p'$ and $q'$ by $p/q = \mu p'/q'$ with $\mu$ fixed by \Eq{eq:fluxdenommu} at $\phi = \frac{2\pi p}{q}$. Note that $q$ and $q'$ are related by $q = q' / \text{gcd}(q', \mu)$ so $q'$ is an integer multiple of $q$ and $q' \geq q$. 

In the $1 \times q'$ magnetic unit cell, the Peierls phases are periodic and we can diagonalize the Hofstadter hamiltonian in momentum space. The $1 \times q'$ unit cell is more convenient for practical purposes, such as determining the spectrum numerically. We emphasize that $\mu$, and hence the magnetic unit cell defined by $q'$, \textit{are both gauge-dependent}. It is only the quantity $\mu p'/q'  = \phi / (2\pi)$ that is gauge-invariant.

We prove \Eq{eq:fluxdenommu} now. For brevity, we denote the Peierls phases $\varphi_{\mbf{R} + \pmb{\delta}_\al, \mbf{R}' + \pmb{\delta}_\be}$ as $\varphi_{r_2 \al, r_2' \be}$, recalling that $\mbf{R} = r_1 \mbf{a}_1 + r_2 \mbf{a}_2$ is the reference point as in \Eq{eq:peierlstel} and $\mbf{R}' = r'_1 \mbf{a}_1 + r'_2 \mbf{a}_2 $ is the unit cell of the second orbital. We have used \Eq{eq:peierlstel} to show that $\varphi_{\mbf{R} + \pmb{\delta}_\al, \mbf{R}' + \pmb{\delta}_\be}$ does not depend on $r_1$. Then we have, under $r_2 \to r_2 + q'$,
\bea
\label{eq:varphildep}
\varphi_{r_2 \al, r_2' \be} &=  ( \rho_{\mbf{r} \mbf{r}'} r_2 + \rho'_{\mbf{r} \mbf{r}'} ) \phi \to  ( \rho_{\mbf{r} \mbf{r}'} (r_2+q') + \rho'_{\mbf{r} \mbf{r}'} ) \phi \\
&= \varphi_{r_2 \al, r_2' \be} + \phi q' \rho_{\mbf{r} \mbf{r}'} \\
&= \varphi_{r_2 \al, r_2' \be}+ \mu \frac{2\pi p'}{q'} q' \rho_{\mbf{r} \mbf{r}'} \\
&= \varphi_{r_2 \al, r_2' \be} + 2\pi p' \ \rho_{\mbf{r} \mbf{r}'} \mu \\
&= \varphi_{r_2 \al, r_2' \be}  \mod 2\pi \ . \\
\eea
Thus $\varphi_{r_2 \al, r_2' \be}$ only depends on $r_2 \text{ mod } q'$, showing that the Peierls paths, and hence the Hofstadter Hamiltonian, has a $1 \times q'$ unit cell. In many simple examples of the Hofstadter Butterfly \cite{PhysRevB.100.245108,PhysRevB.14.2239}, the orbitals are all on the atomic sites, so $\pmb{\delta}_{\al} = 0$ for all $\al$. In this case, \Eq{eq:mudef} trivially gives $\mu =1$, and the flux takes the familiar form $\phi = \frac{2\pi p'}{q'} = \frac{2\pi p}{q}$, so $p=p'$ and $q=q'$.  

We now turn to the example of the square lattice with 1b atoms (see \Fig{fig:1blatticeex}) to discuss $\mu \neq 1$. If we choose the lattice vectors $\mbf{a}_1 = (1,0), \ \mbf{a}_2 = (0,1)$ (noting $\mbf{a}_1 \times \mbf{a}_2 = +1$), then $\mbf{b}_1 = (1,0)$. We calculate $\mu$ using \Eq{eq:mudef} and find
\bea
\mu &= \text{lcd } \{ \mbf{b}_1 \cdot (0,0),\mbf{b}_1 \cdot (1/2,0), \mbf{b}_1 \cdot (0,0) \} \\
&= \text{lcd } \{ (1,0) \cdot (0,0), (1,0) \cdot (1/2,0), (1,0) \cdot (0,0) \} \\
&= 2 \ . \\
\eea
This means that at $\phi = 2 \frac{2\pi p'}{q'}$, we can Fourier transform in the Landau gauge over a $1\times q'$ unit cell. Note that $\mu = 2$ even when the $t'$ hopping is zero, since it arises from the hopping connecting the 1a and 1b atoms (see \Fig{fig:1blatticeex}). Thus we see that when when $t' = 0$ we have $n =1, \mu =2$ and when $t' \neq 0$ we have $n=2, \mu =2$ in this gauge. Alternatively, we could choose a different Landau gauge $\mbf{A}'(\mbf{r}) = - \phi \mbf{b}_1' (\mbf{r} \cdot \mbf{b}_2')$ given by $\mbf{a}'_1 = (0,1), \mbf{a}'_2 = (-1,0)$ (noting $\mbf{a}'_1 \times \mbf{a}'_2 = +1$) where $\mbf{b}'_1 = (0,1)$. In this gauge, we compute
\bea
\mu' &= \text{lcd } \{ \mbf{b}'_1 \cdot (0,0),\mbf{b}'_1 \cdot (1/2,0), \mbf{b}'_1 \cdot (0,0) \} \\
&= \text{lcd } \{ (0,1) \cdot (0,0), (0,1) \cdot (1/2,0), (0,1) \cdot (0,0) \} \\
&= 1 \ .
\eea
Again, this result does not depend on $t'$. Hence in this gauge we have $n=1, \mu' =1$ when $t' = 0$ and $n =2, \mu' =1$ when $t' \neq 0$. The values of $\mu$ are distinct in the different gauges, but $n$ is the same. Indeed, $\Phi = 2\pi n$ is gauge-invariant, whereas $\mu$ simply allowed us to Fourier transform the Peierls phases in a given gauge. 

In some cases (but not all), it is possible to choose lattice vectors so that the Landau gauge $\mbf{A}(\mbf{r}) = - \phi \mbf{b}_1 (\mbf{r} \cdot \mbf{b}_2)$ yields $\mu =1$, which we discuss now.

\subsection{Residual $SL(2,\mathds{Z})$ Gauge Freedom}
\label{app:residualgf}

Despite fixing our gauge to the Landau gauge $\mbf{A}(\mbf{r}) = - \phi \mbf{b}_1 (\mbf{r} \cdot \mbf{b}_2)$ which ensures a $1 \times q'$ magnetic unit cell, there is a residual gauge freedom which arises due to the choice of reciprocal lattice vectors $\mbf{b}_i$.  In deriving \Eq{eq:peierlsphaseform}, the form of the Peierls phases which defines $\mu$ (\Eq{eq:mudef}), we relied on the fact that $\mbf{b}_i \cdot \mbf{a}_j = \delta_{ij}$. For any given set of basis vectors $\mbf{a}_i$, $\mbf{b}_i$ is defined uniquely and our Landau gauge is fixed. However, we can gauge transform to a different Landau gauge $\mbf{A}'(\mbf{r}) = - \phi \mbf{b}_1'(\mbf{r} \cdot \mbf{b}'_2)$ while preserving the form of \Eq{eq:peierlsphaseexp} by choosing new lattice vectors
\bea
\mbf{a}'_i = m_{i1} \mbf{a}_1 + m_{i2} \mbf{a}_2, \qquad m_{ij} \in \mathds{Z} \ .
\eea
For $\mbf{a}'_i$ to preserve the area of the unit cell, we require $\mbf{a}'_1 \times \mbf{a}'_2 = \mbf{b}'_1 \times \mbf{b}'_2 = 1$ which holds iff $\det [m] = m_{11} m_{22} - m_{12} m_{21} = 1$. Thus $[m] \in SL(2, \mathds{Z})$. The corresponding reciprocal lattice is spanned by $\mbf{b}'_i$ satisfying $\mbf{b}'_i \cdot \mbf{a}'_j = \delta_{ij}$ and $\pmb{\nabla} \times \mbf{A}'(\mbf{r}) = \phi \mbf{b}'_1 \times \mbf{b}'_2 = \phi$. It can be checked using $\det [m] = 1$ that the reciprocal vectors are given by
\bea
\mbf{b}'_1 &= m_{22} \mbf{b}_1 - m_{21} \mbf{b}_2, \\
\mbf{b}'_2 &= -m_{12} \mbf{b}_1 + m_{11} \mbf{b}_2 \ . \\
\eea
We find that even after fixing the form of our Landau gauge, there is still a nontrivial residual gauge symmetry
 \bea
\mbf{A}'(\mbf{r}) &= - \phi \mbf{b}'_1 (\mbf{r} \cdot \mbf{b}'_2) = \mbf{A}(\mbf{r}) + \pmb{\nabla} \Lambda 
\eea
where one may verify that $\Lambda = \frac{\phi}{2} \Big( (\mbf{r} \cdot \mbf{b}_1) (\mbf{r} \cdot \mbf{b}_2) - (\mbf{r} \cdot \mbf{b}'_1) (\mbf{r} \cdot \mbf{b}'_2) \Big) $ by direct computation. In the gauge $\mbf{A}'(\mbf{r})$, we find a new rationalization of the flux $\phi = \mu' \frac{2\pi p''}{q''}$ where 
\bea
\mu' &=  \text{lcd } \{ \mbf{b}'_1 \cdot (\pmb{\delta}_\be - \pmb{\delta}_\al) \} \text{ for all hoppings connecting $\mbf{R} + \pmb{\delta}_\al$ and $\mbf{R}' + \pmb{\delta}_\be$} \ .
\eea
This is once again an illustration that $\mu$ is gauge-dependent. It is tempting to think that with an appropriate $SL(2, \mathds{Z})$ gauge transformation, it is possible to find a basis where $\mu' = 1$. We prove a condition on the orbitals $\pmb{\delta}_\al$ demonstrating when it is possible to find such a basis, but we also show that in general it is impossible, i.e. that there is no choice of basis where $\mu = 1$ in our Landau gauge. 

$\mu$ is calculated from the distances between hoppings $\{\pmb{\delta}_\be - \pmb{\delta}_\al \}$. Because we assume $\pmb{\delta}_\be - \pmb{\delta}_\al$ is a rational sum of lattice vectors, we have $\pmb{\delta}_\be - \pmb{\delta}_\al = \omega_{1,\al\be} \mbf{a}_{1} + \omega_{2,\al\be} \mbf{a}_{2}$ for $\omega_{1, \al\be},\omega_{2, \al\be} \in \mathbb{Q}$. By \Eq{eq:mudef}, we have
\bea
\mu' &= \text{lcd } \{ \mbf{b}'_1 \cdot (\omega_{1,\al\be} \mbf{a}_{1} + \omega_{2,\al\be} \mbf{a}_{2}) \}  \text{ for all hoppings connecting $\mbf{R} + \pmb{\delta}_\al$ and $\mbf{R}' + \pmb{\delta}_\be$} \\
&= \text{lcd } \{ m_{22} \omega_{1,\al\be} - m_{21} \omega_{2,\al\be}  \}  \text{ for all hoppings connecting $\mbf{R} + \pmb{\delta}_\al$ and $\mbf{R}' + \pmb{\delta}_\be$} \ . \\
\eea
Let $\mu_i =  \text{lcd } \{ \omega_{i,\al\be} \}$  for all hoppings connecting $\mbf{R} + \pmb{\delta}_\al$ and $\mbf{R}' + \pmb{\delta}_\be$. If $\mu_1$ and $\mu_2$ are coprime, \textit{it is possible to set $\mu' =1$}. We show this by explicitly constructing $[m]$. Set $m_{22} = \mu_1$ and $m_{21} = - \mu_2$ in which case $m_{22} \omega_{1i} = \mu_{1} \omega_{1i} \in \mathds{Z}$ and $-m_{21} \omega_{2i} = \mu_{2} \omega_{2i} \in \mathds{Z}$ for all $i$. Hence $\mu' = 1$. Additionally, there exist $m_{11}, m_{12} \in \mathbb{Z}$ such that $\det [m]  = \mu_1 m_{11} + \mu_2  m_{12} = 1$. Indeed, $\mu_1$ and $\mu_2$ are coprime. We can thus choose $m_{12}$ to be the modular inverse of $\mu_2$, i.e.  $\mu_2  m_{12} = 1 \mod \mu_1$, simultaneously setting $m_{11} \in \mathbb{Z}$ such that $\mu_2 m_{12} = 1 - m_{11} \mu_1$. 

However, when $\mu_1$ and $\mu_2$ are not coprime, this proof fails. We now give an example of such a case where it is impossible to construct a basis where $\mu =1$. Let there be three orbitals $\pmb{\delta}_i = (0,0),(1/2,1/2),(1/2,1/4)$ in the $\mbf{a}_1, \mbf{a}_2$ basis such that $\pmb{\delta}_1$ is connected to $\pmb{\delta}_2$ and $\pmb{\delta}_3$ so that $\{ \omega_{1,\al \be}\} = \{1/2\}$ and $\{ \omega_{2,\al \be}\} = \{1/2, 1/4\}$. Then we find $\mu_1 = 2, \mu_2 = 4$ 
 which are not coprime. Assume for contradiction that coprime $m_{22}$ and $m_{21}$ exist such that $ \mu' =  \text{lcd } \{ \mbf{b}'_1 \cdot (\pmb{\delta}_1 - \pmb{\delta}_2),\mbf{b}'_1 \cdot (\pmb{\delta}_1 - \pmb{\delta}_3) \} $ is 1, recalling $\mbf{b}'_1 = m_{22} \mbf{b}_1 - m_{21} \mbf{b}_2$. (Note that if $m_{22}$ and $m_{21}$ are not coprime, then $\det [m] \neq 1$ and $\pmb{\nabla} \times \mbf{A}'(\mbf{r}) \neq \phi$.) This implies that each component of
 \bea
 \{ \mbf{b}'_1 \cdot (\pmb{\delta}_1 - \pmb{\delta}_2),\mbf{b}'_1 \cdot (\pmb{\delta}_1 - \pmb{\delta}_3) \} &= \{\frac{1}{2}(m_{22} - m_{21}) , \frac{1}{4}(2m_{22} - m_{21}) \} \\
 \eea
 is integer, so $m_{22} - m_{21} \in 2 \mathds{Z}$. We write $m_{22} = m_{21} + 2s, s \in \mathds{Z}$. Then
 \bea
 \frac{1}{4}(2m_{22} - m_{21})  = \frac{1}{4}(m_{21} + 4s)
 \eea
 is an integer only if $m_{21}$ is a multiple of 4. But $m_{22} - m_{21} \in 2 \mathds{Z}$ implies $m_{21}$ is then a multiple of $2$, so $m_{21}$ and $m_{22}$ are not coprime. 
 
In summary, we have shown that at $\phi = \frac{2\pi p}{q} = \mu \frac{2\pi p'}{q'}$ for $p',q'$ coprime, the Peierls phases are periodic in a $1\times q'$ magnetic unit cell \emph{in our Landau gauge}. $\mu$ is calculated from the positions of the orbitals and the choice of lattice vectors. We gave a simple criterion to determine if it is generically possible to find a basis where $\mu =1$, but we showed that there exist lattices where it is impossible to find such a basis. In the remaining sections, we develop our theory for a general $\mu$. 
 
 \subsection{Construction of the Hofstadter Hamiltonian}
\label{app:hofhamconstruct}

We have shown that our choice of Landau gauge preserves translation invariance along $\mbf{a}_1$, but at $\phi = \mu \frac{2\pi p'}{q'}$, there is only translational invariance along $\mbf{a}_2$ given by $r_2 \to r_2+q'$ \emph{even if $\mu$ and $q'$ are not coprime}. By choosing the Landau gauge, we are able to easily calculate the Peierls phases and form the Hofstadter Hamiltonian in a $1 \times q'$ magnetic unit cell. From the magnetic translation group, we know that it is possible to make a gauge transformation and diagonalize $H^\phi$ in the minimal $1 \times q$ unit cell at $\phi = \frac{2\pi p}{q}$, but it is difficult to find such a gauge explicitly. The Landau gauge is much more convenient for numerics, even at the cost of a larger magnetic unit cell. 

Redefining $r_1$ and $r_2$, we write an atomic position in the crystal in the form $r_1 \mbf{a}_1 + q' r_2 \mbf{a}_2 + \ell \mbf{a}_2 + \pmb{\delta}_\al$ where $\ell = 0, \dots, q'-1$ indexes the zero-field unit cells within the magnetic unit cell, and the orbitals are given by $\al = 1, \dots, N_{orb}$ with positions $\pmb{\delta}_\al$. Because the Peierls phases are periodic over the magnetic lattice, we have $\varphi_{q'r_2+\ell, \al, q' r_2' + \ell', \be}  = \varphi_{\ell \al, \ell' \be}$. 

The momentum space operators are defined as usual in the magnetic unit cell. In this notation, they are given by
\bea
\label{eq:eigenstateslandau}
c^\dag_{k_1, k_2, \ell, \al} &= \frac{1}{\sqrt{\mathcal{N}/q'}} \sum_{r_1 r_2} e^{-i \mbf{k} \cdot (r_1 \mbf{a}_1 + r_2q' \mbf{a}_2+ \ell \mbf{a}_2 + \pmb{\delta}_\al)} c^\dag_{r_1,r_2, \ell, \al} \\
\eea 
where the sum is over $r_1 = 1, \dots \mathcal{N}_1, r_2 = 1, \dots \mathcal{N}_2/q'$, and $\mathcal{N} = \mathcal{N}_1 \mathcal{N}_2$ is the number of unit cells in the lattice. We assume that $ \mathcal{N}_2 /q' \in \mathds{Z}$ for the periodic boundary conditions to have an integer number of $1\times q'$ magnetic unit cells. The Hamiltonian can be written in terms of these operators
\bea
\label{eq:magBZandhof}
H^{\phi = \mu \frac{2\pi p'}{q'}} &= \sum_{r_1 r_2,r_1'r_2'} \sum_{ \ell \al, \ell' \be} e^{i\varphi_{\ell \al, \ell' \be}} t_{\al \be}\big( (r_1-r_1') \mbf{a}_1 + (r_2-r_2') q' \mbf{a}_2 + (\ell - \ell') \mbf{a}_2 + (\pmb{\delta}_\al- \pmb{\delta}_\be) \big) c^\dag_{r_1, r_2, \ell, \al} c_{r_1',r_2', \ell', \be}  \\
&= \sum_{k_1,k_2} \sum_{\ell \al, \ell' \be} c^\dag_{k_1, k_2, \ell, \al} \, [\mathcal{H}(k_1,k_2)]_{\ell, \al, \ell', \be} \, c_{k_1,k_2, \ell', \be}, \qquad k_1 \in (-\pi, \pi), \ k_2 \in \lp 0, \frac{2\pi}{q'} \rp \\
\eea
where $\mathcal{H}$ is the Fourier transform over the \emph{magnetic} lattice:
\bea
\label{eq:hofmomentumspace}
\null [\mathcal{H}(k_1,k_2)]_{\ell, \al, \ell', \be} &= \frac{1}{\mathcal{N}/q'} \sum_{r_1, r_1', r_2, r_2' } e^{i\varphi_{\ell \al, \ell' \be} - i \mbf{k} \cdot ((r_1 - r_1') \mbf{a}_1 + (r_2 - r_2') q'\mbf{a}_2 + (\ell - \ell') \mbf{a}_2 + (\pmb{\delta}_\al- \pmb{\delta}_\be) )} \\
&\qquad \qquad \qquad \times  t_{\al \be}\big((r_1-r_1') \mbf{a}_1 + (r_2-r_2') q' \mbf{a}_2 + (\ell - \ell') \mbf{a}_2 + (\pmb{\delta}_\al- \pmb{\delta}_\be) \big) 
\eea 
and the magnetic BZ is defined as $k_1 = \frac{2\pi}{\mathcal{N}_x} (0, \dots, \mathcal{N}_x -1)$, $k_2 = \frac{2\pi}{\mathcal{N}_y} (0, \dots, \mathcal{N}_y /q' -1)$.  We note that for an infinite crystal, i.e. taking $\mathcal{N}_1, \mathcal{N}_2 \to \infty$, the magnetic BZ is defined to be $k_1 \in(-\pi,\pi) , k_2 \in(0, 2\pi /q')$. At a given momentum, the matrix \Eq{eq:hofmomentumspace} can be diagonalized numerically to determine the $q' n_{orb}$ band energies. We have separated the $\ell, \al$ indices because it is useful to think of them in a tensor product basis. 

We comment now on the apparent discontinuity that $\mathcal{H}^\phi(\mbf{k})$ suffers. Because commuting translation operators only exist when $\phi$ is rational, $\mathcal{H}^\phi(\mbf{k})$ only exists at $\phi = \mu \frac{2\pi p'}{q'}$. Additionally, the Fourier transform is only well-defined when the number of lattice sites (on periodic boundary conditions as we have assumed) is a multiple of $q'$. For two arbitrarily close values of $\phi$, their denominators may be arbitrarily different, and for any finite size lattice, $\mathcal{H}^\phi(\mbf{k})$ may not strictly exist. While $\mathcal{H}$ and $\mbf{k}$ have very discontinuous behavior in $\phi$, the spectrum of $H^\phi$ evolves smoothy. \Ref{PhysRevB.14.2239} proves that the spectrum is continuous for the simple Hamiltonian which it considers, but extending this proof to momentum space for a general Hamiltonian is beyond the scope of this work. Instead, we appeal to the position space representation of the Hofstadter Hamiltonian. It is clear that the spectrum evolves smoothly there because each term in $H^\phi$ is an analytic function of $\phi$.

\subsection{Embedding Matrices in Flux}
\label{sec:embedding}

\begin{figure}
 \centering
 \includegraphics[width=10.2cm]{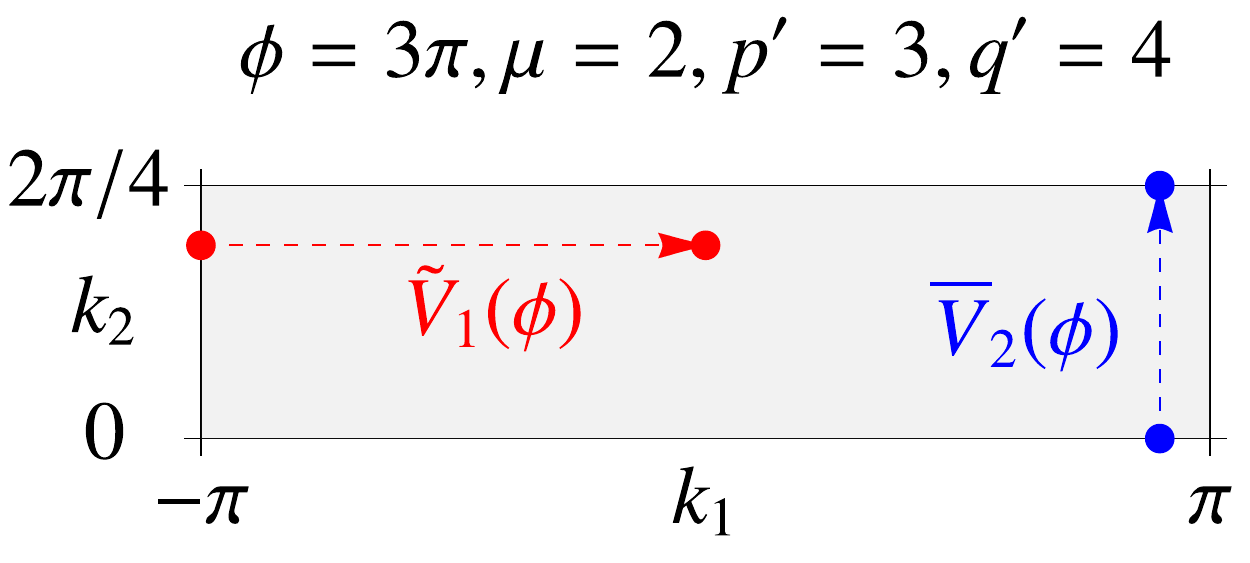} 
\caption{We show the magnetic BZ at $\phi = 2 \frac{2\pi \times 3}{4} = 3\pi$ with $\mu =2$ and $q' = 4$ not coprime, and $p'=3$. There is a $\pi$ periodicity along $k_1$ implemented by $\tilde{V}_1(\phi)$, but a $\pi/2 = 2\pi/q'$ periodicity along $k_2$ implemented by $\overline{V}_2(\phi)$ because the magnetic BZ is defined as $k_1 \in (-\pi,\pi), k_2 \in (0, \pi/2)$. The embedding matrices $\tilde{V}_2(\phi) = [\overline{V}_2(\phi)]^2$ or $V_2(\phi) = [\overline{V}_2(\phi)]^6$ implement periodicities that are larger than the domain of $k_2$.}
\label{fig:magBZ}
\end{figure}

Thus far, we have shown that at $\phi = \mu \frac{2\pi p'}{q'}$, the Hofstadter Hamiltonian can be diagonalized in a $1 \times q'$ magnetic unit cell, and the magnetic BZ may be taken as $k_1 \in (- \pi, \pi), k_2 \in (0, 2\pi/q')$. This establishes a $2\pi/q'$ periodicity in $k_2$. In this section, we will show that the energy spectrum and eigenstates of the Hofstadter Hamiltonian are also periodic in $k_1$ across the magnetic BZ with period $\phi = \mu \frac{2\pi p'}{q'}$, with $p',q'$ coprime. If $\mu$ and $q'$ are coprime, then iterating the $\phi$ periodicity will give a $\frac{2\pi}{q'}$ periodicity in $k_1$, matching the $\frac{2\pi}{q'}$ periodicity along $k_2$. If $\mu$ and $q'$ are \emph{not} coprime, then iterating the $\phi$ periodicity will only yield a $\frac{2\pi}{q} = \frac{2\pi}{q'/ \text{gcd}(q',\mu)}$ periodicity along $k_1$. In contrast, there will always be a $\frac{2\pi}{q'}$ periodicity along $k_2$ because the magnetic unit cell is $1\times q'$ at $\phi= \mu \frac{2\pi p'}{q'}$, so a $\frac{2\pi}{q'}$ periodicity appears due to the definition of the magnetic BZ. We emphasize that the states at $\mbf{k}$ and $\mbf{k} +  \frac{2\pi}{q} \mbf{b}_1$ are independent states in the Hilbert space although they have the same energy, but states at $\mbf{k}$ and $\mbf{k} + \frac{2\pi}{q'} \mbf{b}_2 $ are not independent states due to the definition of the magnetic BZ (see \Eq{eq:magBZandhof}). We sketch an example of the magnetic BZ when $\mu =2$, as is the case for a simple choice of basis for the lattice in \Fig{fig:1blatticeex}, and $\phi = 2 \frac{2\pi \times 3}{4} = 3\pi$ in \Fig{fig:magBZ}, where there is a $\pi$ periodicity along $k_1 \in (-\pi,\pi)$ and a $\frac{2\pi}{q'} = \pi/2$ periodicity along $k_2 \in (0, \frac{\pi}{2})$. 

In our Landau gauge $\mbf{A}(\mbf{r}) = - \phi \mbf{b}_1 (\mbf{r} \cdot \mbf{b}_2)$, we will find expressions for magnetic embedding matrices $V_i(\phi)$ which obey
\bea
\label{eq:embedding}
\mathcal{H}^\phi(\mbf{k} + \phi \mbf{b}_i ) &= V_i(\phi) \mathcal{H}^\phi(\mbf{k}) V^\dag_i(\phi)  \\
\eea
and implement the $\phi$ periodicity in the spectrum. This is the same $\phi$ periodicity that was proved in \Eq{eq:BZper} using the magnetic translation operators. The energy spectrum always has a $\phi$ periodicity along both axes of the magnetic BZ, no matter what magnetic unit cell is chosen to diagonalize the Hamiltonian.

We begin with the expression for $V_2(\phi)$ in our Landau gauge, noting that
\bea
\label{eq:V2calc}
c_{k_1, k_2+ \phi, \ell, \al} &=  \frac{1}{\sqrt{\mathcal{N}/q'}}  \sum_{r_1 r_2} e^{i (\mbf{k} + \mbf{b}_2 \phi) \cdot (r_1 \mbf{a}_1+ q' r_2 \mbf{a}_2 + \ell \mbf{a}_2 + \pmb{\delta}_\al)} c_{r_1,r_2,\ell, \al} \\
&=  \frac{1}{\sqrt{\mathcal{N}/q'}}  \sum_{r_1 r_2} e^{i \mbf{k} \cdot (r_1 \mbf{a}_1+ q' r_2 \mbf{a}_2+ \ell \mbf{a}_2 + \pmb{\delta}_\al)} e^{i \phi(q'r_2 +\ell + \mbf{b}_2 \cdot \pmb{\delta}_\al)} c_{r_1,r_2, \ell, \al}  \\
&=  \frac{1}{\sqrt{\mathcal{N}/q'}}  \sum_{r_1 r_2} e^{i \mbf{k} \cdot (r_1 \mbf{a}_1+ q' r_2 \mbf{a}_2+ \ell \mbf{a}_2 + \pmb{\delta}_\al)} e^{i \phi(\ell + \mbf{b}_2 \cdot \pmb{\delta}_\al)} c_{r_1,r_2, \ell, \al}  \\
&= \sum_{\ell' \be} [V_2(\phi)]_{\ell, \al, \ell', \be} \, c_{k_1, k_2, \ell', \be}   \\
\eea
where we have defined 
\bea
\label{eq:V2defintion}
\null [V_2(\phi)]_{\ell, \al, \ell', \be} = \delta_{\ell \ell'} e^{i \phi \ell} \delta_{\al \be} e^{i \phi \pmb{\delta}_\al \cdot \mbf{b}_2} \ . 
\eea
This allows us to deduce the action of $V_2$ on the Hamiltonian:
\bea
H^\phi &= \sum_{k_1,k_2} \sum_{\ell \al, \ell' \be} c^\dag_{k_1, k_2, \ell, \al} \, [\mathcal{H}(k_1,k_2)]_{\ell, \al, \ell', \be} \, c_{k_1,k_2, \ell', \be} \\
&= \sum_{k_1,k_2} \sum_{\ell \al, \ell' \be} c^\dag_{k_1, k_2+\phi, \ell, \al} \, [\mathcal{H}(k_1,k_2+\phi)]_{\ell, \al, \ell', \be} \, c_{k_1,k_2+\phi, \ell', \be} \\
&= \sum_{k_1,k_2} \sum_{\ell \al, \ell' \be} \sum_{\ell'',\ell''', \al',\be'} c^\dag_{k_1, k_2, \ell''', \al'} \, [V_2^\dag(\phi)]_{\ell''', \al',\ell, \al} [\mathcal{H}(k_1,k_2+\phi)]_{\ell, \al, \ell', \be} \, [V_2(\phi)]_{\ell', \be, \ell'', \be'} c_{k_1,k_2, \ell'', \be'} \ . \\
\eea
Matching terms, we see that \Eq{eq:embedding} is satisfied. We note that $V_2(\phi)$ is a diagonal matrix of phases, and so is unitary. 

 To derive $V_1(\phi)$ in our Landau gauge, we focus without loss of generality on a generic hopping of $H^\phi$ from $r_1 \mbf{a}_1 + (q'r_2+\ell) \mbf{a}_2 + \pmb{\delta}_\al$ to $r_1' \mbf{a}_1 + (q' r_2' + \ell') \mbf{a}_2 + \pmb{\delta}_\be$. Let $(\Delta_1, \Delta_2) = (r_1 - r_1', q' (r_2 -r_2') + \ell - \ell' )$ denote the distance between the unit cells of the orbitals in the hopping, and let $t$ denote the amplitude of the hopping. After Fourier transforming this hopping, we get a term in the Hamiltonian \Eq{eq:hofmomentumspace} given by
\bea
\label{eq:hdeltablochform}
[ \mathcal{H}_t^\phi(\mbf{k}) ]_{\ell, \al, \ell', \be} &= t e^{i \varphi_{\ell \al, \ell' \be} - i k_1\mbf{b}_1 \cdot ( \Delta_1 \mbf{a}_1 + \pmb{\delta}_\al -\pmb{\delta}_\be) - i k_2 \mbf{b}_2 \cdot ( \Delta_2 \mbf{a}_2 + \pmb{\delta}_\al -\pmb{\delta}_\be)  }, \\
\eea
and the whole Hamiltonian $\mathcal{H}^\phi(\mbf{k}) $ consists of a sum over the hoppings $t$ of $\mathcal{H}_t^\phi(\mbf{k})$. We recall that $\varphi_{\ell \al, \ell'\be}$ is the Peierls phase (see \Eq{eq:varphildep} and the following discussion). Using the expression for $\mathcal{H}_t^\phi(k_1, k_2)$ in  \Eq{eq:hdeltablochform}, we see that
\bea
\label{eq:hdeltablochform2}
\null [ \mathcal{H}_t^\phi(k_1 + \phi,k_2) ]_{\ell-1, \al, \ell'-1, \be} &=  t e^{i \varphi_{\ell-1, \al, \ell'-1, \be} - i (k_1+\phi) \mbf{b}_1 \cdot ( \Delta_1 \mbf{a}_1 + \pmb{\delta}_\al -\pmb{\delta}_\be) - i k_2 \mbf{b}_2 \cdot ( \Delta_2 \mbf{a}_2 + \pmb{\delta}_\al -\pmb{\delta}_\be)}  \\
\eea
where $\ell$ and $\ell'$ are defined mod $q'$. Now we calculate that 
\bea
\label{eq:V1hoppingfact}
\varphi_{\ell-1, \al, \ell'-1, \be} - \phi\mbf{b}_1 \cdot ( \Delta_1 \mbf{a}_1 + \pmb{\delta}_\al -\pmb{\delta}_\be) &= -\phi \int_{ r'_1 \mbf{a}_1+(\ell'-1 +q'r_2') \mbf{a}_2 + \pmb{\delta}_\be}^{r_1 \mbf{a}_1 + (\ell-1 + q'r_2) \mbf{a}_2 + \pmb{\delta}_\al} (\mbf{r} \cdot \mbf{b}_2)  \mbf{b}_1 \cdot d \mbf{r} - \phi\mbf{b}_1 \cdot ( \Delta_1 \mbf{a}_1 + \pmb{\delta}_\al -\pmb{\delta}_\be)\\
&= -\phi \int_{ r_1' \mbf{a}_1+(\ell' +q'r_2') \mbf{a}_2 + \pmb{\delta}_\be}^{r_1 \mbf{a}_1 + (\ell+ q'r_2) \mbf{a}_2 + \pmb{\delta}_\al} ( (\mbf{r} - \mbf{a}_2) \cdot \mbf{b}_2)  \mbf{b}_1 \cdot d \mbf{r} - \phi\mbf{b}_1 \cdot ( \Delta_1 \mbf{a}_1 + \pmb{\delta}_\al -\pmb{\delta}_\be)\\
&= \varphi_{\ell, \al, \ell', \be} + \phi \int_{ r'_1 \mbf{a}_1+(\ell' +q'r_2') \mbf{a}_2 + \pmb{\delta}_\be}^{r_1 \mbf{a}_1 + (\ell+ q'r_2) \mbf{a}_2 + \pmb{\delta}_\al} \mbf{b}_1 \cdot d \mbf{r} - \phi\mbf{b}_1 \cdot ( \Delta_1 \mbf{a}_1 + \pmb{\delta}_\al -\pmb{\delta}_\be)\\
&= \varphi_{\ell, \al, \ell', \be} + \phi \mbf{b}_1 \cdot  ( \Delta_1 \mbf{a}_1 + \pmb{\delta}_\al - \pmb{\delta}_\be) - \phi\mbf{b}_1 \cdot ( \Delta_1 \mbf{a}_1 + \pmb{\delta}_\al -\pmb{\delta}_\be) \\
&= \varphi_{\ell, \al,\ell', \be} \\
\eea
where we have used $\mbf{a}_2 \cdot \mbf{b}_2 = 1$. We observe that the cancelation of the $\Delta_1$-dependent terms is due to the structure of our Landau gauge. Comparing \Eqs{eq:hdeltablochform}{eq:hdeltablochform2}, we see that
\bea
\label{eq:V1fact}
\null [ \mathcal{H}_t^\phi(k_1 + \phi,k_2) ]_{\ell-1, \al, \ell'-1, \be} &= [ \mathcal{H}_t^\phi(k_1, k_2) ]_{\ell, \al, \ell', \be} \ . 
\eea
Hence, defining
\bea
\label{eq:V1ansatz}
\null [V_1(\phi)]_{\ell, \al, \ell', \be}  = \delta_{\ell, \ell'-1} \delta_{\al \be} , \\
\eea
emphasizing again that $\ell$ and $\ell'$ are defined mod $q'$ in the Kronecker delta. We find that $\mathcal{H}_t^\phi(k_1 + \phi, k_2 ) = V_1(\phi) \mathcal{H}_t^\phi(k_1, k_2) V^\dag_1(\phi)$. Note that $V_1(\phi)$ is unitary because it is the tensor product of a permutation matrix and the identity. . Thus we have proven \Eq{eq:embedding} holds for $H^\phi$ when focusing on a single generic hopping. It is simple to extend this result for $H^\phi$ having arbitrary hoppings. We observe that
\bea
\mathcal{H}^\phi(k_1, k_2 ) &= \sum_{\text{hoppings } t} \mathcal{H}_t^\phi(k_1, k_2) \\ 
\mathcal{H}^\phi(k_1+ \phi, k_2 ) &= \sum_{\text{hoppings } t} \mathcal{H}_t^\phi(k_1+ \phi, k_2) \\
&= \sum_{\text{hoppings } t}  V_1(\phi) \mathcal{H}_t^\phi(k_1, k_2) V_1^\dag(\phi) \\ 
&= V_1(\phi)  \lp \sum_{\text{hoppings } t}  \mathcal{H}_t^\phi(k_1, k_2) \rp V_1^\dag(\phi) \\ 
&= V_1(\phi) \mathcal{H}^\phi(k_1, k_2 ) V_1^\dag(\phi) \ . \\ 
\eea
By explicit computation, we obtain the braiding relation of the embedding matrices,
\bea
\label{eq:Vbraid}
V_1(\phi) V_2(\phi) &= e^{i \phi} V_2(\phi) V_1(\phi) \ .
\eea
According to \Eq{eq:embedding}, the embedding matrices $V_i(\phi)$ establish a $\phi$ periodicity along $k_1$ and $k_2$ in the energy spectrum. 

At zero flux, we also have an $N_{orb} \times N_{orb}$ embedding matrix corresponding to the $2\pi$ periodicity in the BZ. The embedding matrix $[\mathcal{V}_i]_{\al \be} = \delta_{\al \be} e^{2\pi i \pmb{\delta}_\al \cdot \mbf{b}_i}$ satisfies
\bea
\mathcal{H}^{\phi = 0}(\mbf{k} + 2\pi \mbf{b}_i ) &= \mathcal{V}_i \mathcal{H}^{\phi = 0}(\mbf{k}) \mathcal{V}_i^\dag \ . \\
\eea
We extend this embedding matrix to the nonzero flux case to recover the usual $2\pi$ periodicity in the magnetic BZ by defining its action on the Hofstadter Hamiltonian as the identity on the $\ell$ indices, i.e. $ [\mathcal{V}_i]_{\al \be} \to [\mathcal{V}_i]_{\ell \al, \ell' \be} = \delta_{\ell \ell'} \delta_{\al \be} e^{2\pi i \pmb{\delta}_\al \cdot \mbf{b}_i} $ such that
\bea
\mathcal{H}^{\phi}(\mbf{k} + 2\pi \mbf{b}_i ) &= \mathcal{V}_i \mathcal{H}^{\phi}(\mbf{k}) \mathcal{V}_i^\dag \ . \\
\eea
It is sometimes useful to combine the $2\pi$ periodicity given by $\mathcal{V}_i$ and the $\phi$ periodicity given by $V_i(\phi)$. For example at $\phi = \frac{4\pi}{3}$, the spectrum is actually $\frac{2\pi}{3}$-periodic, as we see from iterating the $\phi = \frac{4\pi}{3}$ periodicity twice, and then applying the $2\pi$ periodicity of $k_1$ given by $\mathcal{V}_i$. The embedding matrix $\mathcal{V}_i^\dag V_i(\phi)^2$ establishes this $\frac{2\pi}{3}$-periodicity. By combining the $\phi$ periodicity and the $2\pi$ periodicity, we will show that there is a $\frac{2\pi}{q}$ periodicity along $k_1$ at $\phi = \frac{2\pi p}{q} = \mu \frac{2\pi p'}{q'}$. Note that the periodicity along $k_1$ depends on $q$ and is gauge-independent. This $2\pi/q$ periodicity along $k_1$ is the same as the periodicity established by the magnetic translation group in \Eq{eq:BZper}.

To find expressions for the embedding matrices that implement the $\frac{2\pi}{q}$ periodicity, we define $\zeta, s \in\mathds{Z}$ by 
\bea
\label{eq:zeta}
\zeta \phi &= \frac{2\pi}{q} + 2\pi s \\
\eea
or equivalently $\zeta \phi = 2\pi /q \! \mod 2\pi$. To prove the $\frac{2\pi }{q}$ periodicity along $k_1$, we first note that
\bea
\label{eq:embed1}
\mathcal{H}^\phi(\mbf{k} + \zeta \phi \mbf{b}_i ) &= [V_1(\phi)]^\zeta \mathcal{H}^\phi(\mbf{k}) [V^\dag_1(\phi)]^\zeta \ . \\
\eea
Additionally, we see
\bea
\label{eq:embed2}
\mathcal{H}^\phi(\mbf{k} + \zeta \phi \mbf{b}_1 )  &=\mathcal{H}^\phi(\mbf{k} + \frac{2\pi}{q} \mbf{b}_1 + 2\pi s \mbf{b}_1)\\
&= [\mathcal{V}_1]^s \mathcal{H}^\phi(\mbf{k} + \frac{2\pi}{q} \mbf{b}_1) [\mathcal{V}_1^\dag]^s \ . \\
\eea
Combining \Eqs{eq:embed1}{eq:embed2}, we obtain
\bea
\label{eq:tildeVprop}
\mathcal{H}^\phi(\mbf{k} + \frac{2\pi}{q} \mbf{b}_1)  &= \tilde{V}_1(\phi) \mathcal{H}^\phi(\mbf{k}) \tilde{V}^\dag_1(\phi), \qquad \tilde{V}_1(\phi) =  [\mathcal{V}_i^\dag]^s [V_1(\phi)]^\zeta \\
\eea
which shows the spectrum is $2\pi/q$-periodic along $k_1$. We remark that $[\mathcal{V}_i, V_i(\phi)] = 0$. The algebra of $\tilde{V}_1$ and $\tilde{V}_2$ is readily obtained by repeated application of \Eq{eq:Vbraid} and \Eq{eq:zeta}. We find
\bea
\label{eq:Vtilde}
\tilde{V}_1(\phi) \tilde{V}_2(\phi) &= e^{i \zeta \frac{2\pi}{q}} \tilde{V}_2(\phi) \tilde{V}_1(\phi)  \ .
\eea
We have thus far shown that $V_1(\phi)$ and $V_2(\phi)$ give a $\phi$ periodicity along $k_1$ and $k_2$. Also we combined the $\phi$ and $2\pi$ periodicity along $k_1$ to show that $\tilde{V}_1(\phi)$ gives a minimal $\frac{2\pi}{q}$ periodicity along $k_1$ at $\phi = \frac{2\pi p}{q}$. Recall from \App{app:fluxrat} that $q'$ is an integer multiple of $q$, and $q=q'$ when $\mu$ and $q'$ are coprime. We emphasize that $k_1$ is periodic in $\frac{2\pi}{q}$. However, we will now show that because the magnetic BZ is given by $k_1 \in (-\pi, \pi), k_2 \in (0, 2\pi/q')$,  $k_2$ is periodic in $\frac{2\pi}{q'}$ which is a finer periodicity, and implies a larger $\frac{2\pi}{q}$ periodicity along $k_2$ when iterated. 
We now construct the embedding matrix $\overline{V}_2(\phi)$ that gives the $\frac{2\pi}{q'}$ periodicity along $k_2$. This periodicity arises because we diagonalized the Hamiltonian in a $1 \times q'$ magnetic unit cell and is gauge dependent. $\overline{V}_2(\phi)$ satisfies 
\bea
\mathcal{H}^{\phi }(\mbf{k} + \frac{2\pi}{q'} \mbf{b}_2 ) = [\overline{V}_2(\phi)] \mathcal{H}^{\phi}(\mbf{k}) [\overline{V}_2(\phi)]^\dag \ . 
\eea
We follow the steps of \Eq{eq:V2calc} and find
\bea
\label{eq:V2primedef}
c_{k_1, k_2+ \frac{2\pi}{q'}, \ell, \al} &= \sum_{r_1 r_2} e^{i (\mbf{k} + \mbf{b}_2 \frac{2\pi}{q'}) \cdot (r_1 \mbf{a}_1 + q' r_2 \mbf{a}_2 + \ell \mbf{a}_2 + \pmb{\delta}_\al)} c_{r_1,r_2,\ell, \al} \\
&= \sum_{r_1 r_2} e^{i \mbf{k} \cdot (r_1 \mbf{a}_1 + q' r_2 \mbf{a}_2 + \ell \mbf{a}_2 + \pmb{\delta}_\al)} e^{i \frac{2\pi}{q'}(q'r_2 +\ell + \mbf{b}_2 \cdot \pmb{\delta}_\al)} c_{r_1,r_2, \ell, \al}  \\
&= [\overline{V}_2(\phi)]_{\ell, \al, \ell', \be} \, c_{k_1, k_2, \ell', \be} , \\
\eea
with 
\bea 
\label{eq:overlineV2}
\null [\overline{V}_2(\phi)]_{\ell, \al, \ell', \be} = \delta_{\ell \ell'} e^{i \frac{2\pi}{q'} \ell} \delta_{\al \be} e^{ i \frac{2\pi}{q'} \pmb{\delta}_\al \cdot \mbf{b}_2} \ .
\eea   
We observe from \Eqs{eq:overlineV2}{eq:V2defintion} that $\overline{V}_2(\phi)^{\mu p'} = V_2(\phi)$. The algebra with with $V_1(\phi)$ can be directly computed and reads
\bea
\label{eq:V1barV2}
V_1(\phi) \overline{V}_2(\phi) = e^{i \frac{2\pi}{q'}} \overline{V}_2(\phi) V_1(\phi) \ . \\
\eea 
In \App{app:wilson}, we will need the algebra of $\overline{V}_2(\phi)$ with $\tilde{V}_1(\phi)$ to study the Wilson loop in the magnetic BZ. Using \Eq{eq:V1barV2} and the definition of $\tilde{V}_1(\phi)$ in \Eq{eq:tildeVprop}, we arrive at
\bea
\label{eq:wilsonloopidentity}
 \tilde{V}_1(\phi)  \overline{V}_2(\phi)&=  e^{i \frac{2\pi}{q'} \zeta}\overline{V}_2(\phi) \tilde{V}_1(\phi)  \ .
\eea
We depict the magnetic BZ and embedding matrices for $\phi = 2\frac{2\pi \times 3}{4} = 3\pi$, with $\mu =2$ and $p=3,q' = 4$ in \Fig{fig:magBZ}. Note that $\mu$ and $q'$ are not coprime, so the periodicities along $k_1$ (which is $\pi$) and $k_2$ (which is  $\pi/2$) are not equal. 

We now give a brief example illustrating the different embedding matrices. In \App{app:fluxrat}, we discussed a model on the square lattice with 1a and 1b atoms (see \Fig{fig:1blatticeex}) where $\mu = 2$ for $\mbf{a}_1 = (1,0), \mbf{a}_2 = (0,1)$. In our Landau gauge defined by $\mbf{A}(\mbf{r}) = - \phi \mbf{b}_1(\mbf{r} \cdot \mbf{b}_2) = \phi(-y,0)$, the Hofstadter Hamiltonian could be formed by taking $\phi = 2 \frac{2\pi p'}{q'}$ for $p'$ and $q'$ coprime, and defining the magnetic BZ as $k_1 \in (-\pi, \pi), k_2 \in (0, 2\pi /q')$. The $\tilde{V}_1(\phi)$ embedding matrix provides a $2\pi/q$ periodicity in the energy spectrum along $k_1$ and $k_2$, where $\phi = \frac{2 \pi p}{q}$, $p$ and $q$ coprime. Due to $\overline{V}_2(\phi)$, there is always a $2\pi/q'$ periodicity along $k_2$. In this example, when $q'$ and $\mu = 2$ are not coprime (i.e. $q'$ is even), we have $\frac{2\pi}{q'} = \frac{2\pi}{2q} $, and so $k_2$ has twice a fine a periodicity as $k_1$. When $q'$ and $\mu = 2$ are coprime, both $k_1$ and $k_2$ have the same periodicity.

\section{Bounded Gaps in the Trivial Limit}
\label{App:atomiclimit}

In this Appendix, we prove that a Hamiltonian with a large enough on-site, per orbital, potential has a gapped Hofstadter Butterfly spectrum for any flux and hence is adiabatically connected to the atomic limit for all flux (\App{App:gershgorin}). Such a property was mentioned in the derivation of the bulk gap closing of an insulator with nonzero Chern number in Sec. II of the main text. We exemplify this feature on the checkerboard lattice model in \App{app:checkerboardham}.

\subsection{Gershgorin circles}
\label{App:gershgorin}

\begin{figure}
 \centering
 \includegraphics[width=5.5cm]{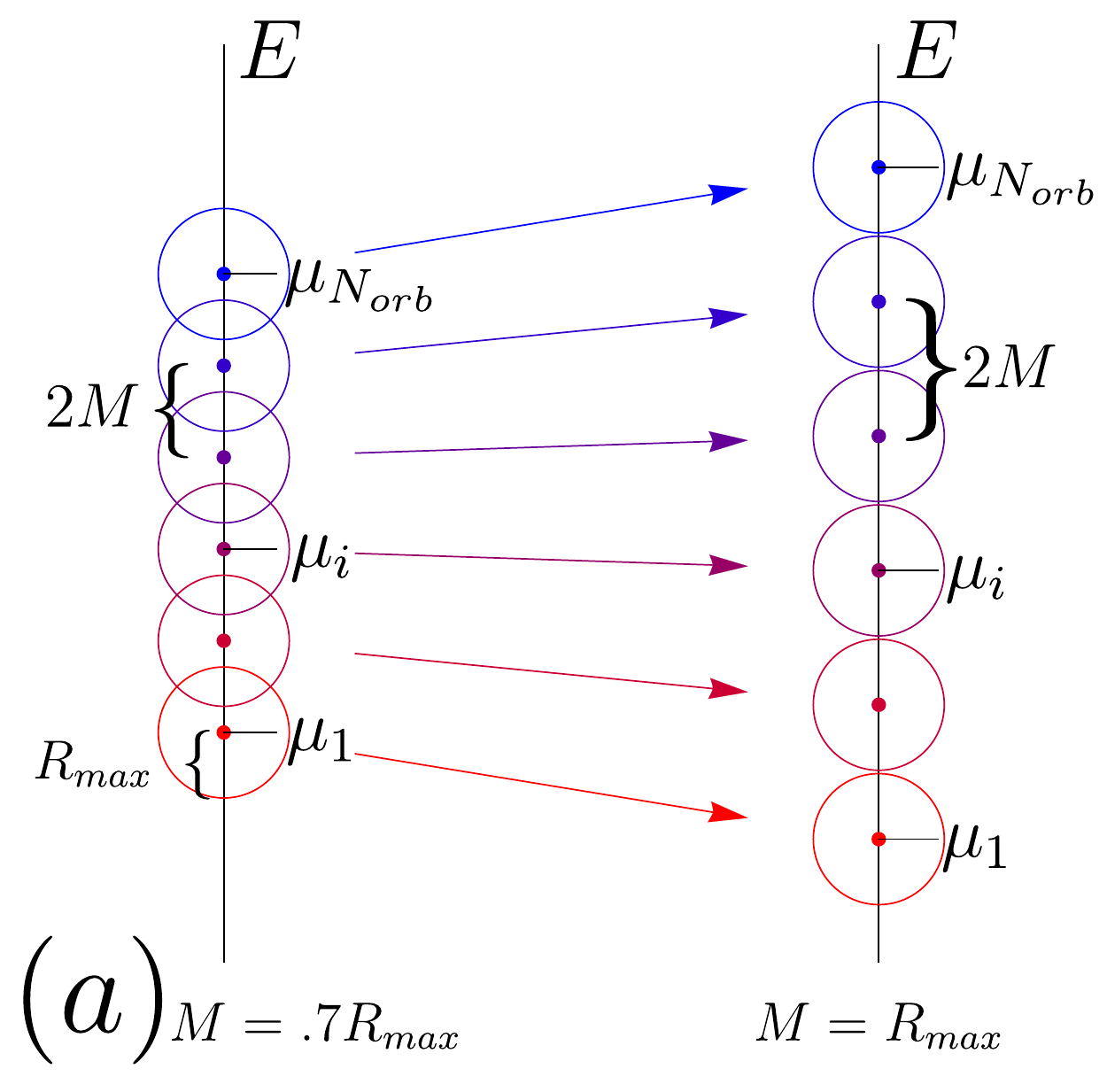} 
 \includegraphics[width=5.5cm]{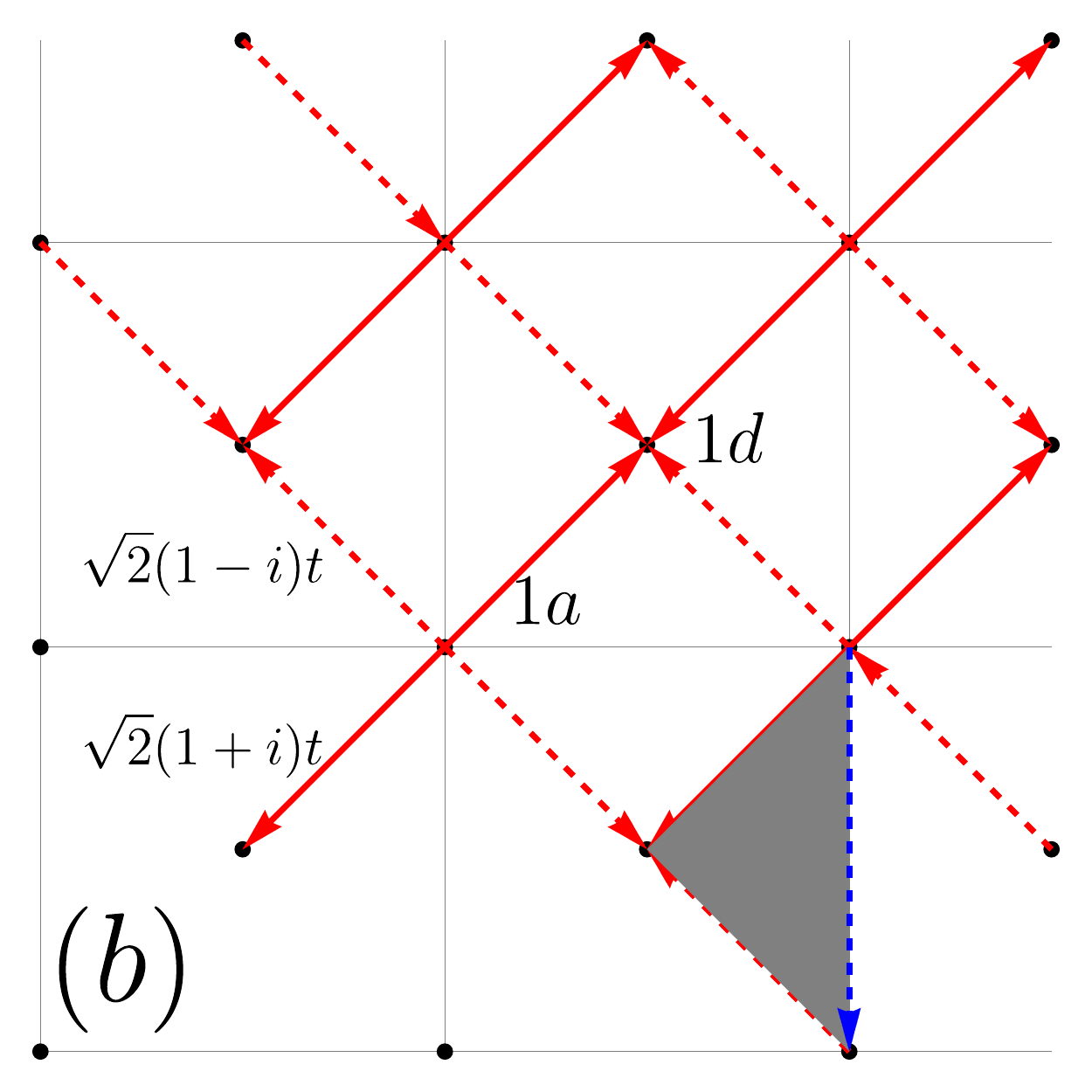}  
  \includegraphics[width=5.5cm]{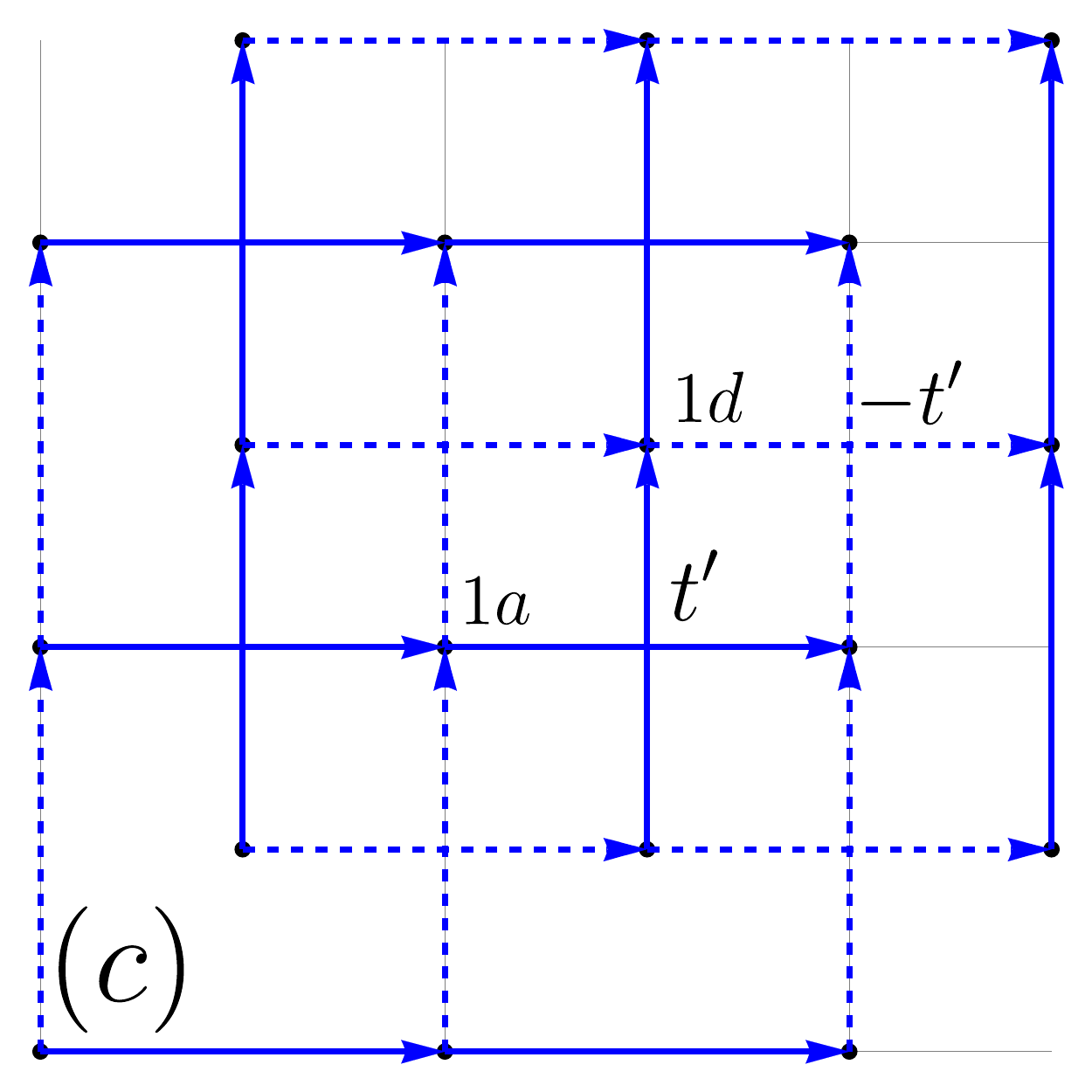}   \\
\caption{(a) We illustrate the Gershgorin circles $\mathcal{D}_\al$ with centers $\mu_\al$ and radii $R_{\al} \leq R_{\max}$. For clarity, we draw all the circles with radius $R_{mas}$. The onsite potential (proportional to $M$) sets the distance between the centers $\mu_\al$ to be $2M$. By increasing $M$ to $M > R_{max}$, we can make all Gershgorin circles disjoint. We depict the two hoppings of the checkerboard model (\Eq{eq:checkerboardham}) in (b), which shows nearest-neighbor hoppings from the $1a$ position to the $1d$ position, and in (c), which shows next nearest-neighbor hoppings from $1a$ to $1a$ and $1d$ to $1d$. In (b), we also provide an example of the minimal area ($1/4$ of the unit cell) enclosed by Peierls paths as a grey triangle. }
\label{fig:gershcheck}
\end{figure}

We first consider the Hamiltonian at zero field. We add to the lattice model described by the hopping amplitudes $t_{\al \be}(\mbf{r}-\mbf{r}')$ an on-site, per-orbital potential
\bea
\mu_\al&=(2\al - N_{orb} - 1) M \label{eqn:GershgorinCenter}
\eea
where $\al = 1, \dots, N_{orb}$ labels the orbitals and $M$ is the on-site amplitude. Let $H$ be the matrix representation of the tight-binding hamiltonian in real space including both $\mu_\al$ and the hopping amplitudes. $H$ is a $(N_{orb} \mathcal{N}) \times (N_{orb} \mathcal{N})$ hermitian matrix where $\mathcal{N}$ is the number of unit cells. We want to focus on the eigenvalues of $H$. Thanks to the translation invariance, we can define $N_{orb}$ Gershgorin circles ${\cal D}_\al$ of center $\mu_\al$ and radius $R_\al$ with
\bea
R_\al&=&\sum_{\be \neq \al}\sum_{\mbf{r}} \left| t_{\al \be}(\mbf{r}) \right| .\label{eqn:GershgorinRadius}
\eea
We assume that these series appearing in \Eq{eqn:GershgorinRadius} do converge, which is true for any system with only finite range hopping terms or exponentially decaying hoppings $|t_{\al \be}(\mbf{r})| \sim \exp (- \kappa |\mbf{r}|), \kappa > 0$, as may arise in the case of flat Chern bands \cite{2013CRPhy..14..816P,Brouder_2007}. From there we define $R_{max}={\rm max }\left\{R_\al; \al = 1, \dots, N_{orb}\right\}$. The Gershgorin circle theorem states that any eigenvalue of $H$ lies within at least one of the Gershgorin circles, and a union of $m$ circles disjoint from all other circles must contain exactly $m$ eigenvalues of $H$. For $|M|$ large, i.e., in the atomic limit, we have exactly $\mathcal{N}$ eigenvalues lying in each ${\cal D}_\al$ circle, forming $N_{orb}$ separated trivial bands. As long as $|M|>R_{max}$, we are guaranteed that all the ${\cal D}_\al$ circles are disjoint and thus all bands are separated by a gap from above and below, as we sketch in \Fig{fig:gershcheck}a.

Turning on the magnetic field, the hopping terms acquire a phase $t_{\al \be}(\mbf{r}-\mbf{r}') \to t_{\al \be}(\mbf{r}-\mbf{r}') \exp \left[ i \int^{\mbf{R} + \pmb{\delta}_\al}_{\mbf{R}'+ \pmb{\delta}_\be} \mbf{A} \cdot d\mbf{r} \right]$ from the Peierls substitution as discussed in \App{app:Peierlspaths}. Note that $\mu_\al$ and $R_\al$ are unchanged under this substitution and so are the ${\cal D}_\al$ circles. This implies that for $|M|>R_{max}$ and irrespective of the magnetic field, the Gershgorin circles remain disjoint and each band is gapped. As a consequence, if we start with $N_{occ}$ occupied bands at $\phi=0$, these bands will stay gapped for any $\phi \neq 0$ if $|M|>R_{max}$. This is valid subject to the convergence of the sum in \Eq{eqn:GershgorinRadius}. This is not in contradiction to Sec. II where we proved that a model with nonzero Chern number necessarily has a gap closing: for large enough $|M|$, i.e. $|M| > R_{max}$, the model cannot host nontrivial bands because it is adiabatically connected to an atomic limit. When $M < R_{max}$, a gap closing is not forbidden in the Hofstadter spectrum. 

\subsection{Hofstadter Hamiltonian on the Checkerboard Lattice}
\label{app:checkerboardham}

We consider a checkerboard lattice with $s$ orbitals at the $1a = (0,0)$ and $1d = (1/2,1/2)$ positions \cite{PhysRevLett.106.236803,PhysRevB.85.075128}. In the basis $c^\dag_{\mbf{R}} = (c^\dag_{\mbf{R},1a}, c^\dag_{\mbf{R},1d})$, the Hamiltonian is given by
\bea
\label{eq:checkerboardham}
H^{\phi = 0} &= \sum_{\mbf{R}} M c^\dag_{\mbf{R}} \sigma_z c_{\mbf{R}} + t \sum_{\mbf{R}} \sum_j c^\dag_{\mbf{R} + \mbf{d}_{j}} (\sigma_1 \cos \theta + (-1)^j \sigma_2 \sin \theta) c_{\mbf{R}} +  t' \sum_{\mbf{R}} c^\dag_{\mbf{R} + \hat{x}} \sigma_z c_{\mbf{R}} - c^\dag_{\mbf{R} + \hat{y}} \sigma_z c_{\mbf{R}} + h.c. \\
\eea
where $\mbf{d}_j$ are defined by $\mbf{d}_0 = 0, \mbf{d}_1 = - \hat{x}, \mbf{d}_2 = - \hat{x} - \hat{y}, \mbf{d}_3 = - \hat{y}$. The hoppings $t \, (t')$ represent nearest (next-nearest) neighbor hoppings as shown in \Fig{fig:gershcheck}b and \Fig{fig:gershcheck}c. Note that the parameter $M$ matches the defintion of the on-site potential $\mu_\al$ Eq.~\ref{eqn:GershgorinCenter}. We use the parameters $t = 1, t' = 1/2, \theta = \pi/4$. At $\phi = 0$,  the model is topological with $C=1$ for $|M| < 2$, and for $|M| > 2$, the model is in a trivial atomic limit.  

We introduce magnetic flux through the lattice in our Landau gauge $\mbf{A} = - \phi( y, 0)$ and take each Peierls integral along the straight-line paths between the endpoints. This is the conventional choice for $s$ orbitals. Our choice for the Peierls paths allows minimal loops that enclose $1/4$ of the unit cell, as depicted for instance in \Fig{fig:gershcheck}b as a grey triangle, so the periodicity of the Hofstadter spectrum is $\Phi = 4 \times 2\pi$. The Hofstadter Hamiltonian is given in position space by
\bea
H^{\phi} &= \sum_{\mbf{R}} M c^\dag_{\mbf{R}} \sigma_z c_{\mbf{R}} + \sum_{\mbf{R}} \sum_j c^\dag_{\mbf{R} + \mbf{d}_{j}} \left[ \bpm 0 & t_j(\phi)^* \\
t_j(\phi) & 0 \\ \epm \cos \theta + (-1)^j \bpm 0 & -i t_j(\phi)^* \\
i t_j(\phi) & 0 \\ \epm\sin \theta \right] c_{\mbf{R}} \\
& \qquad +  \sum_{\mbf{R}} c^\dag_{\mbf{R} + \hat{x}} \bpm t'_{1,a}(\phi) & 0 \\ 0 & - t'_{1,d}(\phi) \epm c_{\mbf{R}} - t' c^\dag_{\mbf{R} + \hat{y}} \sigma_z c_{\mbf{R}} + h.c. \\
\eea
where the Peierls phases are given by
\bea
\label{eq:checkhamphi}
t_j(\phi) &= t e^{ i \int_{\mbf{R}}^{\mbf{R} + \mbf{d}_j} \mbf{A} \cdot d\mbf{r}} = \{t e^{-i \frac{\phi}{8} - i \phi \frac{y}{2}},  t e^{i \frac{\phi}{8} + i \phi \frac{y}{2}}, t e^{-i \frac{\phi}{8} + i \phi \frac{y}{2}}, t e^{i \frac{\phi}{8} - i \phi \frac{y}{2}} \}_j , \\
t'_{1,a}(\phi) &= t' e^{ i \int_{\mbf{R}}^{\mbf{R} + \mbf{x}} \mbf{A} \cdot d\mbf{r}} = t' e^{- i \phi y} , \quad t'_{1,d}(\phi) = t' e^{ i \int_{\mbf{R}+\frac{1}{2}(\hat{x}+\hat{y})}^{\mbf{R} + \frac{1}{2}(3\hat{x}+\hat{y})} \mbf{A} \cdot d\mbf{r}} = t' e^{- i \phi (y+1/2)} , \\
t'_{2,a}(\phi) &= t' e^{ i \int_{\mbf{R}}^{\mbf{R} + \mbf{y}} \mbf{A} \cdot d\mbf{r}} = t', \quad t'_{2,d}(\phi) = t' e^{ i \int_{\mbf{R}}^{\mbf{R} + \mbf{y}} \mbf{A} \cdot d\mbf{r}} = t'  \\
\eea
Using \Eq{eq:mudef}, we calculate $\mu =  \text{lcd } \{ \mbf{b}_1 \cdot (\pmb{\delta}_\be - \pmb{\delta}_\al) \} = \text{lcd } \{1, 1/2\} = 2$ because there are orbitals at $1a = (0,0)$ and $1a = (1/2,1/2)$ connected by a hopping.  Indeed, from the Peierls phases themselves, we see that setting $\phi = 2 \frac{2\pi p'}{q'}$ gives a $1 \times q'$ unit cell, meaning the phases \Eq{eq:checkhamphi} are invariant under $y \to y + q'$.

We define the Hermitian matrices
\bea
\Sigma_j(k_x, k_y) &= \left[ \bpm 0 & t_j(\phi)^* e^{i \mbf{k} \cdot (\pmb{\delta} + \mbf{d}_j) } \\
t_j(\phi)  e^{-i \mbf{k} \cdot (\pmb{\delta} + \mbf{d}_j) }  & 0 \\ \epm \cos \theta + (-1)^j \bpm 0 & - i t_j(\phi)^* e^{i \mbf{k} \cdot (\pmb{\delta} + \mbf{d}_j) }  \\
i t_j(\phi) e^{-i \mbf{k} \cdot (\pmb{\delta} + \mbf{d}_j) }  & 0 \\ \epm\sin \theta \right] \\
&=  \bpm 0 & e^{-i (-1)^j \theta} t_j(\phi)^* e^{i \mbf{k} \cdot (\pmb{\delta} + \mbf{d}_j) } \\
e^{i (-1)^j \theta} t_j(\phi)  e^{-i \mbf{k} \cdot (\pmb{\delta} + \mbf{d}_j) }  & 0 \\ \epm 
\eea
where the location of the $1d$ Wyckoff position is $\pmb{\delta} = (1/2, 1/2)$. We Fourier transform over the magnetic unit cell at $\phi = 2 \frac{2\pi p'}{q'}$ to find 
\bea
\label{eq:hofhamch}
\mathcal{H}^{\phi}_{y,y'} &= \delta_{y, y'} h_y + \delta_{y, y'+1} T_y + \delta_{y+1,y'} T_{y'}^\dag \\
h_y &= M \sigma_z + \Sigma_0(k_x,k_y) + \Sigma_1(k_x, k_y) + 2 t' \bpm \cos(k_x + \phi y)& 0 \\ 0 & -  \cos(k_x + \phi y + \phi/2) \epm  \\
T_{y+1} &= \bpm 0 & e^{-i \theta} t_2(\phi)^*  e^{i \mbf{k} \cdot (\pmb{\delta} + \mbf{d}_2) } + e^{i \theta}  t_3(\phi)^*  e^{i \mbf{k} \cdot (\pmb{\delta} + \mbf{d}_3) }  \\
0 & 0 \\ \epm  - t' \sigma_z e^{- i k_y} \\
\eea
where $y = 0, \dots, q'-1$ at $\phi = 2 \frac{2\pi p'}{q'}$.

\begin{figure*}[h]
 \centering
\begin{overpic}[width=0.6\textwidth,tics=10]{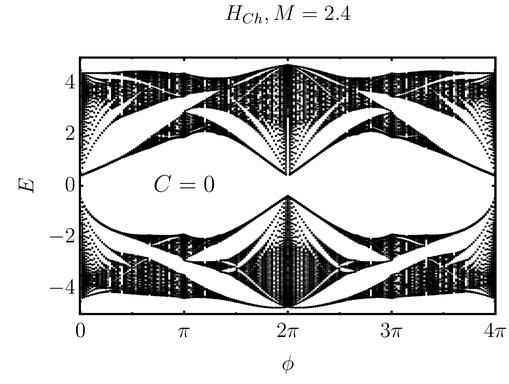}
\end{overpic} 
\caption{We show the the Hofstadter Butterflies for $H_{ch}$ (\Eq{eq:checkerboardham}) at $M = 2.4$ in the trivial phase. Because $H_{ch}^{\phi = 0}$ is trivial, the Hofstadter spectrum may be gapped everywhere. }
\label{fig:Hchgapped}
\end{figure*}

We numerically diagonalize this Hofstadter Hamiltonian \Eq{eq:hofhamch} to compute the Hofstadter Butterfly at $M = 2.4$  in \Fig{fig:Hchgapped}. The two Gershgorin circles for the checkerboard lattice model are defined by the two centers $\mu_{1a}=|M|$ and $\mu_{1d}=-|M|$, and the Gershgorin radius $R_{max}=R_{1a}=R_{1d}=4\left(|t|+|t'|\right)$ (each site being connected to four nearest neighbors and four next nearest neighbors). For the parameters used in \Fig{fig:Hchgapped}, we get $R_{max}=6$. For any $|M|>6$, we are guaranteed that the system at half filling will have a gap irrespective of $\phi$. Note that this value is \emph{only} an upper bound of the minimal $|M|$ value where this gap can appear. As already mentioned previously, the transition from a $C=1$ topological band to a trivial band occurs at $|M| = 2$, and for $M>2$ the Hofstadter Butterfly is gapped everywhere. We show an example of the Hofstadter Butterfly in this phase in \Fig{fig:Hchgapped}. It is hence expected that further analysis can produce tighter bounds. 

\section{Wilson Loops}

In this Appendix, we define the Wilson loop at nonzero flux and determine the constraints on its eigenvalues, the Wannier centers of the magnetic unit cell, due to the magnetic translation group (\App{app:wilson}). In \App{app:chernwilson}, we provide an alternative proof of the major claim in Sec. II, that $C^{\phi = \frac{2\pi p}{q}} \in q \mathds{Z}$, which leads to a proof of gap closing in the bulk of the Hofstadter Butterfly when the $\phi=0$ model has a nonzero Chern number. We also discuss in depth the behavior of the occupied bands as the flux is pumped. In \App{app:mirrorchernwilson}, we show that a Mirror Chern number also enforces a gap closing. We provide an alternative proof of the gap closing for a nonzero Chern number using only inversion eigenvalues (\App{app:invchern}). 

\subsection{Definition of the Wilson Loop at Nonzero Flux}
\label{app:wilson}

The Wilson loop at $\phi = 0$ is calculated from the eigenvectors of $\mathcal{H}^0(\mbf{k})$. Denote the $j$th eigenvector as $\ket{u_j(\mbf{k})}$, which is a vector with $N_{orb}$ components corresponding to the $j$th eigenvalue $\eps_j(\mbf{k})$. (Note that this is a slight abuse of the ket notation, which usually specifies a state in the Hilbert space, as opposed to merely a vector in $\mathds{R}^{N_{orb}}$.) We denote the $i$th element of the $j$th (ordered by energy) eigenvector by $\ket{u_j(\mbf{k})}_i, i = 1, \dots, N_{orb}$. When the bulk spectrum is gapped, we may consider only the $N_{occ} < N_{orb}$ occupied bands which determine the topology of the Hamiltonian at a given filling $\nu = N_{occ} / N_{orb}$. For brevity, we write $[U_{\mbf{k}}]_{ij} = \ket{u_j(\mbf{k})}_i$ as the matrix of occupied eigenvectors. (Here we assume the bulk spectrum is gapped so the occupied spectrum is well-defined.) Then the Wilson loop is defined by
\bea
W^{\phi=0}(k_1) &= U^\dag_{(k_1, 2\pi)} \left[ \prod_{k_2}^{2\pi \leftarrow 0} \mathcal{P}_{(k_1, k_2)} \right] \ U_{(k_1,0)}, \quad \mathcal{P}_{(k_1, k_2)} = U_{(k_1, k_2)} U^\dag_{(k_1, k_2)}
\eea
where the ordered product is understood to be discretized: $\prod_{k_2}^{2\pi \leftarrow 0}  \mathcal{P}_{(k_1, k_2)} \equiv \lim_{\eps \to 0} \mathcal{P}_{(k_1,2\pi)} \mathcal{P}_{(k_1,2\pi - \eps)}  \dots \mathcal{P}_{(k_1,0)} $. It is convenient to fix the $U(1)$ gauge symmetry of the eigenvalues by requiring $U_{(k_1, 2\pi)}  = \mathcal{V}_2 U_{(k_1, 0)}$.

When we add flux to the model in our Landau gauge, the magnetic BZ can be chosen to be $k_1 \in (-\pi,\pi), k_2 \in (0, \frac{2\pi}{q'})$ as discussed in \App{app:hofhamconstruct}. To define the Wilson loop along the $k_2$ direction of the BZ at $\phi = \frac{2\pi p}{q}$, we assume there is a set of $N_{occ}^\phi$ occupied bands which are gapped from the higher energy bands whose eigenvectors form the matrix $[U^\phi_{\mbf{k}}]_{ij} = \ket{u^\phi_j(\mbf{k})}_i$. Then we may calculate the Wilson loop according to
\bea
W_{(k_1,\frac{2\pi}{q'}) \leftarrow (k_1, 0)} \equiv W^{\phi=\frac{2\pi p}{q}}(k_1)  &= [U^\phi_{(k_1, \frac{2\pi}{q'})}]^\dag \left[ \prod_{k_2}^{\frac{2\pi}{q'} \leftarrow 0} \mathcal{P}^\phi_{(k_1, k_2)} \right] \ U^\phi_{(k_1,0)} \ . \\ 
\eea
We have chosen to take the Wilson loop along the $k_2$ direction so the Wilson loop eigenvalues correspond to positions along the extended $\mbf{a}_2$ direction of the magnetic unit cell \cite{Alexandradinata:2012sp}.  We will usually be interested in the Wilson loop at the fixed filling $\nu = N_{occ}/N_{orb}$ of the zero-field Hamiltonian. In this case at filling $\nu$, $N_{occ}^{\phi = \mu \frac{2 \pi p'}{q'}} = q' N_{occ}$ because there are $q' N_{orb}$ bands in $H^\phi$ due to the magnetic unit cell being $1 \times q'$. (Of course, it is possible for gaps to exist at other fillings in the Hofstadter Hamiltonian at a given flux.) As in the zero-field case, it is convenient to fix the gauge by requiring 
\bea
\label{eq:UVperiodicity}
U^\phi_{(k_1, \frac{2\pi}{q'})}  = \overline{V}_2(\phi) U^\phi_{(k_1, 0)}
\eea
 with the embedding matrix $\overline{V}_2(\phi)$ providing the $2\pi/q'$ periodicity in $k_2$ (see \Eq{eq:V2primedef}). 

We study the effect of the magnetic translation group on the Wilson loop following the presentation of \Ref{2017PhRvB..96x5115B} (App. VII D) which develops constraints on the Wilson loop for tight-binding models with the general symmetry
\bea
g_\mbf{k} \mathcal{H}(\mbf{k})g^\dag_{\mbf{k}} &= \mathcal{H}(D_g \mbf{k})
\eea
where $D_g$ is an operator acting on the momentum vector. In the Hofstadter Hamiltonian, 
the magnetic translation group requires $\tilde{V}_1(\phi) \mathcal{H}^\phi(\mbf{k}) \tilde{V}^\dag_1(\phi) = \mathcal{H}^\phi(\mbf{k} + \frac{2\pi}{q} \mbf{b}_i)$ where $\phi = \frac{2\pi p}{q}, \ p,q$ coprime (see \Eq{eq:tildeVprop}). This allows us to form a sewing matrix between the occupied bands at $\mbf{k}$ and $\mbf{k} + \frac{2\pi}{q} \mbf{b}_1$ defined by
\bea
\label{eq:sewing}
B^{ij}_\mbf{k} &= \braket{u^\phi_i(\mbf{k} + \frac{2\pi}{q} \mbf{b}_1) | \tilde{V}_1(\phi) | u^\phi_j(\mbf{k})}, \quad i,j = 1, \dots, q' N_{occ} \ . \\
\eea
This matrix is unitary as long as there is a gap above the occupied bands \cite{Alexandradinata:2012sp} (if there is a gap at $k_1$, there is also a gap at $k_1 + \frac{2\pi}{q}$). \Ref{2017PhRvB..96x5115B} demonstrates that the sewing matrix also obeys
\bea
\ket{u^\phi_j(\mbf{k})} &= \sum_i \tilde{V}_1^\dag(\phi) \ket{u^\phi_i(\mbf{k} + \frac{2\pi}{q} \mbf{b}_1) } B^{ij}_\mbf{k} \ . \\
\eea
This prepares us to calculate a small segment of a Wilson line. We find
\bea
\null [W^\phi_{\mbf{k}_1' \leftarrow \mbf{k}_1}]^{ij} &= \braket{u^\phi_i(\mbf{k}_1') | u^\phi_j(\mbf{k}_1)} \\
&= \sum_{rs} B^{\dag \, ir}_{\mbf{k_2}} \braket{ u^\phi_r(\mbf{k}_1' + \frac{2\pi}{q} \mbf{b}_1) | V_1(\phi) V_1(\phi)^\dag |u^\phi_s(\mbf{k}_1 + \frac{2\pi}{q} \mbf{b}_1) B^{sj}_{\mbf{k}_1} }\\
&= \sum_{rs} B^{\dag \, ir}_{\mbf{k_2}}  [W^\phi_{\mbf{k}_1' + \frac{2\pi}{q} \mbf{b}_1 \leftarrow \mbf{k}_1 + \frac{2\pi}{q} \mbf{b}_1 }]^{rs} B^{sj}_{\mbf{k}_1} \ .
\eea
We are interested in the Wilson loops $W^\phi(k_1) = W^\phi_{(k_1,\frac{2\pi}{q'}) \leftarrow (k_1,0)}$. Piecing together the single Wilson line transformations, we find
\bea
\label{eq:B1}
B_{(k_1, \frac{2\pi}{q'})} W^\phi(k_1) B^\dag_{(k_1, 0)} &= W^\phi(k_1 + \frac{2\pi}{q}).
\eea
Now we must relate $B_{(k_1, \frac{2\pi}{q'})}$ to $B_{(k_1, 0)}$ to establish a unitary relation between the Wilson loops at $k_1$ and $k_1+\frac{2\pi}{q}$. By \Eq{eq:V2primedef}, $\ket{u^\phi_j(k_1, \frac{2\pi}{q'})}  = \overline{V}_2(\phi) \ket{u^\phi_j(k_1, 0)}$, so returning to \Eq{eq:sewing}, we obtain
\bea
\label{eq:Bij}
B^{ij}_{(k_1, \frac{2\pi}{q'})} &= \braket{u_i^\phi(k_1 + \frac{2\pi}{q}, \frac{2\pi}{q'}) | \tilde{V}_1(\phi) | u^\phi_j(k_1, \frac{2\pi}{q'})} \\
&= \braket{u^\phi_i(k_1 + \frac{2\pi}{q}, 0) | {\overline{V}}^\dag_2(\phi) \tilde{V}_1(\phi) \overline{V}_2(\phi) | u^\phi_j(k_1, 0)} \\
\eea
We simplify the product ${\overline{V}}^\dag_2(\phi) \tilde{V}_1(\phi) \overline{V}_2(\phi)$ using the algebra in \Eq{eq:wilsonloopidentity}. We find
\bea
B^{ij}_{(k_1, \frac{2\pi}{q'})} &= \braket{u_i(k_1 + \frac{2\pi}{q}, 0) | {\overline{V}}_2^\dag(\phi) e^{i \zeta \frac{2\pi}{q'}} \overline{V}_2(\phi) \tilde{V}_1(\phi) | u_j(k_1, 0)} \\
&= e^{i \zeta \frac{2\pi}{q'}} B^{ij}_{(k_1, 0)} \ , \\
\eea
relating the sewing matrix $B$ at $k_2 = 0$ and $k_2 = 2\pi/q'$. We remind the reader that $\zeta \in \mathds{Z}$ satisfies $\zeta \phi = 2\pi/q \mod 2\pi$. Plugging in this result to \Eq{eq:B1}, we establish
\bea
W^\phi(k_1 +\frac{2\pi}{q}) = e^{i \frac{2 \pi}{q'}\zeta} B^\dag_{(k_1,0)} W^\phi(k_1) B_{(k_1, 0)} 
\eea
and hence the eigenvalues $e^{i \vartheta_j(k_1)}$ of the Wilson loop, also called the Wannier centers \cite{Alexandradinata:2012sp}, must satisfy
\bea
\label{eq:wilsonshift}
\{ \vartheta_j(k_1 + \frac{2\pi}{q}) \}  &= \{ \vartheta_j(k_1) + \frac{2\pi}{q'} \zeta \} \ . 
\eea
This feature of Wilson spectrum can be used to prove topological properties of the Hamiltonian. Below, we show how these results allow an alternative proof of gap closing in a Chern insulator.

\subsection{Wilson Loop Proof of Gap Closing in a Chern Insulator}
\label{app:chernwilson}

The winding of the determinant of the Wilson loop is equal to the Chern number, and we may leverage this fact to constrain $C^\phi$ \textit{at the filling of} $H^{\phi = 0}$, $\nu = N_{occ}/ N_{orb}$, with the results of \App{app:wilson}. Let us again define $q$ as the denominator of $\phi$, i.e. $\phi = \mu \frac{2\pi p'}{q'} = \frac{2\pi p}{q}$. Then assuming a gap exists so that the Wilson loop is well-defined, we establish that 
\bea
\det \left[ W^\phi(k_1 + \frac{2\pi}{q}) \right] &= \prod_{j=1}^{q' N_{occ}} e^{i \vartheta_j(k_1 + \frac{2\pi}{q})} \\
&= \exp \lp i \sum_{j=1}^{q' N_{occ} } \lp \vartheta_j(k_1) + \zeta \frac{2\pi}{q'} \rp \rp  \\
&= \exp \lp i\sum_{j=1}^{q' N_{occ} } \vartheta_j(k_1) + 2\pi i  \zeta N_{occ}  \rp \\
&=  \prod_{j=1}^{q' N_{occ}} e^{i \vartheta_j(k_1)} \\
&= \det [W^\phi(k_1)] \\
\eea
using \Eq{eq:wilsonshift}. We see that the determinant is periodic in $k_1$ with period $\frac{2\pi}{q} $ and consequently must wind a multiple of $q$ times across the magnetic BZ. This proves $C^{\phi = \frac{2\pi p}{q}} \in q \mathds{Z}$ at the filling $\nu = N_{occ}/N_{orb}$. If we were to consider another filling where $H^\phi$ is gapped, we see from the above calculation that the determinant would not necessarily be $\frac{2\pi}{q}$ periodic, and the Chern number is not required to be quantized in multiples of $q$. 

We now discuss in more detail what it means for $H^{\phi = 0}$ to be an insulator with a gap at filling $\nu$, and yet have a gap closing at $\phi \to 0$ enforced by a nonzero Chern number at filling $\nu$. This apparent discrepancy is due to the discontinuity in the energy spectrum \textit{at a fixed filling} of $H^\phi$. While the entire spectrum evolves continuously in $\phi$, there is no guarantee that the spectrum at a given filling is continuous in $\phi$. This is because the filling is determined by the number of bands, and as we have seen in \App{app:hofhamconstruct}, the number of bands of $H^{\phi = \mu \frac{2\pi p'}{q'}}$ depends on $q'$ and thus is discontinuous everywhere. 

\begin{figure*}[h]
 \centering
\begin{overpic}[width=0.42\textwidth,tics=10]{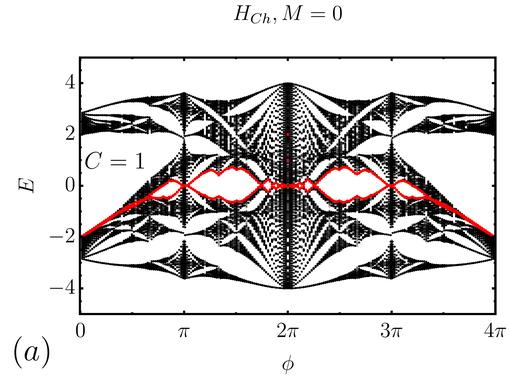}
\end{overpic}  \quad
\begin{overpic}[width=0.42\textwidth,tics=10]{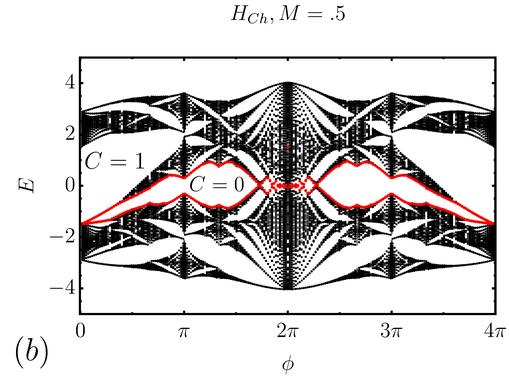}
\end{overpic}  \\
\begin{overpic}[width=0.42\textwidth,tics=10]{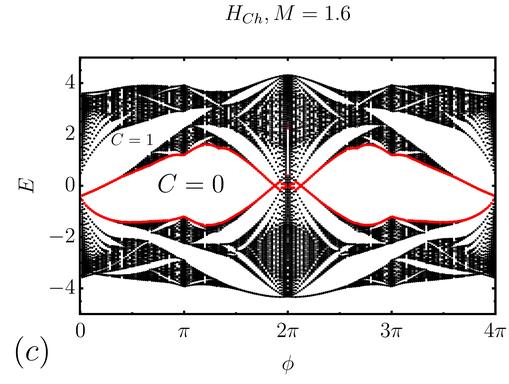}
\end{overpic}  \quad
\begin{overpic}[width=0.42\textwidth,tics=10]{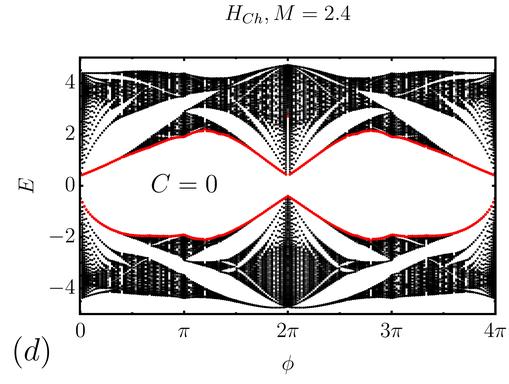}
\end{overpic} 
\caption{We show the Hofstadter Butterflies for $H_{ch}$ (\Eq{eq:checkerboardham}) for $(a)$ the topological phase, $M = 0$, $(b)$ the topological phase, $M = .5$, $(c)$ the topological phase, $M = 1.6$, $(d)$ the trivial phase, $M = 2.4$. We indicate the highest occupied state at half filling and the lowest unoccupied state in red, so the half filling gap occurs between the red lines. In $(a), (b), (c)$, we see that the occupied spectrum at half filling \textit{does} connect smoothly to the occupied spectrum at $\phi = 0$, but the unoccupied spectrum is \textit{discontinuous} at $\phi = 0$. In particular, there are conduction bands (those above filling $\nu$) that connect to the valence bands (those at or below filling $\nu$) as $\phi$ approaches $0$ despite the zero-field Hamiltonian being a gapped insulator at half filling. We have proven that this is required by the nonzero Chern number of the zero-field Hamiltonian. In (a) where $M=0$, we see an extreme case where the half-filling gap closes, as it must, at $\phi$ but remains extremely small up until $\phi \sim \frac{3\pi}{4}$. (b) The Hofstadter Butterfly is visibly gapped at half-filling for $0 < \phi < \pi$, and the gap closing at $\phi = 0^+$ is clearer. This is even more exagerated in $(c)$, closer to the $\phi =0$ phase transition which happens at $M=2$. Between $(c)$ and $(d)$, the $\phi=0$ gap closes, and $H^0$ undergoes a transition from a topological insulator to a trivial insulator. In the Hofstadter Butterfly, this transition manifests itself as the conduction branch detaching from the valence branch at $\phi = 0^+$ at the $M=2$ phase transition and reconnecting with the conduction branch. In the trivial phase, the half filling gap remains open at all $\phi$. }
\label{fig:chernexample}
\end{figure*}

To illustrate this, we show examples of Hofstadter Butterflies in \Fig{fig:chernexample} calculated from the Hamiltonian $H_{ch}$ (\Eq{eq:checkerboardham}). The phases of $H_{ch}^{\phi = 0}$ are controlled by a single parameter $M$. When $|M| < 2$, $H_{ch}^{\phi =0}$ has a Chern number $C^{\phi = 0} = 1$  at half filling, and when $|M| > 2$, the Hamiltonian is trivial. In addition to the bulk spectrum, we color the maximum (minimum) of the valence (conduction) bands red in order to show that a branch of the conduction bands connects to the valence bands as $\phi$ approaches zero. Thus, the conduction spectrum is discontinuous at $\phi = 0$, although the valence spectrum is continuous. A gapped Hamiltonian $H^{\phi_1}$ with nonzero Chern number at filling $\nu$ cannot be adiabatically connected to another gapped Hamiltonian $H^{\phi_2}$ at filling $\nu$ in the Hofstadter Butterfly as we argued in Sec. II. We note that $C^{\phi= \frac{2\pi p}{q}} = 0$ is the only number satisfying $C^{\phi} \in q \mathds{Z}$ for all $\phi$, and hence a gap which exists at filling $\nu$ for a range of $\phi$ can only have $C^\phi=0$. This is illustrated in \Fig{fig:chernexample}. It appears there that there are continuously connected gaps with $C^\phi=1$. However, these gaps are \text{not} at the same filling, and hence can be connected without a gap closing. For instance in \Fig{fig:chernexample}c, the $C=1$ gap occurs at filling $5/8$ at $\phi = \pi/2$ and filling $3/4$ at $\phi = \pi$, which are in accordance with the Streda formula $\nu = C \frac{\phi}{4\pi} + \nu_0$ where $\nu_0 = 1/2$ is the filling of the gap at $\phi = 0$.

\subsection{Discussion of Gap Closing due to a Nonzero Mirror Chern Number}
\label{app:mirrorchernwilson}

We now consider the generalization of this theorem to deduce a gap closing in the bulk spectrum of the Hofstadter Butterfly with a nonzero mirror Chern number at $\phi = 0$.  However, unlike the case of a true Chern number, a mirror Chern number does \textit{not} cause the gap to close immediately at $\phi = 0$, as we see in \Fig{fig:hofs12}b of the Main Text. 

The proof for the bulk gap closing at finite $\phi$ in a model with nonzero mirror Chern number reduces to the Chern number case because the Hamiltonian may be block-diagonalized using the mirror eigenvalues $\pm i$. Additionally, the Hofstadter Hamiltonian may still be block diagonalized by the mirror symmetry $M_z$. Being the product of inversion and a two-fold rotation, $M_z$ is not broken by flux. We assume the total Chern number is zero, or else there would be a gap closing at $\phi = 0^+$. Then the mirror Chern number is defined by
\bea
C_{M_z} &= \frac{C_{m = +i} - C_{m = -i}}{2} \\
\eea
where $C_{m = \pm i}$ is the Chern number calculated over the occupied bands with mirror eigenvalue $m = \pm i$. The total Chern number is $C = C_{m = +i} + C_{m = -i}$ which we assume to be zero (otherwise, we would have a gap closing as $\phi \to 0$), so a nonzero value of $C_{M_z}$ means $C_{m = \pm i}$ must be nonzero. Thus both mirror eigenvalue blocks have opposite and nonzero Chern numbers. Considering each block individually at their own fillings, we use the standard Chern number result of Sec. II to show that a band in each must traverse their individual gaps. Considering the full model, the intersection of the Landau levels of different mirror eigenvalues closes the bulk gap at the $\phi = 0$ filling of the whole model.

\begin{figure}
 \centering
 \includegraphics[width=8.5cm]{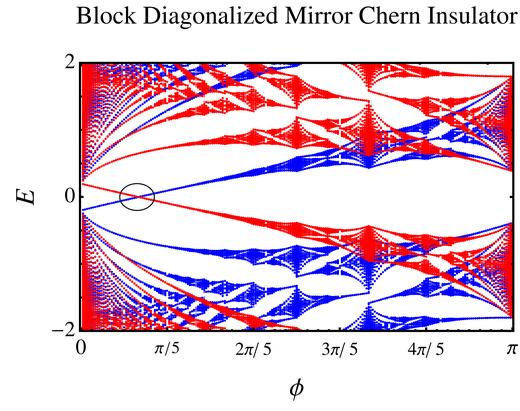} 
\caption{We show a schematic spectrum of a mirror Chern insulator where each state is colored by its mirror eigenvalue, $+i$ (resp. $-i$) for red (resp. blue).  If we consider the red and blue models individually, each has a nonzero and opposite Chern number. Thus the total Chern number of the full Hamiltonian is zero, but there is a nonzero mirror Chern number. The full Hamiltonian, i.e. the red and blue bands, has a gap closing (circled) at $\nu = 1/2$ that occurs at a finite value of the flux where the branches of opposite mirror eigenvalues intersect. The intersection cannot be gapped out by any mirror-preserving perturbation. }
\label{fig:mirrorcartoon}
\end{figure}

\subsection{Alternative Proof of Gap Closing Using Inversion Eigenvalues}
\label{app:invchern}

We proved in Sec. II of the Main Text and independently in \App{app:chernwilson} that a Hamiltonian with a nonzero Chern number had a gap closing at the Fermi level. Here, we show that a weaker version of this statement may be proven with knowledge of only inversion (or $C_{2z}$) eigenvalues. We prove now that an insulator with inversion symmetry and odd Chern number must also have a gap closing.

We will calculate the inversion eigenvalues of $\mathcal{H}^\phi(\mbf{k})$ at $\phi = 0$ and $\phi = \pi$. We are able to do so because inversion is not broken by a magnetic field, i.e.
\bea
\mathcal{I} \mathcal{H}^\phi(\mbf{k}) \mathcal{I} &= \mathcal{H}^\phi(-\mbf{k}), \quad \mathcal{I}^2 = 1, 
\eea
By assumption, the Chern number $C^{\phi = 0}$ (at filling $\nu$) at $\phi = 0$ is odd, and can be diagnosed by an odd number of $-1$ inversion eigenvalues \cite{2011PhRvB..83x5132H}. Hence to show a gap closing, we need only show that the Chern number at $\phi = \pi$ is even (at the same filling $\nu$), since a gap closing must occur between phases with differing Chern number under adiabatic evolution. 

In the magnetic BZ at $\phi = \pi$, we consider the inversion invariant points $\mbf{k}^* = (k_1^*, k_2^*)= (0,0), (0,\frac{\pi}{2}), (\pi, 0), (\pi, \frac{\pi}{2})$  which obey $k^*_1 = -k^*_1 \mod 2\pi$ and $k^*_2 = -k^*_2 \mod \pi$. Now we study the eigenstates of $H^{\phi = \pi}$ at these points. If $\mathcal{I} \ket{\mbf{k}^*} = \xi \ket{\mbf{k}^*}$,  then 
\bea
\mathcal{I} T_2(\pi) \ket{\mbf{k}^*} = \xi T_2^\dag(\pi) \ket{\mbf{k}^*} = \xi T_2(\pi) {T_2^\dag(\pi)}^2\ket{\mbf{k}^*} = \xi e^{-2i k_2} T_2(\pi) \ket{\mbf{k}^*}
\eea
where we have used that the magnetic translation operators satisfy $\mathcal{I} T_i(\phi) \mathcal{I} = T_i(\phi)^\dag$ which follows directly from \Eq{eqn:T}. Recalling that $T_2(\phi) \ket{\mbf{k}^*}$ has momentum $\mbf{k}^* - \phi \mbf{b}_1 = \mbf{k}^* - \pi \mbf{b}_1$, we see that if there are $n_1$ negative inversion eigenvalues at the points $(0,0)$, then there are also $n_1$ negative inversion eigenvalues at $(\pi,0)$. Similarly, if there are $n_2$ negative inversion eigenvalues at $(0,\frac{\pi}{2})$, then there are $n_2$ \textit{positive} ones at $(\pi,\frac{\pi}{2})$. Since there are $2 N_{occ}$ bands in the doubled magnetic unit cell at $\phi = \pi$, we find the total number of negative inversion eigenvalues to be $2n_1 + n_2 + (2 N_{occ} - n_2)$ which is even, so the Chern number must be even (zero included) \cite{2011PhRvB..83x5132H}. Because the $C^{\phi =0}$ is assumed odd, there must be a gap closing between $\phi = 0$ and $\phi = \pi$. 
\section{Hofstadter Topology Protected by Time Reversal symmetry}

In this Appendix, we study the Hofstadter 3D TI phase in generality. First we determine the properties of $U\mathcal{T}$, the effective time-reversal symmetry at $\phi = \Phi/2$ (\App{app:UT}).  We then give two proofs showing that the $\mathds{Z}_2$ index protected by $U\mathcal{T}$ at $\phi = \Phi/2$ is trivial: one in the magnetic BZ (\App{app:genz2proof}) and the other using the Wilson loop (\App{app:wilsonz2}). These proofs are facilitated by a gauge-invariant formalism, introduced in \App{app:UTdetails}, which makes use of the magnetic translation group to construct the Hofstadter Hamiltonian. From this construction, we also find that $U\mathcal{T}$ symmetry at $\phi = \Phi/2$ yields a projective representation of the magnetic space group. 

\subsection{Properties of $U\mathcal{T}$}
\label{app:UT}

First, we discuss time-reversal symmetry (TRS) for the Hofstadter Hamiltonian with a $\phi \to \phi + \Phi$ periodicity in the flux. If the $\phi=0$ model is $\mathcal{T}$-symmetric then the model at $\phi = \Phi/2$ is $U \mathcal{T}$-symmetric. This is because
\bea
U \mathcal{T} H^{\Phi/2} (U \mathcal{T})^\dag  &= U H^{-\Phi/2} U^\dag = H^{\Phi/2} ,
\eea
where $U$ is given by \Eq{eq:hoffluxper}. We now define the anti-unitary operator $\mathcal{T}$ in position space as 
\bea
\label{eq:Tcaldef}
\mathcal{T}^{-1} c_{\mbf{R}, \al} \mathcal{T} &= \sum_{\be'} D_{\al \be}(\mathcal{T}) c_{\mbf{R},\be}
\eea
where the unitary matrix $D_{\al \be}(\mathcal{T})$ is only nonzero when $\pmb{\delta}_\al = \pmb{\delta}_\be$ (and $\al$ and $\be$ are spin-flipped Kramers' pairs), i.e. $\mathcal{T}$ is local, onsite. $D D^* = \pm 1$ for spinful/spin-less electrons ($\mathcal{T}^2 = \pm 1$). The action of $\mathcal{T}$ flips the flux, taking $\phi \to - \phi$ and hence
\bea
\label{eq:UTTU}
\mathcal{T} U \mathcal{T}^{-1} &= U^\dag
\eea
as may be verified using the position space representations of the operators. 
From here we conclude $(U \mathcal{T})^2 = U \mathcal{T} U \mathcal{T} = U U^\dag \mathcal{T}^2 = \mathcal{T}^2$. This result holds for $\Phi = 2\pi n$, $n$ even or odd. Because we study insulators with a nontrivial Kane-Mele invariant, we have $\mathcal{T}^2 = -1$. We recall that $U H^\phi U^\dag = H^{\phi + \Phi}$ and thus $U \mathcal{T}$ also protects a topological classification at $H^{\phi = \Phi/2}$ as per \Eq{eq:mainUT}.

\subsection{Gauge-Invariant Proof of the 3D TI phase}
\label{app:genz2proof}

In the prior sections, we have worked explicitly in the Landau gauge which enables us do explicit, numerical computations such as computing the energy spectrum or the Wilson loop. In this section, we will prove the triviality of the Kane-Mele $\mathds{Z}_2$ index $\delta^{\phi = \Phi/2}$, and do not need to perform explicit computations. Hence we can avoid the cumbersome degeneracies caused by the extended $1\times q'$ magnetic unit cell (see \App{app:hof}) by introducing a new formalism which makes use of the magnetic translation operators to study the Hofstadter Hamiltonian in the $1\times q$ without specifying a gauge. First in \App{app:UTproof}, we will construct the Hofstadter Hamiltonian in this magnetic unit cell in a gauge-invariant manner to facilitate a simple proof that $\delta^{\phi = \Phi/2} = +1$, with the details of the calculations left to \App{app:UTdetails}. 

\subsubsection{Proof of the Trivial Kane-Mele Index due to Gap Closings at the Phase Transition}
\label{app:UTproof}

Recall from Sec. I that we may choose a simultaneous eigenbasis of $T_1(\phi), T_2^q(\phi)$ where $\phi = \frac{2\pi p}{q}$ for $p,q$ coprime. At $\phi = \Phi/2 = \pi n$ for odd $n$, this means $q = 2$. We study the single-particle states $\ket{\al, \ell, k_1, k_2}$ with $\al =1, \dots, N_{orb}, \ell = 0,1$ which are eigenstates of $T_1(\Phi/2)$ and $T_2(\Phi/2)$. Going forward, we write $T_i(\Phi/2) = T_i$ for brevity unless otherwise specified. These states satisfy
\bea
\label{eq:eigenstatesUT}
T_1 \ket{k_1, k_2, \ell, \al} = e^{i k_1} \ket{k_1, k_2, \ell, \al}, \quad  T_2^2 \ket{k_1, k_2, \ell, \al} = e^{i 2k_2} \ket{k_1, k_2, \ell, \al} \\
\eea
where $k_1 \in (-\pi,\pi)$ and we may choose $k_2 \in (-\pi/2, \pi/2)$. Note that these eigenstates differ from the conventional ``momentum" eigenstates in \Eq{eq:eigenstateslandau} which were used to diagonalize the Hofstadter Hamitlonian in \App{app:hof}. Now we can form the Hofstadter Hamiltonian $\mathcal{H}^{\Phi/2}(\mbf{k})$ in this magnetic translation operator eigenbasis directly:
\bea
\label{eq:Hperiodic}
\mathcal{H}^{\Phi/2}_{\ell\al, \ell' \be}(\mbf{k}) \delta_{\mbf{k}, \mbf{k}'} &= \braket{k_1, k_2, \ell, \al| H^{\Phi/2} | k'_1, k'_2, \ell', \be}, \qquad k_1 \in (-\pi,\pi), k_2 \in (-\pi/2,\pi/2)
\eea
which gives a representation of the Hamiltonian in the $1\times 2$ unit cell. We emphasize that the states \Eq{eq:eigenstatesUT} give the minimal, gauge-independent $1\times 2$ unit cell, but we do not have an explicit expression for the Hofstadter Hamiltonian. We can derive the embedding matrices for $\mathcal{H}^{\Phi/2}(\mbf{k})$ that implement the $k_1 \to k_1 + \pi$ periodicity following Sec. I. We prove in \App{app:UTdetails} that there is an embedding matrix $[V_1(\Phi/2, k_2)]_{\ell, \al, \ell', \be}$ satisfying
\bea
\label{eq:genericpishift}
\mathcal{H}^{\Phi/2}(\mbf{k} + \pi \mbf{b}_1) &= V_1(\Phi/2, k_2) \mathcal{H}^{\Phi/2}(\mbf{k}) V^\dag_1(\Phi/2, k_2) \ .
\eea
We will not need an explicit expression for $V_1(\Phi/2, k_2)$ in this section, but we provide one in \App{app:UTdetails}. Note that this embedding matrix depends on $k_2$, which we make explicit in the notation. This differs from the embedding matrices in \App{sec:embedding} where there was no momentum dependence. 

So far we have constructed the Hofstadter Hamiltonian in a gauge-invariant manner that guarantees a magnetic BZ with $k_1 \in (-\pi,\pi), k_2 \in (-\pi/2, \pi/2)$ with a $k_1 \to k_1 + \pi$ periodicity in the spectrum implemented by $V_1(\Phi/2, k_2)$. Now we consider the $U \mathcal{T}$ symmetry that protects a $\mathds{Z}_2$ index, recalling that $(U \mathcal{T})^2 = \mathcal{T}^2 = -1$. We need to determine what properties this symmetry enforces. We will prove in \App{app:UTdetails} that 
\bea
\label{eq:HkappaH}
\sum_{\ell'\al',r'\be'} [D(U \mathcal{T})^\dag]_{\ell\al, \ell' \al' } \mathcal{H}^{\Phi/2 }_{\ell' \al', r' \be'}(\mbf{k}) [D(U \mathcal{T})]_{r'\be' , r \be}  &= \mathcal{H}^{\Phi/2 \ *}_{\ell\al, r \be}(-\mbf{k} - \pmb{\kappa})  \\ 
\eea
where $D(U \mathcal{T})$ is the representation of $U \mathcal{T}$ on the single-particle Hilbert space satisfying $D(U \mathcal{T}) [D(U \mathcal{T})]^* = -1$ and $\pmb{\kappa}$ is a momentum shift. We give a formula for $\pmb{\kappa}$ in \App{app:UTdetails} where we show that it appears as a projective phase in the magnetic space group. 

\begin{figure}
 \centering
 \includegraphics[width=8.cm]{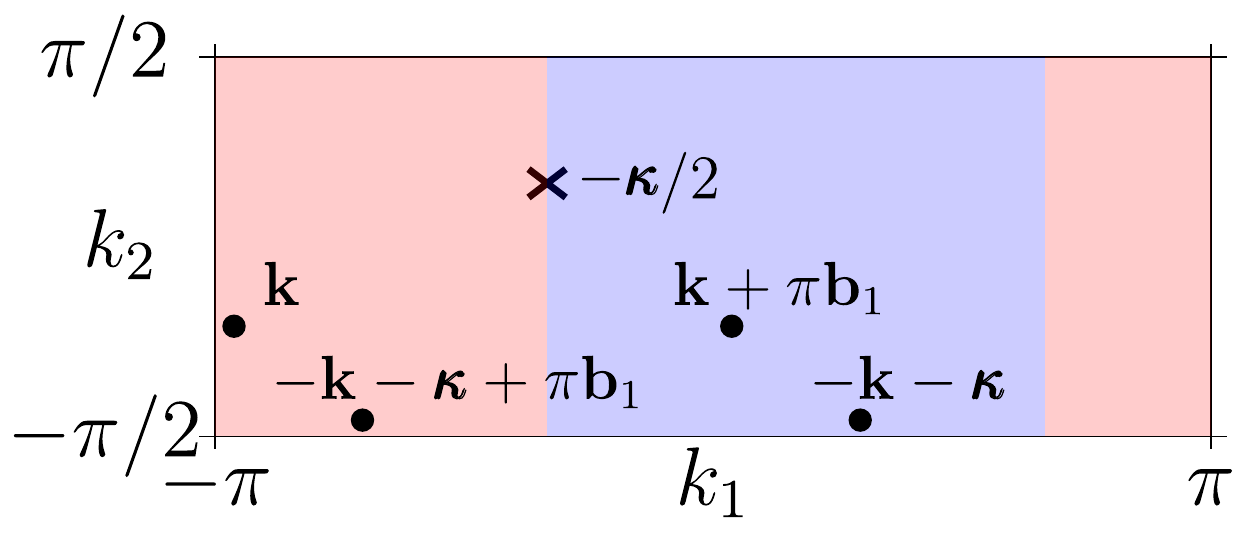} 
\caption{We illustrate an example of $BZ_{1/2, \pmb{\kappa}}$, shown in red, with the rest of the magnetic BZ shown in blue. The point $-\pmb{\kappa}/2$, shown with an ``X" is invariant under $U\mathcal{T}$. If there is a gap closing at $\mbf{k} = -\frac{1}{2} \pmb{\kappa} + \mbf{k}$, then three more gap closings are generated under the application of $U\mathcal{T}$ and $V_1(\Phi/2)$. We denote these points with dots, and observe that exactly two points fall in $BZ_{1/2, \pmb{\kappa}}$. } 
\label{fig:magBZkappa}
\end{figure}

In Sec. III of the Main Text, we proved that the  $\mathds{Z}_2$ invariant was always trivial, $\delta^{\phi = \Phi/2} = +1$, for $\Phi = 2\pi n$, $n$ odd when $\pmb{\kappa} = 0$ by proving that all gap closings come in pairs in half the magnetic BZ. To prove the case with arbitrary $\pmb{\kappa}$, we note that we can rewrite \Eq{eq:HkappaH} in terms of a shifted momentum $\mbf{k} = -\frac{1}{2} \pmb{\kappa} + \mbf{\tilde{k}}$. In matrix notation, this reads
\bea
\label{eq:Hkappaktilde}
\null D(U \mathcal{T})^\dag \mathcal{H}^{\Phi/2 }( -\frac{1}{2} \pmb{\kappa} + \mbf{\tilde{k}}) D(U \mathcal{T}) &= \mathcal{H}^{\Phi/2 \ *}( -\frac{1}{2} \pmb{\kappa} - \mbf{\tilde{k}}) 
\eea
from which we observe that $\mbf{\tilde{k}}$, the momentum measured from $-\frac{1}{2} \pmb{\kappa} $, flips sign under $U\mathcal{T}$. Thus we can replicate the gap closing proof in Sec. IISec. III by choosing the half magnetic BZ defined by $BZ_{1/2, \pmb{\kappa}} = \{ -\frac{1}{2} \pmb{\kappa} + \mbf{\tilde{k}} | \tilde{k}_1 \in (-\pi,0), \tilde{k}_2 \in (-\pi/2,\pi/2) \}$. Now we use the fact that $n$ is odd, so there is a $\pi$ periodicity in the spectrum along $k_1$ given by \Eq{eq:genericpishift}. In addition, due to \Eq{eq:Hkappaktilde}, the spectrum is invariant under $\mbf{k} \to -\mbf{k} - \pmb{\kappa}$. Hence gap closings in the \emph{whole} magnetic BZ come in quartets, two due to the magnetic translation group and two due to $U\mathcal{T}$, at the points
\bea
\label{eq:quartet}
 -\frac{1}{2} \pmb{\kappa} + \mbf{k},  -\frac{1}{2} \pmb{\kappa} - \mbf{k},  -\frac{1}{2} \pmb{\kappa} + \mbf{k} + \pi \mbf{b}_1, \text{ and } -\frac{1}{2} \pmb{\kappa} - \mbf{k} + \pi \mbf{b}_1
\eea
as is shown for instance in \Fig{fig:magBZkappa}. Note that for generic $\mbf{k}$, exactly one of $ -\frac{1}{2} \pmb{\kappa} \pm \mbf{k}$ is in $BZ_{1/2, \pmb{\kappa}}$ and similarly exactly one of $-\frac{1}{2} \pmb{\kappa} \mp \mbf{k} + \pi \mbf{b}_1$ is in $BZ_{1/2, \pmb{\kappa}}$. The degenerate points with $k_1 = 0,\pm \pi/2$ or $\pi$ and $k_2 = 0$ or $\pi/2$ reduce the quartet of points in \Eq{eq:quartet} to pairs of point where \emph{doubly degenerate} gap closings occur. Thus all gap closings in the \emph{half} BZ, $BZ_{1/2, \pmb{\kappa}}$, come in pairs and $\delta^{\phi = \Phi/2} = +1$.

\subsubsection{Construction of the Gauge-invariant Eigenstates and the Projective Symmetry Algebra}
\label{app:UTdetails}

To compute objects like $T_i \ket{ \be, \ell', k_1, k_2}$ and $U \mathcal{T} \ket{ \be, \ell', k_1, k_2}$, we note that there is an explicit expression for the momentum eigenstates given by the magnetic translation operators
\bea
\label{eq:explicitTstates}
 \ket{ k_1, k_2, \ell,\al} &= \frac{1}{\sqrt{\mathcal{N}/2}} \sum_{mn} e^{-i m k_1 - 2i n k_2} T_1^mT_2^{2n} c^\dag_{\ell \mbf{a}_2, \al} \ket{0}, \qquad \ell = 0,1 \text{ and } \ \al = 1, \dots, N_{orb} \\
 \eea
where $T_i^{-m} = [T_i^\dag]^m$. (Recall that have have denoted $T_i(\Phi/2)$ as $T_i$ for brevity). It is straightforward to check that these states have the correct $T_1, T_2^2$ eigenvalues. Here we have assumed $n$ odd so there is a $1\times 2$ magnetic unit cell. 
We will first use \Eq{eq:explicitTstates} to derive the embedding matrix $V_1(\Phi/2, k_2)$ (see \Eq{eq:genericpishift}). We treat the cases of $\ell = 0$ and $\ell = 1$ separately. For $\ell = 0$, we find
\bea
T_2 \ket{ k_1, k_2, 0,\al} &= \frac{1}{\sqrt{\mathcal{N}/2}} \sum_{mn} e^{-i m k_1 - 2i n k_2} T_2 T_1^mT_2^{2n}(\phi) c^\dag_{0, \al} \ket{0} \\
&= \frac{1}{\sqrt{\mathcal{N}/2}} \sum_{mn} e^{-i m k_1 - 2i n k_2} (-1)^m T_1^m  T_2^{2n}  T_2  c^\dag_{0, \al} \ket{0} \\
&= \frac{1}{\sqrt{\mathcal{N}/2}} \sum_{mn} e^{-i m (k_1+\pi) - 2i n k_2} T_1^m  T_2^{2n}  T_2  c^\dag_{0, \al} \ket{0} \\
\eea
using the magnetic translation group and $\exp (i \Phi/2) = -1$ for $n$ odd. Now using \Eq{appeq:T} to compute the action of $T_2 $ on the single-particle state, we find
\bea
T_2  c^\dag_{0 \mbf{a}_2, \al} \ket{0} &=  \exp \Big(i \int_{\pmb{\delta}_\al}^{\mbf{a}_2 + \pmb{\delta}_\al} \mbf{A} \cdot d\mbf{r} \, + \, i \chi_2(\pmb{\delta}_\al ) \Big) c^\dag_{\mbf{a}_2, \al} \ket{0}  \\
\eea
and hence
\bea
\label{eq:T2k0}
T_2  \ket{ k_1, k_2, 0,\al} &=  \exp \Big(i \int_{\pmb{\delta}_\al}^{\mbf{a}_2 + \pmb{\delta}_\al} \mbf{A} \cdot d\mbf{r} \, + \, i \chi_2(\pmb{\delta}_\al ) \Big)  \ket{ k_1+\pi, k_2, 1,\al} 
\eea
For $\ell = 1$, we calculate
\bea
T_2  \ket{ k_1, k_2, 1,\al} &= \frac{1}{\sqrt{\mathcal{N}/2}} \sum_{mn} e^{-i m k_1 - 2i n k_2} T_2 T_1^m T_2^{2n} c^\dag_{\mbf{a}_2, \al} \ket{0} \\
&= \frac{1}{\sqrt{\mathcal{N}/2}} \sum_{mn} e^{-i m (k_1+\pi) - 2i n k_2}  T_1^m  T_2^{2n +1}  c^\dag_{\mbf{a}_2, \al} \ket{0} \\
&= \frac{1}{\sqrt{\mathcal{N}/2}} \sum_{mn} e^{-i m (k_1+\pi) - 2i n k_2}  T_1^m  T_2^{2n +2}  T_2^\dag  c^\dag_{\mbf{a}_2, \al} \ket{0} \\
&= \frac{1}{\sqrt{\mathcal{N}/2}} \sum_{mn} e^{-i m (k_1+\pi) - 2i (n-1) k_2}  T_1^m  T_2^{2n}  T_2^\dag  c^\dag_{\mbf{a}_2, \al} \ket{0} \\
\eea
Now using \Eq{appeq:T} to compute the action of $T^\dag_2$ on the single-particle state, we find
\bea
T^\dag_2 c^\dag_{\mbf{a}_2, \al} \ket{0} &=  \exp \Big(-i \int_{\pmb{\delta}_\al}^{\mbf{a}_2 + \pmb{\delta}_\al} \mbf{A} \cdot d\mbf{r} \, - \, i \chi_2(\pmb{\delta}_\al ) \Big) c^\dag_{0, \al} \ket{0}  \\
\eea
and hence
\bea
\label{eq:T2k1}
T_2 \ket{k_1, k_2,1,\al} &= \exp \Big(2 i k_2 -i \int_{\pmb{\delta}_\al}^{\mbf{a}_2 + \pmb{\delta}_\al} \mbf{A} \cdot d\mbf{r} \, - \, i \chi_2(\pmb{\delta}_\al ) \Big)  \ket{ k_1+\pi, k_2,0,\al} \ . 
\eea
We collect \Eqs{eq:T2k0}{eq:T2k1} into the following formula
\bea
T_2 \ket{ k_1, k_2, \ell,\al} &= \sum_{\ell',\be}\ket{k_1 + \pi , k_2, \ell',\be}  [V_1(\Phi/2, k_2)]_{\ell'\be,\ell\al}  \\
\eea
where 
\bea
\null [V_1(\Phi/2, k_2)]_{\ell'\be,\ell\al} &= \delta_{\al \be} e^{i k_2} \left[ \cos \lp k_2 - \int_{\pmb{\delta}_\al}^{\mbf{a}_2 + \pmb{\delta}_\al} \!\!\!\mbf{A} \cdot d\mbf{r} \, - \chi_2(\pmb{\delta}_\al ) \rp \sigma_1  + \sin \lp k_2 - \int_{\pmb{\delta}_\al}^{\mbf{a}_2 + \pmb{\delta}_\al} \!\!\! \mbf{A} \cdot d\mbf{r} \, - \chi_2(\pmb{\delta}_\al ) \rp \sigma_2\right]_{\ell,\ell'} 
\eea
which by inspection is a unitary matrix. We remark that $V_1(\Phi/2, k_2)$ is periodic along the magnetic BZ, i.e. $V_1(\Phi/2, k_2) = V_1(\Phi/2,k_2 - \pi)$, which we will make use of in \App{app:UTwilsonnew} when considering the Wilson loop. We derive the corresponding transformation on the Hofstadter Hamiltonian to be
\bea
\mathcal{H}^{\Phi/2}_{\ell\al, r \be}(k_1, k_2) 
&=  \sum_{\ell' \al',r',\be}  [V_1^\dag(\Phi/2, k_2)]_{\ell\al,\ell'\al'} \mathcal{H}^{\Phi/2}_{\ell'\al', r' \be'}(k_1+\pi, k_2) [V_1(\Phi/2, k_2)]_{r'\be',r\be} \ . \\
\eea
In matrix notation, this reads
\bea
 \mathcal{H}^{\Phi/2}(\mbf{k}) &= V^\dag_1(\Phi/2, k_2) \mathcal{H}^{\Phi/2}(\mbf{k} + \pi \mbf{b}_1) V_1(\Phi/2, k_2)  %,
\eea
which proves \Eq{eq:genericpishift}.

\begin{figure}
 \centering
 \includegraphics[width=5.5cm]{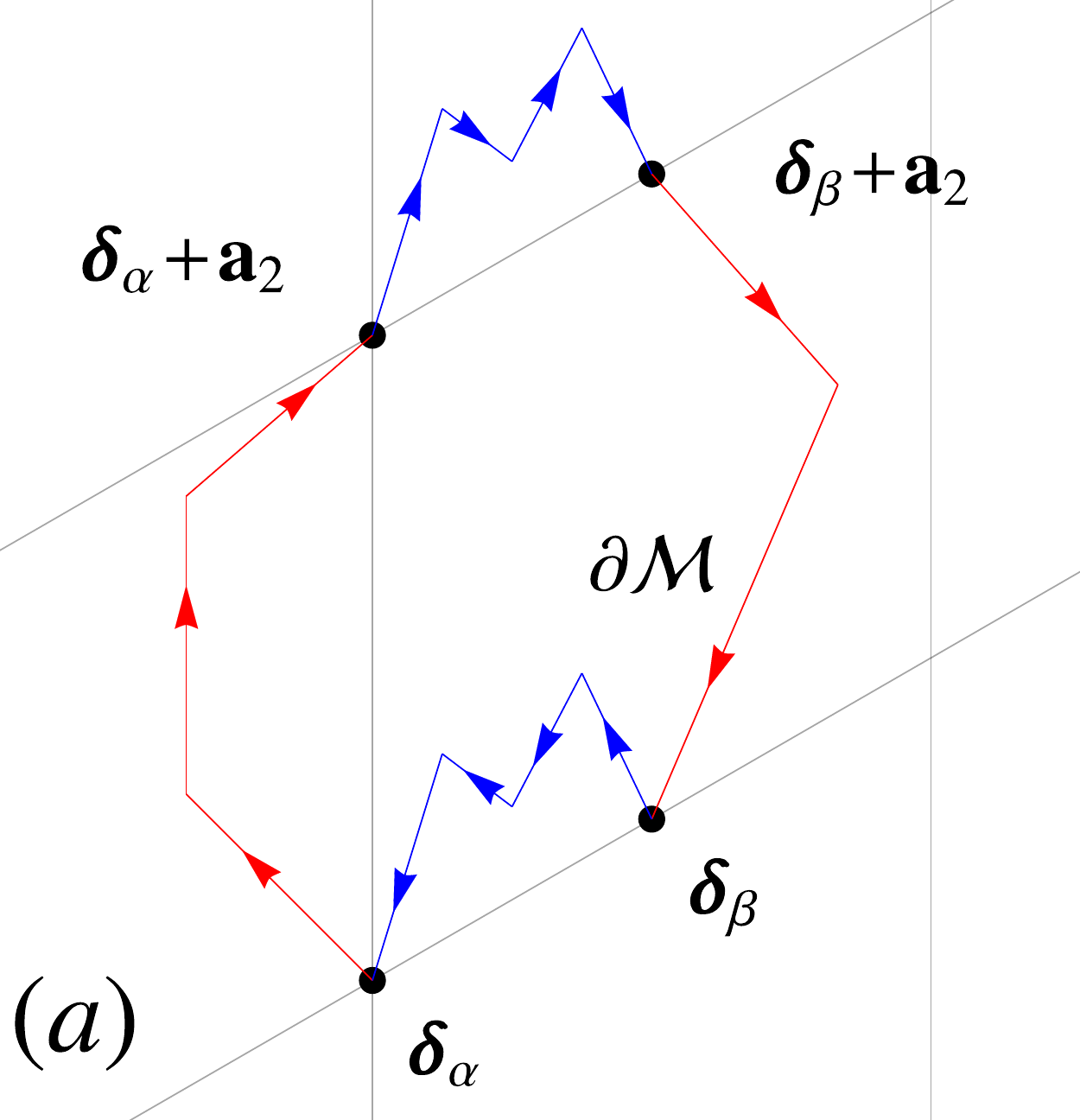} 
  \includegraphics[width=5.5cm]{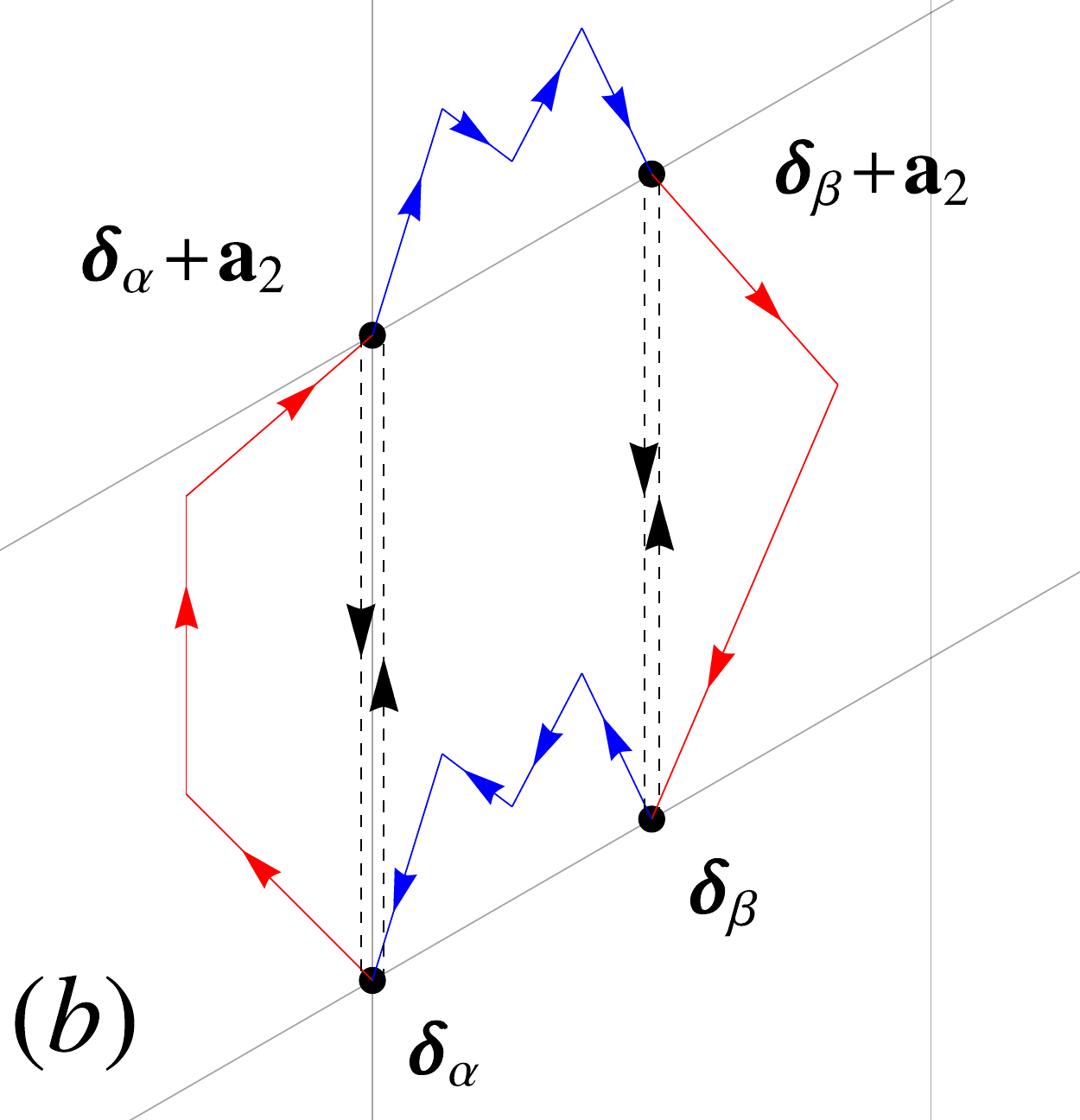}
   \includegraphics[width=5.5cm]{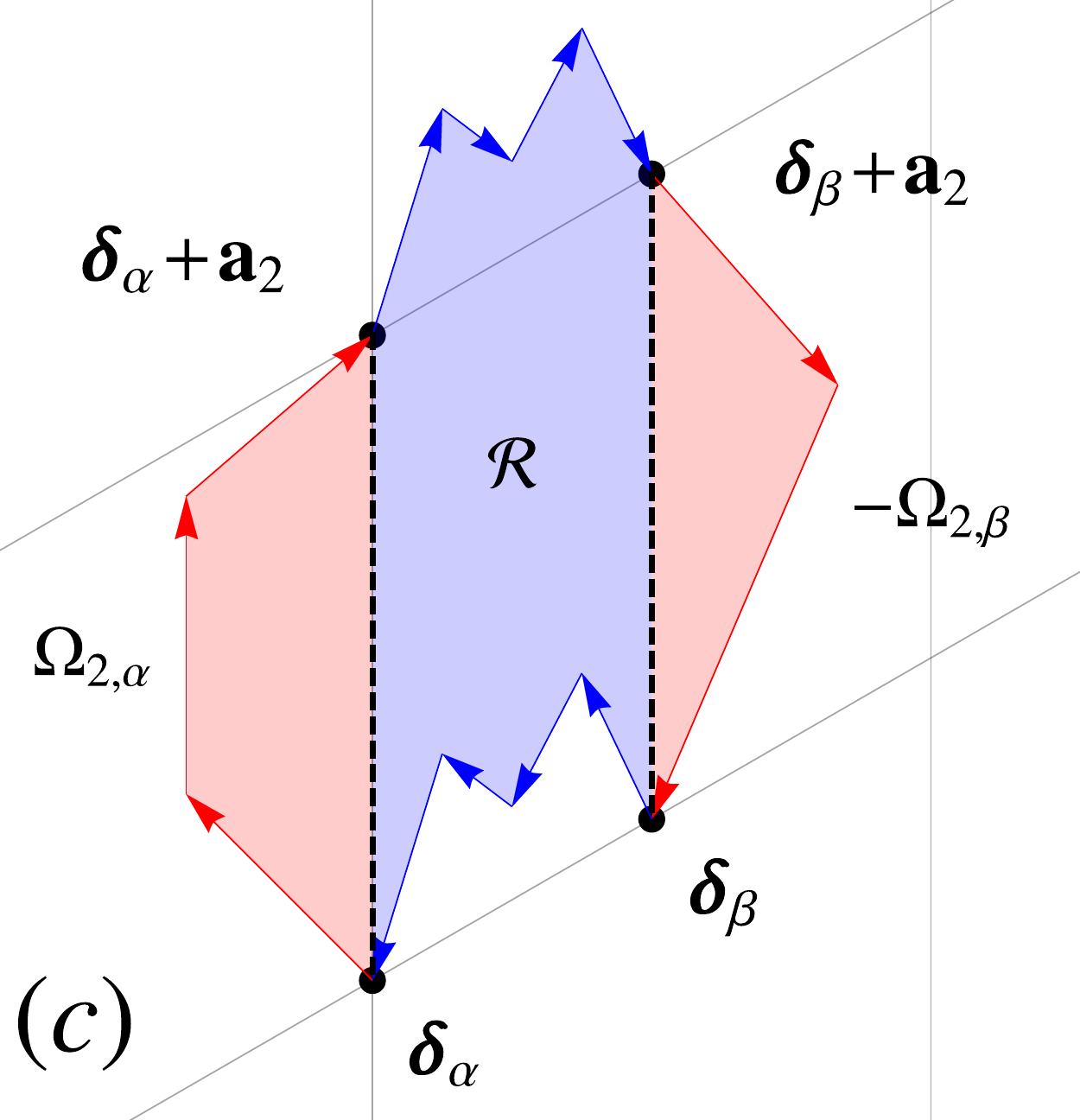} 
\caption{We illustrate the manipulations of the integrals from \Eq{eq:ppeqstart} to \Eq{eq:intUTdeform} with $\mbf{a}_i = \mbf{a}_2$. $(a)$ We show the closed loop along Peierls paths $\del \mathcal{M}$. The loop is composed of (1) the Peierls paths connecting $\pmb{\delta}_{\al} \to \pmb{\delta}_{\al} + \mbf{a}_i$ and $ \pmb{\delta}_{\be} + \mbf{a}_i \to \pmb{\delta}_{\be}$ shown in red which in general have different shapes, and (2) the Peierls paths connecting $\pmb{\delta}_{\al} + \mbf{a}_i \to \pmb{\delta}_{\be} + \mbf{a}_i $ and  $\pmb{\delta}_{\be} \to \pmb{\delta}_{\al}$ shown in blue, which may be chosen to have the same shape because they are related by a lattice vector. $(b)$ Without changing the value of the integral, we add and subtract straight-line paths, dashed in black, that connect $\pmb{\delta}_{\al}, \pmb{\delta}_{\al} + \mbf{a}_i$ and  $\pmb{\delta}_{\be}, \pmb{\delta}_{\be} + \mbf{a}_i$. $(c)$ The line integral can be divided along the straight-line contours and separated into the areas $\Omega_{i,\al}, \mathcal{R}, -\Omega_{i,\be}$ where the minus sign is due to the opposite orientation. We see and can easily prove that $\mathcal{R}$ can be deformed to a parallelogram with corners at $\pmb{\delta}_{\al}, \pmb{\delta}_{\al} + \mbf{a}_2, \pmb{\delta}_{\be} + \mbf{a}_2$, and $\pmb{\delta}_{\be}$. 
}
\label{fig:UTareafigs}
\end{figure}

In a similar manner, we can compute the action of $U\mathcal{T}$ on the eigenstates \Eq{eq:explicitTstates} once we know the algebra of $U\mathcal{T}$ and $T_i(\Phi/2)$. Using \Eq{appeq:T}, we compute
\bea
\mathcal{T} T_i  \mathcal{T}^{-1} &= \mathcal{T}  \sum_{\mbf{R} \al} e^{i \int_{\mbf{R} + \pmb{\delta}_\al}^{\mbf{R} + \pmb{\delta}_\al + \mbf{a}_i} \mbf{A} \cdot d\mbf{r} \, + \, i \chi_i(\mbf{R}+\pmb{\delta}_\al )} c^\dag_{\mbf{R}+\mbf{a}_i, \al} c_{\mbf{R},\al} \mathcal{T}^{-1} \\
\eea
where the integral of $\mbf{A}$ is taken \emph{over a straight-line path}, which is not necessarily a Peierls path. Using \Eq{eq:Tcaldef}, we compute
\bea
\mathcal{T} T_i(\Phi/2) \mathcal{T}^{-1} &= \sum_{\mbf{R} \al} e^{-i \int_{\mbf{R} + \pmb{\delta}_\al}^{\mbf{R} + \pmb{\delta}_\al + \mbf{a}_i} \mbf{A} \cdot d\mbf{r} \, - \, i \chi_i(\mbf{R}+\pmb{\delta}_\al )} \mathcal{T}  c^\dag_{\mbf{R}+\mbf{a}_i, \al} c_{\mbf{R},\al} \mathcal{T}^{-1} \\
 &= \sum_{\mbf{R} \al, \al', \al''} e^{-i \int_{\mbf{R} + \pmb{\delta}_\al}^{\mbf{R} + \pmb{\delta}_\al + \mbf{a}_i} \mbf{A} \cdot d\mbf{r} \, - \, i \chi_i(\mbf{R}+\pmb{\delta}_\al )} c^\dag_{\mbf{R}+\mbf{a}_i, \al'} D^\dag_{\al' \al}(\mathcal{T}) D_{\al \al''}(\mathcal{T}) c_{\mbf{R},\al''} \\
  &= \sum_{\mbf{R} \al} e^{-i \int_{\mbf{R} + \pmb{\delta}_\al}^{\mbf{R} + \pmb{\delta}_\al + \mbf{a}_i} \mbf{A} \cdot d\mbf{r} \, - \, i \chi_i(\mbf{R}+\pmb{\delta}_\al )} c^\dag_{\mbf{R}+\mbf{a}_i, \al} c_{\mbf{R},\al} \\
&=  T_i(-\Phi/2) \\
\eea
where we have used the fact that $\mathcal{T}$ is onsite as discussed in \App{app:UT}. We see that the effect of $\mathcal{T}$ is merely to complex conjugate $T_i(\phi)$, which is equivalent to reversing the flux. We now study $U \mathcal{T} T_i(\phi) (U \mathcal{T})^{-1} = U T_i(-\phi) U^\dag$. Using \Eq{eq:hoffluxper}, we find
\bea
U T_i(-\Phi/2) U^\dag &= \sum_{\mbf{R} \al} e^{-i \int_{\mbf{R} + \pmb{\delta}_\al}^{\mbf{R} + \pmb{\delta}_\al + \mbf{a}_i} \mbf{A} \cdot d\mbf{r} \, - \, i \chi_i(\mbf{R}+\pmb{\delta}_\al )} U c^\dag_{\mbf{R}+\mbf{a}_i, \al} U^\dag U c_{\mbf{R},\al} U^\dag \\
 &= \sum_{\mbf{R} \al} \exp \lp -i \int_{\mbf{R} + \pmb{\delta}_\al}^{\mbf{R} + \pmb{\delta}_\al + \mbf{a}_i} \mbf{A} \cdot d\mbf{r} \, - \, i \chi_i(\mbf{R}+\pmb{\delta}_\al ) + i \int_{\mbf{r}_0}^{\mbf{R} + \mbf{a}_i + \pmb{\delta}_\al} \mbf{\tilde{A}} \cdot d\mbf{r} - i \int_{\mbf{r}_0}^{\mbf{R}+ \pmb{\delta}_\al} \mbf{\tilde{A}} \cdot d\mbf{r} \rp c^\dag_{\mbf{R}+\mbf{a}_i, \al} c_{\mbf{R},\al}  \\
\eea
where we emphasize that the integral of $\mbf{\tilde{A}}$ (generating $\Phi$ flux) is taken over a Peierls path (see \Eq{eq:Usinglepart} ) and the integral of $\mbf{A}$ (generating $\phi = \Phi/2$ flux) is taken over a straight-line path (see \Eq{eqn:T}). We can re-sum the Peierls integrals using \Eq{eq:peierlsdeform}
 \bea
 \label{eq:phaseUT}
U T_i(-\Phi/2) U^\dag   &= \sum_{\mbf{R} \al} \exp \lp -i \int_{\mbf{R} + \pmb{\delta}_\al}^{\mbf{R} + \pmb{\delta}_\al + \mbf{a}_i} \mbf{A} \cdot d\mbf{r} \, - \, i \chi_i(\mbf{R}+\pmb{\delta}_\al ) + i \int_{\mbf{R} + \pmb{\delta}_\al}^{\mbf{R} + \pmb{\delta}_\al+ \mbf{a}_i } \mbf{\tilde{A}} \cdot d\mbf{r} \rp c^\dag_{\mbf{R}+\mbf{a}_i, \al} c_{\mbf{R},\al} \ . \\
\eea
To proceed, we want to relate the gauge fields $\mbf{A}$ and $\mbf{\tilde{A}}$ in a convenient way. We  have $2\mbf{A} = \mbf{\tilde{A}}$ up to a gauge transformation because $\curl \mbf{A} = \Phi/2$ and $\curl \mbf{\tilde{A}} = \Phi$. It is most convenient to take $2\mbf{A} = \mbf{\tilde{A}}$. We now note that the integrals can be combined. For clarity, we denote integrals over Peierls paths with a $P.P.$ label, and the straight-line path with $straight$, and we also denote $\mbf{R} + \pmb{\delta}_\al = \mbf{r}_{\al}$ as shorthand. Then the integrals in \Eq{eq:phaseUT} can be manipulated to yield
\bea
- \int_{straight}^{\mbf{r}_{\al} \to \mbf{r}_{\al} + \mbf{a}_i} \mbf{A} \cdot d\mbf{r} + \int_{P.P}^{\mbf{r}_{\al} \to \mbf{R} + \mbf{a}_i + \pmb{\delta}_\al} \mbf{\tilde{A}} \cdot d\mbf{r} &= \int_{straight}^{\mbf{r}_{\al} \to \mbf{r}_{\al} + \mbf{a}_i} \mbf{A} \cdot d\mbf{r} - 2 \int_{straight}^{\mbf{r}_{\al} \to \mbf{r}_{\al} + \mbf{a}_i} \mbf{A} \cdot d\mbf{r} + \int_{P.P.}^{\mbf{r}_{\al} \to \mbf{r}_\al + \mbf{a}_i} \mbf{\tilde{A}} \cdot d\mbf{r}  \\
&=\int_{straight}^{\mbf{r}_{\al} \to \mbf{r}_{\al} + \mbf{a}_i} \mbf{A} \cdot d\mbf{r}+2\int_{straight}^{ \mbf{r}_{\al} + \mbf{a}_i \to \mbf{r}_{\al} } \mbf{A} \cdot d\mbf{r}+  \int_{P.P.}^{\mbf{r}_{\al} \to \mbf{r}_\al + \mbf{a}_i}\mbf{\tilde{A}} \cdot d\mbf{r}  \\
&=\int_{straight}^{\mbf{r}_{\al} \to \mbf{r}_{\al} + \mbf{a}_i} \mbf{A} \cdot d\mbf{r}+ \lp   \int_{P.P.}^{\mbf{r}_{\al} \to \mbf{r}_\al + \mbf{a}_i} + \int_{straight}^{ \mbf{r}_{\al} + \mbf{a}_i \to \mbf{r}_{\al} } \rp \mbf{\tilde{A}} \cdot d\mbf{r} \ . \\
\eea
Notice that the first integral is of the same form as in $T_i(+\Phi/2)$, and the second is over a loop given by a Peierls path from $\mbf{r}_{\al}$ to $\mbf{r}_{\al} + \mbf{a}_i$ (which may be taken along an arbitrary sequence of Peierls paths because $\curl \mbf{\tilde{A}} = \Phi$) and then a straight-line path back from $\mbf{r}_{\al} + \mbf{a}_i$ to $\mbf{r}_{\al} $. We denote the area enclosed by such a path as $\Omega_{i, \al}$, as shown for example in \Fig{fig:UTareafigs}a. Continuing, we find 
\bea
\label{eq:ppeqstart}
- \int_{straight}^{\mbf{r}_{\al} \to \mbf{r}_{\al} + \mbf{a}_i} \mbf{A} \cdot d\mbf{r} + \int_{P.P}^{\mbf{r}_{\al} \to \mbf{R} + \mbf{a}_i + \pmb{\delta}_\al} \mbf{\tilde{A}} \cdot d\mbf{r} &= \int_{straight}^{\mbf{r}_{\al} \to \mbf{r}_{\al} + \mbf{a}_i} \mbf{A} \cdot d\mbf{r} +\oint_{\del \Omega_{i,\al}} \mbf{\tilde{A}} \cdot d\mbf{r} \\ % 
&= \int_{straight}^{\mbf{r}_{\al} \to \mbf{r}_{\al} + \mbf{a}_i} \mbf{A} \cdot d\mbf{r} +\oint_{\Omega_{i, \al}} \Phi dS \\ 
&=\int_{straight}^{\mbf{r}_{\al} \to \mbf{r}_{\al} + \mbf{a}_i} \mbf{A} \cdot d\mbf{r} +\Phi \Omega_{i, \al} \\ 
\eea
and thus we find
\bea
U T_i(-\Phi/2) U^\dag  &= \sum_{\mbf{R} \al} \exp \lp  i \Phi \Omega_{i, \al} + i \int_{\mbf{R} + \pmb{\delta}_\al}^{\mbf{R} + \pmb{\delta}_\al + \mbf{a}_i} \mbf{A} \cdot d\mbf{r} \, - \, i \chi_i(\mbf{R}+\pmb{\delta}_\al ) % 
\rp c^\dag_{\mbf{R}+\mbf{a}_i, \al} c_{\mbf{R},\al} \ . \
\eea
We want to recover an expression proportional to the magnetic translation operator $T_i(\Phi/2)$. To make progress, we note that at $\phi = \Phi/2$, \Eq{appeq:T} gives $\chi_i(\mbf{r}) = \Phi/2 (\mbf{a}_i \times \mbf{r})$ and hence
\bea
- \chi_i(\mbf{R}+\pmb{\delta}_\al) &= \chi_i(\mbf{R}+\pmb{\delta}_\al) - \Phi \mbf{a}_i \times (\mbf{R}+\pmb{\delta}_\al) \\
&= \chi_i(\mbf{R}+\pmb{\delta}_\al) - \Phi \mbf{a}_i \times \pmb{\delta}_\al \mod 2\pi \\
\eea
and hence
\bea
\label{eq:UTbaralmost}
U T_i(-\Phi/2) U^\dag  &= \sum_{\mbf{R} \al} \exp \lp  i \Phi (\Omega_{i, \al} + \pmb{\delta}_\al \times \mbf{a}_i ) 
+ i \int_{\mbf{R} + \pmb{\delta}_\al}^{\mbf{R} + \pmb{\delta}_\al + \mbf{a}_i} \mbf{A} \cdot d\mbf{r} \, + \, i \chi_i(\mbf{R}+\pmb{\delta}_\al ) \rp c^\dag_{\mbf{R}+\mbf{a}_i, \al} c_{\mbf{R},\al} \ . \
\eea
We can now define the momentum shift $\pmb{\kappa}_{\al}$ by
\bea
\label{eq:kappadef}
\kappa_{i,\al} &= \Phi (\Omega_{i, \al} + \pmb{\delta}_\al \times \mbf{a}_i ) \ . \\
\eea
It appears that the momentum shift depends on the orbital location. However, we will now prove that, in fact
\bea
\label{eq:diffkappa0}
\pmb{\kappa}_{\al} -\pmb{\kappa}_{\be} &= \Phi \big( \Omega_{i, \al} - \Omega_{i, \be } + (\pmb{\delta}_\al  - \pmb{\delta}_{\be} ) \times \mbf{a}_i \big)  = 0 \mod 2\pi  \\
\eea
To prove \Eq{eq:diffkappa0}, we recall that any integral of $\mbf{\tilde{A}}$ over a closed loop of Peierls paths is equal a multiple of $2\pi$. We consider a closed loop $\del \mathcal{M}$ that is formed of Peierls paths connecting $\pmb{\delta}_{\al}, \pmb{\delta}_{\al} + \mbf{a}_i, \pmb{\delta}_{\be} + \mbf{a}_i, \pmb{\delta}_{\be}$ in sequence, which we depict in \Fig{fig:UTareafigs}a. Importantly, the Peierls paths connecting $\pmb{\delta}_{\al}, \pmb{\delta}_{\be}$ and $ \pmb{\delta}_{\al} + \mbf{a}_i, \pmb{\delta}_{\be} + \mbf{a}_i$ can be chosen to be the same shape since they are related by a lattice vector. Now, as shown in \Fig{fig:UTareafigs}b we can separate the line integral over $\del \mathcal{M}$ with \emph{straight-line} paths connecting $\pmb{\delta}_{\al}, \pmb{\delta}_{\al} + \mbf{a}_i$ and $\pmb{\delta}_{\be}, \pmb{\delta}_{\be} + \mbf{a}_i$:
\bea
\label{eq:intUTdeform}
0 &= \oint_{\del \mathcal{M}} \mbf{\tilde{A}} \cdot d\mbf{r} \mod 2\pi \\
&= \oint_{\del \Omega_{i,\al}} \mbf{\tilde{A}} \cdot d\mbf{r} +  \oint_{\del \mathcal{R}} \mbf{\tilde{A}} \cdot d\mbf{r} +  \oint_{-\del \Omega_{i,\be}} \mbf{\tilde{A}} \cdot d\mbf{r} \mod 2\pi \\
&= \Phi \oint_{\Omega_{i,\al}} dS + \Phi \oint_{\mathcal{R}} dS - \Phi \int_{\Omega_{i,\be}} dS \mod 2\pi \\
&= \Phi ( \Omega_{i,\al} - \Omega_{i,\be} ) + \Phi \oint_{\mathcal{R}} dS \mod 2\pi \\
\eea
where $\mathcal{R}$, shown in \Fig{fig:UTareafigs}c, is the area formed by the straight-line paths and the Peierls paths connecting $\pmb{\delta}_{\al}, \pmb{\delta}_{\be}$ and $ \pmb{\delta}_{\al} + \mbf{a}_i, \pmb{\delta}_{\be} + \mbf{a}_i$. Because these Peierls paths are the same shape, $\mathcal{R}$ can be deformed to a parallelogram without changing its area. Hence the area of $\mathcal{R}$ is that of a parallelogram with sides given by $\mbf{a}_i$ and $\pmb{\delta}_{\be} - \pmb{\delta}_{\al}$. Thus, \Eq{eq:intUTdeform} reads
\bea
 \Phi ( \Omega_{i,\al} - \Omega_{i,\be} ) + \mbf{a}_i \times (\pmb{\delta}_{\be} - \pmb{\delta}_{\al}) &= 0 \mod 2\pi, 
\eea
proving \Eq{eq:diffkappa0} . 
Hence we can calculate $\pmb{\kappa} = \pmb{\kappa}_{\al}$ at \emph{any} orbital location. However, we must be consistent with the choice of origin which defines $\pmb{\delta}_{\al}$. Shifting the origin changes both $\pmb{\kappa}$ and the overall phase of $T_i(\Phi/2)$, which we have fixed throughout. Note however that \Eq{eq:diffkappa0} remains invariant under such a shift. 

With $\pmb{\kappa} = \kappa_1 \mbf{b}_1 + \kappa_2 \mbf{b}_2 $ determined, we return to \Eq{eq:UTbaralmost} to obtain 
\bea
\label{eq:UTalgtemp}
U \mathcal{T} T_i(\Phi/2) (U \mathcal{T})^{-1}  &=  \left[ \sum_{\mbf{R} \al} \exp \lp  i \kappa_i + i \int_{\mbf{R} + \pmb{\delta}_\al}^{\mbf{R} + \pmb{\delta}_\al + \mbf{a}_i} \mbf{A} \cdot d\mbf{r} \, + \, i \chi_i(\mbf{R}+\pmb{\delta}_\al ) \rp c^\dag_{\mbf{R}+\mbf{a}_i, \al} c_{\mbf{R},\al} \right]  \\
& = \exp \lp i \pmb{\kappa} \cdot \mbf{b}_i \rp  T_i(\Phi/2)  \ .
\eea
We remark that this result, the algebra of $U\mathcal{T}$ and $T_i(\Phi/2)$ holds for $n$ even or odd, and regardless of the sign of $\mathcal{T}^2$. Now assuming $n$ odd so there is a $1\times 2$ unit cell, (\Eq{eq:UTalgtemp}) lets us calculate
\bea
\label{eq:UTstates}
U \mathcal{T} \ket{k_1, k_2,\ell,\al} &=  \frac{1}{\sqrt{\mathcal{N}/2}} \sum_{mn} e^{+i m k_1 + 2i n k_2} U \mathcal{T} T_1^m T_2^{2n} (U \mathcal{T})^{-1} U \mathcal{T} c^\dag_{\ell \mbf{a}_2, \al} \ket{0} \\
&= \frac{1}{\sqrt{\mathcal{N}/2}}  \sum_{mn} e^{+i m k_1 + 2i n k_2} e^{i m \kappa_{1} + i 2n  \kappa_{2}} T_1^m T_2^{2n}  U \mathcal{T}  c^\dag_{\ell \mbf{a}_2, \al}  \ket{0} \\
&= \frac{1}{\sqrt{\mathcal{N}/2}}  \sum_{mn} e^{+i m (k_1 + \kappa_{1} ) + 2i n (k_2 + \kappa_{2})} T_1^m T_2^{2n} U \mathcal{T}   c^\dag_{\ell \mbf{a}_2, \al}  \ket{0} \\
&= \sum_{\al',\ell'} \ket{-(k_1 + \kappa_{1}) , -(k_2 + \kappa_{2}), \ell', \al'} K [D^\dag(U \mathcal{T})]_{\ell'\al',\ell \al}\\
\eea
where in the last line we defined $D^\dag(U \mathcal{T})$, the single-particle representation of $U \mathcal{T}$ (with $K$ denoting complex conjugation). This follows from 
\bea
U \mathcal{T}c^\dag_{\ell \mbf{a}_2, \al} \ket{0} &= U \mathcal{T}c^\dag_{\ell \mbf{a}_2, \al} \mathcal{T}^{-1} U^\dag U \mathcal{T} \ket{0} \\
&= U \mathcal{T}c^\dag_{\ell \mbf{a}_2, \al} \mathcal{T}^{-1} U^\dag  \ket{0} K \\
&= \sum_{\al'} U c^\dag_{\ell \mbf{a}_2, \al'} U^\dag  [D_{\al \al'}(\mathcal{T})]^*  \ket{0} K \\
&=   \sum_{\al' \ell'} c^\dag_{\ell' \mbf{a}_2, \al'} \delta_{\ell\ell'} e^{i \int_{\mbf{r}_0}^{\ell \mbf{a}_2 +  \pmb{\delta}_{\al'}} \mbf{\tilde{A}} \cdot d \mbf{r}} \ket{0} K D_{\al \al'}(\mathcal{T}) \\
&=   \sum_{\al' \ell'} c^\dag_{\ell' \mbf{a}_2, \al'}  \ket{0} K \delta_{\ell\ell'} e^{-i \int_{\mbf{r}_0}^{\ell \mbf{a}_2 +  \pmb{\delta}_{\al'}} \mbf{\tilde{A}} \cdot d \mbf{r}} D_{\al \al'}(\mathcal{T}) \\
&= \sum_{\al' \ell'} c^\dag_{\ell' \mbf{a}_2, \al'}  \ket{0} K  [D^\dag(U \mathcal{T}) ]_{\ell' \al', \ell \al}, \qquad  [D^\dag(U \mathcal{T}) ]_{\ell' \al', \ell \al} = \delta_{\ell\ell'} e^{-i \int_{\mbf{r}_0}^{\ell \mbf{a}_2 +  \pmb{\delta}_{\al'}} \mbf{\tilde{A}} \cdot d \mbf{r}} D_{\al \al'}(\mathcal{T})  \\
\eea
with $\ell = 0,1$ and $\al = 1, \dots, N_{orb}$ in a tensor product basis. 

\begin{figure}
 \centering
  \includegraphics[width=5.6cm, trim = 0 -.9cm 0 0, clip]{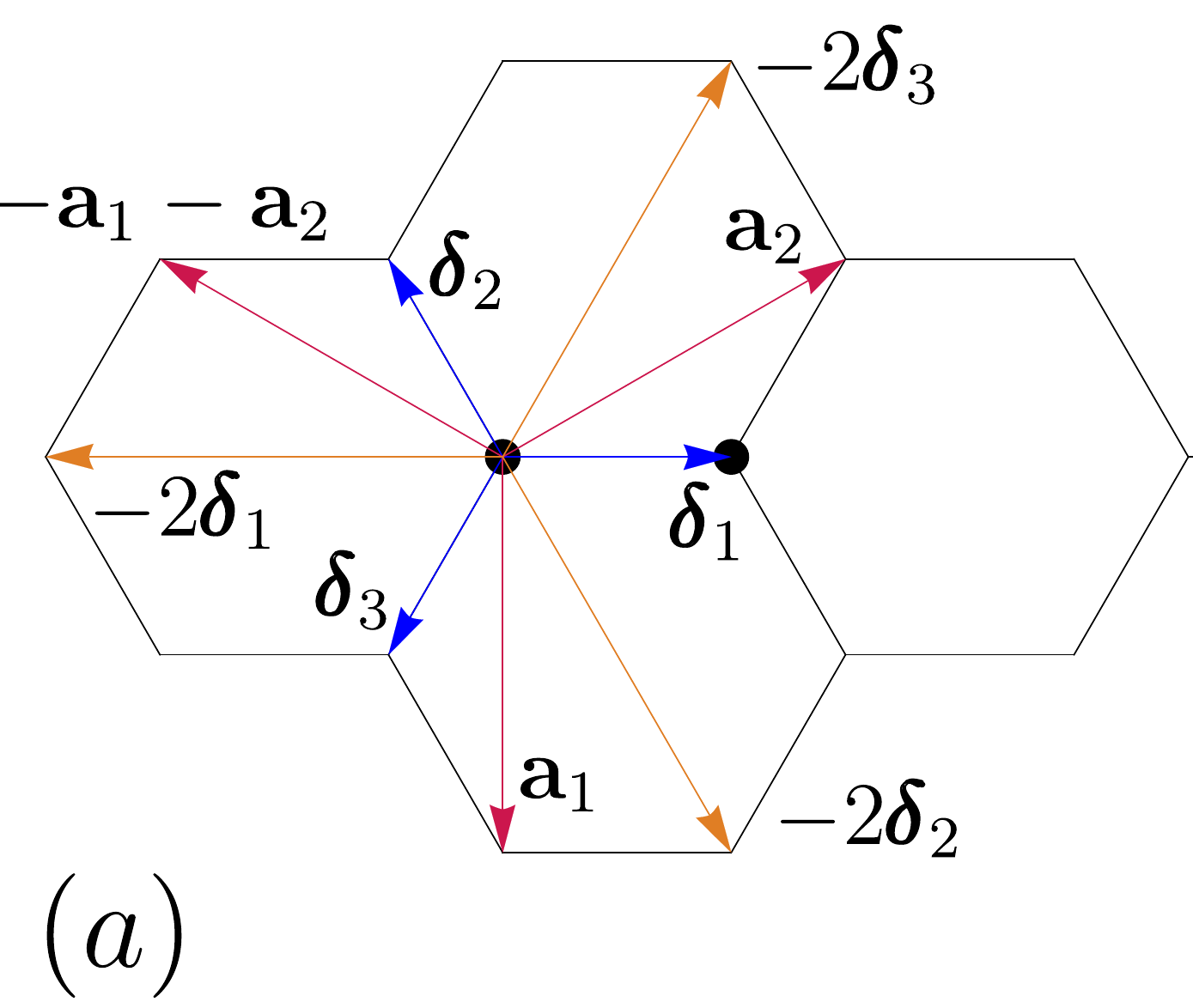}  
 \includegraphics[width=6.cm]{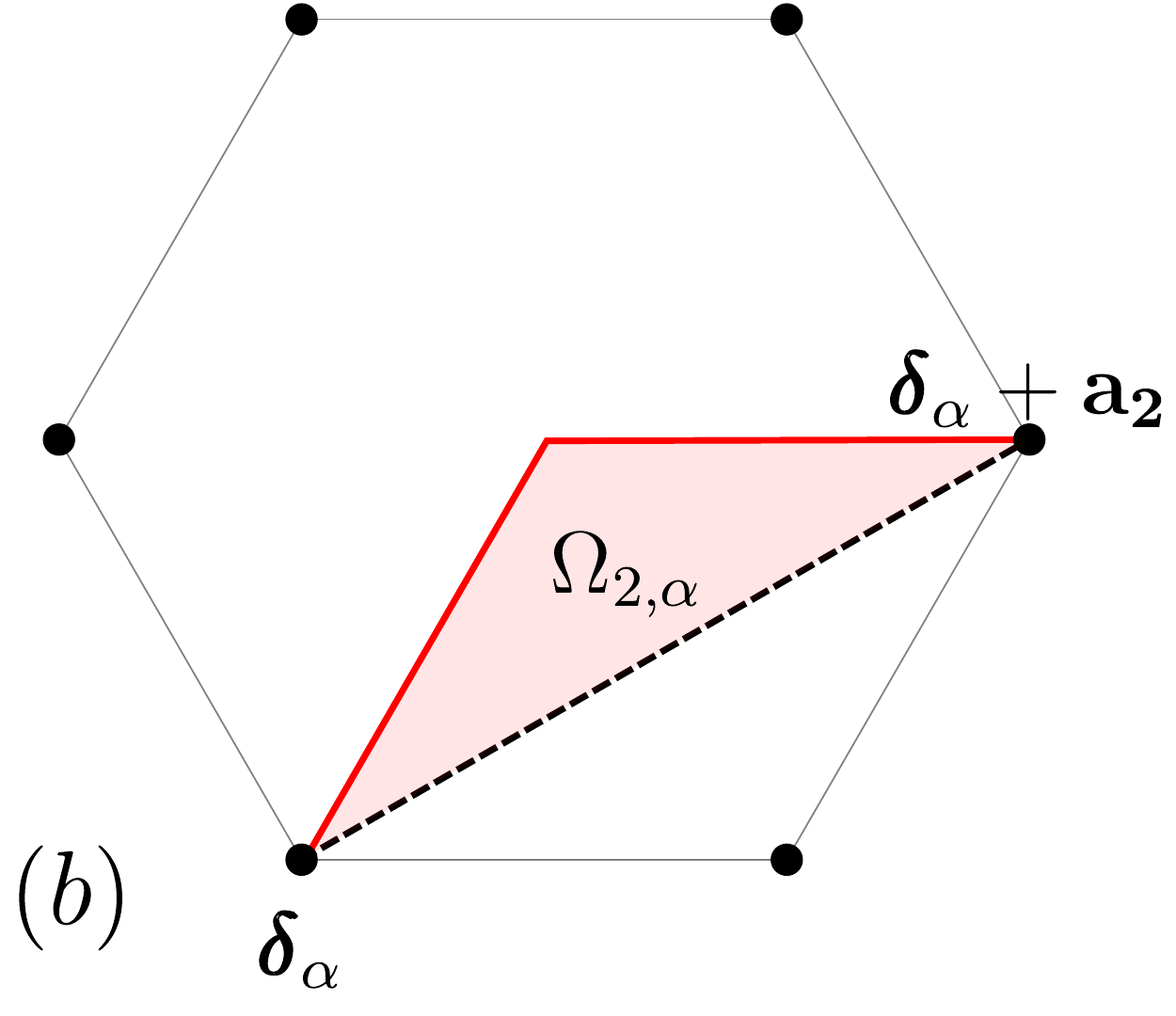}  
  \includegraphics[width=5.35cm]{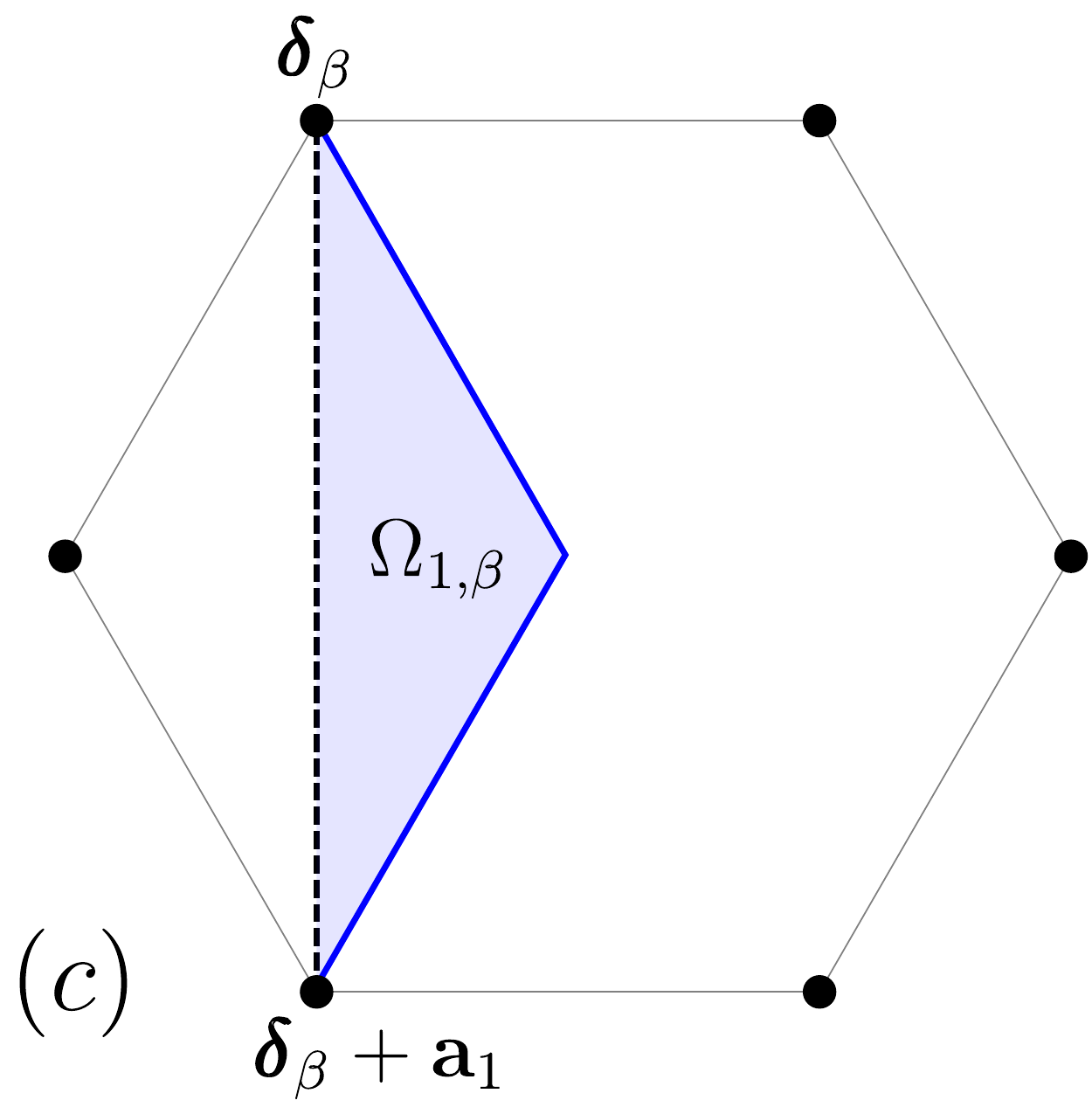} 
 \caption{$(a)$ We show the lattice vectors and nearest neighbor vectors of the TBG model, replicated from \Fig{fig:magtranspic}c. $(b)$ We show the area $\Omega_{1,\al} = 1/6$ enclosed by the Peierls Paths (shown in solid red) and the straight-line path in dashed black for a given orbital $\pmb{\delta}_{\al}$. We recall that $\Phi = 6\pi$, as discussed in detail in \App{app:TBG}, so $\Phi \Omega_{1,\al} = \pi$. In the convention where the origin is at the center of the honeycomb, we calculate $\Phi \pmb{\delta}_{\al} \times \mbf{a}_1 = 6\pi (-\pmb{\delta}_1/2 + \mbf{a}_1/2) \times \mbf{a}_1 = 6\pi \times 1/3 = 0 \mod 2\pi$, and hence from \Eq{eq:kappadef}, we find $\kappa_1 = \pi$. $(c)$ We show the area $\Omega_{2,\be} = 1/6$ enclosed by the Peierls Paths (shown in solid blue) and the straight-line path in dashed black for a given orbital $\pmb{\delta}_{\be}$. We recall that $\Phi = 6\pi$, as discussed in detail in \App{app:TBG}, so $\Phi \Omega_{2,\be} = \pi$.  In the convention where the origin is at the center of the honeycomb, we calculate $\Phi \pmb{\delta}_{\be} \times \mbf{a}_2 = 6\pi (-\pmb{\delta}_1/2 - \mbf{a}_1/2) \times \mbf{a}_2 = 6\pi \times -2/3 = 0 \mod 2\pi$, and hence from \Eq{eq:kappadef}, we find $\kappa_2 = \pi$.}
\label{fig:kappatbg}
\end{figure}

To determine the action of $U \mathcal{T}$ on the Hofstadter Hamiltonian, we use the fact that $U\mathcal{T}$ commutes with $H^{\Phi/2}$, giving
\bea
\mathcal{H}^{\Phi/2}_{\ell\al, \ell' \be}(\mbf{k}) 
&= \sum_{\ell' \al',r' \be'}  [D(U \mathcal{T}) ]_{\ell \al,\ell' \al'}   \mathcal{H}^{\Phi/2 \ *}_{\ell'\al', r' \be'}(-\mbf{k} - \pmb{\kappa})  [D^\dag(U \mathcal{T}) ]_{r' \be', r \be}  \ .  \\ 
\eea
We recognize the usual form of an anti-unitary symmetry acting on a single-particle Hamiltonian, but with a momentum shift that comes from the algebra of $U\mathcal{T}$ and $T(\Phi/2)$. We mention that the full symmetry algebra of $H^{\Phi/2}$, 
\bea
U \mathcal{T}, T_i(\Phi/2)  
\eea
forms a \emph{projective} representation of the magnetic space group $1'$ at $\phi = 0$ with the projective phases $\phi$ and $\pmb{\kappa}$ given by
\bea
\label{eq:UTprojective}
T_1(\Phi/2) T_2(\Phi/2) = e^{i \Phi/2} T_2(\Phi/2) T_1(\Phi/2), \qquad U \mathcal{T} T_i(\Phi/2) = e^{ i \pmb{\kappa} \cdot \mbf{b}_i} T_i(\Phi/2) U \mathcal{T} \ .
\eea
For instance, we can consider the model of twisted bilayer graphene introduced in \App{app:TBG}. In \Fig{fig:kappatbg}, we show that, due to the Peierls paths being taken through the center of the honeycomb, we find nonzero expression for $\Omega_{i,\al}$. Choosing a convention where the origin is fixed at the center of the honeycomb, we find that \Eq{eq:kappadef} yields $\pmb{\kappa} = (\pi,\pi)$. 

\subsection{Wilson Loop Proof of a Nontrivial $\mathds{Z}_2$ Index with $\mathcal{T}$ symmetry}
\label{app:wilsonz2}

The crux of our proof in Sec. III is the establishment that the $\mathds{Z}_2$ Kane-Mele invariant of $H^{\Phi/2}$ is trivial. \App{app:genz2proof} demonstrated this using the action of the magnetic translation group on the BZ. Here, we prove the same result using the Wilson loop. We assume that the magnetic periodicity is given $\Phi = 2\pi n$, $n$ odd and we will prove that $\delta^{\phi= \Phi/2} = +1$. In \App{app:TRissue}, we show when $n$ is even, $\delta^{\phi =\Phi/2}$  is \emph{not} fixed to be $+1$. 

\subsubsection{$U\mathcal{T}$ Constraints on the Wilson Loop}
\label{app:UTwilsonnew}

To begin, we study the constraint imposed by $U \mathcal{T}$ on the Wilson loop at $\phi = \Phi/2$ in the $1 \times 2$ gauge-invariant magnetic unit cell. To begin, we need to work in an eigenbasis of $H^{\Phi/2}, T_1(\Phi/2), T_2^2(\Phi/2)$ which is given by
\bea
\ket{m, \mbf{k}} &= \sum_{\ell \al}  \ket{\mbf{k}, \ell, \al} [\mathcal{U}^\dag(\mbf{k})]_{\ell \al, m} , \quad \ell = 0,1 \ , \al = 1, \dots, N_{orb}, \text{ and } m = 1, \dots, 2 N_{orb} \\
\eea
where $[\mathcal{U}(\mbf{k})]_{\ell \al, m}$ is a $2 N_{orb} \times 2 N_{orb}$ unitary matrix which relates the $\ell, \al$ basis (in \Eq{eq:eigenstatesUT}) to the $m$ energy eigenbasis of $\mathcal{H}^{\Phi/2}(\mbf{k})$. To follow the method of \App{app:wilson}, we write define $\ket{u_j^{\Phi/2}(\mbf{k})}$, the eigenstate of $\mathcal{H}^{\Phi/2}(\mbf{k})$, by
\bea
\braket{\mbf{k} ,\ell,\al|u_j^{\Phi/2}(\mbf{k})} = [U(\mbf{k})]_{\ell \al, j}
\eea 
where $j = 1, \dots, 2 N_{occ}$, assuming the energy spectrum is gapped for $\phi \in (0, \Phi/2)$ and we can separate the occupied bands. For convenience, we define the unitary matrix $Q = D(U\mathcal{T})$ which, by \Eq{eq:HkappaH}, satisfies
\bea
Q K \mathcal{H}^{\Phi/2}(\mbf{k}) K Q^\dag =  \mathcal{H}^{\Phi/2}(-\mbf{k} - \pmb{\kappa}) \\
\eea
where $K$ is complex conjugation. Following \App{app:wilson}, we now form the unitary sewing matrix
\bea
\label{eq:BcalUT}
\mathcal{B}^{ij}_\mbf{k} &= \braket{u^{\Phi/2}_i(-\mbf{k} - \pmb{\kappa}) | Q K |u^{\Phi/2}_j(\mbf{k})} \\
&= \braket{u^{\Phi/2}_i(\mbf{-k}- \pmb{\kappa}) | Q | u^{\Phi/2 \, * }_j(\mbf{k}) }  \\ 
\eea
which connects eigenstates at TRS momenta \cite{2017PhRvB..96x5115B}:
\bea
\ket{u^{\Phi/2}_j(\mbf{k})} &= \sum_i Q \ket{u^{\Phi/2 \, *}_i(\mbf{-k}- \pmb{\kappa})} [\mathcal{B}^\dag]^{ij}_{-\mbf{k}- \pmb{\kappa}} \ .
\eea
Using this relation, we determine that a small segment of a Wilson loop obeys
\bea
\label{eq:wilsonUTsymmetry}
\null [W^{\Phi/2}_{\mbf{k}_1' \leftarrow \mbf{k}_1}]^{ij} &= \braket{u^{\Phi/2}_i(\mbf{k}_1') | u^{\Phi/2}_j(\mbf{k}_1)} \\
&= \sum_{rs} \mathcal{B}^{ir}_{-\mbf{k}_1'- \pmb{\kappa}} \braket{ u^{\Phi/2 \, *}_r(-\mbf{k}_1'- \pmb{\kappa}) | Q^\dag Q |u^{\Phi/2 *}_s(-\mbf{k}_1- \pmb{\kappa} ) }[\mathcal{B}^\dag_{-\mbf{k}_1- \pmb{\kappa}}]^{sj} \\
&= \sum_{rs} \mathcal{B}^{ir}_{-\mbf{k}_1'- \pmb{\kappa}}  [W^{\Phi/2 \, * }_{-\mbf{k}_1' - \pmb{\kappa} \leftarrow -\mbf{k}_1- \pmb{\kappa}}]_{rs} [\mathcal{B}^\dag_{-\mbf{k}_1- \pmb{\kappa}}]^{sj} \ .
\eea
We now want to consider a full Wilson loop. Because of the shift due to $\pmb{\kappa}$, we will find it convenient to shift the origin of the Wilson loop. Hence we define $\mbf{\tilde{k}}$ by $\mbf{k} = - \pmb{\kappa}/2 + \mbf{\tilde{k}}$ and define a Wilson loop $\tilde{W}^{\Phi/2}(\tilde{k}_1)$ by integrating along $\tilde{k}_2$:
\bea
\tilde{W}^{\Phi/2}(\tilde{k}_1) \equiv W^{\Phi/2}_{(-\kappa_1/2 + \tilde{k}_1, -\kappa_2/2 +\pi/2) \leftarrow (-\kappa_1/2 + \tilde{k}_1, -\kappa_2/2 -\pi/2)} \ . \\
\eea
 Using \Eq{eq:wilsonUTsymmetry}, we find
\bea
\label{eq:UTwilson}
\tilde{W}^{\Phi/2}(\tilde{k}_1) &= \mathcal{B}_{(-\kappa_1/2 - \tilde{k}_1, -\kappa_2/2 - \pi/2)} \tilde{W}^{\Phi /2 *}_{(-\kappa_1/2 - \tilde{k}_1, -\kappa_2/2 -\pi/2) \leftarrow (-\kappa_1/2 - \tilde{k}_1, -\kappa_2/2 +\pi/2)} \mathcal{B}^\dag_{(-\kappa_1/2 - \tilde{k}_1, -\kappa_2/2 +\pi/2)} \\ 
&= \mathcal{B}_{(-\kappa_1/2 - \tilde{k}_1, -\kappa_2/2 - \pi/2)} \tilde{W}^{\Phi /2 * \, \dag}_{ (-\kappa_1/2 - \tilde{k}_1, -\kappa_2/2 +\pi/2) \leftarrow (-\kappa_1/2 - \tilde{k}_1, -\kappa_2/2 -\pi/2)} \mathcal{B}^\dag_{(-\kappa_1/2 - \tilde{k}_1, -\kappa_2/2 +\pi/2)} \\ 
&= \mathcal{B}_{(-\kappa_1/2 - \tilde{k}_1, -\kappa_2/2 - \pi/2)} [\tilde{W}^{\Phi /2}(-\tilde{k}_1)]^T \mathcal{B}^\dag_{(-\kappa_1/2 - \tilde{k}_1, -\kappa_2/2 +\pi/2)} \\ 
\eea
We now observe that from the construction of the translation eigenstates in \Eq{eq:explicitTstates}, we have $\ket{k_1, k_2, \ell, \al} = \ket{k_1+2\pi, k_2, \ell, \al} = \ket{k_1, k_2+\pi, \ell, \al}$, and thus from \Eq{eq:Hperiodic}, we find that $\mathcal{H}^{\Phi/2}(k_1, k_2) = \mathcal{H}^{\Phi/2}(k_1 + 2\pi, k_2) = \mathcal{H}^{\Phi/2}(k_1, k_2+\pi)$. In other words, the embedding matrices that enforce the magnetic BZ periodicity are trivial. This is due to the choice of the eigenbasis in \Eq{eq:explicitTstates}, which differ from the bases we considered in \Eq{eq:eigenstateslandau} in the Landau gauge. Hence, we have 
\bea
\mathcal{B}^{ij}_{(-\kappa_1/2 - \tilde{k}_1, -\kappa_2/2 +\pi/2)} 
&= \mathcal{B}^{ij}_{(-\kappa_1/2 - \tilde{k}_1, -\kappa_2/2 -\pi/2)} \ . 
\eea
Returning to \Eq{eq:UTwilson} with this result, we find
\bea
\label{eq:WilsonUTresult}
\tilde{W}^{\Phi/2}(\tilde{k}_1) &= \mathcal{B}_{(-\kappa_1/2 - \tilde{k}_1, -\kappa_2/2 -\pi/2)}[\tilde{W}^{\Phi/2}(-\tilde{k}_1)]^T \mathcal{B}^\dag_{(-\kappa_1/2 - \tilde{k}_1, -\kappa_2/2 -\pi/2)} \ .
\eea
We define the eigenvalues of $\tilde{W}^{\Phi/2}(\tilde{k}_1)$ as $\{ \exp i  \tilde{\vartheta}_j(\tilde{k}_1) \} $. Using \Eq{eq:WilsonUTresult}, we find the usual constraint on the Wannier centers: 
\bea
\label{eq:thetaUT}
\{ \tilde{\vartheta}_j(\tilde{k}_1) \} = \{ \tilde{\vartheta}_j(-\tilde{k}_1) \}
\eea
emphasizing that $\tilde{k}_1$ is the coordinate measured from $-\pmb{\kappa}/2$, i.e. $ \tilde{\vartheta}_j(\tilde{k}_1) $ is the Wilson eigenvalue of the loop at $-\pmb{\kappa}/2 + \tilde{k}_1 \mbf{b}_1$. By re-centering the magnetic BZ around $-\pmb{\kappa}/2$, we have found that the $U\mathcal{T}$ symmetry behaves like $\mathcal{T}$ at $\phi = 0$ on the shifted Wilson loop $\tilde{W}(\tilde{k}_1)$.
We now study the constraint on the Wilson loop spectrum due to the $\Phi/2$ periodicity of $\tilde{k}_1$ (recalling that we take $n$ odd here). From \Eq{eq:genericpishift}, we have
\bea
V_1(\Phi/2, k_2) \mathcal{H}^{\Phi/2}(\mbf{k}) V_1^\dag(\Phi/2, k_2) =  \mathcal{H}^{\Phi/2}(\mbf{k} +\pi \mbf{b}_1) \ . \\
\eea
Following \App{app:wilson}, we now form the unitary sewing matrix 
\bea
\label{eq:BV1gaugeinv}
B^{ij}_\mbf{k} &= \braket{u^{\Phi/2}_i(\mbf{k} + \pi \mbf{b}_1 ) | V_1(\Phi/2, k_2) |u^{\Phi/2}_j(\mbf{k})} \\
\eea
which is not to be confused with $\mathcal{B}_{\mbf{k}}$, the sewing matrix for $U\mathcal{T}$ defined in \Eq{eq:BcalUT}. $B_{\mbf{k}}$ which connects eigenstates that are $\pi \mbf{b}_1$ apart \cite{2017PhRvB..96x5115B}:
\bea
\ket{u^{\Phi/2}_j(\mbf{k})} &=  \sum_i V_1(\Phi/2, k_2) \ket{u^{\Phi/2}_i(\mbf{k} + \pi \mbf{b}_1)} [B^\dag]^{ij}_{\mbf{k} + \pi \mbf{b}_1} \ .
\eea
Following the same discussion in \Eq{eq:wilsonUTsymmetry}, we find
\bea
\label{eq:WgaugeinvV2}
\tilde{W}^{\Phi/2}(\tilde{k}_1) &= B_{(-\kappa_1/2 + \tilde{k}_1 + \pi, -\kappa_2/2 + \pi/2)} \tilde{W}^{\Phi /2}(\tilde{k}_1 + \pi) B^\dag_{(-\kappa_1/2 + \tilde{k}_1 +\pi, -\kappa_2/2 -\pi/2)} \ . \\ 
\eea
The two embedding matrices $B_{\mbf{k}}$ and $B_{\mbf{k} - \pi \mbf{b}_2}$ are identical:
\bea
B^{ij}_{\mbf{k}-\pi\mbf{b}_2} &= \braket{u^{\Phi/2}_i(\mbf{k} + \pi \mbf{b}_1 - \pi \mbf{b}_2 ) | V_1(\Phi/2,k_2- \pi ) |u^{\Phi/2}_j(\mbf{k}- \pi \mbf{b}_2)} \\
&= \braket{u^{\Phi/2}_i(\mbf{k} + \pi \mbf{b}_1) | V_1(\Phi/2, k_2) |u^{\Phi/2}_j(\mbf{k})} \\
&= B^{ij}_{\mbf{k}} 
\eea
where we have used that the eigenstates (\Eq{eq:explicitTstates}) and the embedding matrix $V_1(\Phi/2, k_2)$ are periodic in $\pi \mbf{b}_2$, as we can see from \App{app:UTdetails}. Thus in the basis of \Eq{eq:explicitTstates}, we find that \Eq{eq:WgaugeinvV2} enforces 
\bea
\label{eq:thetaV2}
\{ \tilde{\vartheta}_j(\tilde{k}_1) \} = \{ \tilde{\vartheta}_j(\tilde{k}_1 + \pi) \}
\eea
which is in contrast to \Eq{eq:wilsonshift}, which was performed in the Landau gauge. This is due to a different choice of eigenstates. In the Landau gauge, we chose to diagonalize the Hamiltonian in the basis of \Eq{eq:eigenstateslandau} on which $T_1(0)$ was diagonal (see \App{app:hofhamconstruct}). In contrast, the basis constructed in \Eq{eq:explicitTstates} is diagonal under $T_1(\Phi/2)$. 

\subsubsection{$\mathds{Z}_2$ Index from the Wilson Loop}

We will now see that, given \Eqs{eq:thetaUT}{eq:thetaV2}, the $\mathds{Z}_2$ invariant must be trivial. \Ref{2011PhRvB..84g5119Y} provides the following method of calculating $\delta^{\phi = \Phi/2}$ from the Wilson loop eigenvalues. Draw an arbitrary line of arbitrary constant $\tilde{\vartheta}(\tilde{k}_1) = \tilde{\vartheta}^*$ through the \emph{half} Wilson spectrum $\{\tilde{\vartheta}_j(\tilde{k}_1)\}, \ \tilde{k}_1 \in(-\pi,0)$ and count the number of times it crosses the Wilson bands; $\delta^{\phi = \Phi/2}$ is trivial if there are an even number of crossings, and non-trivial is there are are odd number. We emphasize $\mbf{\tilde{k}}$ refers to the momentum $\mbf{k} = - \pmb{\kappa}/2 + \mbf{\tilde{k}}$ in the magnetic BZ. We will prove that all crossings of the Wilson loop spectrum with the arbitrary line $\tilde{\vartheta}'$ occur in pairs, and hence $\delta^{\phi = \Phi/2}$ is trivial. 

Suppose a crossing occurs at the point $(\tilde{k}_1^*, \tilde{\vartheta}^*)$ with $\tilde{k}_1^* \in (-\pi,0)$, so there is a Wilson band satisfying $\tilde{\vartheta}(\tilde{k}_1^*) = \tilde{\vartheta}^*$. By $U \mathcal{T}$, \Eq{eq:thetaUT} ensures there is also a band satisfying $\tilde{\vartheta}(-\tilde{k}_1^*) = \tilde{\vartheta}^*$. Then, using \Eq{eq:thetaUT}, there are bands satisfying $\tilde{\vartheta}(-\tilde{k}_1^*-\pi) = \tilde{\vartheta}^*$ and $\tilde{\vartheta}(\tilde{k}_1^*+\pi) = \tilde{\vartheta}^*$. Note that generically, $\tilde{k}_1^*$ and $-\tilde{k}_1^* - \pi$ are distinct points and $-\tilde{k}_1^* - \pi \in (-\pi, 0)$ (see \Fig{fig:magBZkappa}). Thus each crossing at $\tilde{k}_1^*$ comes with a partner at $-\tilde{k}_1^* - \pi$ and the total number of crossing in the half spectrum $\tilde{k}_1 \in (-\pi,0)$ must be even. We can always avoid a crossing at the degenerate point $\tilde{k}_1^* = - \pi/2 = -\tilde{k}_1^* - \pi$ where this argument breaks down by changing $\tilde{\vartheta}^*$, since it is arbitrary.

\section{ $C_{2z} \mathcal{T}$ symmetry}

In this Appendix, we discuss the Hofstadter topological phases protected by $C_{2z} \mathcal{T}$ symmetry. First, we demonstrate that $(U C_{2z} \mathcal{T})^2 = \pm (C_{2z} \mathcal{T})^2$ where the sign is determined by an integral along Peierls paths (\App{app:UCT}). We discuss the topological invariants in both cases. If $(U C_{2z} \mathcal{T})^2 = +1$, the $w_2$ invariant at $\Phi/2$ can be calculated in an expanded unit cell where $U$ is diagonal in momentum space (\App{app:w2calc}). When $(U C_{2z} \mathcal{T})^2 = -1$, we argue the phase must be trivial in real space, and then calculate the nested Wilson loop to show the triviality explicitly (\App{app:UCTtrivial}).

\subsection{Symmetry Properties}
\label{app:UCT}

In \Sec{sec:fragile}, we consider Hamiltonians with the symmetry $C_{2z} \mathcal{T}$ at $\phi = 0$ which satisfies $(C_{2z} \mathcal{T})^2 = +1$. At $\phi = \Phi/2$, the symmetry of $H^{\Phi/2}$ is $U C_{2z} \mathcal{T}$, which may square to either $\pm1$. We derive a formula for this sign as follows. First, the action of $C_{2z}$ on the annihilation operator reads
\bea
C_{2z}^\dag c_{\mbf{R},\al}  C_{2z} &= \sum_{\be} D_{\al \be}(C_{2z}) c_{-\mbf{R} - \pmb{\delta}_\al - \pmb{\delta}_\be,\be}
\eea
where $D_{\al \be}(C_{2z}) = 0$ if $\pmb{\delta}_{\al} - C_{2z} \pmb{\delta}_\be = \pmb{\delta}_{\al} + \pmb{\delta}_\be$ is not a lattice vector. We let $U = e^{i \mathcal{O}}$ as defined by \Eq{eq:hoffluxper}, and then compute 
\bea
C_{2z} \mathcal{O} &= C_{2z}  \sum_{\mbf{R} \be} c^\dag_{\mbf{R}, \be} c_{\mbf{R}, \be}  \int^{\mbf{R}+ \pmb{\delta}_\be}_{\mbf{r}_0}  \mbf{\tilde{A}}  \cdot d\mbf{r}, \\
&= \sum_{\mbf{R} \be}  C_{2z} c^\dag_{\mbf{R}, \be}  C_{2z}^\dag  C_{2z}c_{\mbf{R}, \be} C_{2z}^\dag C_{2z} \int^{\mbf{R}+ \pmb{\delta}_\be}_{\mbf{r}_0}  \mbf{\tilde{A}}  \cdot d\mbf{r} \\
&= \lp \sum_{\mbf{R} \be \be' \be''}  c^\dag_{-\mbf{R} - \pmb{\delta}_\be - \pmb{\delta}_{\be'}, \be'} [D(C_{2z})^\dag]_{\be' \be } [D(C_{2z})_{\be \be''}] c_{-\mbf{R} - \pmb{\delta}_\be - \pmb{\delta}_{\be''}, \be''}  \int^{\mbf{R}+ \pmb{\delta}_\be}_{\mbf{r}_0}  \mbf{\tilde{A}}  \cdot d\mbf{r} \rp C_{2z}  \\
\eea
where $\pmb{\nabla} \times \tilde{\mbf{A}} = \Phi$ and the integral is taken along Peierls paths. Now we take $-\mbf{R} - \pmb{\delta}_\be - \pmb{\delta}_{\be'} \to \mbf{R}$ by relabeling the sum. This takes $c_{-\mbf{R} - \pmb{\delta}_\be - \pmb{\delta}_{\be''}, \be''} \to c_{\mbf{R} + \pmb{\delta}_{\be'} - \pmb{\delta}_{\be''}, \be''}$. But because $[D(C_{2z})^\dag]_{\be' \be } [D(C_{2z})_{\be \be''}]$ is only nonzero when $\pmb{\delta}_{\be'} - \pmb{\delta}_{\be''} =0$ by unitarity, we have
\bea
C_{2z} \mathcal{O}  &=  \lp \sum_{\mbf{R} \be \be' \be''}  c^\dag_{\mbf{R}, \be'} [D(C_{2z})^\dag]_{\be' \be } [D(C_{2z})]_{\be \be''} c_{\mbf{R}, \be''}  \int^{-\mbf{R} - \pmb{\delta}_\be - \pmb{\delta}_{\be'}+ \pmb{\delta}_\be}_{\mbf{r}_0}  \mbf{\tilde{A}}  \cdot d\mbf{r} \rp C_{2z}  \\
&=  \lp \sum_{\mbf{R} \be' \be''}  c^\dag_{\mbf{R}, \be'} \lp \sum_\be [D(C_{2z})^\dag]_{\be' \be } [D(C_{2z})]_{\be \be''} \rp c_{\mbf{R}, \be''}  \int^{-\mbf{R} - \pmb{\delta}_{\be'} }_{\mbf{r}_0}  \mbf{\tilde{A}}  \cdot d\mbf{r} \rp C_{2z}  \\
&=  \lp \sum_{\mbf{R}  \be' } c^\dag_{\mbf{R}, \be'} c_{\mbf{R}, \be'}  \int^{-\mbf{R}- \pmb{\delta}_{\be'}}_{\mbf{r}_0}  \mbf{\tilde{A}}  \cdot d\mbf{r} \rp C_{2z} \ . \\
\eea
We recall for the reader that $\mbf{\tilde{A}}$ is a gauge field generating $\Phi$ flux. Hence, from \Eq{eq:peierlsdeform}, we can deform the integral along Peierls paths to find
\bea
 \int^{-\mbf{R}- \pmb{\delta}_{\be'}}_{\mbf{r}_0}  \mbf{\tilde{A}}  \cdot d\mbf{r} &=  \lp \int^{-\mbf{R}- \pmb{\delta}_{\be'}}_{-\mbf{r}_0} +  \int_{\mbf{r}_0}^{-\mbf{r}_0} \rp \mbf{\tilde{A}}  \cdot d\mbf{r} \mod 2\pi \ . \\
\eea
By a change of variables in the integral $\mbf{r} = - \mbf{s}$, we establish
\bea
\int^{-\mbf{R}- \pmb{\delta}_{\be'}}_{-\mbf{r}_0} \mbf{\tilde{A}}(\mbf{r})  \cdot d\mbf{r}  = \int^{\mbf{R} + \pmb{\delta}_{\be'}}_{\mbf{r}_0} \mbf{\tilde{A}}(-\mbf{s})  \cdot d (-\mbf{s}) = \int^{\mbf{R}+ \pmb{\delta}_{\be'}}_{\mbf{r}_0} \mbf{\tilde{A}}(\mbf{s}) \cdot d\mbf{s} 
\eea
where we have used that $\mbf{\tilde{A}}(\mbf{r})$ is an odd function of $\mbf{r}$. This allows us to write
\bea
\label{eq:U0}
C_{2z} \mathcal{O} &= \lp\mathcal{O} + \left[  \int_{\mbf{r}_0}^{-\mbf{r}_0} \mbf{\tilde{A}}  \cdot d\mbf{r} \right]  \sum_{\mbf{R} \be} c^\dag_{\mbf{R}, \be} c_{\mbf{R}, \be} \rp C_{2z}, \\
&= \lp\mathcal{O} + \left[ \int_{\mbf{r}_0}^{-\mbf{r}_0} \mbf{\tilde{A}}  \cdot d\mbf{r} \right] N \rp C_{2z} , \\
\eea
where $N$ is the total number of electrons in the many-body state. In this work, we study single-particle physics where $N =1$. In this case, we exponentiate \Eq{eq:U0} to find
\bea
\label{eq:C2zTOexp}
C_{2z} U &= C_{2z} \sum_{n=0}^\infty \frac{1}{n!} (i \mathcal{O})^n \\
&= \sum_{n=0}^\infty \frac{1}{n!} i^n \lp \mathcal{O} + \left[ \int_{\mbf{r}_0}^{-\mbf{r}_0} \mbf{\tilde{A}}  \cdot d\mbf{r} \right] \rp^n C_{2z}  \\
&= \exp \lp i \mathcal{O} + i \left[ \int_{\mbf{r}_0}^{-\mbf{r}_0} \mbf{\tilde{A}}  \cdot d\mbf{r} \right] \rp C_{2z}  \\
&= \exp \lp i \int_{\mbf{r}_0}^{-\mbf{r}_0} \mbf{\tilde{A}}  \cdot d\mbf{r} \rp  U C_{2z} 
\eea
where in the last line we use that $\mathcal{O}$ commutes with the $c$-number $ i \int_{\mbf{r}_0}^{-\mbf{r}_0} \mbf{\tilde{A}}  \cdot d\mbf{r}$. Using \Eq{eq:C2zTOexp} and the relation $\mathcal{T} U = U^\dag \mathcal{T}$ (see \Eq{eq:UTTU}), we find 
\bea
\label{eq:UCT}
(U C_{2z} \mathcal{T})^2 &= U C_{2z}  \mathcal{T} U C_{2z} \mathcal{T} \\
&= U C_{2z} U^\dag \mathcal{T} C_{2z} \mathcal{T} \\
&= \exp \lp - i \int_{\mbf{r}_0}^{-\mbf{r}_0} \mbf{\tilde{A}}  \cdot d\mbf{r} \rp C_{2z} U  U^\dag \mathcal{T} C_{2z} \mathcal{T} \\
&= \exp \lp - i \int_{\mbf{r}_0}^{-\mbf{r}_0} \mbf{\tilde{A}}  \cdot d\mbf{r} \rp C_{2z} \mathcal{T} C_{2z} \mathcal{T} \\
&= \exp \lp -i \int_{\mbf{r}_0}^{-\mbf{r}_0} \mbf{\tilde{A}}  \cdot d\mbf{r} \rp (C_{2z} \mathcal{T})^2 \ . \\
\eea
We recall that the integrals must be taken along the Peierls paths, and $\mbf{r}_0$ is a fixed but arbitrary orbital \emph{of the Hamiltonian} (see \Eq{eq:hoffluxper}). Additionally, in defining $C_{2z}$, we have fixed the origin of the lattice to coincide with the $C_{2z}$-symmetric point of $\tilde{\mbf{A}}$, i.e. $\tilde{\mbf{A}}(C_{2z}\mbf{r}) = - \tilde{\mbf{A}}(\mbf{r})$. This fixes the origin of the lattice, so we cannot redefine $\mbf{r}_0$. \Eq{eq:UCT} demonstrates that $(U C_{2z} \mathcal{T})^2$ may differ from $(C_{2z} \mathcal{T})^2 = 1$ by a phase
\bea
\label{eq:gammadef}
\gamma_2 &= \int_{\mbf{r}_0}^{-\mbf{r}_0} \mbf{\tilde{A}}  \cdot d\mbf{r} \mod 2\pi 
\eea
that depends only on the Peierls paths and orbitals of the Hamiltonian. It is simple to determine $\gamma_2$ by direct computation for a given model. But we show first that $\gamma_2$ may only take the value $0$ or $\pi$, on the condition that the Peierls paths of the model are themselves $C_{2z}$-symmetric. Let the path $\mathcal{C}_1$ consist of Peierls paths connecting $\mbf{r}_0$ to $C_{2z} \mbf{r}_0 = - \mbf{r}_0$, and define $\mathcal{C}_2 = C_{2z} \mathcal{C}_1$ which connects $-\mbf{r}_0$ to $\mbf{r}_0$. By the same change of variables, we have that
\bea
\int_{\mathcal{C}_2} \mbf{\tilde{A}}  \cdot d\mbf{r} =  \int_{C_{2z} \mathcal{C}_1} \mbf{\tilde{A}}  \cdot d\mbf{r} =  \int_{\mathcal{C}_1} -\mbf{\tilde{A}}  \cdot d(-\mbf{r}) = \int_{\mathcal{C}_1} \mbf{\tilde{A}}  \cdot d\mbf{r} \ .
\eea
With this result, it is also true that
\bea
\oint_{\mathcal{C}_1 + \mathcal{C}_2} \mbf{\tilde{A}}  \cdot d\mbf{r} &= \int_{\mathcal{C}_1} \mbf{\tilde{A}}  \cdot d\mbf{r}  + \int_{\mathcal{C}_2} \mbf{\tilde{A}}  \cdot d\mbf{r}  = 2 \int_{\mathcal{C}_1} \mbf{\tilde{A}}  \cdot d\mbf{r} \ . \\
\eea
Because $\mathcal{C}_1$ and $\mathcal{C}_2$ are both Peierls paths, $\mathcal{C}_1 + \mathcal{C}_2 = \del \mathcal{R}$ is a closed loop taken along Peierls paths, and thus
\bea
2 \int_{\mbf{r}_0}^{- \mbf{r}_0} \mbf{\tilde{A}}  \cdot d\mbf{r} &= \oint_{\del \mathcal{R}} \mbf{\tilde{A}}  \cdot d\mbf{r}  =  \int_{\mathcal{R}} \Phi dS \in 2\pi \mathbb{Z}
\eea
where we have used that $\curl \mbf{\tilde{A}} = \Phi$ and, by the definition of $\Phi$, all closed loops along Peierls' paths enclose an integer multiple of $2\pi$ flux. Hence find that $\int_{\mbf{r}_0}^{-\mbf{r}_0} \mbf{\tilde{A}}  \cdot d\mbf{r}$ is a multiple of $\pi$ and thus the phase is quantized to be $0$ or $\pi$. 

\begin{figure*}
 \centering
\includegraphics[width=6cm]{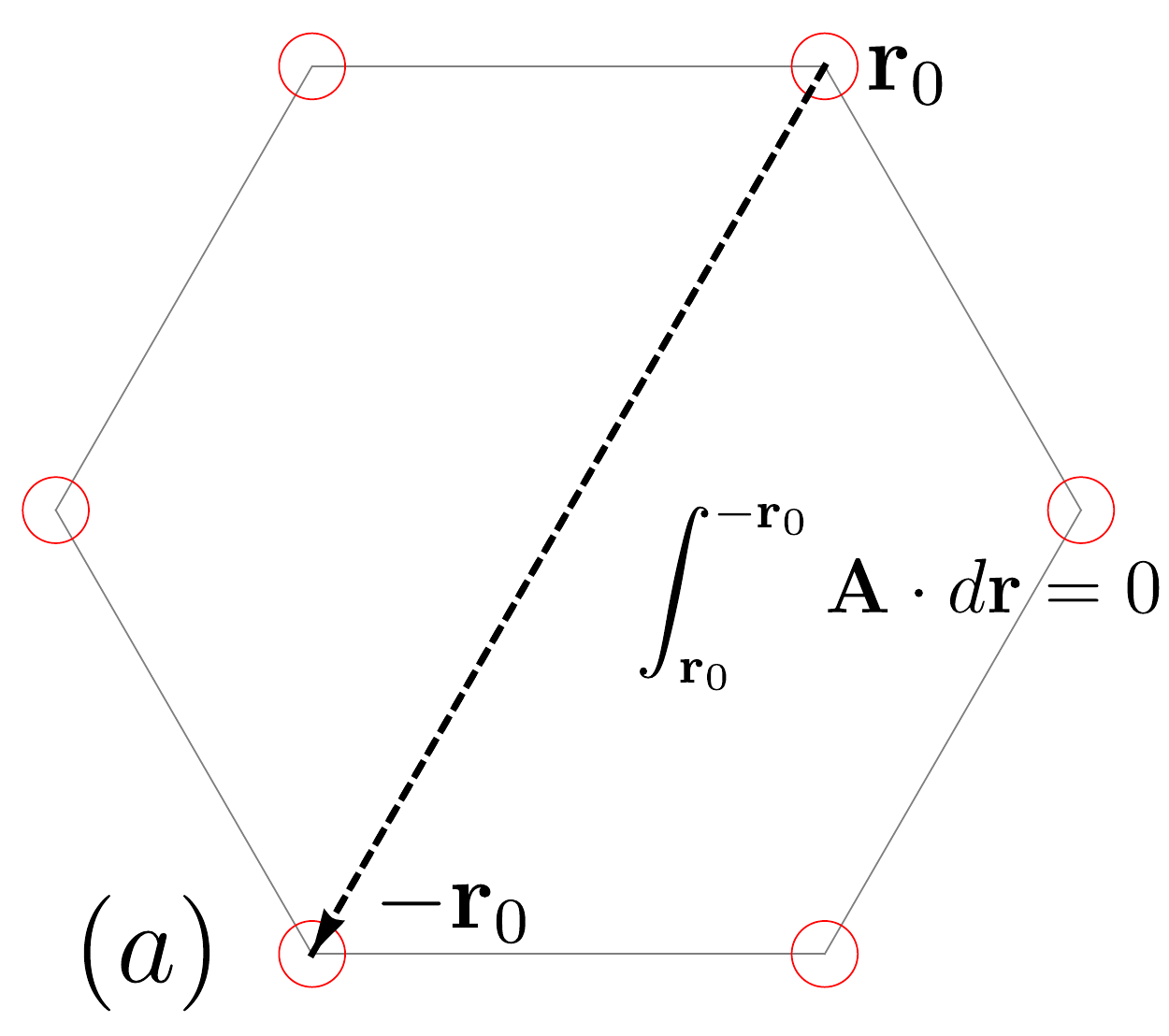}  \quad 
\includegraphics[width=6cm]{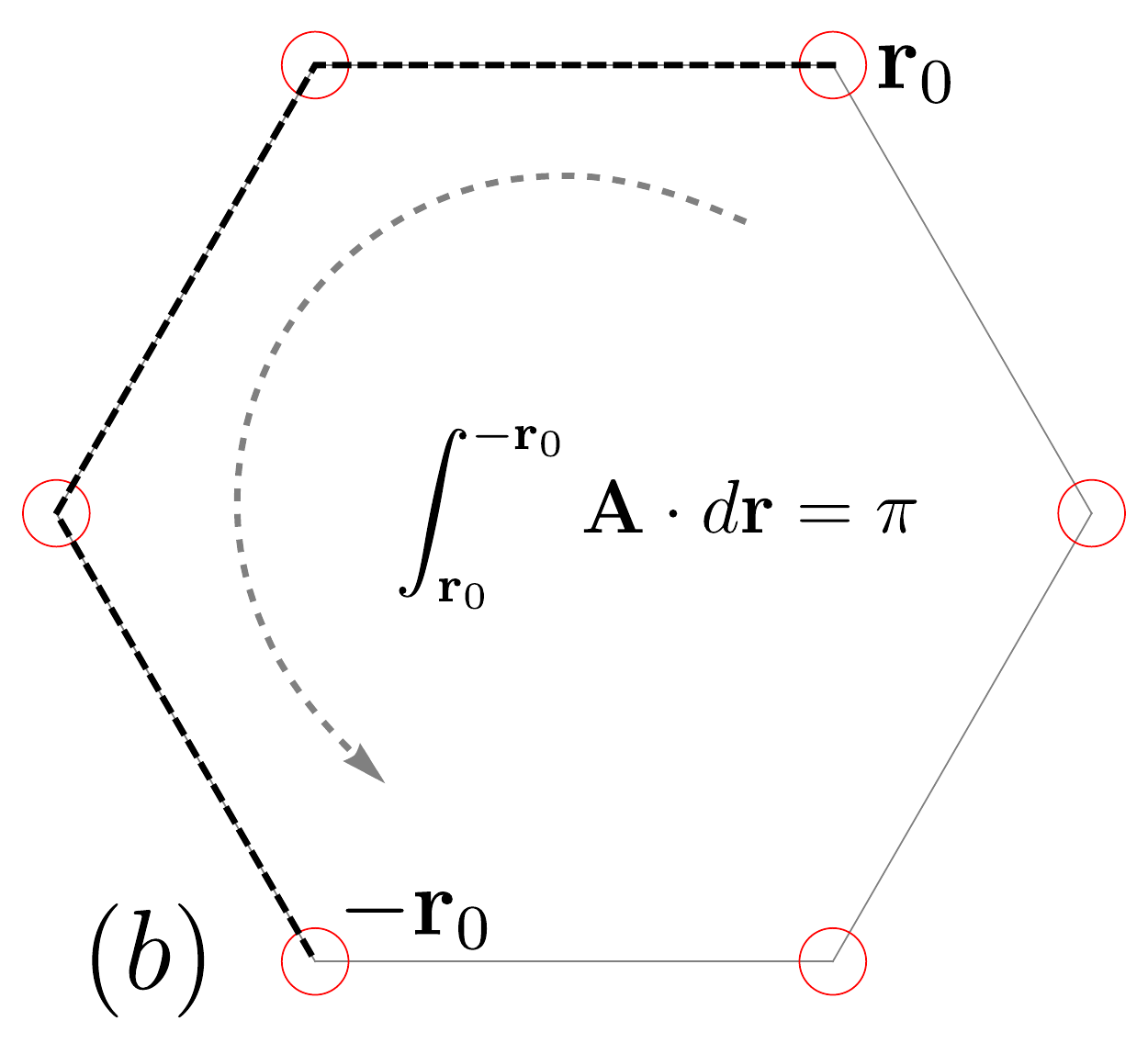} 
\caption{We show the Peierls paths joining $\mbf{r}_0$ and $-\mbf{r}_0$ in $(a)$ our model of twisted bilayer graphene and $(b)$ an alternative model discussed in \Ref{2018arXiv181111786L} where Peierls paths are taken along the bonds of the lattice. The different Peierls paths lead to different values of $\gamma_2$: $+1$  ( resp. $-1$) in the case of our model (resp. the alternative model). } 
\label{fig:C2zTphase}
\end{figure*}

The phase $\gamma_2$ may be calculated to determine the sign of $(U C_{2z} \mathcal{T})^2$. In some cases however, it may be determined more simply. For instance, if there is an orbital on the $1a = (0,0)$ position, we may choose $\mbf{r}_0 = (0,0)$ in which case the integration path vanishes and $\gamma_2$ must be zero. More generally, if there are \emph{any} Peierls paths that connect $\mbf{r}_0$ to the origin, then we can break up the integral into $C_{2z}$-symmetric parts:
\bea
\label{eq:toorigin}
\gamma_2 &= \int_{\mbf{r}_0}^{-\mbf{r}_0} \mbf{\tilde{A}}  \cdot d\mbf{r} 
= \int_{\mbf{r}_0}^{0} \mbf{\tilde{A}}  \cdot d\mbf{r} + \int_0^{-\mbf{r}_0} \mbf{\tilde{A}}  \cdot d\mbf{r} 
= - \int^{\mbf{r}_0}_{0} \mbf{\tilde{A}}  \cdot d\mbf{r} + \int_0^{-\mbf{r}_0} \mbf{\tilde{A}}  \cdot d\mbf{r} 
= - \int^{\mbf{r}_0}_{0} \mbf{\tilde{A}}  \cdot d\mbf{r} + \int_0^{\mbf{r}_0} \mbf{\tilde{A}}  \cdot d\mbf{r} 
= 0 \ . \\
\eea
We find that if the origin may be reached along Peierls path, then the phase is also forced to be zero. (We emphasize that at $\phi = \Phi$, integrals may be arbitrarily deformed along Peierls paths.) For the the QSH model of \App{app:QSH}, orbitals lie on the $1a$ position so the phase is trivial, and in the model of twisted bilayer graphene discussed in \App{app:TBG}, the Peierls paths are taken through the origin as shown in \Fig{fig:C2zTphase}a, so the phase must also be trivial. To furnish an example where the phase is $\pi$, we consider an alternative model of TBG in \Ref{2018arXiv181111786L} that is identical to our model of TBG \emph{except} for the choice of Peierls' paths. In the alternative model, the Peierls paths are taken along the bonds, so that $\Phi = 2\pi$, as shown in \Fig{fig:C2zTphase}b. We can calculate $\gamma_2$ directly via \Eq{eq:gammadef}. We choose $\mbf{r}_0 = \frac{1}{2} \pmb{\delta}_1 - \frac{1}{2} \mbf{a}_1$, the same as in \Fig{fig:Utbgexample}, and calculate
\bea
\gamma_2 &= \int_{\mbf{r}_0}^{- \mbf{r}_0} \tilde{\mbf{A}} \cdot m \mbf{r} \mod 2\pi \\
&= \pi 
\eea
where the path of integration (which is arbitrary as long as it is taken along Peierls paths) is shown in \Fig{fig:C2zTphase}b. This emphasizes once more that the Peierls paths are physical. Different Peierls paths in the TBG model lead to different values of $\gamma_2$, giving $(U C_{2z} \mathcal{T})^2 = \pm (C_{2z} \mathcal{T})^2$. In general, $\gamma_2$ may be simply calculated from \Eq{eq:gammadef} for any model. In coming work (\Ref{hofsymtoappear}), we will shown that similar phases $\gamma_n$ characterize the algebra of general $UC_n\mathcal{T}$-symmetric point groups at high symmetry Wyckoff positions. 

The sign of $(U C_{2z} \mathcal{T})^2$ has important physical consequences. If $(U C_{2z} \mathcal{T})^2 = + 1$, then $U C_{2z} \mathcal{T}$ protects a $w_2$ invariant at $\phi = \Phi/2$ in the same way as $C_{2z} \mathcal{T}$ protects a $w_2$ invariant at $\phi = 0$. We discuss how to compute this invariant in \App{app:w2calc}. If $(U C_{2z} \mathcal{T})^2 = - 1$, then there is no $w_2$ invariant \cite{2019PhRvX...9b1013A}. In this case, we show in \App{app:UCTtrivial} using the nested Wilson and the 10 Fold Way that there is \emph{no nontrivial phase} protected by $UC_{2z} \mathcal{T}$, and hence $H^{\phi = \Phi/2}$ is trivial. 

\subsection{Calculation of the $w_2$ invariant at $(UC_{2z} \mathcal{T} )^2 = +1$}
\label{app:w2calc}

In this section, we show how to use the Wilson loop to compute the $w_2$ index of $H^{\phi = \Phi/2}$. We work in the Landau gauge, which is best suited for explicit calculations. Recall from \Ref{2019PhRvX...9b1013A} that $w_2$ may be computed at $\phi = 0$ from the Wilson loop spectrum, which is ``particle-hole" symmetric thanks to $C_{2z} \mathcal{T}$ \cite{2018arXiv180710676S}. We can follow the same protocol at $\phi = \Phi/2$, but we will find that $U$ acts off-diagonally in momentum space and changes the symmetries of the Wilson spectrum. 

\subsubsection{Construction of the Extended unit cell}
\label{app:Umom}

To compute $w_2$ in the conventional way, we must construct $U$ in an \emph{extended} unit cell where it is diagonal in momentum space. We begin with the expression for $U$ in a single-particle Hilbert space:
\bea
\label{eq:Uhere}
U &= \sum_{\mbf{R} \al} \exp \lp i \int^{\mbf{R}+ \pmb{\delta}_\al}_{\mbf{r}_0}  \mbf{\tilde{A}}  \cdot d\mbf{r} \rp \ket{\mbf{R}, \al} \bra{\mbf{R},\al}, \quad \pmb{\nabla} \times \mbf{\tilde{A}} = \Phi ,
\eea
as is rewritten from \Eq{eq:Usinglepart} of the Main Text. Recall that the integral is taken along Peierls paths and is single-valued mod $2\pi$. Here we have used the single-particle bases $\ket{\mbf{R},\al} = c^\dag_{\mbf{R},\al} \ket{0}$. To derive the action of $U$ on the Hofstadter Hamiltonian $\mathcal{H}^{\phi = \Phi/2}(\mbf{k})$ in the Landau gauge, we could Fourier transform $U$ over the $1\times q'$ unit cell at $\phi = \Phi/2$, where $\phi = \mu \frac{2\pi p'}{q'}$ as discussed in \App{app:hofhamconstruct}. (The gauge-independent formalism introduced in \App{app:UTdetails} which ensures a $1\times q$ magnetic unit cell is not suited to doing numerical calculations. Thus we will rely on the Landau gauge (see \App{app:hofhamconstruct}) to numerically evaluate the Wilson loop and $w_2^{\phi = \Phi/2}$.) However, this generically results in an expression for $U$ that is off-diagonal in momentum space because the positionally-dependent phase in \Eq{eq:Uhere} does not generically share the $1 \times q'$ periodicity of $H^{\phi = \Phi/2}$. Our tactic is to find an extended $\la_1 \times \la_2 $ unit cell where $U$ is diagonal in momentum space. Then $UC_{2z}\mathcal{T}$ acts diagonally on the eigenstates of $\mathcal{H}^{\phi = \Phi/2}$ when we  work in a \emph{new} $\la_1 \times \frac{q' \la_2}{\text{gcd}(q',\la_2)}$ magnetic unit cell which is commensurate with the $1 \times q'$ magnetic unit cell and the $\la_1 \times \la_2$ unit cell where $U$ is diagonal in momentum space. 

First, we will determine in what unit cell $U$ is diagonal in momentum space \emph{irrespective} of $\phi$. Note that $\exp ( i \int_{\mbf{r}_0}^{\mbf{R}  + \pmb{\delta}_{\al}} \tilde{\mbf{A}} \cdot d\mbf{r} )$ is a single-valued function of $\mbf{R}$ on the lattice by the definition of $\Phi$ (see \App{app:magper}) and is periodic in $\mbf{R}$ due to the assumption of commensurate orbitals (see \App{app:ppA}). Hence it has some minimal spatial periodicity $(\la_1, \la_2)$ with $\la_1, \la_2 \in \mathbb{N}$ given by
\bea
\label{eq:kshiftexpanded}
\int_{\mbf{r}_0}^{\la_1 \mbf{a}_1 + \mbf{R} + \pmb{\delta}_{\al}} \tilde{\mbf{A}} \cdot d\mbf{r} &= \int_{\mbf{r}_0}^{\la_2 \mbf{a}_2 + \mbf{R} + \pmb{\delta}_{\al}} \tilde{\mbf{A}} \cdot d\mbf{r} = \int_{\mbf{r}_0}^{\mbf{R} + \pmb{\delta}_{\al}} \tilde{\mbf{A}} \cdot d\mbf{r} \mod 2\pi, \quad \forall \al = 1, \dots, N_{orb}
\eea
with $\mbf{R}$ being any lattice vector. (Although it is not necessary for any of the following calculations, we can show that $\la_i$ is the denominator of $\kappa_i$, defined in \Eqs{eq:kappadef}{eq:diffkappa0}, i.e. $\la_i \kappa_i = 0 \mod 2\pi$. From this perspective, the magnetic unit cell is extended so translations by $\la_i \mbf{a}_i$ commute with $U\mathcal{T}$, as we see from \Eq{eq:UTprojective}.)  When we Fourier transform over the Bravais lattice composed of the enlarged $\la_1 \times \la_2$ unit cells, the BZ shrinks to  $k_1 \in  (- \frac{\pi}{\la_1}, \frac{\pi}{\la_1}), k_2 \in (- \frac{\pi}{\la_2}, \frac{\pi}{\la_2})$ which we label as $BZ_{\la}$. In this unit cell, the position bases are defined
\bea
\ket{r_1,r_2,\ell_1, \ell_2, \al} &= c^\dag_{(\la_1 r_1 + \ell_1)\mbf{a}_1 + (\la_2 r_2 + \ell_2) \mbf{a}_2, \al} \ket{0}, \qquad \ell_1 =0,\dots, \la_1 -1, \ \ell_2 =0,\dots, \la_2 -1 \ . \\
\eea
The momentum eigenstates are then defined by
\bea
\ket{r_1,r_2, \ell_1, \ell_2, \al} = \frac{1}{\sqrt{\mathcal{N}/\la_1 \la_2}} \sum_{k_1, k_2 \in BZ_\la} e^{i (k_1 \mbf{b}_1 + k_2 \mbf{b}_2) \cdot ( (\la_1 r_1 + \ell_1) \mbf{a}_1 +( \la_2 r_2 +  \ell_2 ) \mbf{a}_2  + \pmb{\delta}_\al)} \ket{k_1, k_2, \ell, \al}, \\
\eea
where $\mathcal{N}$ is the number of $1\times 1$ unit cells in the lattice. We may now compute
\bea
U &=  \sum_{r_1,r_2, \ell_1, \ell_2,\al} \exp \lp i \int^{ (\la_1 r_1 + \ell_1) \mbf{a}_1 +( \la_2 r_2 +  \ell_2 ) \mbf{a}_2 + \pmb{\delta}_\al}_{\mbf{r}_0}  \mbf{\tilde{A}}  \cdot d\mbf{r} \rp \ket{r_1, r_2, \ell_1, \ell_2, \al} \bra{r_1, r_2, \ell_1,\ell_2, \al}  \\
&=  \frac{1}{\mathcal{N}/\la_1 \la_2 } \sum_{\mbf{k},\mbf{k'} \in BZ_\la, \ell_1,\ell_2, \al} \ket{\mbf{k}, \ell_1, \ell_2, \al} \bra{\mbf{k}', \ell_1, \ell_2, \al}  \times \\
& \qquad \left[ \sum_{r_1,r_2} \exp \lp i (\mbf{k} - \mbf{k}') \cdot ((\la_1 r_1 + \ell_1) \mbf{a}_1 +( \la_2 r_2 +  \ell_2 ) \mbf{a}_2 + \pmb{\delta}_\al) \rp  \exp \lp i \int^{(\la_1 r_1 + \ell_1) \mbf{a}_1 +( \la_2 r_2 +  \ell_2 ) \mbf{a}_2 + \pmb{\delta}_\al}_{\mbf{r}_0}  \mbf{\tilde{A}}  \cdot d\mbf{r} \rp \right]  \\
&=  \frac{1}{\mathcal{N}/\la_1 \la_2 } \sum_{\mbf{k},\mbf{k'} \in BZ_\la, \ell_1,\ell_2, \al} \ket{\mbf{k}, \ell_1, \ell_2, \al} \bra{\mbf{k}', \ell_1, \ell_2, \al}  \times \\
& \qquad \left[ \sum_{r_1,r_2} \exp \lp i (\mbf{k} - \mbf{k}') \cdot ((\la_1 r_1 + \ell_1) \mbf{a}_1 +( \la_2 r_2 +  \ell_2 ) \mbf{a}_2 + \pmb{\delta}_\al) \rp  \exp \lp i \int^{ \ell_1  \mbf{a}_1 +  \ell_2 \mbf{a}_2 + \pmb{\delta}_\al}_{\mbf{r}_0}  \mbf{\tilde{A}}  \cdot d\mbf{r} \rp \right]  \\
\eea
where we have used \Eq{eq:kshiftexpanded} to remove the $r_1, r_2$ dependence of the integral. Then the sum in brackets can be explicitly evaluated and we find
\bea
\label{eq:UandS}
U &= \frac{1}{\mathcal{N}/(\la_1 \la_2)}  \sum_{\mbf{k},\mbf{k'} \in BZ_\la, \ell_1,\ell_2, \al} \ket{\mbf{k}, \ell_1, \ell_2, \al} \bra{\mbf{k}', \ell_1, \ell_2, \al}  e^{i \int_{\mbf{r}_0}^{\ell_1 \mbf{a}_1 + \ell_2 \mbf{a}_2  + \pmb{\delta}_{\al}} \tilde{\mbf{A}} \cdot d\mbf{r} + i (\mbf{k}- \mbf{k}' )\cdot ( \ell_1 \mbf{a}_1 + \ell_2 \mbf{a}_2 + \pmb{\delta}_{\al})} \times \\
& \qquad  \sum_{r_1,r_2 } e^{i (\mbf{k} - \mbf{k}') \cdot (\la_1 r_1 \mbf{a}_1 + \la_2 r_2 \mbf{a}_2) } \\ 
&= \sum_{\mbf{k} \in BZ'} \sum_{\ell_1, \ell_2, \al}  \mathcal{U}_{\ell_1,\ell_2, \al} \ket{\mbf{k},\ell_1,\ell_2, \al} \bra{\mbf{k}, \ell_1,\ell_2, \al}, \qquad \mathcal{U}_{\ell_1,\ell_2, \al} = e^{i \int_{\mbf{r}_0}^{\ell_1 \mbf{a}_1 + \ell_2 \mbf{a}_2  + \pmb{\delta}_{\al}} \tilde{\mbf{A}} \cdot d\mbf{r} } \ . \\
\eea
We see explicitly that $U$ is diagonal in momentum space in the $\la_1 \times \la_2$ unit cell. 

\begin{figure}
 \centering
 \includegraphics[width=6.2cm]{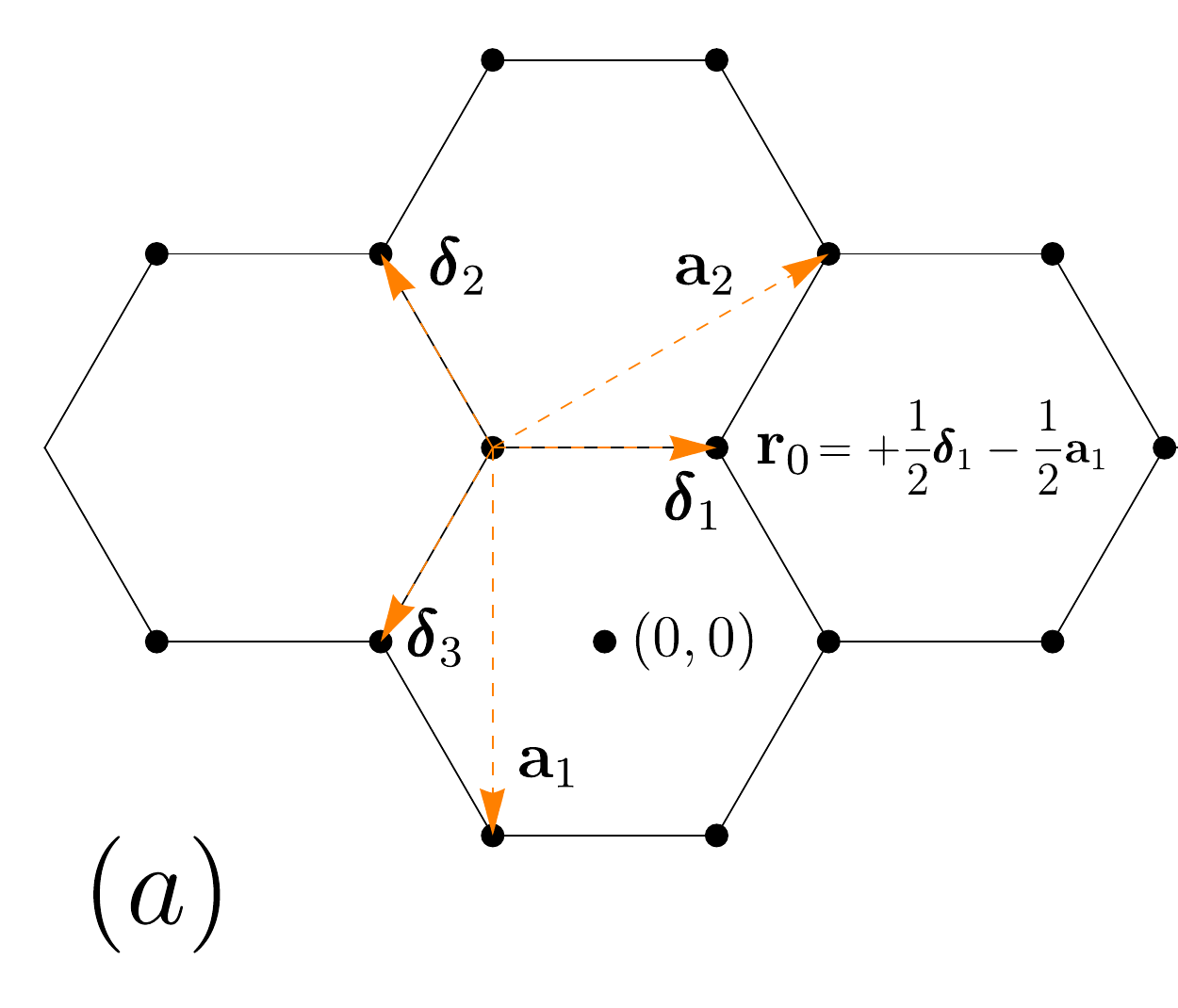} \quad
\includegraphics[width=5.8cm]{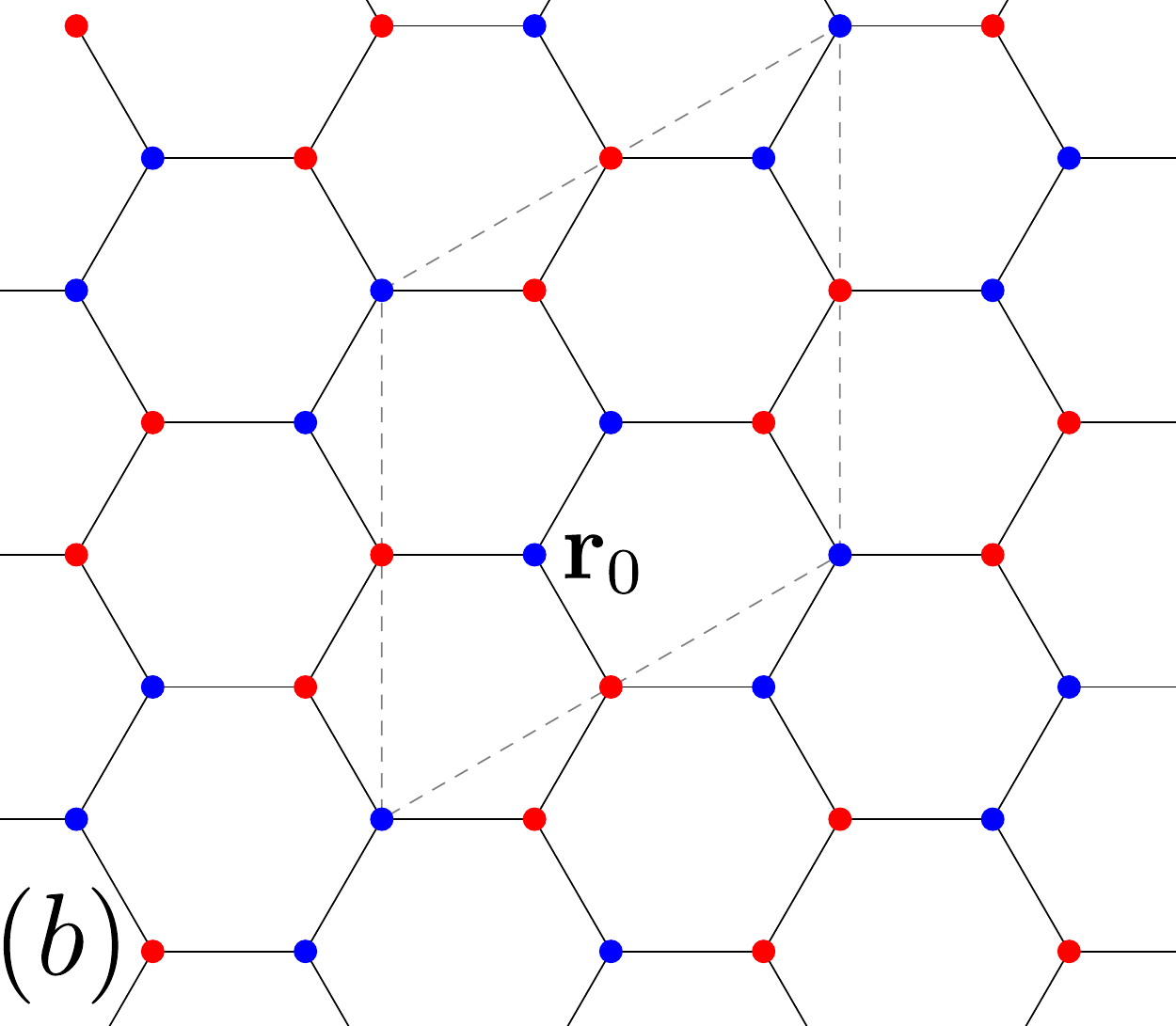}
\caption{(a) We show the lattice vectors $\mbf{a}_i$ and the nearest-neighbor vectors $\pmb{\delta}_i$ for the TBG model of \App{app:TBG}. The atomic sites are located at $\pm \frac{1}{2} \pmb{\delta}_1 - \frac{1}{2} \mbf{a}_1$ modulo lattice vectors, and we fix $\mbf{r}_0 = \frac{1}{2} \pmb{\delta}_1 - \frac{1}{2} \mbf{a}_1$. (b) We depict the value of \Eq{eq:TBGS} on the lattice, with a value of $+1$ ($-1$) corresponding to blue (red). We outline the $2\times 2$ unit cell in dashed, gray lines which indicates the periodicity of $U$. }
\label{fig:Utbgexample}
\end{figure}

We now give an example of this calculation for model of twisted bilayer graphene (introduced in detail in \App{app:TBG}). In fact, we will use this expression for $U$ in momentum space later for Wilson loop calculations in \App{app:C2zTwilson}. We recall that the nearest-neighbor vectors of the honeycomb lattice are $\pmb{\delta}_\al, \al = 1,2,3$ (as shown for convenience in \Fig{fig:Utbgexample}a), and that there are $s$ and $p_z$ orbitals on the atomic sites which are located at $\pm\frac{1}{2} \pmb{\delta}_1 - \frac{1}{2} \mbf{a}_1$. We refer to the choice of $+$ (resp. $-$) as the A (resp. B) sublattice.  The Peierls paths of the model (\Fig{fig:TBGhoppings}) enclosed multiples of a third of the unit cell, so $n=3$ and $\Phi = 6\pi$. Using our Landau gauge for $\mbf{\tilde{A}}$, it is straightforward to calculate the following integrals along Peierls' paths (see \Fig{fig:TBGhoppings})
\bea
\label{eq:usefulints}
\exp \lp i \int_{\mbf{R} + \pmb{\delta}_\al}^{\mbf{R} + \pmb{\delta}_\al \pm \mbf{a}_i} \tilde{\mbf{A}} \cdot d \mbf{r} \rp = -1, \qquad \exp \lp i \int_{\frac{1}{2} \pmb{\delta}_1 - \frac{1}{2} \mbf{a}_1}^{-\frac{1}{2} \pmb{\delta}_1 - \frac{1}{2} \mbf{a}_1} \tilde{\mbf{A}} \cdot d \mbf{r} \rp = -1 \ .
\eea
$U$ also requires a choice of $\mbf{r}_0$, an arbitrary but fixed position of an orbital \emph{of the Hamiltonian}, e.g. $\mbf{r}_0$ may \emph{not} be a honeycomb center for $H_{TBG}$ because there are no orbitals there. We choose $\mbf{r}_0 = +\frac{1}{2} \pmb{\delta}_1 - \frac{1}{2} \mbf{a}_1$. Then by repeated application of \Eq{eq:usefulints}, we find 
\bea
\label{eq:TBGS}
\exp \lp i \int_{\frac{1}{2} \pmb{\delta}_1 - \frac{1}{2} \mbf{a}_1}^{r_1 \mbf{a}_1 + r_2 \mbf{a}_2 \pm \frac{1}{2} \pmb{\delta}_1 - \frac{1}{2} \mbf{a}_1} \tilde{\mbf{A}} \cdot d\mbf{r} \rp &= \exp \lp i \int_{\frac{1}{2} \pmb{\delta}_1 - \frac{1}{2} \mbf{a}_1}^{\pm \frac{1}{2} \pmb{\delta}_1 - \frac{1}{2} \mbf{a}_1} \tilde{\mbf{A}} \cdot d\mbf{r} \rp \exp \lp i \int_{\pm\frac{1}{2} \pmb{\delta}_1 - \frac{1}{2} \mbf{a}_1}^{r_1 \mbf{a}_1 + r_2 \mbf{a}_2 \pm \frac{1}{2} \pmb{\delta}_1 - \frac{1}{2} \mbf{a}_1} \tilde{\mbf{A}} \cdot d\mbf{r} \rp \\
&= \pm (-1)^{r_1 + r_2} 
\eea
which we show pictorially in \Fig{fig:Utbgexample}b. According to \Eq{eq:kshiftexpanded}, we can choose a $2 \times 2$ unit cell because, using \Eq{eq:TBGS}, 
\bea
\exp \lp i \int_{\mbf{r}_0}^{2r_1 \mbf{a}_1 + 2r_2 \mbf{a}_2 \pm \frac{1}{2} \pmb{\delta}_1 - \frac{1}{2} \mbf{a}_1} \tilde{\mbf{A}} \cdot d\mbf{r} \rp &= \pm (-1)^{2r_1 + 2r_2}  = \pm 1
\eea 
where the $r_1, r_2$ dependence has disappeared.  We index the unit cell by $\ell_1, \ell_2 \in \{0,1\}$ and define $BZ_{\la}$ as $k_1 \in (- \pi/2, \pi/2), k_2 \in (-\pi/2, -\pi/2)$. Then following \Eq{eq:UandS}, we calculate
\bea
U &= \sum_{\mbf{k} \in BZ'} \sum_{\ell_1, \ell_2, \al}  \mathcal{U}_{\ell_1,\ell_2, \al} \ket{\mbf{k},\ell_1,\ell_2, \al} \bra{\mbf{k}, \ell_1,\ell_2, \al}, \\
 \mathcal{U}_{\ell_1,\ell_2, \al} &= \exp \lp i \int_{\mbf{r}_0}^{\ell_1 \mbf{a}_1 + \ell_2 \mbf{a}_2  + \pmb{\delta}_{\al}} \tilde{\mbf{A}} \cdot d\mbf{r} \rp \\
 &= (-1)^{\ell_1 + \ell_2} \exp \lp i \int_{\mbf{r}_0}^{\pmb{\delta}_{\al}} \tilde{\mbf{A}} \cdot d\mbf{r} \rp \\
 &= \text{sub}(\al) (-1)^{\ell_1 + \ell_2} 
\eea
where we have used \Eq{eq:usefulints}, and $\text{sub}(\al) = \pm 1$ where the sign is positive (resp. negative) for $\pmb{\delta}_{\al}$ on the A (resp. B) sublattice. We will use this result in \App{app:C2zTwilson}.

\subsubsection{Constraints on the Wilson loop at $\phi = \Phi/2$}
\label{app:conwilphi}

Now we study how $U C_{2z} \mathcal{T}$ symmetry constrains the Wilson loop eigenvalues. We will demonstrate that in a unit cell where $U$ is diagonal in momentum space, $U C_{2z} \mathcal{T}$ creates a ``particle-hole" symmetry in the spectrum of the Wilson Hamiltonian at $\phi = \Phi/2$, i.e. $\{ \vartheta(k_1) \} = \{- \vartheta(k_1) \}$. This allows the $w_2$ invariant to be calculated by counting the number of points where $\vartheta(k_1) = 0$ and $\vartheta(k_1) = \pi$ 
as described in \Ref{2019PhRvX...9b1013A}. We emphasize that $w_2^{\phi = \Phi/2}$ is a topological invariant and is not affected by the choice of unit cell. This is in contrast to the relative winding of the Wilson loop eigenvalues which may change when the unit cell is expanded.

To begin, we can construct the extended unit cell to be $\la_1 \times \la_2 q' / \text{gcd}(\la_2, q')$ which explicitly ensures that the $\la_1 \times \la_2$ unit cell in which $U$ is diagonal and the $1 \times q'$ unit cell in which we can diagonalize $H^{\phi = \Phi/2 = \mu \frac{2\pi p}{q}}$ are commensurate. For brevity, we let $\la_2' = \la_2 q' / \text{gcd}(\la_2, q')$. In this extended unit cell, we can take $k_1 \in (-\pi/\la_1, \pi/\la_1), k_2 \in (0, 2\pi/\la_2')$ which we define as $BZ_{\la'}$. We define the $i$th energy eigenstate of $H^{\phi = \Phi/2 = \mu \frac{2\pi p}{q}}$ as $\ket{u^{\Phi/2}_i}$ which is an $N_{orb} \times \la_1 \la_2'$ vector.  We call the representation of $U C_{2z} \mathcal{T}$ on the $\ket{u^{\Phi/2}_i}$ eigenstates $D[U C_{2z} \mathcal{T}] = \mathcal{Q} K$ for brevity. We can derive an explicit expression by acting $U C_{2z} \mathcal{T}$ on the momentum eigenstates:
\bea
\label{eq:Qconnect}
U C_{2z} \mathcal{T} \ket{\mbf{k}, \ell_1, \ell_2, \al} &= U C_{2z} \mathcal{T} \frac{1}{\sqrt{\mathcal{N}/\la_1 \la_2'}} \sum_{r_1, r_2} e^{- i \mbf{k} \cdot (\, (\la_1 r_1 + \ell_1) \mbf{a}_1 + (\la_2' r_2 + \ell_2) \mbf{a}_2 + \pmb{\delta}_{\al})}  \ket{r_1, r_2, \ell_1, \ell_2, \al}  \\
&=\frac{1}{\sqrt{\mathcal{N}/\la_1 \la_2'}} \sum_{r_1, r_2} e^{-i (-\mbf{k}) \cdot (\, (\la_1 r_1 + \ell_1) \mbf{a}_1 + (\la_2' r_2 + \ell_2) \mbf{a}_2 + \pmb{\delta}_{\al})}  U C_{2z} \mathcal{T}  \ket{r_1, r_2, \ell_1, \ell_2, \al}  \\
&=\frac{1}{\sqrt{\mathcal{N}/\la_1 \la_2'}} \sum_{r_1, r_2,\be}e^{-i (-\mbf{k}) \cdot (\, (\la_1 r_1 + \ell_1) \mbf{a}_1 + (\la_2' r_2 + \ell_2) \mbf{a}_2 + \pmb{\delta}_{\al})}  \mathcal{U}_{-\ell_1,-\ell_2, \be} [D(C_{2z} \mathcal{T})]_{\al \be}  \ket{-r_1, -r_2, -\ell_1, -\ell_2, \be} K \\
\eea
where we have used \Eq{eq:UandS} to determine the action of $U$ on the states. Now we define 
\bea
\label{eq:Qdef}
\null[\mathcal{Q}]_{\ell_1\ell_2 \al, \ell'_1 \ell'_2 \be} = \mathcal{U}_{-\ell_1,-\ell_2, \al} \delta_{- \ell_1,\ell_1'}  \delta_{- \ell_2,\ell_2'} [D(C_{2z} \mathcal{T})]_{\al \be} \ .
\eea
Note from \Eq{eq:Qconnect} that $\mathcal{Q}_{\ell_1\ell_2 \al, \ell'_1 \ell'_2 \be} $ is only nonzero when $\ell_1 \mbf{a}_1 + \ell_2 \mbf{a}_2 + \pmb{\delta}_\al = -(\ell_1' \mbf{a}_1 + \ell_2' \mbf{a}_2 + \pmb{\delta}_\be)$ mod $\la_1 \mbf{a}_1 + \la_2' \mbf{a}_2$. This is the usual constraint on spatial symmetries, but in the expanded unit cell. Using this property, we continue from \Eq{eq:Qconnect} to find
\bea
U C_{2z} \mathcal{T} \ket{\mbf{k}, \ell_1, \ell_2, \al} &= \frac{1}{\sqrt{\mathcal{N}/\la_1 \la_2'}} \sum_{r_1, r_2} \sum_{\ell_1',\ell_2',\be}e^{-i (-\mbf{k}) \cdot (\, (\la_1 r_1 - \ell'_1) \mbf{a}_1 + (\la_2' r_2 - \ell'_2) \mbf{a}_2 - \pmb{\delta}_{\be})}  [Q]_{\ell_1\ell_2 \al, \ell'_1 \ell'_2 \be}  \ket{-r_1, -r_2,\ell_1', \ell_2', \be} K \\
&= \frac{1}{\sqrt{\mathcal{N}/\la_1 \la_2'}} \sum_{r_1, r_2} \sum_{\ell_1',\ell_2',\be}e^{-i \mbf{k} \cdot (\, (\la_1 r_1 + \ell'_1) \mbf{a}_1 + (\la_2' r_2 + \ell'_2) \mbf{a}_2 + \pmb{\delta}_{\be})}  [Q]_{\ell_1\ell_2 \al, \ell'_1 \ell'_2 \be}  \ket{r_1, r_2,\ell_1', \ell_2', \be} K \\
&=  \sum_{\ell_1',\ell_2',\be}  [Q]_{\ell_1\ell_2 \al, \ell'_1 \ell'_2 \be} \ket{\mbf{k},\ell_1', \ell_2', \be} K \ . \\
\eea
We assume that $(U C_{2z} \mathcal{T} )^2 = +1$ in this section, so $ \mathcal{Q}  \mathcal{Q}^* = + 1$. Note that in the $\la_1 \times \la_2'$ unit cell, $U C_{2z} \mathcal{T}$ is also diagonal in momentum space because both $C_{2z}$ and $\mathcal{T}$ reverse the momentum of a state. Then we can define a unitary sewing matrix
\bea
\mathcal{B}^{ij}_\mbf{k} = \braket{u^{\Phi/2}_i(\mbf{k}) | \mathcal{Q} |u^{\Phi/2 \, *}_j(\mbf{k})} \\ 
\eea
which obeys
\bea
\label{eq:invertBuct}
\ket{u^{\Phi/2}_j(\mbf{k})} &= \mathcal{Q} \ket{u^{\Phi/2 \, *}_i(\mbf{k})} [\mathcal{B}^\dag]^{ij}_{\mbf{k}} \ .
\eea
Using \Eq{eq:invertBuct}, we determine that a small segment of a Wilson loop obeys
\bea
\null [W^{\Phi/2}_{\mbf{k}_1' \leftarrow \mbf{k}_1}]^{ij} &= \braket{u^{\Phi/2}_i(\mbf{k}_1') | u^{\Phi/2}_j(\mbf{k}_1)} \\
&= \sum_{rs} \mathcal{B}^{ir}_{\mbf{k}_1'} \braket{ u^{\Phi/2 \, *}_r(\mbf{k}_1') | \mathcal{Q}^\dag \mathcal{Q} |u^{\Phi/2 \, *}_s(\mbf{k}_1 ) [\mathcal{B}^\dag]^{sj}_{\mbf{k}_1} }\\
&= \sum_{rs} \mathcal{B}^{ir}_{\mbf{k}_1'}  [W^{\Phi/2 \, *}_{\mbf{k}_1' \leftarrow \mbf{k}_1}]_{rs} [\mathcal{B}^\dag]^{sj}_{\mbf{k}_1} \ .
\eea
Extending this to a full Wilson loop along $BZ_{\la'}$ where we integrate along $k_2 \in (0, 2\pi/\la_2')$, we find
\bea
\label{eq:UCTwilson}
W^{\Phi/2}(k_1) &= \mathcal{B}_{(k_1, \frac{2\pi}{\la_2'})} W^{\Phi/2 \, *}_{(k_1, \frac{2\pi}{\la_2'}) \leftarrow (k_1, 0)} \mathcal{B}^\dag_{(k_1, 0)} \\ 
&= \mathcal{B}_{(k_1, \frac{2\pi}{\la_2'})} W^{\Phi/2 \, *}(k_1) \mathcal{B}^\dag_{(k_1, 0)} \ . \\ 
\eea
To related the sewing matrices $\mathcal{B}$ at points across the $BZ_{\la'}$, we need to use the embedding matrix $V_2'(\Phi/2)$ that implements the $2\pi/\la_2'$ periodicity. Following identically the calculation of the the embedding matrix $\overline{V}_2(\phi)$ in the $1\times q'$ unit cell in \Eq{eq:V2calc}, we find
\bea
\label{eq:V2laUCT}
\null [V_2'(\Phi/2)]_{\ell_1 \ell_2 \al, \ell'_1 \ell'_2 \be} = \delta_{\ell_1 \ell'_1} \delta_{\ell_2 \ell'_2} e^{i \frac{2\pi}{\la_2'} \ell_2} \delta_{\al \be} e^{i \frac{2\pi}{\la_2'} \pmb{\delta}_\al \cdot \mbf{b}_2} 
\eea
where $\ell_1 =0, \dots, \la_1-1$ and $\ell_2 =0, \dots, \la_2'-1$ index the unit cells within the extended $\la_1 \times \la_2'$ unit cell, respectively. We now observe that
\bea
\mathcal{B}^{ij}_{(k_1, \frac{2\pi}{\la_2'})} &= \braket{u^{\Phi/2}_i(k_1, \frac{2\pi}{\la_2'}) | \mathcal{Q} K  | u^{\Phi/2}_j(k_1,\frac{2\pi}{\la_2'})} \\
&= \braket{u^{\Phi/2}_i(k_1, 0) | [V'_2(\Phi/2)]^\dag \mathcal{Q} K V'_2(\Phi/2) | u^{\Phi/2}_j(k_1,0)} \\
\eea
By direct calculation with \Eqs{eq:Qdef}{eq:V2laUCT}, we compute
\bea
\null [\, V'_2(\Phi/2)^\dag \mathcal{Q} K V'_2(\Phi/2) ]_{\ell_1 \ell_2 \al,\ell'_1 \ell'_2 \be} &= e^{-i \frac{2\pi}{\la_2'} (\ell_2 \mbf{a}_2 + \pmb{\delta}_{\al} + \ell_2' \mbf{a}_2 + \pmb{\delta}_\be) \cdot \mbf{b}_2} [\mathcal{Q} K ]_{\ell_1 \ell_2 \al,\ell'_1 \ell'_2 \be} \\
&=  [\mathcal{Q} K ]_{\ell_1 \ell_2 \al,\ell'_1 \ell'_2 \be} \\
\eea
where in the last line we have used that $ [\mathcal{Q} K ]_{\ell_1 \ell_2 \al,\ell'_1 \ell'_2 \be}$ is only nonzero if $\ell_1 \mbf{a}_1 + \ell_2 \mbf{a}_2 + \pmb{\delta}_\al = -(\ell_1' \mbf{a}_1 + \ell_2' \mbf{a}_2 + \pmb{\delta}_\be)$ mod $\la_1 \mbf{a}_1 + \la_2' \mbf{a}_2$, so $ \ell_2 + \pmb{\delta}_{\al} \cdot \mbf{b}_2 + \ell_2' + \pmb{\delta}_{\be} \cdot \mbf{b}_2 = 0 \mod \la_2'$. Hence we obtain
\bea
\mathcal{B}^{ij}_{(k_1, \frac{2\pi}{\la_2'})}  &= \braket{u^{\Phi/2}_i(k_1, 0) | \mathcal{Q} K | u^{\Phi/2}_j(k_1,0)} \\
&= \mathcal{B}^{ij}_{(k_1, 0)} \ .
\eea
Returning to \Eq{eq:UCTwilson}, we find the desired ``particle-hole" symmetry
\bea
W^{\Phi/2}(k_1) &= \mathcal{B}^{\dag}_{(k_1, 0)} W^{\Phi/2 \, *} (k_1) \mathcal{B}_{(k_1, 0)}, \\ 
\{ \vartheta_j(k_1) \} &= \{ -\vartheta_j(k_1) \} \\
\eea
as previously stated. This is identical to the Wannier center constraint arising from $C_{2z} \mathcal{T}$ at $\phi=0$ and hence we can calculate $w_2$ directly from the Wilson loop spectrum following the discussion of \Ref{2019PhRvX...9b1013A}.

\subsection{Proof of a Trivial Phase when $(UC_{2z} \mathcal{T} )^2 = -1$}
\label{app:UCTtrivial}

Now we consider the case where $(U C_{2z} \mathcal{T})^2 = - 1$ and there is no $w_2$ invariant defined. We remark that no $2D$ crystalline system at $\phi = 0$ can have such a symmetry because $(C_{2z} \mathcal{T})^2 = C_{2z}^2 \mathcal{T}^2 = (\pm 1)^2 = 1$. From the 3D perspective where $\phi$ is interpreted as $k_z$ and $U$ is the embedding matrix along the $z$ direction, the symmetry algebra at $k_z = 0$ is always the same as at $k_z = \pi$, so having $(C_{2z} \mathcal{T})^2 = + 1$ and $(U C_{2z} \mathcal{T})^2 = - 1$ is also impossible. The possibility of such projective symmetry algebras (see also \Eq{eq:UTprojective}) is a novel feature of Hofstadter physics. 

The Higher Order Topological Insulator (HOTI) phase is still characterized by Wannier flow between $\phi = 0$ and $\phi = \Phi/2$. Because we assume a non-trivial $w_2$ index at $\phi = 0$ that protects corner states, we need only show corner states are not stable if $(UC_{2z} \mathcal{T})^2 = -1$. This establishes that $H^{\Phi/2}$ is in a trivial atomic limit, and so pumping must occur between $\phi = \Phi/2$ and $\phi = 0$. 

First we provide a heuristic argument that $(U C_{2z} \mathcal{T})^2 = -1$ trivializes the phase. We consider a finite crystal that preserves $U C_{2z} \mathcal{T}$ symmetry, where $C_{2z}$ is a rotation around the origin. Consider a high symmetry Wyckoff position $\mbf{w}$ with $U C_{2z} \mathcal{T}$ in its magnetic point group. If the state has a center exactly at $\mbf{w}$, then by Kramers Theorem it must have a Kramers partner at $\mbf{w}$ because $(UC_{2z} \mathcal{T})^2 =-1$. If a state has a Wannier center at $\mbf{w} + \pmb{\delta}$, perturbed slightly from $\mbf{w}$, then by $U C_{2z} \mathcal{T}$ there is another state at $\mbf{w} - \pmb{\delta}$. Hence any pair of states at $\mbf{w}$ can be moved adiabatically away from $\mbf{w}$ along $U C_{2z} \mathcal{T}$-symmetric paths. This indicates a trivial (Real Space Invariant) RSI \cite{2020Sci...367..794S}. In coming work (\Ref{hofsymtoappear}), we derive expressions for the RSIs of all the 2D magnetic point groups. This indicates that the bulk is trivialized, and by the Bulk-Boundary correspondence, we expect the boundary states to be trivial \cite{2020Sci...367..794S,2018PhRvX...8c1070K,Geier_2018,PhysRevX.9.011012,Khalaf_2018}.

In absence of the $w_2$ invariant, corner states can still be diagnosed using the nested Berry phase formalism of \Refs{2017PhRvB..96x5115B}{2017Sci...357...61B}. Hence, we can compute the determinant of the nested Wilson loop $\mathcal{W}$ to confirm the heuristic argument given before. We remark that \Ref{2018arXiv180611116W} demonstrated that $C_{2z} \mathcal{T}$ symmetry quantizes $\det \mathcal{W} = \pm 1$ (where $-1$ is the nontrivial value of the topological phase) when $(C_{2z} \mathcal{T})^2 = \pm1$ and the nested Wilson loop is taken over a particle-hole symmetric configuration of Wannier bands. Here we study the specific case where $(UC_{2z} \mathcal{T})^2 = -1$, and we show that the determinant of the nested Wilson loop over particle-hole symmetric Wannier bands is fixed to be $+1$, trivial. 

We define the Wannier Hamiltonian $H_W(k_1)$ by $W^{\Phi/2}(k_1) = e^{ i H_W(k_1)}$, and the Wannier bands as
\bea
H_W(k_1) \ket{w_j(k_1)} = \vartheta_j(k_1) \ket{w_j(k_1)} \ . \\
\eea
We call $\mathcal{K} = \mathcal{B}_{(k_1,0)} K$ the representation of $U C_{2z} \mathcal{T}$ on $H_W(k_1)$ such that
\bea
\label{eq:uc2thamalg}
\mathcal{K} H_W(k_1) \mathcal{K}^{-1} &= - H_W(k_1) \ . \\
\eea
The essential difference between the $\mathcal{K}^2 = +1$ case and the $\mathcal{K}^2 = -1$ case is that the latter creates ``anti-Kramers' pairs", i.e. the states $\ket{w_j(k_1)}$ and $\mathcal{K}\ket{w_j(k_1)}$ have opposite Wilson eigenvalues, $\pm \vartheta(k_1)$ , but are necessarily distinct even when $\vartheta(k_1) = - \vartheta(k_1) \mod 2\pi$. First we prove that the two states $\ket{w(k_1)}$ and $\mathcal{K} \ket{w(k_1)}$ have opposite eigenvalues. Let $H_W(k_1) \ket{w(k_1)} = \vartheta(k_1) \ket{w(k_1)}$. Then
\bea
\label{eq:Kkramers}
H_W(k_1) \mathcal{K} \ket{w(k_1)} &= - \mathcal{K} H_W(k_1) \ket{w(k_1)} \\
&= - \mathcal{K}  \vartheta(k_1) \ket{w(k_1)} \\
&= - \vartheta(k_1)  \mathcal{K}  \ket{w(k_1)} \\
\eea
where we have used \Eq{eq:uc2thamalg} to anti-commute $\mathcal{K}$ and $H_W(k_1)$. Now we prove that anti-Kramers' pairs represent distinct states by contradiction. Suppose $\vartheta(k_1) = 0,\pi$  and $\mathcal{K}^{-1} \ket{w_j(k_1)}$ and $ \ket{w_j(k_1)}$ represent the same state, so $ \mathcal{K} \ket{w_j(k_1)} = e^{i \al}  \ket{w_j(k_1)}$. Then by acting $\mathcal{K}$ again, we find
\bea
\mathcal{K}^2 \ket{w_j(k_1)} &= \mathcal{K} e^{i \al}  \ket{w_j(k_1)} \\
- \ket{w_j(k_1)} &=  e^{-i \al} \mathcal{K} \ket{w_j(k_1)} \\
- \ket{w_j(k_1)} &=  e^{-i \al + i \al} \ket{w_j(k_1)} \\
- \ket{w_j(k_1)} &= \ket{w_j(k_1)} \\
\eea
so we reach a contradiction because $\ket{w_j(k_1)} \neq 0$. Importantly, we see that for $\vartheta = 0$ or $\vartheta = \pi$, where $\vartheta = - \vartheta$, $\ket{w(k_1)}$ and $\mathcal{K} \ket{w(k_1)}$ are true Kramers' partners, meaning they are distinct states with the same eigenvalue. 

Following \Ref{2018arXiv180611116W}, the Wilson loop bands selected to compute the nested Wilson loop should preserve $U C_{2z} \mathcal{T}$. By \Eq{eq:Kkramers}, the states $\ket{w(k_1)}$ and $\mathcal{K} \ket{w(k_1)}$ are automatically particle-hole symmetric and hence respect $U C_{2z} \mathcal{T}$. We now argue that the bands of a generic nested Wilson loop can be gapped into pairs related by $\mathcal{K}$ because there are no other unitary symmetries to protect crossings, so in general the selected bands decompose into two sets, one above $\vartheta = 0$ and the other below $\vartheta = 0$. Note that there must always be an even number of bands because $(U C_{2z} \mathcal{T})^2 = -1$. For now, we assume that there at least 4 bands so we can calculate the nested Wilson loop on a $U C_{2z} \mathcal{T}$-symmetric pair of bands which is a subset of full Wilson Hamiltonian spectrum. We treat the special case of two bands later and show it is also trivial. 

We now study a generic $2\times2$ block to show that its determinant is always $+1$, which is sufficient to establish that the determinant of a generic nested Wilson loop (with at least four bands) is $+1$ by block diagonalization. At every $k_1$, we can diagonalize the Wilson Hamiltonian $H_W(k_1)$ to find its two eigenvectors which, due to the $UC_{2z} \mathcal{T}$ symmetry, can be written as $\ket{w(k_1)}$ and $\mathcal{K}\ket{w(k_1)}$. The nested Wilson loop in the $k_1 \in(-\pi/\la_1, \pi/\la_1), k_2 \in (0, 2\pi/\la_2')$ BZ of the extended unit cell is written
\bea
\label{eq:nestedWdef}
\mathcal{W} &= \tilde{U}^\dag(2\pi/\la_1) \left[ \prod_{k_1}^{2\pi/\la_1 \leftarrow 0} \tilde{\mathcal{P}}_{k_1} \right] \, \tilde{U}(0), \qquad \tilde{\mathcal{P}}_{k_1} = \tilde{U}(k_1) \tilde{U}^\dag(k_1)
\eea
where $\tilde{U}(k_1)$ is the $\la_1 \la_2' N_{orb} \times 2$ matrix of the eigenvectors of the Wilson loop. We choose a conventional ordering where the columns of $\tilde{U(k_1)}$ are ordered such that $\tilde{U}(k_1) = [w_1(k_1), w_2(k_1)]$ where $w_1(k_1)$ is the column vector of length $\la_1 \la_2' N_{orb}$ corresponding to $\ket{w(k_1)}$ and $w_2(k_1)$ is the column vector corresponding to $\mathcal{K}\ket{w(k_1)}$. Note that in this ordering, 
\bea
\mathcal{K} \tilde{U}(k_1) &= \mathcal{K} [w_1(k_1), w_2(k_1)] \\
&=  [\mathcal{K} w_1(k_1), \mathcal{K} w_2(k_1)] K \\
&=  [w_2(k_1), \mathcal{K}^2 w_1(k_1)] K \\
&=  [w_2(k_1),  -w_1(k_1)] K \\
&= [w_1(k_1), w_2(k_1)] \bpm 0 & -1 \\ 1 & 0 \epm K \\
&= \tilde{U}(k_1) \, (-i \sigma_2K)  \\
\eea
where $\sigma_2$ is a Pauli matrix. Acting on the projectors, we find that they commute with $\mathcal{K}$:
\bea
\mathcal{K} \tilde{\mathcal{P}}_{k_1} &= \tilde{U}(k_1) (-i \sigma_2 K) U^\dag(k_1) \\
&=  \tilde{U}(k_1) (U(k_1) \cdot i \sigma_2 K)^\dag \\
&=  \tilde{U}(k_1) (- \mathcal{K} U(k_1))^\dag \\
&= - \tilde{U}(k_1) U^\dag(k_1) \mathcal{K}^{-1} \\
&=  \tilde{U}(k_1) U^\dag(k_1) \mathcal{K} \\
&= \tilde{\mathcal{P}}_{k_1} \mathcal{K}  \\
\eea
so $[\tilde{\mathcal{P}}_{k_1}, \mathcal{K}] = 0$. Using this identity on the nested Wilson loop, we find
$(-i \sigma_2K) \mathcal{W}  (-i \sigma_2K)^\dag =  \mathcal{W}$. Writing 
\bea
\label{eq:Whamanti}
\mathcal{W} = e^{i \mathcal{H}_W}, \qquad \mathcal{H}_W = \sum_{i=0}^3 d_i \sigma_i, 
\eea
defining $\sigma_0$ as the $2\times 2$ identity matrix, we find that $\mathcal{H}_W$ must obey 
\bea
\label{eq:reality}
- \sigma_2 \mathcal{H}^*_W \sigma_2 = \mathcal{H}_W \ .
\eea 
\Eq{eq:reality} requires $d_0 = 0$ or $d_0= \pi$, but the $d_i$ for $i = 1,2,3$, are free. We emphasize that the anti-symmetric $\sigma_2$ matrix appearing in \Eq{eq:reality} is due to $\mathcal{K}^2 = -1$. In the other case where $\mathcal{K}^2 = +1$, we could choose a symmetric matrix, like $\sigma_1$, and we would have a different reality condition. We can now compute the determinant from \Eq{eq:Whamanti}:
\bea
\det[\mathcal{W}] = \exp \lp i \Tr[\mathcal{H}_W] \rp &= \exp (2 i d_0), 
\eea
so because $d_0$ is quantized by $U C_{2z} \mathcal{T}$ to be $0$ or $\pi$, the determinant must equal $+1$. Thus for a Wilson loop with four or more bands, the nested Wilson loop indicates a trivial phase. 

We now consider the special case of a Wilson loop $W^{\Phi/2}(k_1) = e^{ i H_W(k_1)}$ with only two bands, so it is not possible to take a $UC_{2z} \mathcal{T}$-symmetric \emph{subset} of bands. The $2 \times 2$ Wilson Hamiltonian $H_W(k_1)$ obeys  $\mathcal{K} H_W(k_1) \mathcal{K}^{-1} = - H_W(k_1)$ with $\mathcal{K}^2 = -1$. Without loss of generality, we make take $\mathcal{K} = i \sigma_2 K$. We can apply the same reasoning of \Eqs{eq:Whamanti}{eq:reality} but for the Wilson Hamiltonian. We have that $H_W(k_1)$ must be in the form
\bea
\label{eq:ucttrivham}
H_W(k_1) = \sum_{i=0}^3 h_{W,i}(k_1) \sigma_i \\
\eea
where $h_{W,0}(k_1) = 0$ or $h_{W,0}(k_1) = \pi$ but $h_{W,i}(k_1)$ for $i = 1,2,3$ are free. Because the $h_{W,i}$ are free, there are no protected crossings (Weyl nodes) at $\vartheta = 0$ or $\vartheta = \pi$ and no protected winding number, unlike in the $(C_{2z} \mathcal{T})^2 = +1$ case studied in \Ref{2018arXiv180710676S}. Indeed, the Wilson Hamiltonian \Eq{eq:ucttrivham} must be topologically trivial as a map from $k_1 \in S^1$ to $h_{W,i} \in S^3$ because $\pi_1(S^3) = 0$. Thus we have shown that with $U C_{2z} \mathcal{T}$ symmetry satisfying $(U C_{2z} \mathcal{T})^2 = -1$, the Wilson loop with two bands is topologically trivial. 

There is also a more abstract way to understand that $(U C_{2z} \mathcal{T})^2 = -1$ ensures the Wilson Hamiltonian is trivial. Returning to \Eq{eq:uc2thamalg}, we recognize that the Wilson Hamiltonian (assuming $\mathcal{K}^2 = -1$ is the only symmetry of the model) is in the symmetry class $C, d=1$ of the Ten-fold Way \cite{2009AIPC.1134...22K,2010NJPh...12f5010R}. Therefore it has no topological index and is necessarily trivial. In comparison, when we consider $U C_{2z} \mathcal{T}$ symmetry squaring to $+1$, the Wilson Hamiltonian would be in the symmetry class $D, d=1$ in which case there is a $\mathds{Z}_2$ topological invariant which we identify as the quantized determinant of the nested Wilson loop \cite{2018arXiv180611116W, 2018arXiv180710676S}.

\section{The Quantum Spin Hall Model}
\label{app:6}

In this Appendix, we introduce the Quantum Spin Hall (QSH) model of Bernevig, Hughes, and Zhang (BHZ) as a simple lattice model to exemplify our proofs of the Hofstadter topological phase in  \Secs{sec:cherngap}{sec:3DTI} \cite{2006Sci...314.1757B}. In \App{app:QSH}, we recap the essentials of the model before explicitly writing down the Hofstadter Hamiltonian. We enumerate the symmetries of the model and introduce onsite symmetry-breaking perturbations that isolate different topological phases: a mirror Chern insulator, a Kane-Mele insulator, and a fragile $w_2$ insulator. We then add a next-nearest neighbor term to the Hamiltonian to break the flux periodicity from $\Phi = 2\pi$ $(n=1)$ to $\Phi = 4\pi$ $(n=2)$. Using this model, we demonstrate that when $n$ is even, the Hofstadter phase of a zero-field $\mathds{Z}_2$ insulator may be a weak TI or 3D (strong) TI (\App{app:TRissue}). 

\subsection{Hofstadter Hamiltonian}
\label{app:QSH}

\begin{figure}
 \centering
\includegraphics[height=4cm]{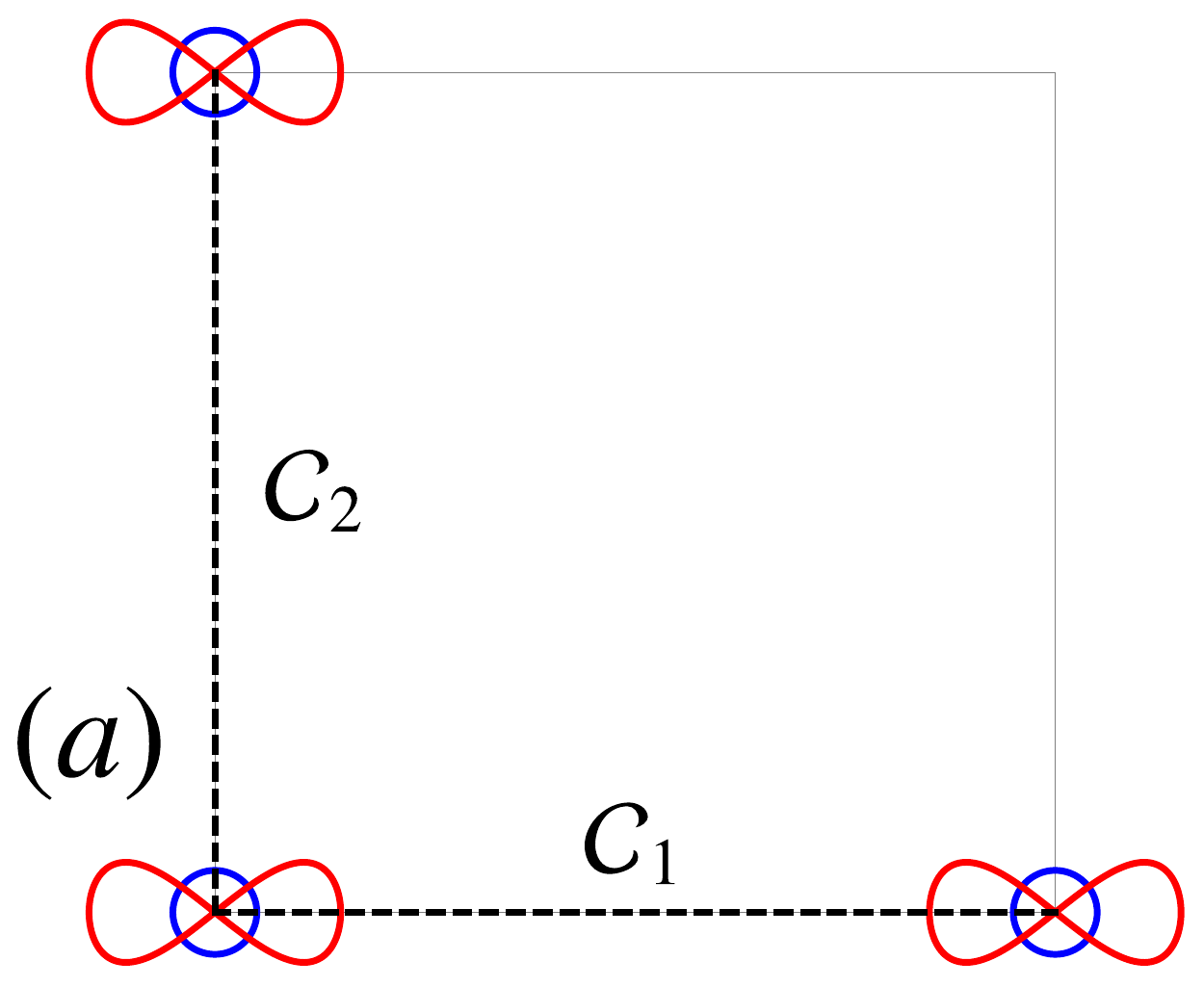} 
\includegraphics[height=3.75cm, trim = 0 -1cm 0 0 , clip]{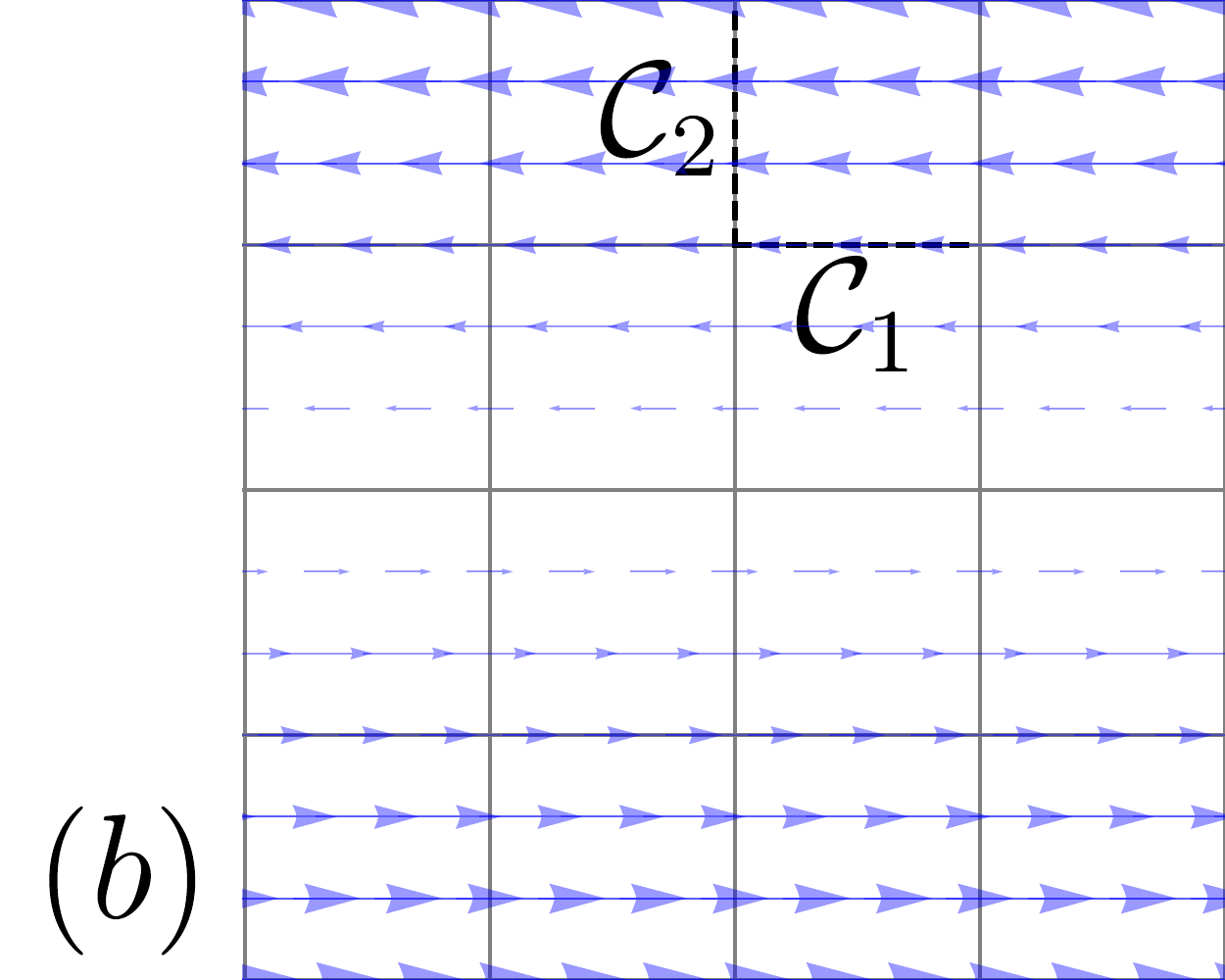} 
\caption{(a) We show the Peierls paths for the QSH model of \App{app:QSH} with the hopping amplitudes. All orbitals are on the $1a $ Wyckoff position and, being $s$ (blue)  and $p_z$ (red) orbitals, are localized near the atoms. and the only hoppings are nearest neighbor. The Peierls paths are taken along the lattice vectors, indicated by $\mathcal{C}_1, \mathcal{C}_2$. (b) We show the lattice with the vector field $\mbf{A} = \phi (-y ,0)$ and the two hoppings in an arbitrary unit cell. We see that along $\mathcal{C}_1$, there is a nonzero phase accumulated along the hopping, but there is no phase accumulated along $\mathcal{C}_2$ because $\mbf{A}$ is always perpendicular to the path of integration.}
\label{fig:QSHpeierls}
\end{figure}

The BHZ Hamiltonian is a model with spin-$1/2$ electrons $\sigma$ and $s,p$ orbitals $\tau$, all at the $1a$ Wyckoff position \cite{2006Sci...314.1757B}. The momentum space Hamiltonian is
\bea
\label{eq:qsh}
H_{QSH}(k_x, k_y) &= (M- \cos k_x  - \cos k_y )I\otimes \tau_3 + \sin k_x \, \sigma_3 \otimes \tau_1 +  \sin k_y \, I \otimes \tau_2 
\eea
and realizes a Quantum Spin Hall effect with a mirror Chern number equal to $-1$ (resp. $+1$) for $0 < M < 2$ (resp. $-2 < M < 0$). We define this topological invariant using the mirror symmetry $M_z = -i \sigma_3 \otimes I$, the product of a rotation $C_{2z} = -i \sigma_3 \otimes \tau_3 $ and inversion $\mathcal{I} = I \otimes \tau_3$ which satisfies $[H_{QSH}(\mbf{k}), M_z] = 0$ for any $\mbf{k}$.  Because $M_z$ is local in the BZ, all occupied bands can be labeled by a mirror eigenvalue $m = \pm i$. Then we define the usual Chern number $C$ and mirror Chern number $C_{M_z}$
\bea
C = C_{m=+i} + C_{m=-i}, \quad C_{M_z} = \frac{C_{m=+i} - C_{m=-i}}{2} \ .
\eea
From these expressions, one can calculate that at half filling, the Chern number $C$ is identically zero (owing to TRS) but that
\bea
C_{M_z} &= \begin{cases}
-1 , & M \in (0, 2) \\
+1 , & M \in (-2, 0) \\
0 , & |M| > 2 \\
\end{cases} \ .
\eea
As per the discussion in \Sec{sec:cherngap}, the Hofstadter Butterfly must have a gap closing when $|M| < 2$ since each mirror block of the Hamiltonian has its own non-zero Chern number. To verify this, we construct the Hofstadter Hamiltonian explicitly. 

The Hofstadter Hamiltonian is simple to construct because all hoppings are nearest neighbor and we take the Peierls path directly along the bonds for all orbitals in the model, as shown in \Fig{fig:QSHpeierls}a. In our Landau gauge $\mbf{A} = (-\phi y,0)$, the only term that acquires a phase the $y$-hopping: $c^\dag_{x+1,y} c_{x,y} \to e^{- i \phi y} c^\dag_{x+1,y} c_{x,y}$ as we depict \Fig{fig:QSHpeierls}b. All closed loops enclose an integer area, so $n=1$, and we can take $\phi = \frac{2 \pi p}{q}$  to recover a $1 \times q$ magnetic unit cell. (Note that because $\mu =1$, there is no distinction between $q$ and $q'$ in this gauge.) At $\Phi = 2\pi$, all the Peierls phases of the model are manifestly zero mod $2\pi$, so it can be trivially shown that the $H^{\phi + \Phi} = H^\phi$, and hence $U = 1$ (\Eq{eq:hoffluxper}). The Hofstadter Hamiltonian is
\bea
\mathcal{H}^\phi_{y,y'} &= \delta_{y,y'} h_y + \delta_{y, y'+1} T + \delta_{y+1, y'} T^\dag \\
h_y &= (M- \cos(k_x- \phi y))I \otimes \tau_3 + \sin(k_x - \phi y) \sigma_3 \otimes \tau_1,\\
T &=  -  \frac{1}{2} e^{-i k_y} I \otimes (\tau_3 - i \tau_2 )
\eea
We promote the Mirror symmetry to $M_z = \delta_{y,y'} (-i \sigma_3 \otimes I)$ since it commutes with each block of the Hofstadter Hamiltonian. Thus we see explicitly that $M_z$ remains a symmetry at all $\phi$. 

As discussed in \Sec{sec:cherngap}, we add symmetry-breaking terms to the QSH model \Eq{eq:qsh} to remove symmetries from the Hamiltonian. This is accomplished by enumerating all of the symmetries of the model. The elementary symmetries are $\mathcal{I}, C_{2z}, C_{4z}, \mathcal{T}$, and $C_{2x}$, and $M_z = \mathcal{I} C_{2z}$. We collect their representations and products in \Tab{tab:qshsym}. For brevity, we do not include the additional products of other symmetries with $C_{4z}$ because they are broken if $C_{4z}$ is broken. Similarly, $C_{4z}$ is broken if $C_{2z}$ is broken. 

Inversion symmetry $\mathcal{I}$ and $C_{4z}$ also require composition with other gauge-dependent unitary operators to remain symmetries at $\phi \neq 0$. The Landau gauge preserves inversion symmetry. However, the $1\times q$ magnetic unit cell at $\phi = \frac{2\pi p}{q}$ breaks inversion symmetry. Hence we find that $\mathcal{I}$ acting on the Hofstadter Hamiltonian must be multiplied by a permutation matrix $\mathcal{D}$ that inverts the order of unit cells within the $1 \times q$ magnetic unit cell. Explicitly, $\mathcal{D}_{\ell, \ell'} = \delta_{q-\ell,\ell'} $. The Landau gauge explicitly breaks the $C_{4z}$ symmetry, so a gauge transformation is necessary. In \Ref{hofsymtoappear}, we will develop the full theory of space group symmetries in the Hofstadter Butterfly. Because we do not use the $C_{4z}$ symmetry in this work, we refer the reader to \Ref{hofsymtoappear} for further details. Considering only the spatial structure and not the orbital character, a generic term in the $H_{QSH}$ (\Eq{eq:qsh}) transforms as
\bea
\label{eq:c4transform}
\sum_{\mbf{R}}C_{4z} \left[ \exp \lp i \int_{\mbf{R}}^{\mbf{R} + \mbf{a}_i} \mbf{A}(\mbf{r}) \cdot d \mbf{r} \rp c^\dag_{\mbf{R} + \mbf{a}_i} c_{\mbf{R} }\right] C_{4z}^\dag &= \sum_{\mbf{R}} \exp \lp i \int_{\mbf{R}}^{\mbf{R} + \mbf{a}_i} \mbf{A}(\mbf{r}) \cdot d \mbf{r} \rp c^\dag_{C_{4z} (\mbf{R} + \mbf{a}_i)} c_{C_{4z} \mbf{R}} \\
&= \sum_{\mbf{R}} \exp \lp i \int_{C_{4z}^{-1} \mbf{R}}^{C_{4z}^{-1} (\mbf{R} + C_{4z} \mbf{a}_i)} \mbf{A}(\mbf{r}) \cdot d \mbf{r} \rp c^\dag_{\mbf{R} + C_{4z} \mbf{a}_i} c_{\mbf{R}} \\
&= \sum_{\mbf{R}} \exp \lp i \int_{\mbf{R}}^{\mbf{R} + C_{4z} \mbf{a}_i} C_{4z}^{-1} \mbf{A}(C_{4z}\mbf{r}) \cdot d \mbf{r} \rp c^\dag_{\mbf{R} + C_{4z} \mbf{a}_i} c_{\mbf{R}} \ . \\
\eea
This shows us that, because the Peierls phase acquired from the hopping $\mbf{R} \to \mbf{R} + \mbf{a}_i$ is different from the phase acquired from $C_{4z} \mbf{R} \to C_{4z} (\mbf{R} + \mbf{a}_i)$, a gauge transformation $G$ is required. We define
\bea
G c_{\mbf{R}, \al} G^{-1} = e^{- i \la(\mbf{R})} c_{\mbf{R},\al}, \quad G &= \exp \lp i \sum_{\mbf{R}} c^\dag_{\mbf{R}, \al} c_{\mbf{R}, \al} \la(\mbf{R}) \rp , 
\eea
which acts as a gauge transformation of the vector potential in the Peierls substitution: $
G c^\dag_{\mbf{R}+\mbf{a}_i} c_{\mbf{R}} G^{-1} = \exp ( i \int_{\mbf{R}}^{\mbf{R}+\mbf{a}_i} \pmb{\nabla} \la \cdot d\mbf{r} ) c^\dag_{\mbf{R}+\mbf{a}_i} c_{\mbf{R}}$. $\la(\mbf{r})$ must satisfy
\bea
\label{eq:lambdac4}
C_{4z}^{-1} \mbf{A}(C_{4z}\mbf{r})  - \pmb{\nabla} \la(\mbf{r}) &=  \mbf{A}(\mbf{r}) \ .
\eea
It is trivial to check that $\la(\mbf{r}) = \phi x y$ satisfies \Eq{eq:lambdac4}. This completes our discussion of the symmetries. 

\begin{table}
\caption{Symmetries of the BHZ Model} % title of Table
\centering % used for centering table
\begin{tabular}{|c |c | c| c| } % centered columns (4 columns)
\hline
%heading
Symmetry &$ \phi=0 $&$ \phi\neq 0$& Mapping of $\mbf{k}, \phi$  \\
\hline
$\mathcal{I} $ & $ I \otimes \tau_3	$ & $ \mathcal{I} \mathcal{D} $ & $ (-k_x, -k_y, \phi)$ \\
\hline
$\mathcal{T}  $ & $ i\sigma_2 \otimes I K$ & $\mathcal{T}$ & $(-k_x, -k_y, - \phi)$\\
\hline
$C_{2z}  $ & $ - i \sigma_3 \otimes \tau_3$ & $(C_{4z} G)^2 $ & $(-k_x, -k_y, \phi)$ \\
\hline
$C_{4z}$ & $ 1/\sqrt{2}(I + C_{2z})$ & $C_{4z} G$ & $(-k_y,k_x, - \phi)$ \\
\hline
$M_z $ & $ - i \sigma_3 \otimes I	$ & $M_z$ & $(k_x, k_y, \phi)$ \\
\hline
$C_{2x} $ & $	i \sigma_1 \otimes \tau_3$ & $ G^{-1} C_{2x} G$  & $(k_x, -k_y, - \phi)$ \\
\hline
$ M_z C_{2x}  $ & $i \sigma_2 \otimes \tau_3	$ & $G^{-1} M_z C_{2x} G $ & $(k_x, -k_y, - \phi)$\\
\hline
$\mathcal{I}\mathcal{T} $ & $i \sigma_2 \otimes \tau_3 K	$ & $\mathcal{I}T  \mathcal{D}  $  & $(k_x, k_y, - \phi)$ \\
\hline
$C_{2z}\mathcal{T} $ & $-i \sigma_1 \otimes \tau_3 K$ & $(C_{4z} G)^2 \mathcal{T} $ & $(k_x, k_y, -\phi)$ \\
\hline
$M_{z} \mathcal{T} $ & $	-i \sigma_1 \otimes I K$ &$M_z \mathcal{T}$ & $(-k_x, -k_y, - \phi)$ \\
\hline
$\mathcal{I} C_{2x} $ & $i \sigma_1 \otimes I$ & $\mathcal{I} C_{2x}$ & $(-k_x, k_y, - \phi)$\\
\hline
$\mathcal{I} M_z C_{2x} $ & $i \sigma_2 \otimes I$ & $\mathcal{I} M_z C_{2x}$ & $(-k_x, k_y, - \phi)$ \\
\hline
$\mathcal{T} C_{2x} $ & $-i \sigma_3 \otimes \tau_3 K$ & $ \mathcal{T} G^{-1} C_{2x} G $ & $(-k_x, k_y, \phi)$\\
\hline
$\mathcal{T} M_z C_{2x} $ & $- I \otimes \tau_3 K$ & $\mathcal{T} G^{-1} M_z C_{2x} G $ & $(-k_x, k_y, \phi)$\\
\hline
$\mathcal{I} \mathcal{T} C_{2x} $ & $- i \sigma_3 \otimes I K$ & $\mathcal{I} \mathcal{T} C_{2x} $ & $(k_x, -k_y, \phi)$ \\
\hline
$\mathcal{I} \mathcal{T} M_z C_{2x} $ & $K$ & $\mathcal{I} \mathcal{T} M_z C_{2x} $ & $(k_x, -k_y, \phi)$\\
\hline\end{tabular} \\
\justify
We list the symmetries of the BHZ model in column 1 and their representations on the Bloch Hamiltonian in column 2. In column 3, we provide their representations on the Hofstadter Hamiltonian $\mathcal{H}_{QSH}$ in the presence of nonzero flux. The symmetries listed in column 3 refer to the $4\times 4$ representations defined in column 2. Note that for $\phi \neq 0$, some of the symmetries are broken, and take $\phi \to - \phi$. The mapping of $\mbf{k}$ and $\phi$ under the action of the symmetries is shown in column 4.
\label{tab:qshsym}
\end{table}

\begin{figure*}
 \centering
\includegraphics[width=8cm]{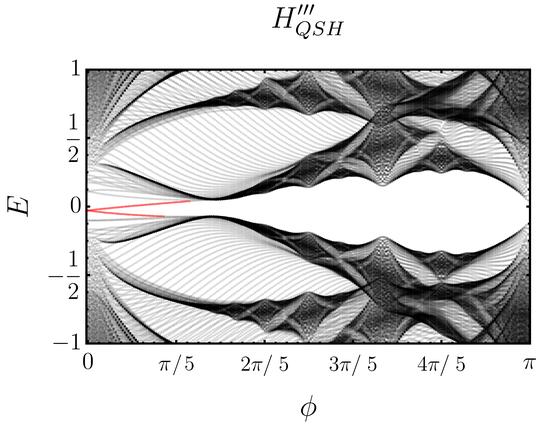} 
\caption{We show the Hofstadter Butterfly for $H'''_{QSH}$ computed on a $25 \times 25$ unit cell lattice with open boundary conditions along both directions. The model as defined in \Tab{tb:qshhams} is in the topological phase and exhibits corner modes (red) that are pumped into the bulk (grey) as $\phi$ is increased.} 
\label{fig:QSHhoti}
\end{figure*}

From \Tab{tab:qshsym} , it can be checked that the Hamiltonian with onsite perturbation 
\bea
H'_{QSH} = H_{QSH} + \eps_1 I\otimes \tau_1 + \eps_2 \sigma_3 \otimes \tau_2 
\eea
breaks all symmetries except $M_z, \mathcal{T}, M_z \mathcal{T}$. Thus $H'_{QSH}$ can have no Chern number that would enforce a gap closing, but does still have a Mirror Chern number. If we also add a mirror-breaking term, defining 
\bea
H''_{QSH} = H'_{QSH} + \eps_3 \sigma_2 \otimes \tau_2,
\eea
then we lose the mirror Chern number and a bulk gap may open (see \choose{\Fig{fig:hofs12}}{\Fig{fig:hofs12} of the Letter}). As demonstrated in Sec. III, $H''_{QSH}$ -- which has only $\mathcal{T}$ symmetry -- can be classified as a 3D TI with gapless surface states in $\phi$ for open boundary conditions, but it need not have gapless bulk states. Finally, we also can break all symmetries except $C_{2z} \mathcal{T}$. Explicitly, we let 
\bea
H_{QSH}''' = H_{QSH} + \eps_4 I  \otimes \tau_2 + \eps_5 (\sigma_1 + \sigma_2) \otimes I + \eps_6 (\sigma_1 \otimes \tau_2 + \sigma_2 \otimes \tau_2 + \sigma_1 \otimes \tau_3)
\eea
which preserves $C_{2z} \mathcal{T}$ only and opens a gap at all $\phi$. We show the Hofstadter Butterfly for this model in \Fig{fig:QSHhoti} and confirm the pumping of corner states that characterizes a HOTI. Note that because $U=1$ (\Eq{eq:hoffluxper}) in our gauge and $H^{\phi = \Phi}$ is identical to $H^{\phi=0}$, we trivially have $(U C_{2z} \mathcal{T})^2 =  (C_{2z} \mathcal{T})^2 = 1$, which agress with our general calculation in \Eq{eq:toorigin}. We list the values of the parameters in these perturbed Hamiltonians in \Tab{tb:qshhams}. 

\begin{table}
\caption{QSH Hamiltonians and their Symmetries} % title of Table
\centering % used for centering table
\begin{tabular}{|c |c | c|} % centered columns (4 columns)
\hline
%heading
Hamiltonian  & Symmetries & Parameters \\
\hline
$H'_{QSH} = H_{QSH}+ \eps_1 I \otimes \tau_1 + \eps_2 \sigma_3 \otimes \tau_2 $ & $ M_z, \mathcal{T}$ & $\eps_1 = .1, \eps_2 = .11, M = 1.5 $\\
\hline
$H''_{QSH} = H_{QSH}+ \eps_1 I \otimes \tau_1 + \eps_2 \sigma_3 \otimes \tau_2 + \eps_3 \sigma_2  \otimes \tau_2 $ & $ \mathcal{T} $ & $\eps_1 = .1, \eps_2 = .11, \eps_3 = .12, M = 1.5 $\\
\hline
$H'''_{QSH} =H_{QSH}+  \eps_4 I  \otimes \tau_2 + \eps_5 (\sigma_1 + \sigma_2) \otimes I + \eps_6 (\sigma_1 \otimes \tau_2 + \sigma_2 \otimes \tau_2 + \sigma_1 \otimes \tau_3)$ & $ C_{2z} \mathcal{T}$ & $\eps_4 = .1, \eps_5 = .11, \eps_6 = .05, M = 1.6 $\\
\hline
\end{tabular}
\label{tb:qshhams}
\justify
We list the variations of the BHZ model (column 1), the symmetries they retain from the perturbations (column 2), and the values of the parameters that realize their nontrivial Hofstadter topology (column 3). The Hofstadter Butterflies of the models may be found in \choose{\Fig{fig:hofs12} and \Fig{fig:QSHhoti}}{\Fig{fig:hofs12} of the Letter and \Fig{fig:QSHhoti}}. 
\end{table}

\subsection{Discussion of the Proof  for $\mathcal{T}$-symmetric TIs with General Flux Periodicity $\phi \to \phi + 2 \pi n$}
\label{app:TRissue}

\begin{figure}[h!]
 \centering
\includegraphics[width=5.cm, trim = 0 -3.1cm 0 0 , clip]{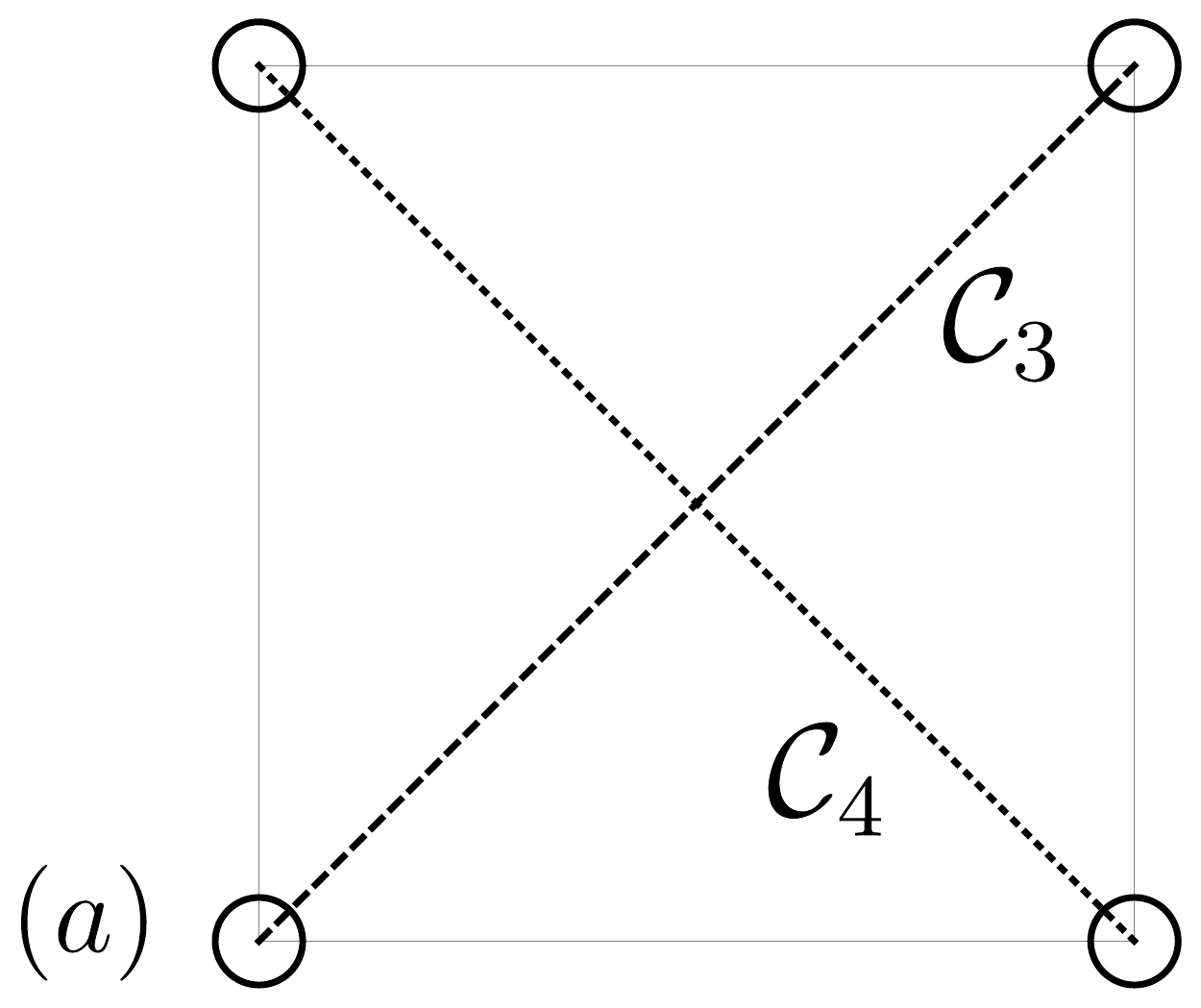}  \qquad \qquad
\includegraphics[height=6.5cm]{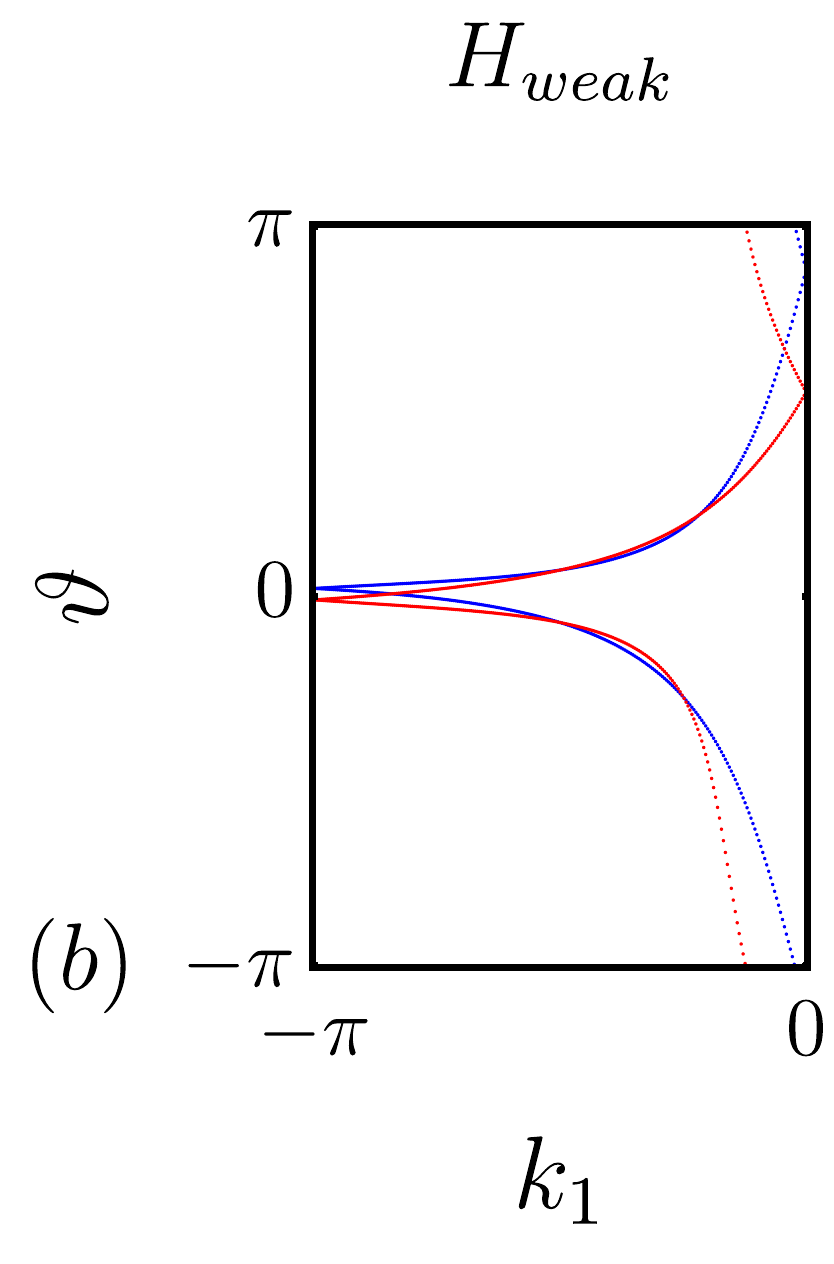} \qquad \qquad
\includegraphics[height=6.5cm]{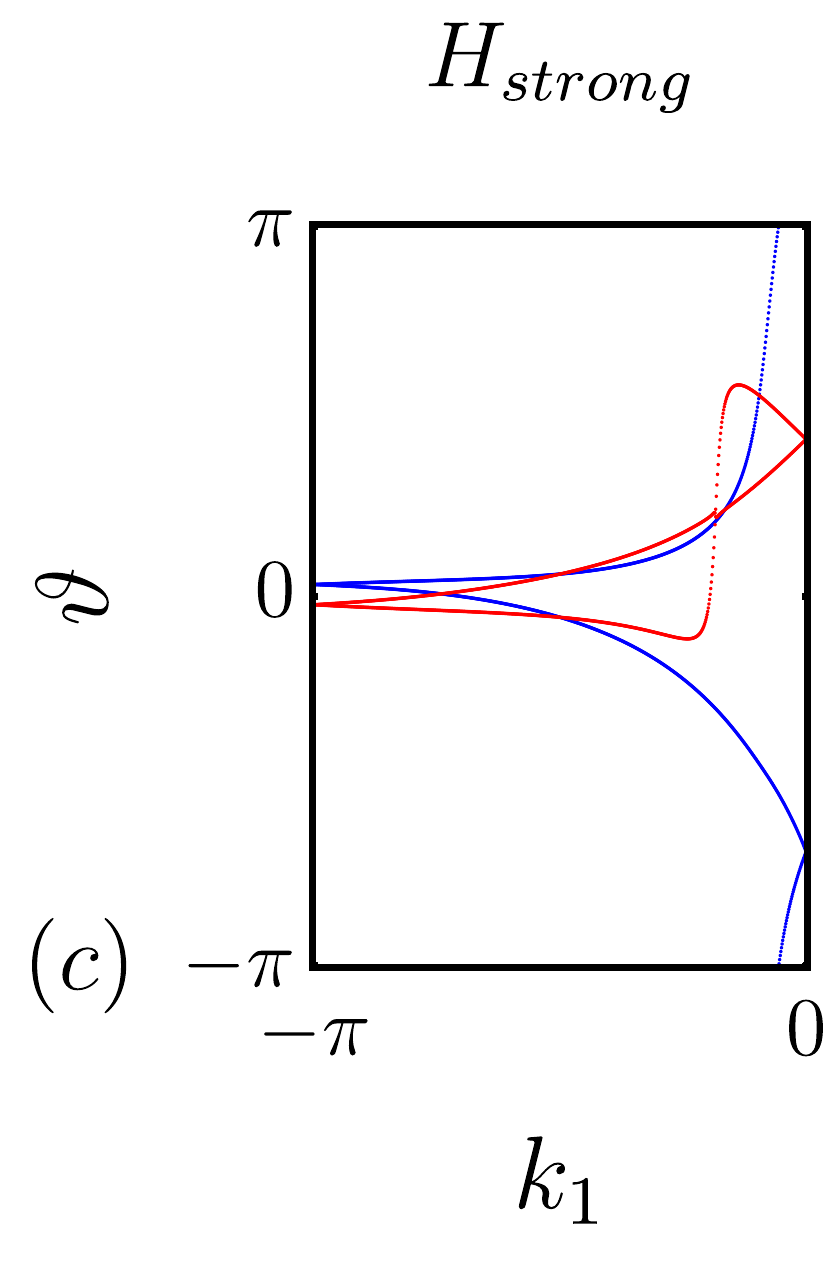} 
\caption{(a) In \App{app:TRissue}, we introduce two next-nearest neighbor hoppings of amplitude $\la$ whose Peierls paths are taken along the diagonals and are indicated by $\mathcal{C}_3, \mathcal{C}_4$. We calculate the Peierls phase in \Eq{eq:QSHbreakpeierls}. (b) We show the Wilson loop spectra for the model $H_{weak}$ of \Tab{tb:weakstrong} which is $4\pi$-periodic in flux at $\phi = 0$ (blue) and $\phi = 2\pi$ (red).  We calculate the $\mathds{Z}_2$ invariant by observing that any line of constant $\vartheta$ intersects an odd number of eigenvalues. We find that both phases at $0$ and $2\pi$ are nontrivial, and the model is a weak TI. (c) By tuning the parameters to $H_{strong}$ of \Tab{tb:weakstrong}, the model can reach a strong 3D TI phase where the $\phi = 0$ model is $\mathds{Z}_2$ nontrivial, but the $\phi = 2\pi$ model is trivial because lines of constant $\vartheta$ intersect an even number of eigenvalues. In both cases, the 3D topology is nontrivial. 
}
\label{fig:QSHpeierls2}
\end{figure}

In this section, we discuss how our proof of the trivial $\mathds{Z}_2$ invariant at $\phi =\Phi/2, \Phi = 2 \pi n$ (see Sec. III) relies on $n \in \mathbb{N}$ being odd. This leads to a $k_1 \to k_1 + \pi$ periodicity in the magnetic BZ which ensures every gap closing comes in pairs. If $n$ is even, this proof fails because there is no increased periodicity along $k_1$. In this case, it is unclear a priori whether the $\phi = \Phi/2$ model may be trivial or nontrivial. We show an example of a model with $n=2$ that can realize either a trivial or nontrivial phase at $\phi = \Phi/2$, confirming that the Hofstadter topology is not uniquely determined by the zero-field topology when $n$ is even.

We begin with the QSH model of \App{app:QSH}. To break the $\Phi = 2\pi$ periodicity to a $\Phi = 4\pi$ periodicity, we add next nearest-neighbor diagonal hoppings as shown in \Fig{fig:QSHpeierls2}a. This term breaks the $\phi \to \phi+2\pi$ periodicity because now half a unit cell can be encircled along Peierls paths. However, it does not change $\mu$ because the new hoppings obey $\mbf{b}_1 \cdot ( (\mbf{R} + \pmb{\delta}_{\al}) - (\mbf{R}' + \pmb{\delta}_{\be})) =  \mbf{b}_1 \cdot (\mbf{a}_1 + \mbf{a}_2) = 1$, which does not affect \Eq{eq:mudef}. We couple $s$ orbitals to $p$ orbitals, preserving $\mathcal{T}$ symmetry, and find that in momentum space, the appropriate term is 
\bea
H_{diag}(\mbf{k}) &=  \la \, I \otimes \tau_1 (\cos(k_x + k_y) + \cos(k_x - k_y) ) \ . \\
\eea
Consulting \App{app:QSH}, we see that this term breaks a number of the zero-field symmetries ($\mathcal{I}, C_{2z}, C_{4z}, C_{2x}, M_z C_{2x}, \mathcal{I} \mathcal{T}, C_{2z} \mathcal{T}, C_{2x} \mathcal{T}, M_{z} C_{2x} \mathcal{T}$) but preserves $\mathcal{T}, M_z, M_z \mathcal{T}, \mathcal{I} C_{2x}, M_z \mathcal{I} C_{2x}, \mathcal{I} \mathcal{T} C_{2x}, M_z  \mathcal{I} \mathcal{T} C_{2x}$. We break all the remaining symmetries except $\mathcal{T}$ using the perturbations shown in \Tab{tb:weakstrong}.  The Peierls phase for the two diagonal hoppings are found to be
\bea
\label{eq:QSHbreakpeierls}
\la \to \la(\phi) =  \la \exp \lp i \int_{(x,y)}^{(x+1, y\pm1)} \mbf{A} \cdot d\mbf{r} \rp = t \exp \lp -i \phi y \pm i \frac{\phi}{2} \rp \ . \\
\eea
At $\phi = 2\pi$ where the new model is $\mathcal{T}$ symmetric, the magnetic unit cell is the same as at $\phi=0$. Indeed, adding the new hopping has not changed $\mu =1$, thus \emph{not} requiring an enlarged magnetic unit cell at $\phi = 2\pi$, and the Hamiltonian is the same except for the diagonal coupling which obeys $\la(2\pi) = - \la$. We now construct two models: $H_{weak}$ is characterized by the parameters $\la = -0.1, M = 1.7$ and and $H_{strong}$ is characterized by $\la = -0.2, M = 1.85$ (see \Tab{tb:weakstrong}), each of which has a nontrivial $\mathds{Z}_2$ invariant at $\phi = 0$. We show in \Fig{fig:QSHpeierls2}b that the first model $H_{weak}$ exhibits a weak Hofstadter TI phase where both the $\phi = 0$ and $\phi = \Phi/2 = 2\pi$ Hamiltonians have nontrivial $\mathds{Z}_2$ invariants, so $\theta = 0$. On the contrary, we show in \Fig{fig:QSHpeierls2}c that the second model $H_{strong}$ exhibits a strong Hofstadter TI phase where the $\mathds{Z}_2$ invariant at $\phi = \Phi/2$ is trivial, and hence $\theta = \pi$. From these two examples, we see that the topology at $\Phi/2 = n \pi$ cannot be uniquely determined from the $\phi = 0$ topology when $n$ is even, and thus the Hofstadter Hamiltonian may be either a weak or strong TI. 

\begin{table}
\caption{Weak and Strong TI Hamiltonians} % title of Table
\centering % used for centering table
\begin{tabular}{|c |c | c|} % centered columns (4 columns)
\hline
%heading
Hamiltonian  & Symmetries & Parameters \\
\hline
$H_{weak} = H_{QSH}+ \eps_1 I \otimes \tau_1 + \eps_2 \sigma_3 \otimes \tau_2 + \eps_3 \sigma_2  \otimes \tau_2 + \la H_{diag} $ & $ \mathcal{T} $ & $\eps_1 = .2, \eps_2 = .21, \eps_3 = .22, \la = -.1, M = 1.7 $\\
\hline
$H_{strong} = H_{QSH}+ \eps_1 I \otimes \tau_1 + \eps_2 \sigma_3 \otimes \tau_2 + \eps_3 \sigma_2  \otimes \tau_2 + \la H_{diag} $ & $ \mathcal{T} $ & $\eps_1 = .2, \eps_2 = .21, \eps_3 = .22, \la = -.2, M = 1.85 $\\\hline
\end{tabular}
\label{tb:weakstrong}
\justify
We list two variants of BHZ model which realize a weak and strong 3D TI phase (column 1). Each has only $\mathcal{T}$ symmetry (column 2). The values of the parameters for these phases are found in column 3. 
\end{table}

\section{A Model of Twisted Bilayer Graphene}
\label{app:7}

In this Appendix, we introduce a model of twisted bilayer graphene on the Moir\'e lattice. Briefly, we review the fragile topology in zero field, and then we move on to construct the Hofstadter Hamiltonian ( \App{app:TBG}). In \App{app:tbgsymbreak}, we introduce terms that isolate various symmetries of the model, and use these perturbed Hamiltonians to illustrate the results of \Sec{sec:fragile}. We also provide a detailed discussion of the $U C_{2z} \mathcal{T}$ symmetry at $\phi = \Phi/2$ and its constraints on the Wilson loop in both the Landau gauge magnetic unit cell and an expanded unit cell (\App{app:C2zTwilson}). We discuss the bulk gap closings enforced by $C_{2z}$ as mentioned in \Sec{sec:fragile} and the Wannier flow protected by $C_{2z} \mathcal{T}$ from a real space perspective (\App{app:symeig}). Finally, we also argue that $C_{2x} \mathcal{T}$ can protect a bulk gapless point in \App{app:c2xt}. 

\subsection{The Hofstadter Hamiltonian}
\label{app:TBG}

\begin{figure}
 \centering
\includegraphics[width=10cm]{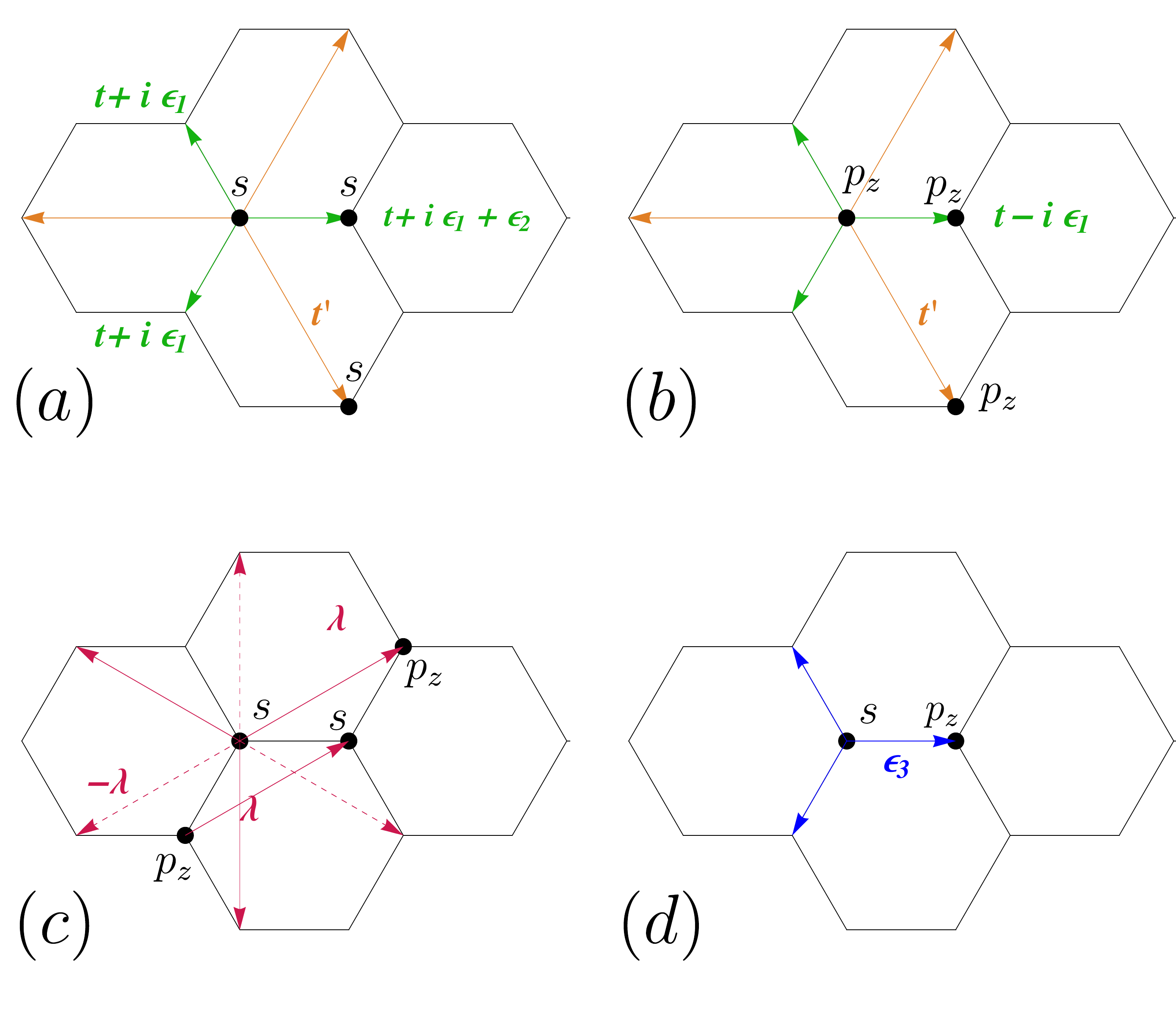} 
\includegraphics[height=6.6cm, trim = 0 -1cm 0 0, clip]{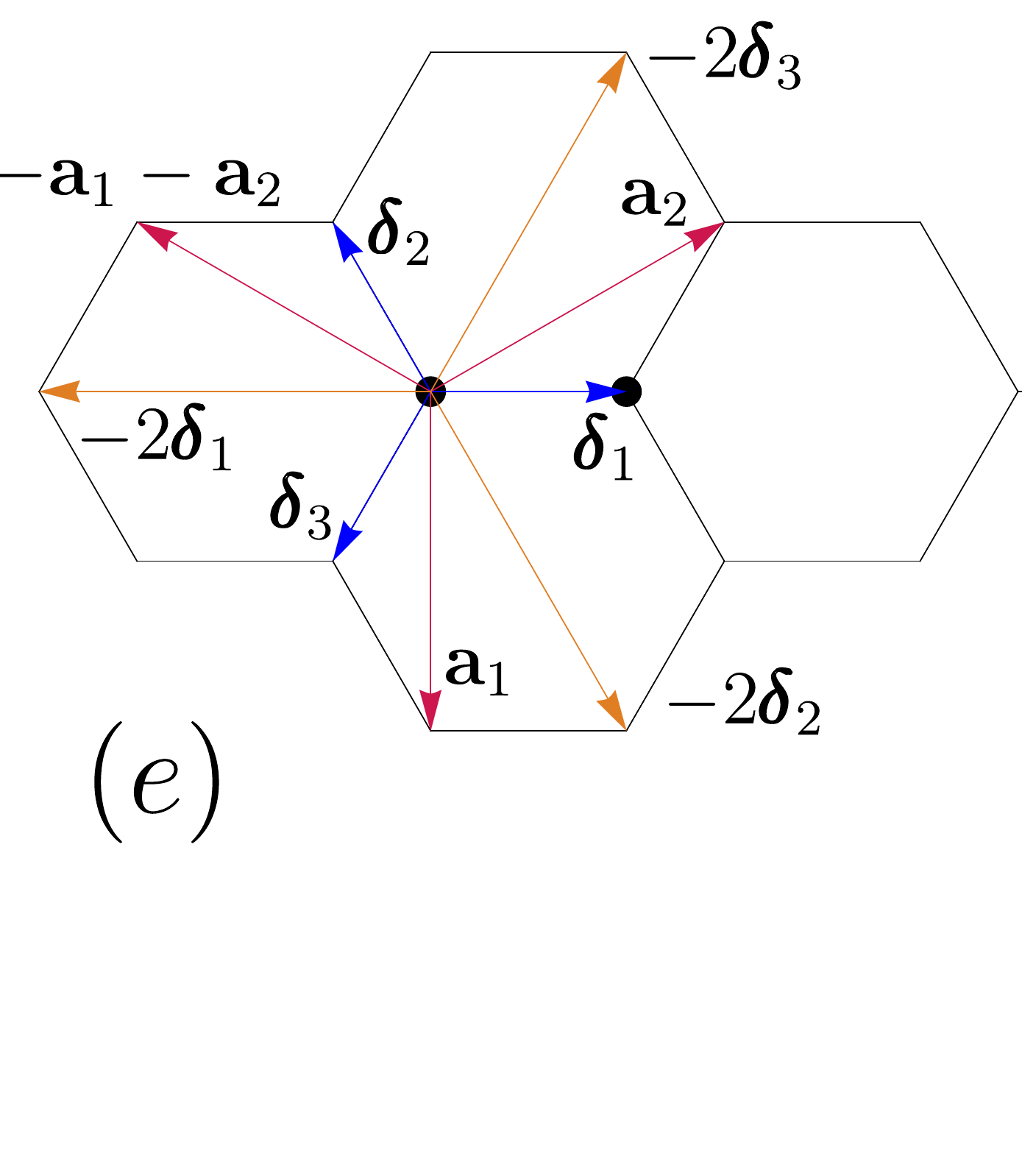}   \\
{\larger[2] (f)} \includegraphics[width=4.5cm]{tbgpeierls1} 
{\larger[2] (g)} \includegraphics[width=5cm]{tbgpeierls2} 
{\larger[2] (h)} \includegraphics[width=5cm]{tbgpeierls3} 
\caption{$(a)-(d)$ We depict the hoppings $t, t', $ and $\la$ for the $s$ and $p_z$ orbitals of the model in \Ref{2018arXiv180710676S}. We also show the amplitudes $\eps_1, \eps_2$, and $\eps_3$ of the symmetry breaking terms which are discussed in \App{app:tbgsymbreak}. For visual clarity, the arrows denoting the hoppings do \emph{not} correspond to the Peierls paths which are instead shown in \Fig{fig:TBGhoppings}. $(e)$ We also depict the first-, second-, and third-nearest neighbor vectors of the model. (f) We show the two Peierls paths of the nearest-neighbor hopping, $\mathcal{C}_1$ and $\mathcal{C}_2$, we are taken in superpositon as we discuss in \Fig{fig:TBGhoppings}. (g) We show the second nearest-neighbor path $\mathcal{C}_3$. (f) We show the second nearest-neighbor path $\mathcal{C}_4$. 
}
\label{fig:TBGfigs}
\end{figure}

In \Ref{2018arXiv180710676S}, a 4-band model of twisted bilayer graphene was constructed on the Moir\'e lattice to capture the phenomenology of the fragile topology inherent to the system. In momentum space, the model is written
\bea
H_{TBG}(\mbf{k}) &= \Delta \mu_3 \otimes \sigma_0 + \mu_0 \otimes \sigma_1 \sum_{i=1}^3 \ t \cos (\pmb{\delta}_i \cdot \mbf{k}) + t' \cos (- 2 \pmb{\delta}_i \cdot \mbf{k})  \\&-\mu_0 \otimes\sigma_2  \sum_{i=1}^3 \ t \sin (\pmb{\delta}_i \cdot \mbf{k})  + t' \sin (- 2 \pmb{\delta}_i \cdot \mbf{k}) \ - 2 \la \mu_2 \otimes \sigma_3 \sum_{i=1}^3 \sin ( \mbf{d}_i \cdot \mbf{k} ) \ . 
\eea
Here,  $\mu$ (resp. $\sigma$) are the $s,p$ orbital (resp. sublattice) Pauli matrices, and the Moir\'e lattice vectors are $\mbf{a}_1 = \sqrt{3} (0,-1) D, \mbf{a}_2 = (\frac{3}{2} , \frac{\sqrt{3}}{2} )D$ where $D$ is the length of the Moir\'e superlattice unit cell edge. The nearest-neighbor vectors are $\pmb{\delta}_1= \frac{1}{3} \mbf{a}_1 + \frac{2}{3} \mbf{a}_2, \pmb{\delta}_2 = -\frac{2}{3} \mbf{a}_1 - \frac{1}{3} \mbf{a}_2, \pmb{\delta}_3 = \frac{1}{3} \mbf{a}_1 - \frac{1}{3} \mbf{a}_2$ and second-nearest-neighbors are $\mbf{d}_1 = \mbf{a}_1, \mbf{d}_2 = \mbf{a}_2, \mbf{d}_3 = -\mbf{a}_1 - \mbf{a}_2$. We depict the hopping amplitudes in \Fig{fig:TBGfigs}a-\Fig{fig:TBGfigs}d and the vectors in \Fig{fig:TBGfigs}e. From here forward, we normalize $\mbf{a}_1$ and $\mbf{a}_2$ to have unit area matching the convention established earlier, so we take $\mbf{a}_i \to \mbf{a}_i / \Omega, \ \Omega = \mbf{a}_1 \times \mbf{a}_2 = \frac{3 \sqrt{3}}{2} D^2 $. The Peierls paths $\mathcal{C}_1,\mathcal{C}_2,\mathcal{C}_3$ and $\mathcal{C}_4$ of the model are sketched in \Fig{fig:TBGfigs}.

We define the lattice with the center of the honeycomb at the origin, so the orbitals are located at $(r_1 - \frac{1}{2}) \mbf{a}_1 + r_2 \mbf{a}_2 \pm \frac{1}{2} \pmb{\delta}_1$. Following \Ref{2018arXiv180710676S}, we choose parameters $t' = -t/3, \la = \sqrt{2/27} t$ in which case the onsite splitting $\Delta$ determines the topology. For $|\Delta| < 2t$, there is a fragile pair winding of 1 in the Wilson loop of the occupied bands at half filling, which we plot in \Fig{fig:TBGcorner}b. The winding is protected by a $C_{2z} \mathcal{T} = I \otimes \sigma_1 K$ symmetry, and the model also has physical $C_{3z} = I, C_{2x} = \mu_3 \otimes I$ symmetries \cite{2018arXiv180710676S}. As written, this model also exhibits the ``accidental" symmetries $C_{2z} =  I \otimes \sigma_1, \mathcal{T} = K$ which are separately preserved in the model but need not be preserved in the physical system. Note that $C_{2z}^2 = \mathcal{T}^2 = +1$, indicating spinless particles. 

\begin{figure}
 \centering
\includegraphics[width=5.7cm]{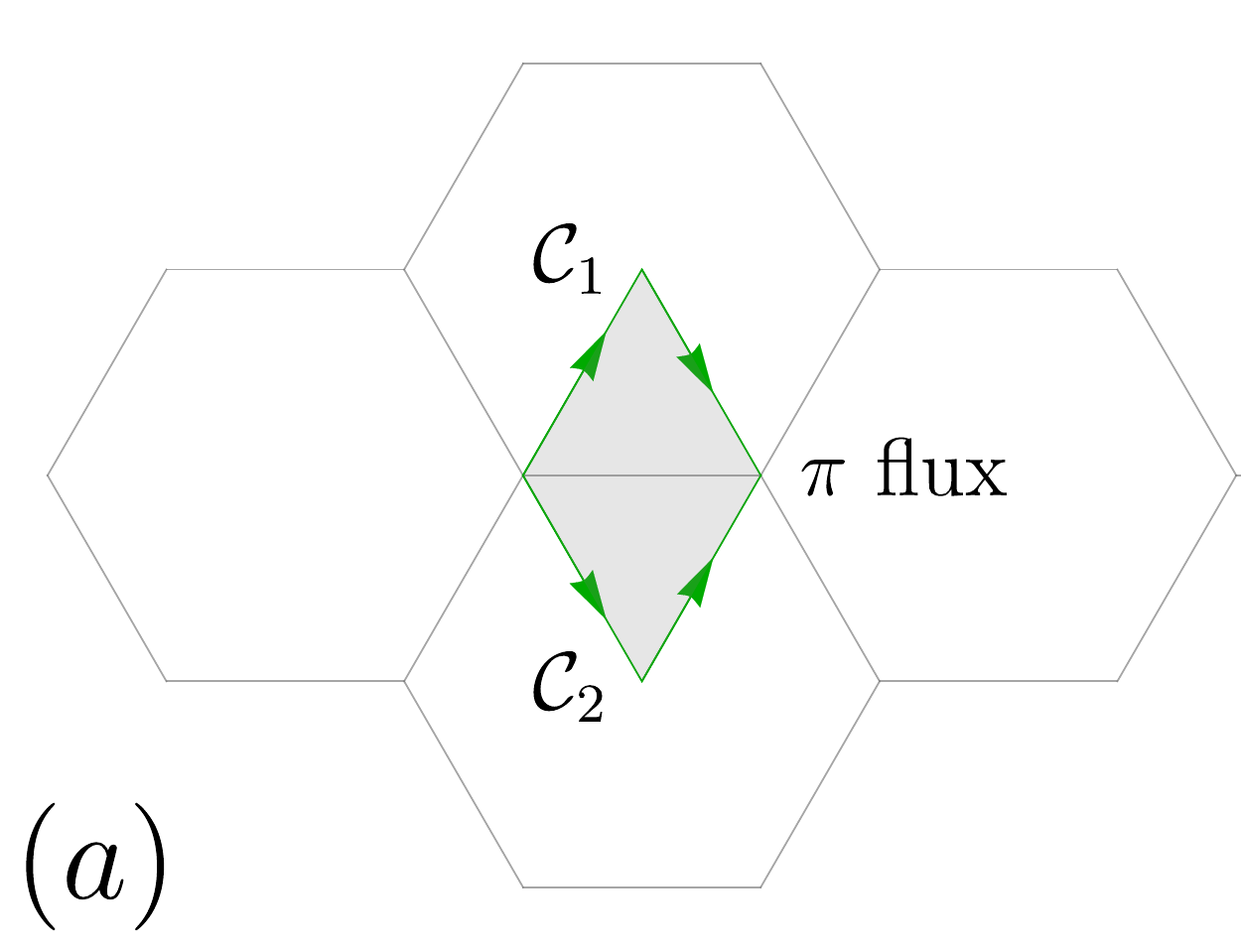}   
\includegraphics[width=5.7cm]{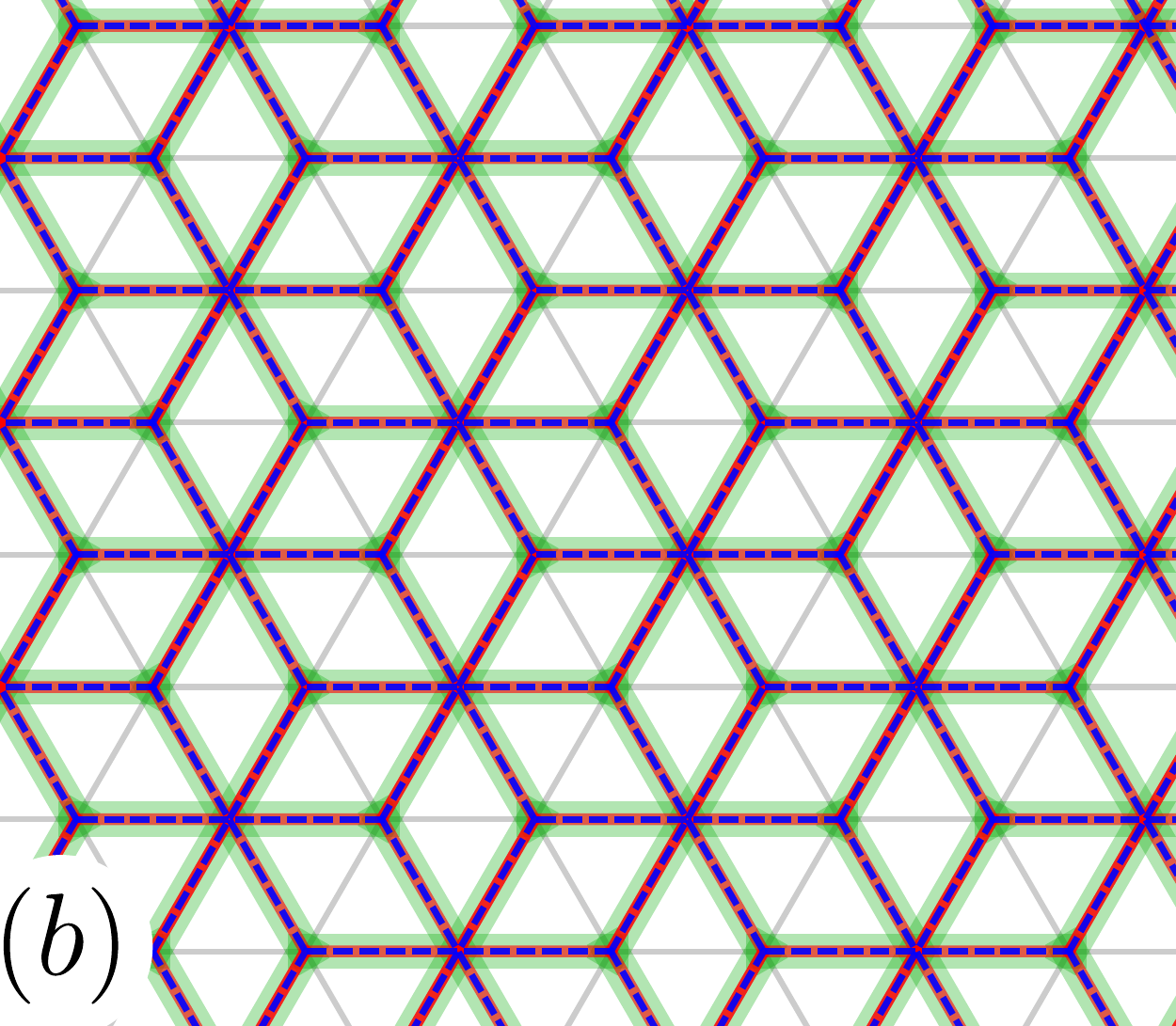}  \quad 
\includegraphics[width=5.7cm]{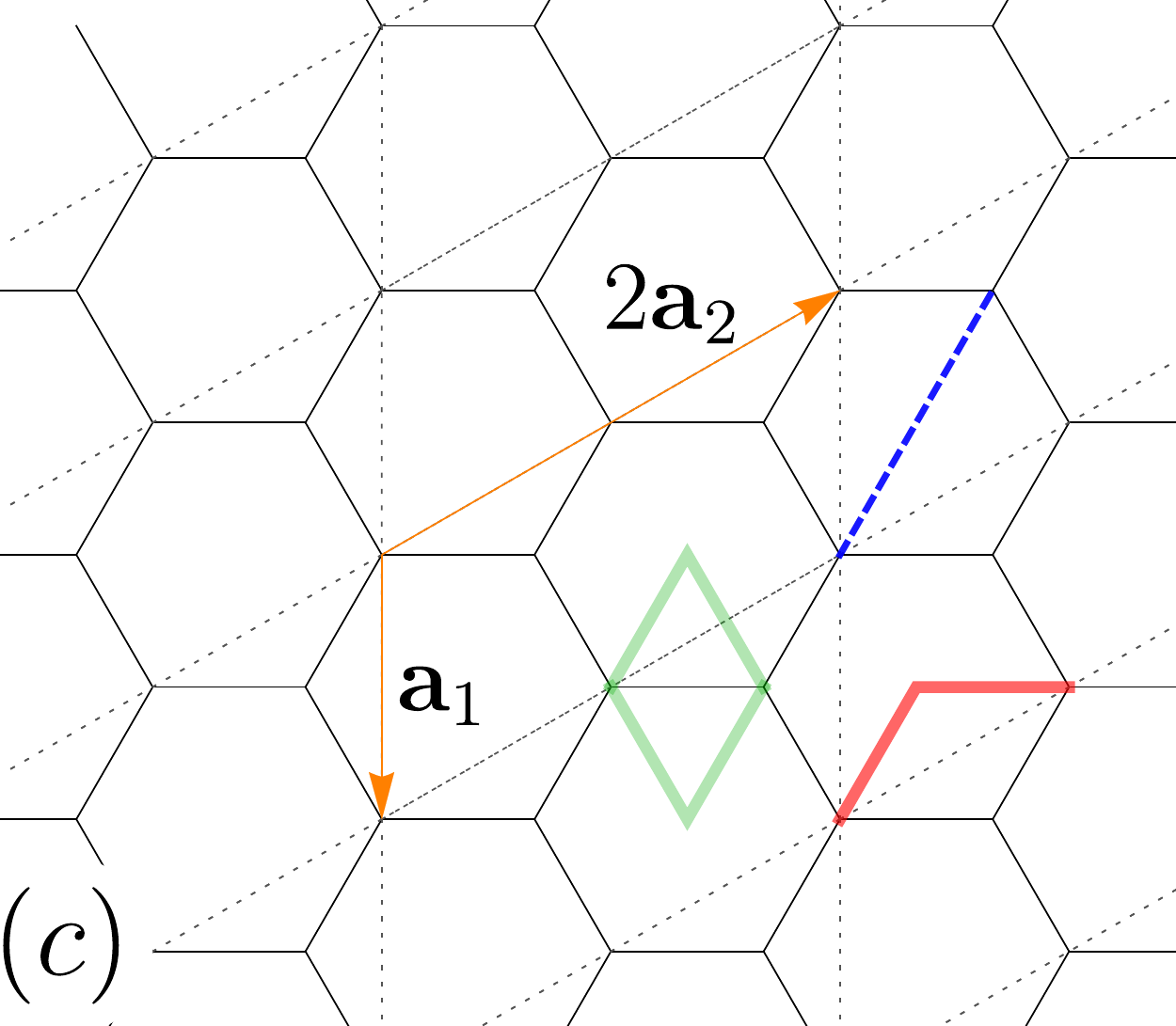}   
\caption{  $(a)$ We sketch the two paths ($\mathcal{C}_1$ and $\mathcal{C}_2$) of a nearest-neighbor hopping which are taken in superposition. At $\phi = \Phi/2 = 3\pi$, the phase difference between $\mathcal{C}_1$ and $\mathcal{C}_2$ is equal to $3\pi \times \frac{1}{3} = \pi$ because the paths enclose $1/3$ of a unit cell. Thus, their contributions to the amplitude add with exactly opposite signs, and the hopping vanishes.  $(b)$ We depict the Peierls paths of the model with the nearest-neighbor hoppings in green, the second nearest-neighbor hoppings in red, and the third nearest-neighbor hoppings in dashed blue. The honeycomb is shown in light grey. We see that any closed loop along the Peierls' paths encloses an integer number of rhombuses, each of which is $1/3$ of the unit cell. Hence $n= 3$ and $\Phi = 6\pi$.  $(c)$ We show the $1 \times 2$ magnetic unit cell of the model at $\phi = \mu \frac{2\pi p'}{q'} = 3 \frac{2\pi}{2}$ outlined in dotted grey lines. In addition for visual clarity, we also show examples of the first-, second-, and third nearest neighbor paths shown with the same colors as in $(b)$. 
}
\label{fig:TBGpeierls}
\end{figure}

Now we discuss the construction of the Hofstadter Hamiltonian. We proceed in the Landau gauge $\mbf{A}(\mbf{r}) = - \phi \mbf{b}_1 (\mbf{r} \cdot \mbf{b_2})$ which we emphasize is centered at the $1a$ position in the center of the honeycomb although there is no orbital there. As argued in \Ref{2018arXiv181111786L}, the path of integration of the Peierls phases should be taken through the centers of the honeycomb cells because, from microscopics, the orbital overlap is greatest there. This is depicted in \Fig{fig:TBGhoppings}. Calculating the Peierls phases along those paths, we write the model in position space with the magnetic field as
\bea
\label{eq:TBGmodelham}
H_{TBG}^{\phi} &= \sum_{\mbf{R}}  \Big(c^\dag_{\mbf{R}}  \Delta \mu_3 \otimes \sigma_0 c_\mbf{R} +  \sum_{i=1}^3 \frac{t_i}{2} c^\dag_{\mbf{R} + \pmb{\delta}_i} \mu_0 \otimes \sigma_1 c_\mbf{R} + \frac{t'_i}{2} c^\dag_{\mbf{R} - 2 \pmb{\delta}_i}\mu_0 \otimes \sigma_1  c_\mbf{R}  \\& \quad -  i \sum_{i=1}^3 \frac{t_i}{2} c^\dag_{\mbf{R} + \pmb{\delta}_i} \mu_0 \otimes \sigma_2 c_\mbf{R} + \frac{t'_i}{2} c^\dag_{\mbf{R} - 2 \pmb{\delta}_i} \mu_0 \otimes \sigma_2 c_\mbf{R} -  i \sum_{i=1}^3  c^\dag_{\mbf{R} + \mbf{d}_i} \mu_2 \otimes \text{diag}(\la_i,-\la'_i) c_\mbf{R} +h.c. \Big) \\
\eea
where the new hopping elements at each $\mbf{R} = r_1 \mbf{a}_1 + r_2 \mbf{a}_2$ are calculated using \Eq{eq:peierlsdefphase} and the Peierls paths are given in \Fig{fig:TBGfigs}f,g,h. We compute the Peierls phases along these paths to be
\bea
\label{eq:tbgphases}
t_i &= t \cos \lp \frac{\phi}{6} \rp \left\{e^{-i \phi \frac{r_2}{3} },e^{i \phi (\frac{2 r_2}{3} - \frac{1}{3})}, e^{i \phi (-\frac{r_2}{3} + \frac{1}{6})} \right\}, t'_i=  t' \left\{ e^{i \phi (\frac{2 r_2}{3} - \frac{2}{3})}, e^{-i \phi \frac{4 r_2}{3} },  e^{i \phi \frac{2 r_2}{3}} \right\} ,\\
\la_i & = \la \left\{  e^{-i \phi (r_2 - \frac{1}{6})},  e^{-i \frac{\phi}{6}},  e^{i \phi (r_2 - 1)} \right\} , \la'_i  = \la \left\{  e^{-i \phi (r_2 + \frac{1}{6})},  e^{i \frac{\phi}{6}},  e^{i \phi r_2} \right\} .\\
\eea
In particular, the nearest-neighbor hoppings have two Peierls paths taken in superposition (see \Fig{fig:TBGfigs}f). For example, the paths $\mathcal{C}_1, \mathcal{C}_2$ for the $t_1$ hopping are shown in \Fig{fig:TBGpeierls}a, and the Peierls substitution reads
\bea
t_1 &\to t_1(\phi) = \frac{t}{2} \lp e^{i \int_{\mathcal{C}_1} \mbf{A} \cdot d\mbf{r} } + e^{i \int_{\mathcal{C}_2} \mbf{A} \cdot d\mbf{r} } \rp  \\
&= \frac{t}{2} \lp 1 + e^{i \int_{\mathcal{C}_2 - \mathcal{C}_1} \mbf{A} \cdot d\mbf{r} } \rp e^{i \int_{\mathcal{C}_1} \mbf{A} \cdot d\mbf{r} } \ . \\
\eea
Noting that $\mathcal{C}_2 - \mathcal{C}_1 = \del \mathcal{R}$, where $\mathcal{R}$ is marked as the grey rhombus in \Fig{fig:TBGpeierls}a, is a closed loop of area $1/3$ (recalling that have normalized the area of the unit cell to 1), we find
\bea
t_1(\phi) &= \frac{t}{2} \lp 1 + e^{i \phi /3 } \rp e^{i \int_{\mathcal{C}_1} \mbf{A} \cdot d\mbf{r} }  \\
&= \frac{t}{2} \lp e^{- i \phi/6} + e^{i \phi /6} \rp e^{ i \frac{\phi}{6} + i \int_{\mathcal{C}_1} \mbf{A} \cdot d\mbf{r} }  \\
&= t \cos \lp \frac{\phi}{6} \rp e^{ i \frac{\phi}{6} + i \int_{\mathcal{C}_1} \mbf{A} \cdot d\mbf{r} } \ . \\
\eea
Computing the remaining integral gives \Eq{eq:tbgphases}. We emphasize in the $\la$ hopping term, $\sigma_z \to \text{diag}(\la, - \la')$ because the different sublattice sites have different hopping paths. We refer to the phases of the hoppings as $\arg t_i = \varphi_{t i}(\ell),\arg t'_i = \varphi_{t' i}(\ell), \arg \la_i = \varphi_{\la i}(\ell), \arg \la'_i = \varphi_{\la' i}(\ell)$.

To Fourier transform \Eq{eq:TBGmodelham} in momentum space in the Landau gauge, we compute $\mu$ using \Eq{eq:mudef}. To do so, we only need the vectors between orbitals connected by hoppings, which for nearest neighbors are $\pmb{\delta}_i$, second nearest neighbors are $\mbf{a}_i$, and third nearest neighbors are $-2 \pmb{\delta}_i$. From \Eq{eq:mudef}, we calculate $\mu =  \text{lcd } \{ \mbf{b}_1 \cdot \pmb{\delta}_i, \mbf{b}_1 \cdot \mbf{a}_i, \mbf{b}_1 \cdot (-2\pmb{\delta}_i) \}_{i=1}^3 = \text{lcd } \left\{\frac{1}{3},-\frac{2}{3},\frac{1}{3},1,0,-\frac{2}{3},\frac{4}{3},-\frac{2}{3}\right\}= 3$. Hence, we can form the Hofstadter Hamiltonian by taking $\phi = 3 \frac{2\pi p'}{q'}$ for $p',q'$ coprime, and define the magnetic BZ as $k_1 \in (-\pi, \pi), k_2 \in (0, 2\pi /q')$ with a $1 \times q'$ magnetic unit cell. Indeed, one can check that $\phi = 3 \frac{2\pi p'}{q'}$ explicitly gives a $r_2 \to r_2 + q'$ periodicity in the Peierls phases from \Eq{eq:tbgphases}. Note that in \Ref{2018arXiv181111786L}, the Hofstadter of the model was constructed in the square lattice gauge $\mbf{A}_{sq}(\mbf{r}) = \phi (0, x)$, which is not the form of our Landau gauge, $\mbf{A}(\mbf{r}) = - \phi \mbf{b}_1 (\mbf{r} \cdot \mbf{b}_2)$. In the square lattice gauge, \Ref{2018arXiv181111786L} found it was necessary to rationalize the flux as $\phi = 2 \frac{2\pi p'}{q'}$. There is no contradiction because the rationalization is gauge-dependent.  

The flux periodicity $\Phi$ is gauge invariant and is determined by the possible loops along Peierls paths. We overlay all possible Peierls paths in \Fig{fig:TBGpeierls}b, from which we can see that all paths enclose a multiple of $1/3$ of a unit cell. Hence $n=3$ and $\Phi = 6\pi$. Note that $\mu$ and $n$ being identical is coincidental, as can be seen by considering the gauge choice of \Ref{2018arXiv181111786L} where this explicitly does not hold. 

We now construct the Hofstadter Hamiltonian. As required, there is no dependence on $r_1$ and the Hamiltonian can be Fourier transformed over $\mbf{a}_1$ immediately. At $\phi = 3 \frac{2\pi p'}{q'}$, we choose the unit cell to be indexed by $\ell = 0, \dots, q'-1$. An example of the $\phi = 3\pi$ ($q' = 2$) magnetic unit cell is shown in \Fig{fig:TBGpeierls}c. Keeping track of how the hoppings connect different atoms in the magnetic unit cell, we derive the Hofstadter Hamiltonian
\bea
\null [\mathcal{H}^{\phi}(\mbf{k})]_{\ell, \ell'} &= \delta_{\ell \ell'} h_{\ell} + \delta_{\ell-1, \ell'} T^\dag_{\ell} + \delta_{\ell'-1, \ell} T_{\ell'} + \delta_{\ell -2,\ell'} S^\dag_{\ell} + \delta_{\ell'-2, \ell} S_{\ell'} \\
\eea
where the $4\times 4$ blocks are given by
\bea
h_{\ell} &= \Delta \mu_3 \otimes \sigma_0 + \mu_0 \otimes \sigma_1 \big( |t_i| \cos(\mbf{k} \cdot \pmb{\delta}_1 -\varphi_{t1}(\ell) ) +  t'\cos( 2\mbf{k} \cdot \pmb{\delta}_2 + \varphi_{t'2}(\ell) ) +t'\cos(2\mbf{k} \cdot \pmb{\delta}_3 + \varphi_{t'3}(\ell) ) \big) \\ & \quad  - \mu_0 \otimes \sigma_2 \big(|t_i| \sin( \mbf{k} \cdot \pmb{\delta}_1 - \varphi_{t1}(\ell))   + t'\sin( -2\mbf{k} \cdot \pmb{\delta}_2 - \varphi_{t'2}(\ell)) +t'\sin( -2\mbf{k} \cdot \pmb{\delta}_3 - \varphi_{t'3}(\ell) ) \big) \\ & \quad - \la \mu_2 \otimes \text{diag}(\sin (\mbf{k} \cdot \mbf{d}_1 - \varphi_{\la1}(\ell)),-\sin (\mbf{k} \cdot \mbf{d}_1 - \varphi_{\la'1}(\ell))) \\
T_{\ell} &= \frac{|t_i|}{2}  \lp \mu_0 \otimes \sigma_1  - i \mu_0 \otimes \sigma_2 \rp (e^{i \varphi_{t2}(\ell) - i \mbf{k} \cdot \pmb{\delta}_2} +e^{i \varphi_{t3}(\ell) - i \mbf{k}\cdot \pmb{\delta}_3} ) \\
&  \quad - i \la \mu_2 \otimes \big(-\text{diag}(e^{-i \varphi_{\la2}(\ell-1)},-e^{-i \varphi_{\la'2}(\ell-1)}) e^{i \mbf{k} \cdot \mbf{d}_2} + \text{diag}(e^{i \varphi_{\la3}(\ell)},-e^{i \varphi_{\la'3}(\ell)}) e^{- i \mbf{k} \cdot \mbf{d}_3} \big) \\
S_{\ell} &=\frac{t'}{2} \lp \mu_0 \otimes \sigma_1 - i \mu_0 \otimes \sigma_2  \rp e^{i \varphi_{t'1}(\ell) + 2 i \mbf{k} \cdot \pmb{\delta}_1} \ . 
\eea
This completes the construction of the Hofstadter Hamiltonian. We can now discuss the symmetries of the $H^\phi_{TBG}$ at nonzero flux. The $\phi=0$ symmetries are promoted to those in \Tab{tb:tbsym} (third column) when $\phi \neq 0$ with the additional $\mathcal{D}$ and $G'$ factors. For brevity, we do not include product symmetries with $C_{3z}$ and the other symmetries since they are broken when we break $C_{3z}$. Again, $\mathcal{D}_{\ell \ell'} = \delta_{\ell, q'-\ell'}$ acts on the (enlarged) magnetic unit cell indices to flip the magnetic unit cell under $C_{2z}$. Additionally, $C_{3z}$ requires a gauge transformation because our Landau gauge explicitly breaks $C_{3z}$ symmetry. To determine the appropriate gauge transformation $G'$, we refer to the discussion around \Eq{eq:c4transform} which shows that
\bea
G' = \exp \lp i \sum_{\mbf{R} \al} c^\dag_{\mbf{R} \al} c_{\mbf{R} \al} \chi(\mbf{R} + \pmb{\delta}_{\al}) \rp \\
\eea
where $\chi(\mbf{r}) = - \phi xy$ is the solution to 
\bea
C_{3z}^{-1} \mbf{A}(C_{3z}\mbf{r})  - \pmb{\nabla} \chi(\mbf{R}) &=  \mbf{A}(\mbf{r}) \ .
\eea
We refer the reader to \Ref{hofsymtoappear} for a detailed treatment of space group symmetries in the presence of magnetic fields. We note that there is also a particle-hole symmetry $Ph = \mu_1 \otimes \sigma_3$ that obeys 
\bea
\label{eq:PHtbg}
Ph^\dag H^\phi_{TBG}(\mbf{k}) Ph = - H^{-\phi \, *}_{TBG}(-\mbf{k}) 
\eea
which exists at $\phi =0$ and $\phi = \Phi/2$. 

Using these expressions, one can check explicitly that the symmetries in \Tab{tb:tbsym} are preserved. For symmetries that take $\phi \to - \phi$, the zero-field symmetry is restored at $\Phi/2 = 3\pi, p'/q' = 1/2$. At this point, $\cos \frac{\Phi/2}{6} = 0 $, so nearest neighbor couplings vanish due to the interference of their Peierls paths. \Fig{fig:TBGpeierls}a illustrates that the phase difference between the two paths is $\pi$, and they destructively interfere.

\begin{table}
\caption{Symmetries of the TBG Model} % title of Table
\centering % used for centering table
\begin{tabular}{|c |c | c| c| } % centered columns (4 columns)
\hline
%heading
Symmetry &$ \phi=0 $&$ \phi\neq 0$& Mapping of $\mbf{k}, \phi$ \\
\hline
$C_{3z} $ & $ I $ & $ C_{3z}G_{2\pi/3} $ & $ R_{2\pi/3} \mbf{k}, \phi $ \\
\hline
$C_{2x}$ & $\mu_3 \otimes I $ & $C_{2x}$& $ (k_x,-k_y) ,-\phi $\\
\hline
$C_{2z}$ & $ I \otimes \sigma_1 $ & $ C_{2z} \mathcal{D}  $ & $ -\mbf{k}, \phi $ \\
\hline
$\mathcal{T}$ & $ K $ & $\mathcal{T}$ & $ -\mbf{k}, -\phi$ \\
\hline
$C_{2z} \mathcal{T}$ & $ I \otimes \sigma_1 K $ & $  C_{2z}  \mathcal{D}  T $ & $ \mbf{k}, -\phi$ \\
\hline
$C_{2x} \mathcal{T}$ & $ -i \mu_3 \otimes I K $ & $C_{2x} \mathcal{T}$ & $ (-k_x, k_y), \phi $ \\
\hline
$C_{2x} C_{2z}$ & $ i \mu_3 \otimes \sigma_1 $ & $ C_{2x} C_{2z}  \mathcal{D}  $ & $ (-k_x, k_y), -\phi $ \\
\hline
$C_{2x} C_{2z} \mathcal{T}$ & $ \mu_3 \otimes \sigma_1 K$ & $C_{2x} T C_{2z} \mathcal{D} $ & $ (k_x, -k_y), \phi $ \\
\hline
\end{tabular}
\justify
We list the symmetries of the TBG model \Eq{eq:TBGmodelham} in column 1 and their representations on the Bloch Hamiltonian in column 2. In column 3, we provide their representations on the Hofstadter Hamiltonian $\mathcal{H}_{TBG}$ in the presence of nonzero flux. The symmetries listed in column 3 refer to the $4\times 4$ representations defined in column 2. Note that for $\phi \neq 0$, some of the symmetries are broken, and take $\phi \to - \phi$. The mapping of $\mbf{k}$ and $\phi$ under the action of the symmetries is shown in column 4.
\label{tb:tbsym}
\end{table}

\subsection{Breaking Symmetries}
\label{app:tbgsymbreak}

Like in the case of the QSH model, we may add symmetry-breaking terms to $H_{TBG}$ in order to isolate the effects of various symmetries. In \Sec{sec:fragile}, we showed that $C_{2z} \mathcal{T}$ symmetry alone was responsible for the HOTI classification of the Hofstadter phase, and thus we build perturbations to destroy all other symmetries of the model. By coupling nearest neighbor atoms with an imaginary coupling that takes opposite signs for the $s$ and $p_z$ orbitals, we break $C_{2z}$ and $\mathcal{T}$ symmetry individually. In momentum space at $\phi = 0$, this term reads
\bea
\label{eq:TBGbreakC2z}
H_{1}(\mbf{k}) = \mu_3 \otimes \sum_{i} (\sigma_1 \sin  \pmb{\delta}_i \cdot \mbf{k} + \sigma_2 \cos  \pmb{\delta}_i \cdot \mbf{k} ) \ .
\eea
One may check that this term breaks $C_{2z}, \mathcal{T}, C_{2x} \mathcal{T}, C_{2x} C_{2z}$. To break $C_{3z} = I \otimes I$, we add an anisotropic term that alters only the $\pmb{\delta}_1$ hopping, 
\bea
H_2(\mbf{k}) &= \mu_0 \otimes ( \sigma_1 \cos \pmb{\delta}_1 \cdot \mbf{k} - \sigma_2 \sin \pmb{\delta}_1 \cdot \mbf{k} ) \ .
\eea
To break $C_{2x}$ and all the remaining product symmetries, we use the term
\bea
H_3(\mbf{k}) &= \mu_1 \otimes \sum_i (\sigma_1 \cos  \pmb{\delta}_i \cdot \mbf{k} - \sigma_2 \sin \pmb{\delta}_i \cdot \mbf{k} )\ . 
\eea
One may check that all these perturbations preserve the particle-hole symmetry \Eq{eq:PHtbg}. This is desirable for stabilizing the corner modes at zero energy, although this is not essential to the physics of the model. We will show in \App{app:symeig} that if $C_{2z}$ and $\mathcal{T}$ are maintained, then a bulk gap closing is enforced by the $C_{2z}$ eigenvalues. To show this, we remove $H_{1}$ to create the Hamiltonian $H_{TBG}''$ which possesses $C_{2z}, \mathcal{T}$, and their product $C_{2z} \mathcal{T}$. $H_{TBG}''$ has a gapless bulk due to $C_{2z}$ (see \Fig{fig:TBGwannier}). Note that although $H''_{TBG}$ has a $\mathcal{T}$ symmetry, this model is spinless and thus $\mathcal{T}^2 = +1$ so there is no Kane-Mele invariant. The models in this section are summarized in \Tab{tb:tbghams}. 

\begin{table}
\caption{TBG Hamiltonians and their Symmetries} % title of Table
\centering % used for centering table
\begin{tabular}{|c |c | c|} % centered columns (4 columns)
\hline
%heading
Hamiltonian  & Symmetries & Parameters \\
\hline
$H'_{TBG} = H_{TBG}+ \eps_1 H_{1,im}(\mbf{k}) + \eps_2 H_2(\mbf{k}) + \eps_3 H_3(\mbf{k}) $ & $C_{2z} \mathcal{T} $ & $\eps_1 = .12, \eps_2 = .11, \eps_3 = .1, \Delta = 1.6 $\\
\hline
$H''_{TBG} = H_{TBG}+ \eps_2 H_2(\mbf{k}) + \eps_3 H_3(\mbf{k}) $ & $C_{2z}, \mathcal{T}, C_{2z} \mathcal{T} $ & $ \eps_2 = .11, \eps_3 = .1, \Delta = 1.6 $\\
\hline
$H'''_{TBG} = H_{TBG} + \eps_2 H_2(\mbf{k}) + \eps_4 H_4(\mbf{k}) $ & $C_{2x} \mathcal{T}$ & $\eps_2 = .11, \eps_4 = .1, \Delta = 1.6 $ \\
\hline
\end{tabular}
\justify
We list the variations of the TBG model (column 1), the symmetries they retain from the perturbations (column 2), and the values of the parameters that realize their nontrivial Hofstadter topology (column 3). The Hofstadter Butterflies of the models may be found in \choose{\Fig{fig:TBGcorner}, \Fig{fig:TBGwannier}, and \Fig{fig:TBGwannierflow}}{\Fig{fig:TBGcorner} of the Letter, \Fig{fig:TBGwannier}, and \Fig{fig:TBGwannierflow}}. 
\label{tb:tbghams}
\end{table}

\subsection{Computing the $w_2$ index protected by $U C_{2z} \mathcal{T}$ symmetry}
\label{app:C2zTwilson}

\begin{figure}
 \centering
\includegraphics[width=7cm]{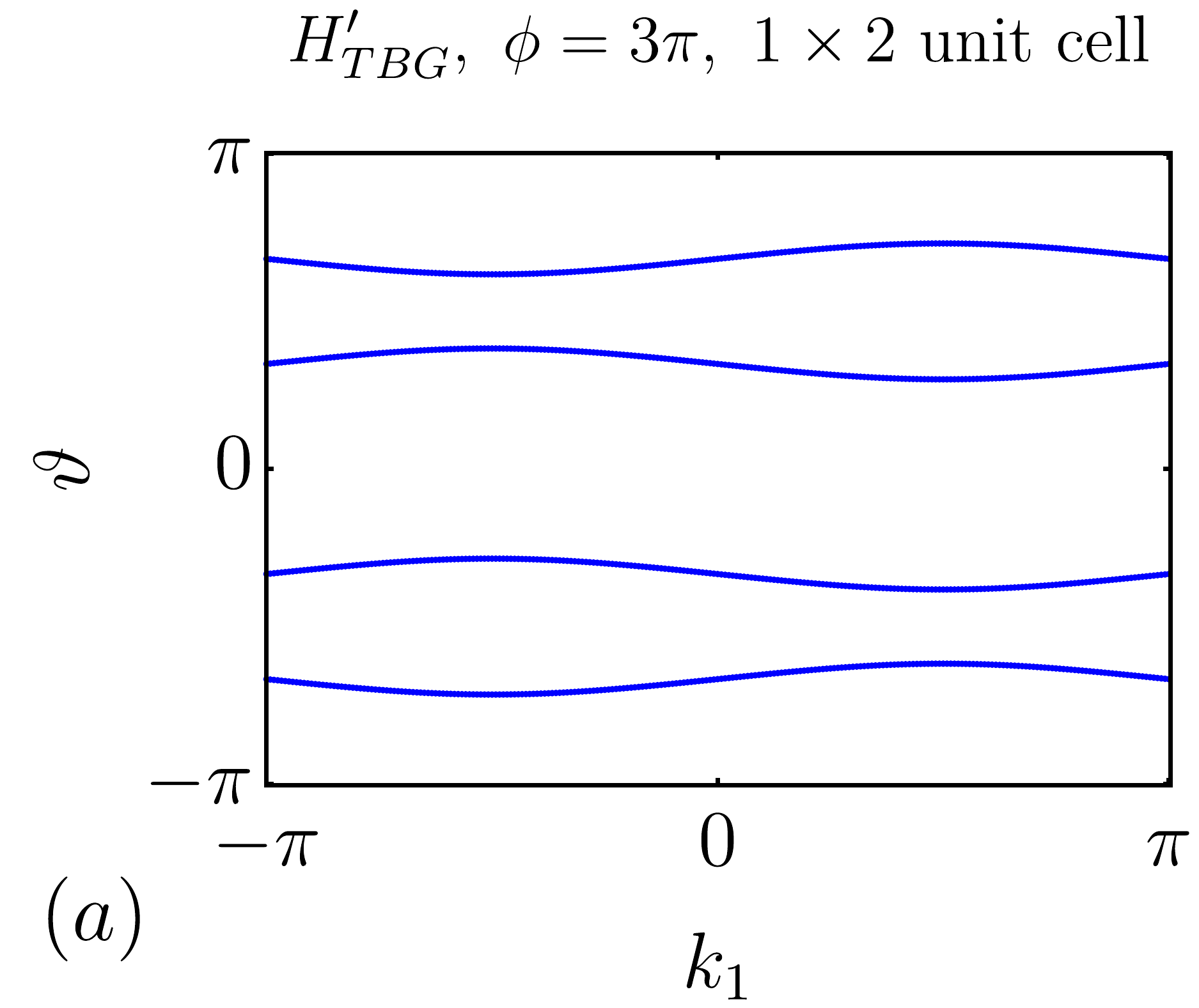} 
\includegraphics[width=7.2cm]{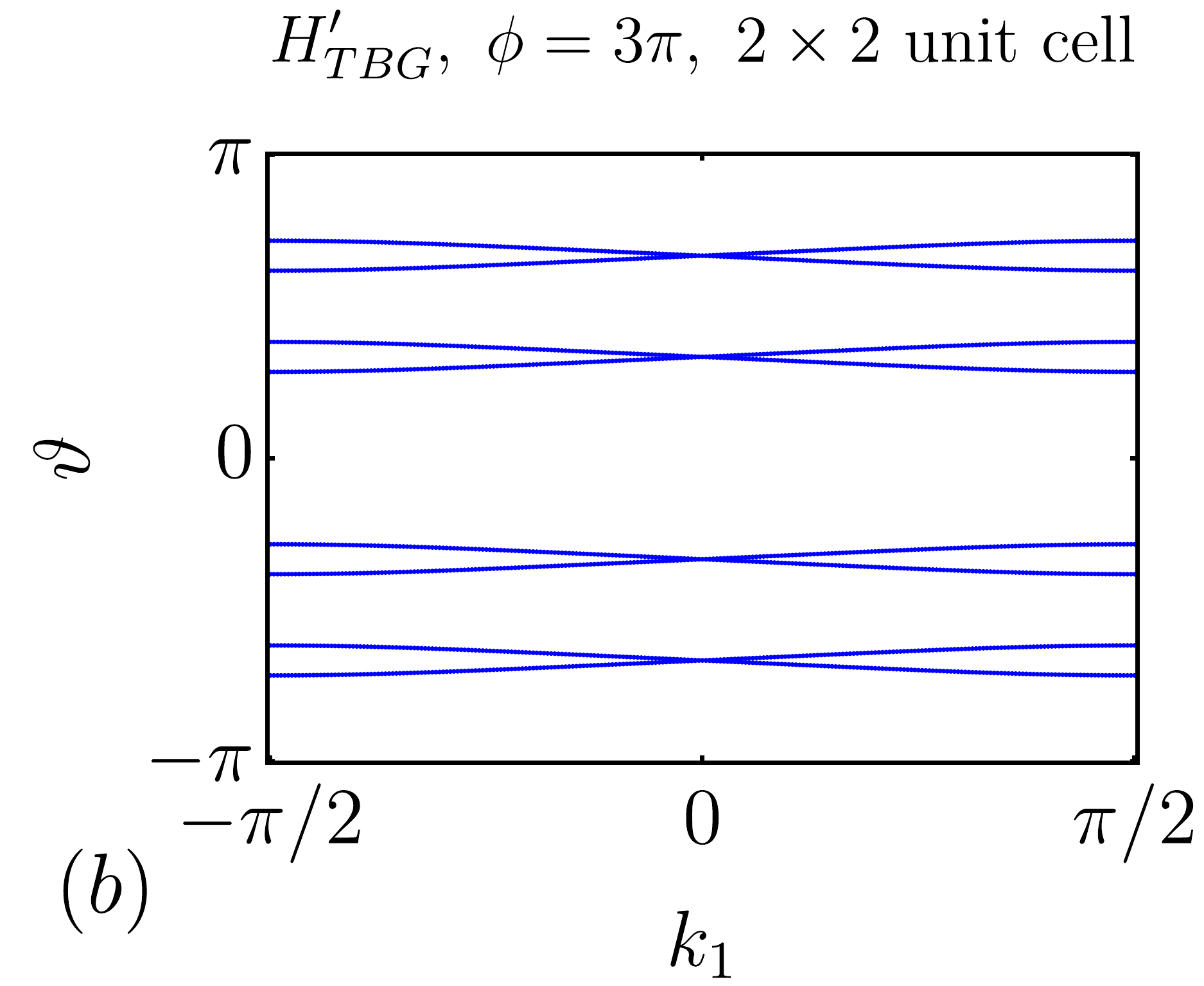} 
\caption{(a) We show the Wilson Loop spectrum for $H'_{TBG}$ in the $1\times 2$ magnetic unit cell of our Landau gauge at $\phi = \Phi/2$ where $UC_{2z} \mathcal{T}$ is not diagonal in momentum space. Hence there is no ``particle-hole" symmetry. (b) To make $U$ diagonal in momentum space such that the Wilson spectrum realizes the ``particle-hole" symmetry of a conventional $w_2$ insulator, we diagonalize $H^{\Phi/2}$ in a $2\times 2$ unit cell where $k_1$ is defined in $(-\pi/2, \pi/2)$. We calculate the Wilson loop for this Hamiltonian and observe that it matches the spectrum of $(a)$ with the Wilson bands folded so $k_1$ is defined only mod $\pi$. Note that $(U C_{2z} \mathcal{T})^2 = +1$, so we may calculate $w_2^{\phi = \Phi/2}$ as described in \App{app:w2calc} without calculating the nested Wilson loop. 
}
\label{fig:TBGwilsoncomp}
\end{figure}

From \Tab{tb:tbghams}, we see that $H'_{TBG}$ has a $C_{2z} \mathcal{T}$ symmetry and hence has $UC_{2z} \mathcal{T}$ symmetry at $\phi = \Phi/2$. As shown in \Fig{fig:C2zTphase}a, $(UC_{2z} \mathcal{T})^2 = e^{i \gamma_2} (C_{2z} \mathcal{T})^2 = +1$, so the phase at $\phi = \Phi/2$ is characterized by the $w_2$ invariant. Following the discussion of \App{app:w2calc} where we used this model as an example, we can compute this invariant in a $2 \times 2$ unit cell where $UC_{2z} \mathcal{T}$ acts diagonally in momentum space. For pedagogical purposes, we also compute the Wilson loop in the $1\times 2$ unit cell at $\phi = \Phi/2$ where $U C_{2z} \mathcal{T}$ is not diagonal and there is no particle-hole symmetry in the Wilson spectrum. We plot the Wilson loop spectra using these two possible unit cells in \Fig{fig:TBGwilsoncomp}. 
From the spectrum of \Fig{fig:TBGwilsoncomp}b, we observe that there are no crossings at $\vartheta = 0$ or $\vartheta = \pi$ and thus $w^{\phi = \Phi/2}_2 = 0$. Note that the number of crossings at $\vartheta = 0$ must equal the number at $\vartheta = \pi$ because $\{\vartheta(k_1 + \phi) \} = \{\vartheta(k_1) + \pi \}$ (see \Eq{eq:wilsonshift}). From here, we can calculate the Hofstadter HOTI invariant
\bea
\theta &= w^{\phi=0}_2 -w^{\phi=\Phi/2}_2 = 1 \\
\eea
recalling that $H^{\phi=0}$ is nontrivial. Thus corner states are pumped into the bulk as the flux is tuned from 0 to $\Phi/2$ as we see in \Fig{fig:TBGcorner}a of the Main Text. We discuss the signatures of this phase in the following section \App{app:symeig} from a real-space perspective. 

\subsection{$C_{2z}$ Symmetry Eigenvalues}
\label{app:symeig}

\begin{figure}
 \centering
 \begin{overpic}[width=0.375\textwidth,tics=10]{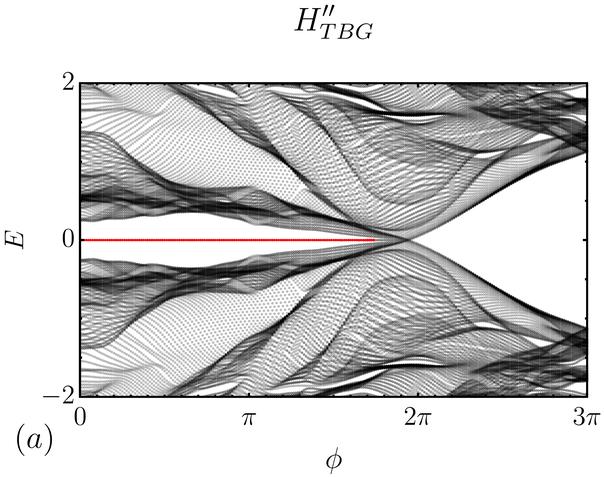}
\end{overpic}  
\begin{overpic}[width=0.30\textwidth,tics=10]{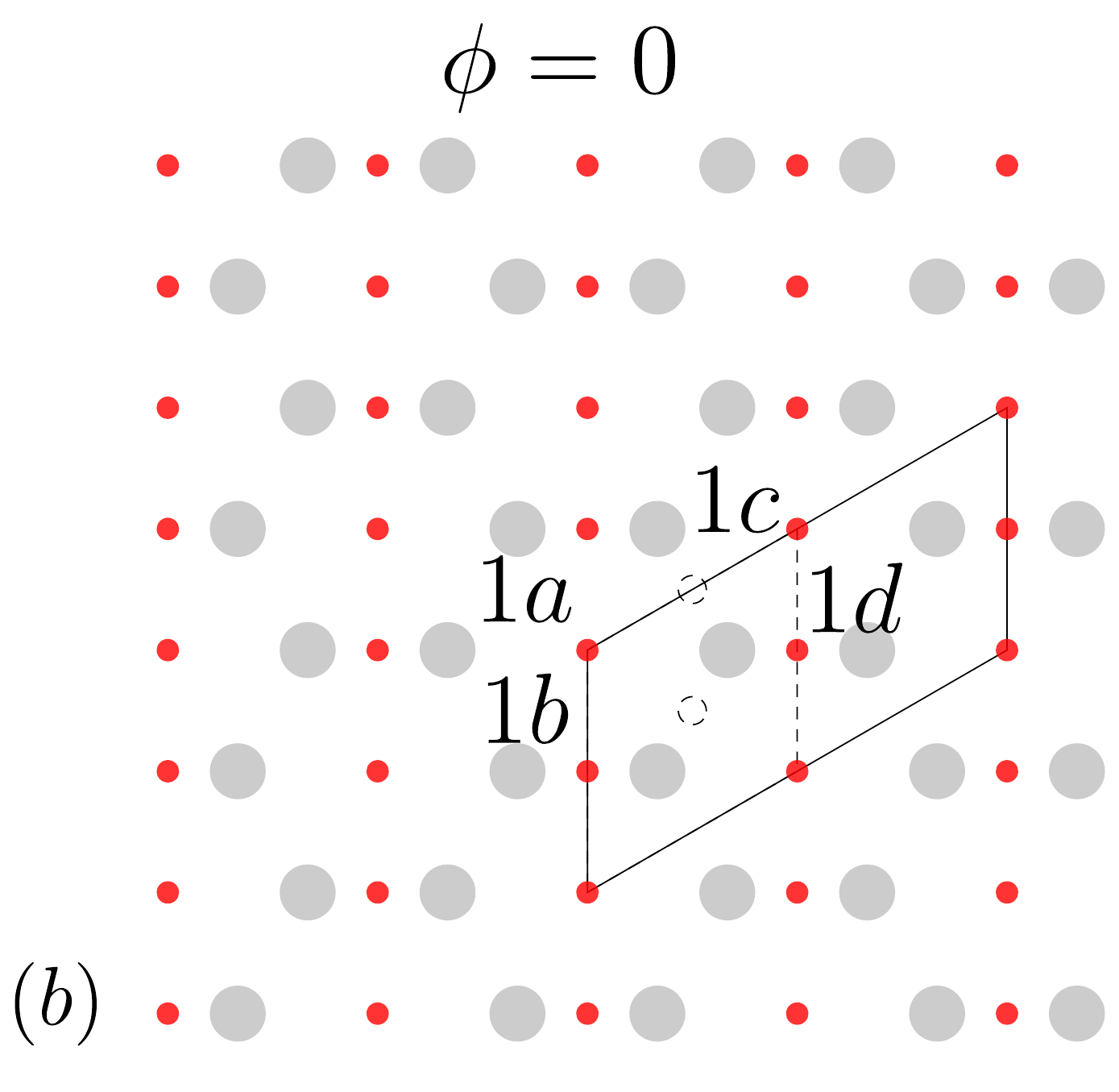}
\end{overpic} 
\begin{overpic}[width=0.3\textwidth,tics=10]{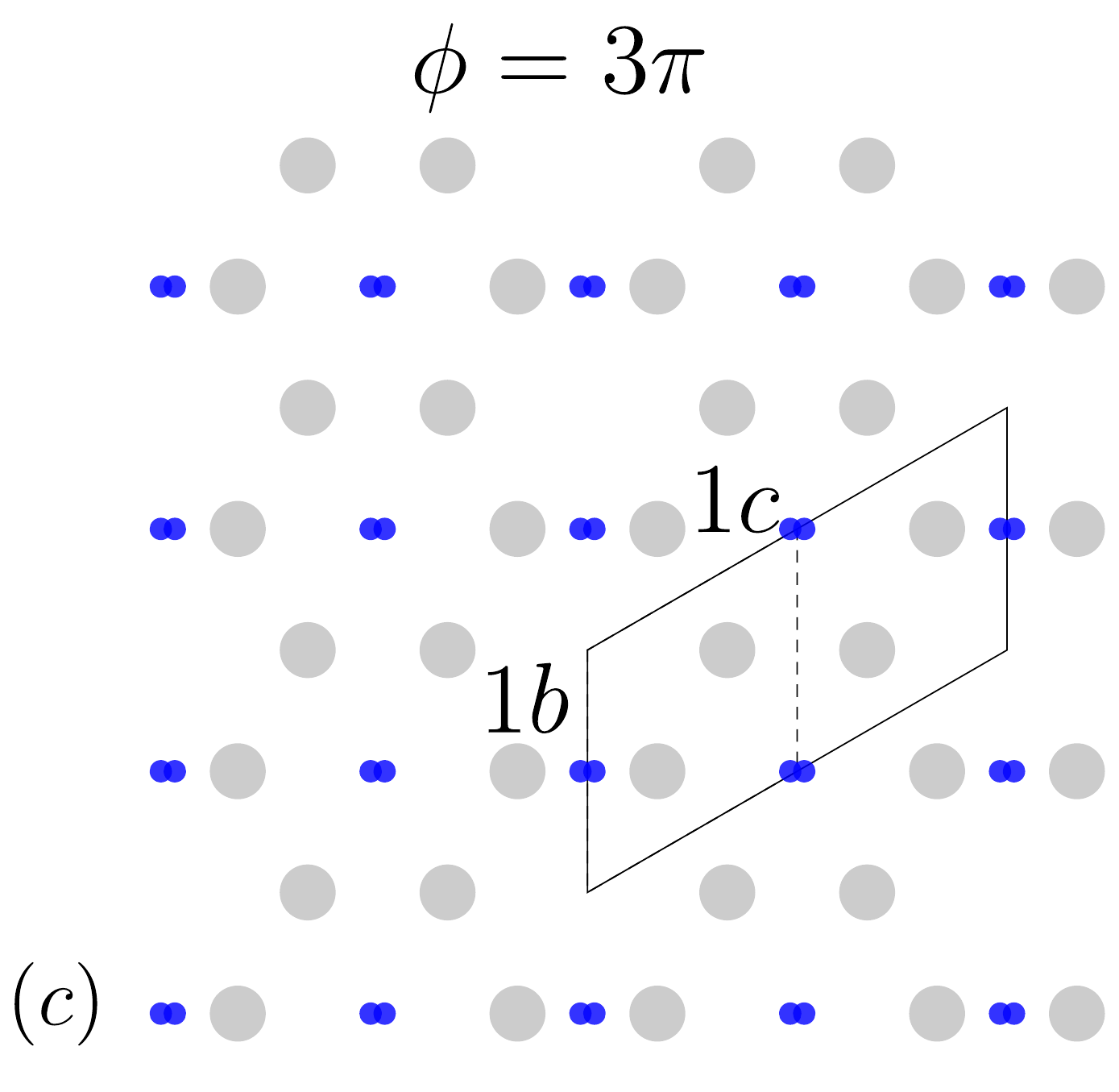}
\end{overpic}
\caption{The Hofstadter Butterfly is calculated on a $20 \times 20$ lattice in the topological phase for $H_{TBG}''$, described in \App{app:TBG}, which has both $C_{2z}$ and $\mathcal{T}$. Because $C_{2z} \mathcal{T}$ remains, there is still corner mode pumping, but the model is not a HOTI due to the bulk gap closing enforced by $C_{2z}$ eigenvalues. We demonstrate this in $(b)$ and $(c)$, where the Wannier centers of the phases at $\phi = 0, 3\pi$ are shown in red and blue respectively in \emph{magnetic} unit cell. (We show  the boundary of the \emph{standard} unit cell and its $1c,1d$ orbitals with dashed lines.) A possible Wannierization is shown in $(b)$, with $s$ and $p_z$ orbitals at the $1b$ and $1c$ positions. Without breaking $C_{2z}$ symmetry, pairs of orbitals may be moved to other high symmetry Wyckoff positions, but it is impossible to reach the state at $\phi = 0$.}
\label{fig:TBGwannier}
\end{figure}

When $C_{2z}$ and $\mathcal{T}$ remain symmetries of the model as in $H_{TBG}'$, a gap closing may be enforced by the symmetry eigenvalues of $C_{2z}$, as we observe from the Hofstadter Butterfly in \Fig{fig:TBGwannier}a. We will prove this gap closing for $H''_{TBG}$ (see \Tab{tb:tbghams}) using the framework of Topological Quantum Chemistry \cite{2017Natur.547..298B,2018PhRvB..97c5139C,PhysRevE.96.023310}. To obtain the band representations of this model, we calculate the $C_{2z}$ eigenvalues directly from $H_{TBG}''$ (see \Tab{tb:tbghams}) for $\phi = 3\pi$ at the 4 inversion invariant points in the $\phi = 3\pi$ BZ. The calculation is straightforward at $\phi = 3\pi$ using the embedding matrices and symmetries defined in \Sec{sec:embedding} and \App{app:TBG} respectively. At $\phi =0$, note that we must artificially extend the $1\times 1$ unit cell to the $1\times 2$ magnetic unit cell to compare the atomic limits of the band representations at $\phi = 0$ and $\phi = 3\pi$ in the same unit cell. 

We collect the eigenvalues of the occupied bands in \Tab{tb:tbgeigs}. We note that at $\phi = 0$ and $\phi = 3\pi$, all bands are connected energetically. We emphasize that Wychoff positions correspond to the \emph{magnetic} unit cell, i.e. $1a = (0,0), 1b =  \mbf{a}_1/2, 1c = 2 \mbf{a}_2 /2, 1d = \mbf{a}_1/2 + 2 \mbf{a}_2 /2 $ taking the center of the hexagonal plaquette ($1a$) as the origin. 

\begin{table}[h]
\caption{TBG $C_{2z}$ eigenvalues} % title of Table
\centering % used for centering table
\begin{tabular}{|c|c |c | c| c| c |} % centered columns (4 columns)
\hline
&$ (0,0) $&$ \pi \mbf{b}_1 $&$ \frac{\pi}{2} \mbf{b}_2  $& $ \pi \mbf{b}_1  + \frac{\pi}{2}  \mbf{b}_2  $ & Band Representation \\
\hline
$\xi^{\phi = 0} $&$ -1, -1,+1,+1 $&$ +1, +1,+1, +1 $&$ -1, -1, +1, +1$& $ -1, +1, -1, +1$ & $BR^{\phi = 0} = 2\Gamma_1 +2 \Gamma_2 + 4 B_1 + 2Y_1 + 2Y_2 + 2A_1 + 2A_2$ \\
\hline \hline
&$  (0,0) $&$ \pi \mbf{b}_1 $&$ \frac{\pi}{2} \mbf{b}_2  $& $ \pi \mbf{b}_1  + \frac{\pi}{2}  \mbf{b}_2 $ & Band Representation \\
\hline
$\xi^{\phi = 3\pi} $ &$ -1,-1,+1,+1 $&$ -1,-1,+1,+1 $&$ -1,-1,+1,+1$& $ -1,-1,+1,+1$ & $BR^{\phi = 3\pi} = 2\Gamma_1 + 2 \Gamma_2 + 2 B_1 + 2 B_2 + 2 Y_1 + 2 Y_2 + 2 A_1 + 2 A_2 $ \\
\hline
\end{tabular}
\justify
In columns 1-4, we show the $C_{2z}$ eigenvalues calculated from $H_{TBG}''$ (see \Tab{tb:tbghams}) at half filling in the $1 \times 2$ magnetic unit cell at both $\phi = 0$ and $\phi = 3\pi$. Column 5 shows the (spinless) band representations determined from the eigenvalues. 
\label{tb:tbgeigs}
\end{table}
We see that we may obtain the momentum space band representations of \Tab{tb:tbgeigs} by inducing atomic orbitals from high symmetry Wyckoff positions to the full space group \cite{Aroyo:firstpaper,Aroyo:xo5013}. First we consider the $\phi = 0$ band representation. Because no $B_2$ irreps appear, the possible atomic orbitals are $A_{1a}, B_{1b}, A_{1c}, B_{1d}$, recalling that $A (B)$ is the even (odd) irrep under $C_{2z}$. There is only one Wannierization possible:
\bea
BR^{\phi = 0 } &= (A_{1a} \oplus B_{1b} \oplus A_{1c} \oplus B_{1d}) \uparrow G \ . \\
\eea
We show the location of the Wannier centers in \Fig{fig:TBGwannier}b. 

Next, we consider the bands at $\phi = 3\pi$ forming $BR^{\phi = 3\pi}$. In this case, there are many non-unique atomic limits that recover the band structure. Note that any pair of locally even irreps ($s$ orbitals) and locally odd irreps ($p_z$ orbitals) at the same Wyckoff position $w$ yields the band representation $(A_{w} \oplus B_w ) \uparrow G = \Gamma_1 + \Gamma_2 + B_1 + B_2 + Y_1 + Y_2 + A_1 + A_2$ when induced to the space group. Thus placing both an $s$ and $p_z$ orbital at any two high symmetry Wyckoff positions $w_1, w_2$ will yield $(A_{w_1} \oplus B_{w_1} \oplus A_{w_2} \oplus B_{w_2}) \uparrow G =  BR^{\phi = 3\pi}$. In fact, these limits $A_{w_1} \oplus B_{w_1} \oplus A_{w_2} \oplus B_{w_2}$ are the only possible atomic limits. This may be shown by exhaustion. In \Fig{fig:TBGwannier}b, we show one such Wannierization to the $1b$ and $1c$ Wyckoff positions within the magnetic unit cell. 

We conclude that there must be an odd number of irreps at the $1a$ position for $\phi = 0$, but there must be an even number of irreps at the $1a$ position for $\phi = 3\pi$. Under $C_{2z}$-preserving perturbations that preserve the gap, electrons may be moved off the high symmetry Wyckoff positions into the $2e = (x,y), (1-x, 1-y)$ position \emph{in pairs}, so necessarily only an even number of electrons can be deformed to another high symmetry Wyckoff position. We see that $BR^{\phi = 0}$ and $BR^{\phi = 3\pi}$ are incompatible in this manner, and a gap closing must occur at an intermediate $\phi$ while $C_{2z}$ is preserved, as we see in \Fig{fig:TBGwannier}a. We underscore that although the magnetic unit and the magnetic BZ do not evolve smoothly as the flux is increased, the Wannier centers do evolve smoothly, which our argument relies on. 

\begin{figure}
 \centering
\includegraphics[width=6.5cm, trim = 0 -.5cm 0 0 , clip]{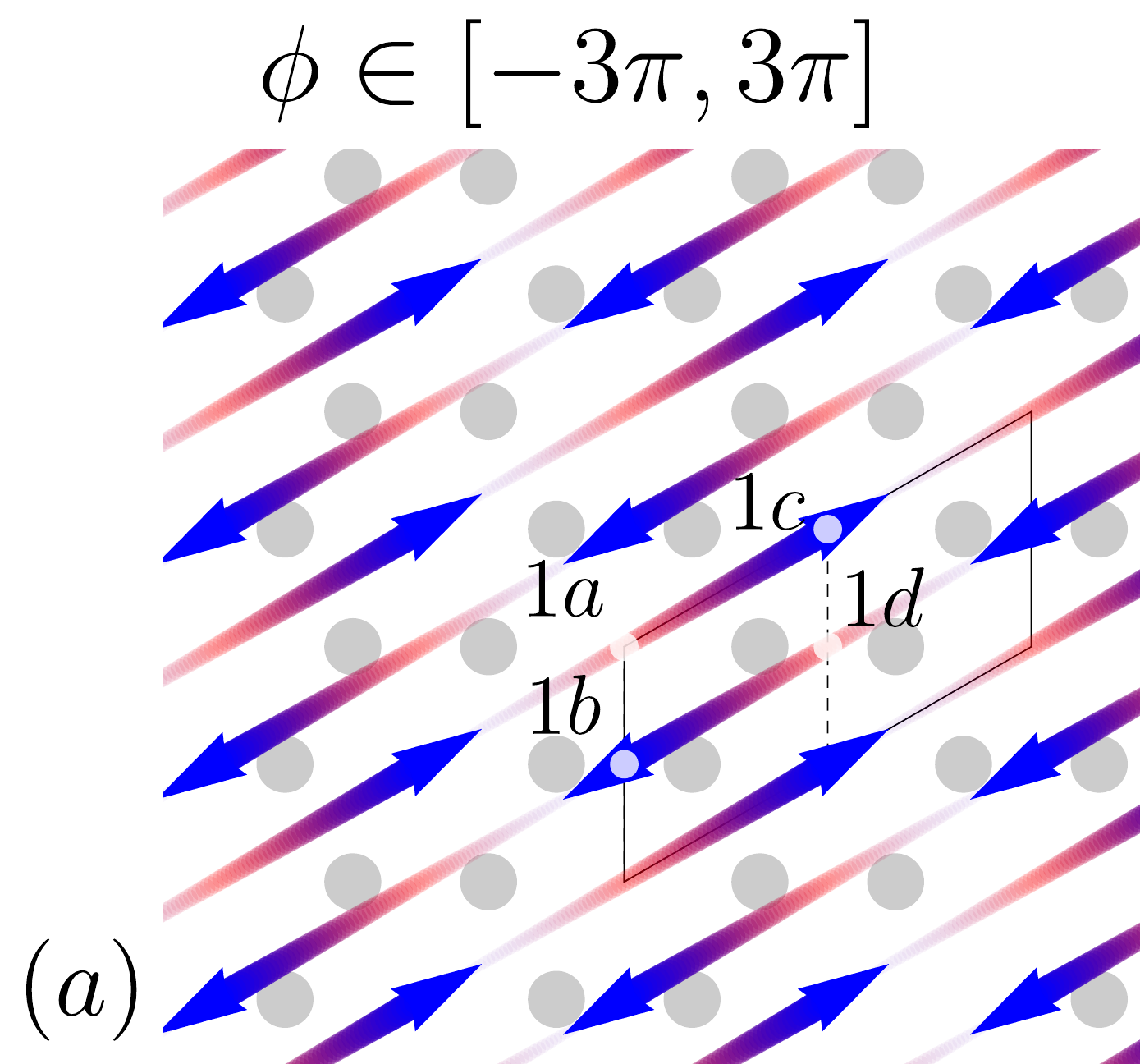} \quad 
\includegraphics[width=9cm]{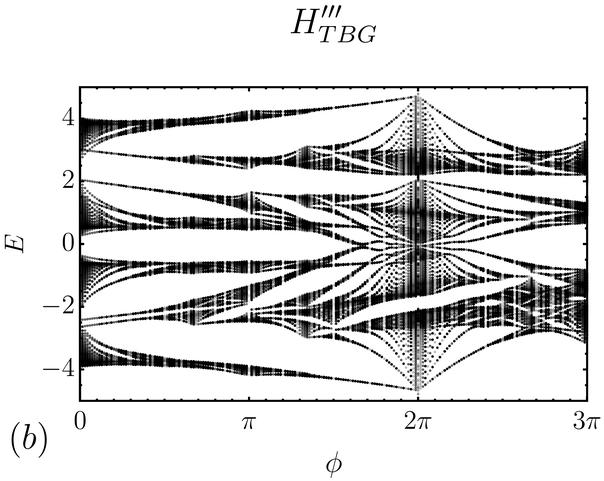} 
\caption{(a) We show an example of $C_{2z} \mathcal{T}$ symmetric Wannier flow from $\phi = - 3\pi$ to $\phi = 3\pi$, going from the $1b,1c$ positions at $\phi = -3\pi$ (blue) to the $1a, 1d$ positions (red), and back to the $1b,1c$ positions at $\phi = 3\pi$. When we break $C_{2z}$, the orbitals may move slightly from the maximal Wyckoff positions. However, with $C_{2z} \mathcal{T}$ intact, the Wannier flow is still protected. (b) We show the Hofstadter Butterfly for $H'''_{TBG}$ on periodic boundary with only $C_{2x} \mathcal{T}$ symmetry, and observe the bulk gap closing. The exact Hamiltonian is given in \Tab{tb:tbghams}. Because $C_{2z} \mathcal{T}$ is broken, there is also no $w_2^{\phi = 0}$ invariant to protect winding in the Wilson loop. Thus the Hofstadter Topology is trivial, although the the bulk gap closing at finite flux is locally protected by $C_{2x} \mathcal{T}$. 
}
\label{fig:TBGwannierflow}
\end{figure}

We can also use this position-space argument to understand the Wannier flow that characterizes the Hofstadter HOTI phase when $C_{2z}$ is broken but $C_{2z} \mathcal{T}$ is preserved. For instance, we can take $H_{TBG}''$ and break its $C_{2z}$ symmetry by adding a small perturbation \Eq{eq:TBGbreakC2z}, resulting in the new model $H'_{TBG}$ which only has $C_{2z} \mathcal{T}$ symmetry. Let us first consider the Wannier centers of $H'_{TBG}$ at $\phi = 0$ and $\phi = 3\pi$ where $U C_{2z} \mathcal{T}$ still pins Wannier centers to the high-symmetry Wyckoff positions (which are invariant under the magnetic point group $2'$ (a proper subset of the point group $2$) symmetries). Then as argued in \App{app:C2zTwilson}, the Wannier centers must flow nontrivially between $\phi = 0$ and $\phi = 3\pi$ where there are different Wannierizations. However, with $C_{2z}$ broken, this Wannier flow does \textit{not} require a gap closing as the flux is tuned from $0$ to $3\pi$. This is because the Wannier centers are not constrained to obey $C_{2z}$ symmetry at all $\phi$. Instead, the Wannier flow is $C_{2z} \mathcal{T}$-symmetric: if there is a Wannier center at $\mbf{r}$ at flux $\phi$, then by $C_{2z} \mathcal{T}$, there is a Wannier center at $-\mbf{r}$ and $-\phi$. So each trajectory $\mbf{r}(\phi)$ must obey $\mbf{r}(\phi) = - \mbf{r}(-\phi)$. Thus as we tune the flux through its full period, we find nontrivial Wannier flow that pumps electrons between unit cells. We depict an example in \Fig{fig:TBGwannierflow}a of two electrons interpolating between the $1a, 1d$ Wyckoff positions at $\phi=0$ and the $1c, 1b$ positions at $\phi= \Phi/2$. On open boundary conditions, this induces corner state flow. The corner states must be degenerate in energy at $\phi=0$ because they are $C_{2z} \mathcal{T}$ partners, and are pinned to zero energy when particle-hole symmetry exists. As the flux is increased, the pumping converts an occupied energy level to a hole on one of the boundaries with the reverse process happening on the other boundary by $C_{2z} \mathcal{T}$, matching the energy splitting in \Fig{fig:TBGcorner}. 

\subsection{$C_{2x} \mathcal{T}$-Protected Gap Closing}
\label{app:c2xt}

While we have not paid much attention to the $C_{2x}$ and $C_{3z}$ symmetries of the model, it is important to break the composite symmetry $C_{2x} \mathcal{T}$ to realize the HOTI phase. (Note $C_{2x} \mathcal{T}$ exists in $H_{TBG}$ due to the accidental $\mathcal{T}$ symmetry.) We argue now that because both $C_{2x}$ and $\mathcal{T}$ take $\phi \to - \phi$ in the Hofstadter Hamiltonian (see \Tab{tb:tbsym}), $C_{2x} \mathcal{T}$ is preserved at all flux and may protect a gap closing \cite{2019arXiv190810976O}. 

Let $\mathcal{H}^\phi(\mbf{k})$ be the Hofstadter Hamiltonian of the $H_{TBG}$. It obeys
\bea
\label{eq:c2xthameq}
C_{2x} \mathcal{T} \mathcal{H}^\phi(k_x, k_y) (C_{2x} \mathcal{T})^{-1} &= \mathcal{H}^\phi(-k_x, k_y), \qquad  C_{2x} \mathcal{T} = I_{q'} \otimes (- i \mu_3 \otimes I_2 K),
\eea
with $I_{q'}$ the $q'\times q'$ identity matrix. Our strategy for understanding the bulk gap closing is to study an effective two band Hamiltonian that models the low energy behavior near the Fermi level. We will show that generically, $C_{2x} \mathcal{T}$ symmetry is sufficient to prove a gap closing between the two bands at some $\phi$, but this is only a proof of local stability. In particular, the assumption of an effective two band model assumes the two bands are are close to each other and well separated from all other bands. This is certainly false when $\Delta$, the onsite potential,  is large and the conduction bands are far from the valence bands. As such no gap closing need exist. This is similar to the local stability of a Weyl node, which is locally protected but can be gapped out in pairs. 

Because $C_{2x} \mathcal{T}$ acts as the identity on the sublattice index but as $-i \mu_3$ on the spin indices (see \Eq{eq:c2xthameq}), the minimal low energy Hamiltonian consists of two bands which adequately models the highest energy valence band and lowest energy conduction band. The most general effective Hamiltonian is then
\bea
\label{eq:22hamc2xt}
\mathcal{H}_{eff}(k_x, k_y, \phi) &= d_0(\mbf{k}, \phi) \mu_0 + d_1(\mbf{k}, \phi) \mu_1 + d_2(\mbf{k}, \phi) \mu_2 + d_3(\mbf{k}, \phi) \mu_3
\eea
and acting on this Hamiltonian, $C_{2x} = -i \mu_3 K$. At $k_x = 0$, \Eq{eq:c2xthameq} mandates that $C_{2x} \mathcal{T}$ commute with $H_{eff}$ at all $k_y, \phi$. Using \Eq{eq:22hamc2xt}, we compute
\bea
C_{2x} \mathcal{T} \mathcal{H}_{eff}(0, k_y, \phi) (C_{2x} \mathcal{T})^{-1} = d_0(0, k_y, \phi) \mu_0 - d_1(0,k_y, \phi) \mu_1 + d_2(0,k_y,\phi) \mu_2 + d_3(0,k_y, \phi) \mu_3 \ .
\eea
But $[C_{2x} \mathcal{T}, \mathcal{H}_{eff}(0, k_y, \phi) ] = 0$, so $d_1 = 0$. There are two free parameters, $k_y \in(-\pi,\pi), \phi \in (0, 6\pi)$, which we fix by the requirement that $d_2(0, k_y^*, \phi^*) = d_3(0, k_y^*, \phi^*) = 0$. Generically, such a $k_y^*$ and $\phi^*$ will exist because the $d_1, d_2$ space is of codimension 0. In this case, $\mathcal{H}_{eff}(0, k_y^*, \phi^*) \propto \mu_0$, and the two bands must be degenerate in energy. 

We substantiate this argument with a numerical calculation of the Hofstadter Butterfly. We build $H'''_{TBG}$ by adding terms to $H_{TBG}$ that break all symmetries except $C_{2x} \mathcal{T}$. We make use of the new perturbation
\bea
H_4(\mbf{k}) &= \mu_0 \otimes \sigma_3 + \mu_2 \otimes \sigma_0 + \mu_3 \otimes \sigma_3 \\
\eea
which is onsite and preserves $C_{2x} \mathcal{T}$. Explicitly, it breaks $C_{2z}, C_{2x}, \mathcal{T}, C_{2z} \mathcal{T}, C_{2x} C_{2z}, C_{2x} C_{2z} \mathcal{T}$, and particle-hole symmetry. Using these perturbations, we build $H'''_{TBG} = H_{TBG} + \eps_2 H_2(\mbf{k}) + \eps_4 H_4(\mbf{k}) $ and include precise values of the parameters in 
 \Tab{tb:tbghams}. We show the spectrum in \Fig{fig:TBGwannierflow}b and observe that the bulk is gapless due to the Weyl node at $(k_x, k_y, \phi) = (0, k_y^*, \phi^*)$.

\setcitestyle{numbers,square}

\end{document}